\documentclass[aps,prb,onecolumn, nofootinbib, 10pt]{revtex4-2}

\usepackage{amsmath,amssymb,bm}
\usepackage{graphicx,comment}
\usepackage[tight]{subfigure}
\usepackage[dvipsnames]{xcolor}
\usepackage[papersize={8.5in,11in}]{geometry}
\usepackage[colorlinks=true]{hyperref}
\hypersetup{
    bookmarks=true,
    unicode=false,
    pdftoolbar=true,
    pdfmenubar=true,
    pdffitwindow=false,
    pdfstartview={FitH},
    pdfnewwindow=true,
    colorlinks=true,
    linkcolor=magenta,
    citecolor=blue,
    filecolor=magenta,
    urlcolor=blue
}

\usepackage{graphicx}
\usepackage{dcolumn}
\usepackage{color}
\usepackage{amssymb,amsmath,makecell}
\usepackage{tabularx,graphicx}
\usepackage{epstopdf}
\usepackage{latexsym}
\usepackage{colortbl}
\usepackage{psfrag}
\usepackage{bbm,bm,array,physics}
\usepackage{dsfont}
\usepackage{float,mathrsfs}

\def \nn{\nonumber \\}
\newcommand{\bs}[1]{\boldsymbol{#1}}

\begin{document}

\title{Direction-dependent magnetoelectric conductivity from dipolar topological semimetals}

\author{Firdous Haidar}
\author{Ipsita Mandal}
\email{ipsita.mandal@snu.edu.in}
\affiliation{Department of Physics, Shiv Nadar Institution of Eminence (SNIoE), Gautam Buddha Nagar, Uttar Pradesh 201314, India}

\begin{abstract}
We compute the linear magnetoelectric conductivity of a two-band vortex nodal-ring (VNR) semimetal and a three-band Hopf semimetal (HSM) within the semiclassical Boltzmann formalism, in the relaxation-time approximation, for three inequivalent planar-Hall configurations, with the electric ($\boldsymbol E$) and magnetic ($\boldsymbol B$) fields oriented differently relative to the nodal-ring plane and the Berry-curvature (BC) dipole axis. Unlike Weyl and multifold semimetals with point-monopole BC flux, the HSM carries a single, point-like BC-dipole fixed at its node, while the VNR hosts a continuous ring of such sources around the nodal loop. Both have vanishing Chern numbers. Working consistently to cubic order in $\boldsymbol B$, we evaluate the Drude, BC, orbital-magnetic-moment (OMM), anomalous-Hall, and Lorentz-force contributions. In both systems, the dipolar BC and OMM activate out-of-plane and odd-in-$\boldsymbol B$ in-plane response-channels that are symmetry-forbidden for a BC-monopole. Mirror symmetry confines the odd longitudinal terms to $B_z$, the field component along the dipole axis. The internode-scattering conductivity of the HSM vanishes identically, since the dipolar BC and OMM are even in momentum and no charge is pumped between two conjugate Hopf nodes, which carry oppositely oriented BC-dipoles. Comparisons with a $\mathcal{PT}$-broken gapped nodal-ring and a pseudospin-1 triple-point semimetal confirm the dipolar-BC origin of these features.
\end{abstract}

\maketitle

\tableofcontents

%======================================================================
\section{Introduction}
\label{secintro}
%======================================================================

The interplay between topology and quantum geometry has emerged as a central
theme in the study of three-dimensional (3d) semimetals. Their low-energy
excitations are often governed by symmetry-protected band-crossings in the
Brillouin zone (BZ), which occur either at isolated
points~\cite{burkov11_Weyl, yan17_topological, bernevig, grushin-multifold}
or along extended one-dimensional
manifolds~\cite{balents-nodal, vortex-nrsm}. These nodal structures are
characterised by topological invariants, which in turn dictate a variety of
measurable electronic responses. An important objective of modern topological
transport is therefore to establish direct connections between experimentally
accessible response functions and the underlying topological indices.

A unifying geometric quantity underlying these phenomena is the Berry curvature
(BC)~\cite{xiao_review,sundaram99_wavepacket,graf-Nband,graf_thesis}, which is
the gauge-invariant field strength associated with the Berry connection of Bloch
electrons. The influence of the BC on transport first came to prominence in Weyl
semimetals (WSMs), where the nodal points act as monopole sources and sinks of
the BC in momentum space. The associated monopole charge is determined by the
Chern number ($\mathcal C$) evaluated over any closed oriented two-dimensional surface
enclosing the node and serves as the corresponding topological invariant. This
momentum-space monopole picture accounts for several hallmark transport
signatures, including the intrinsic anomalous Hall
effect~\cite{haldane04_berry,goswami13_axionic,burkov14_anomolous} and the
unconventional longitudinal and transverse conductivities observed in planar-Hall
geometries~\cite{zhang16_linear,chen16_thermoelectric, amit_magneto,das20_thermal,ips-kush-review,  ips-rsw-ph, ips-spin1-ph, timm, ips-exact-spin1, ips-exact-kwn,ips-exact-rsw, phe_nlsm, ips-nlsm-ph}.

\begin{figure}[t!]
    \centering
    \includegraphics[width=0.9\textwidth]{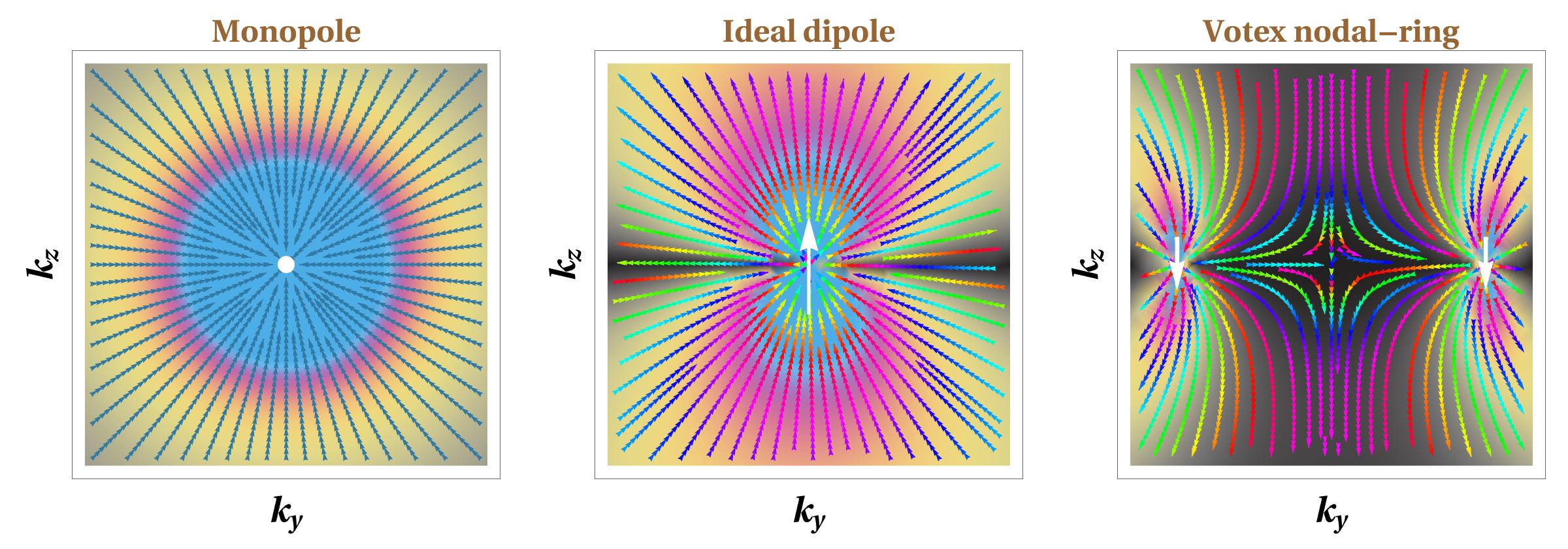}
\caption{BC textures of a monopole, ideal dipole, and vortex nodal-ring projected in the $k_y k_z$-plane.
Unlike Weyl and multifold semimetals, whose BC flux originates from point monopoles, a three-band Hopf semimetals carries a point-like BC-dipole fixed at its node. A vortex nodal-ring's dipolar character comes from a continuous ring of such sources threading around the nodal loop.}
    \label{fig:bc-dip-mono}
\end{figure}

Although the tenfold-way classification~\cite{kitave-10fold,Schnyder2008,Schnyder2009} provides a comprehensive framework for symmetry-protected topological phases, it does not exhaust all manifestations of topology in solids. Topological phases may also be characterised by invariants originating from the global geometry of the Bloch wavefunctions, rather than from symmetry alone~\cite{neupert2021}. A prominent example is furnished by Hopf systems~\cite{graf-hopf,hopf2, hopf3,ips-fa, ips-fa-lett}, whose topology is encoded in the Hopf invariant, an integer-valued linking number associated with the preimages of the Bloch-state map. Because a globally defined Bloch frame exists, the occupied-band bundle has trivial first Chern class and remains topologically trivial in the stable tenfold-way sense. Nevertheless, the Bloch wavefunctions exhibit a nontrivial Hopf invariant, placing these systems in a distinct class beyond the conventional paradigm.
This broader perspective has revealed that momentum-space topology need not be restricted to BC-monopoles. Topological band-crossings may instead generate higher-order multipolar distributions of the BC, including ideal dipoles, quadrupoles, and more intricate textures~\cite{graf-hopf,hopf2,hopf3, ips-fa, ips-fa-lett}.
Fig.~\ref{fig:bc-dip-mono} depicts this. These are the momentum-space counterparts of the multipole expansion in electromagnetism and give rise to transport phenomena fundamentally distinct from those driven by monopole charges.

\begin{figure*}[t!]
\subfigure{\includegraphics[width=0.3 \textwidth]{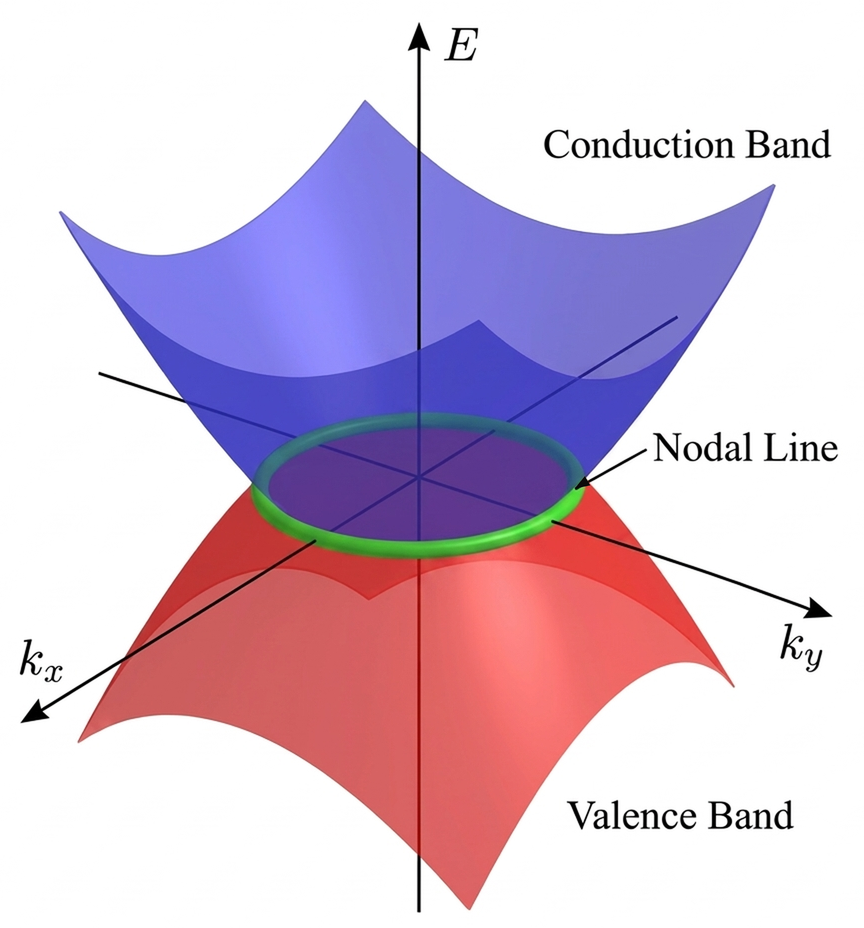}} \hspace{ 1 cm}
\subfigure{\includegraphics[width=0.6 \textwidth]{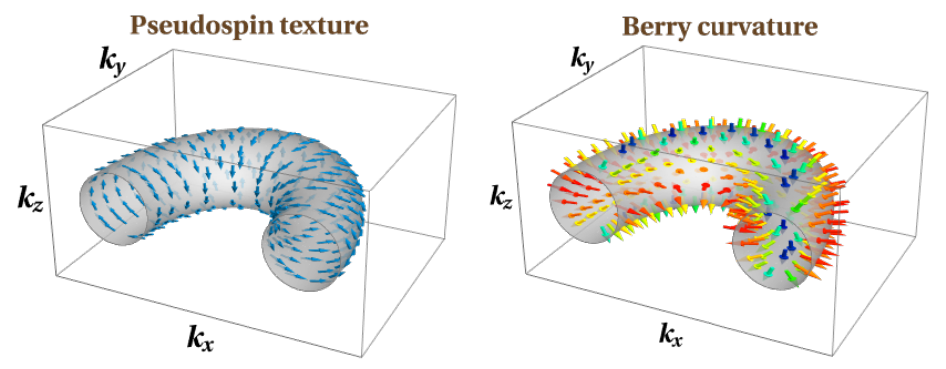}  }
\caption{\label{figfs} Pictorial depiction of some unique features of VNRs:
(a) Dispersion $E$ against the $k_x k_y $-plane showing two bands crossing along a circle (highlighted in green), setting $k_z = 0 $. (b) Nature of the vector fields ${\bs d} (\bs k)$ [or, equivalently, ${\bs d}_0 (\bs k)$ in the vicinity of the VNR] and $\bs \Omega_{s=2} (\bs k)$ in the vicinity of a section of the toroidal Fermi surface (gray). The former displays a smoke-ring pattern.}
\end{figure*}

The first class of systems studied here consists of pseudospin vortex-nodal-rings (VNRs)~\cite{vortex-nrsm}. Unlike conventional $\mathcal{PT}$-symmetric nodal-rings, whose topology is characterised by a $\pi$-quantised Zak phase and a singular BC~\cite{schnyder_nodal, biao_nodal, zak_nodal},\footnote{Introducing a mass term of magnitude $\Delta$ gaps the spectrum and induces a finite BC proportional to $\Delta$, which possesses only in-plane components~\cite{schnyder_nodal,yang1,yang_review_nlsm,chen_nlsm,ips-magnus,flores,phe_nlsm,claudia_nlsm,enke,ips-nlsm-ph,ips-gnr-strain}.} the VNR phase is characterised by a vortex-like winding (or smoke-ring configuration) of the pseudospin texture in momentum space [cf.\ Fig.~\ref{figfs}(b)]. Equivalently, the nodal-ring itself can be viewed as a vortex line in momentum space, with the pseudospin texture encircling it forming a toroidal vector field about its axis~\cite{chang_nlsm_review}. For each fixed $k_z$ slice away from the ring's plane, the map $\hat{\bs d}(\bs k)$ defines a Pontryagin (skyrmion) number $Q(k_z)\in\mathbb{Z}$~\cite{vortex-nrsm}, with $Q(k_z)=+1$ on one side of the ring plane and $Q(k_z)=-1$ on the other. The VNR topology is therefore a family of $\pi_2(S^2)$ degrees that change sign across the ring plane, rather than a single $\pi_3(S^2)$-valued index. The cancellation of contributions from opposite sides of the ring plane is consistent with the absence of a net monopole charge. In contrast to conventional nodal-rings, the VNR hosts a smooth and finite BC distribution~\cite{ips-dipole-vnr}, making it a particularly attractive platform for investigating BC-induced transport. Unlike the single, point-like BC-dipole of an isolated Weyl-type or Hopf node, this distribution is carried by every point of the nodal loop, so the VNR realises a continuous ring of dipole-like BC sources rather than one dipole fixed at a single momentum. Throughout, we will refer to the conventional nodal-ring model and its gapped descendants collectively as the non-vortex nodal-ring (NVNR). VNRs constitute the second class of systems studied here.

In parallel, another route towards unconventional Berry geometry is offered by three-band massless HSMs, which host an ideal (point-like) BC-dipole at each node, despite possessing vanishing monopole Chern charge. The band structure and the contrasting monopolar versus dipolar BC textures are illustrated in Fig.~\ref{fig:hopf}. The anisotropic BC-dipole texture produces characteristic signatures such as BC-dipole-induced transport anisotropies~\cite{hopf-plasmon,ips-dipole-vnr}, and, experimentally, a polar-angle-dependent zeroth Landau level together with helical zero modes at a Berry-dipole node has been observed in an acoustic metamaterial~\cite{dipole-helical-arc}. This observation concerns the gapless Berry-dipole point itself, rather than a bulk-boundary correspondence of a gapped Hopf phase~\cite{hopf-ins-surface-states}. Three-band HSMs therefore provide an ideal platform for exploring how momentum-space BC multipoles manifest themselves in magnetotransport.

Besides the BC, Berry geometry gives rise to an orbital magnetic moment (OMM)~\cite{xiao_review, sundaram99_wavepacket}, which plays a crucial role in magnetic-field-induced transport~\cite{ips-ruiz, ips-rsw-ph, ips-spin1-ph, ips-exact-spin1, ips-exact-kwn, ips-exact-rsw,ips-nlsm-ph}. Although the magnetotransport properties of conventional nodal-ring semimetals have been studied extensively~\cite{schnyder_nodal,yang1,yang_review_nlsm,chen_nlsm,ips-magnus,flores,phe_nlsm, claudia_nlsm,enke, ips-nlsm-ph, ips-gnr-strain}, corresponding studies for VNRs remain scarce, and analogous analyses for BC-dipolar HSMs are largely absent. In this work, we analytically investigate the linear magnetoelectric conductivity in planar-Hall geometries for both systems. Such configurations involve static and uniform electric ($\bs E$) and magnetic ($\bs B$) fields, with $\bs B$ not necessarily perpendicular to $\bs E$. Our primary objective is to elucidate the emergence of an in-plane conductivity driven jointly by the BC and the OMM, as well as the nature of the out-of-plane conductivity sourced by the anomalous-Hall (AH) effect and the Lorentz-force operator (LFO). For VNRs, the orientation of the nodal-ring plane relative to the $\bs E \bs B$-plane allows several distinct planar-Hall configurations~\cite{ips-nlsm-ph, ips-gnr-strain}. For HSMs, the orientation of the BC-dipole axis leaves equally distinctive fingerprints.

%%%%%%%%%%%
%%%%%%%%%%%%%%%%%%

The paper is organised as follows. Sec.~\ref{secmodel} introduces the effective Hamiltonians for the VNR and the three-band HSM, together with their BC and OMM structures. Both have vanishing Chern numbers.
Sec.~\ref{secmethod} presents the semiclassical Boltzmann formalism and the three inequivalent planar-Hall set-ups, including the argument for their completeness. Sec.~\ref{secvnr} presents the full conductivity for the VNR in all three set-ups. Sec.~\ref{sechopfres} presents the corresponding results for the HSM, including the internode-scattering contribution. Sec.~\ref{seccompare} compares the VNR and HSM results against the NVNR and the pseudospin-1 TSM, highlighting how the dipolar geometry governs the qualitative structure of the response. Sec.~\ref{secsum} summarises our findings and outlines future directions. Technical details and intermediate steps are detailed in the appendices. Throughout, we use natural units with $\hbar=c=k_B=1$. The elementary charge $e$ is dimensionless in these units, but we retain it explicitly in all expressions to keep track of the charge dependence.

%======================================================================
\section{Model Hamiltonians and their quantum geometry}
\label{secmodel}
%======================================================================

\subsection{Vortex nodal-ring (VNR)}
\label{secVNR}

%%%%%%%%%%%%%%%%%%%%%%%%%%%%%%%%%%%%%%%%%%%%%%%%%%%%%%%%%
%%%%%%%%%%%%%%%%%%%%

The minimal model of a VNR, comprising two bands and a circular nodal loop lying in the $k_x k_y$-plane at $k_z=0$, is captured by~\cite{vortex-nrsm}
%%%%%%%%%%%%%%%%%%%%%%%
\begin{align}
\mathcal{H}_0 (\bs k ) = {\bs d}_0 (\bs k) \cdot \bs{\sigma} \,,
\quad {\bs d}_0 (\bs k)
= \left \lbrace  \frac{-\,k_x\,k_z} {m_z}, \,\frac{-\,k_y\, k_z} {m_z},
\, \frac{ k_\perp^2-k_0^2 - k_z^2} {2\,m_\perp}  \right \rbrace, \quad
k_\perp = \sqrt{k_x^2 + k_y^2 }\,,
\end{align}
%%%%%%%%%%%%%%%%%%
where $k_0$, $m_z$, and $m_\perp$ are material-dependent parameters. At the $k_z=0$ plane, the two bands cross at $k_\perp=k_0$, defining a nodal-ring of radius $k_0$. The Fermi velocities perpendicular and parallel to the nodal plane are $v_z\equiv k_0/m_z$ and $v_\perp\equiv k_0/m_\perp$. For simplicity, we set $v_z=v_\perp=v_0$ in what follows.

For low-energy excitations near the Fermi surface, the dominant transport signatures are captured by a linearised Hamiltonian $\mathcal{H}$ that deviates from the nodal line only at linear order in momentum~\cite{linearize-nlsm,ips-nlsm-ph}. This is achieved by transforming to cylindrical coordinates,
\begin{align}
\label{eqtrs}
k_x = k_\perp \cos\phi\,,\quad
k_y = k_\perp \sin\phi\,,\quad k_\perp = k_0+\kappa_\perp\,,
\end{align}
with $\kappa_\perp\in[-k_0,\infty)$. Low-energy processes require $|\kappa_\perp|\ll k_0$, since $k_0$ is an ultraviolet-energy scale. Linearising in $\kappa_\perp$ and $k_z$, we obtain
\begin{align}
\label{eqham}
\mathcal{H}_\ell (\bs k) = {\bs d}(\bs k)\cdot\bs{\sigma}\,,
\quad
{\bs d}(\bs k) = v_0\left\lbrace -k_z\cos\phi,\,-k_z\sin\phi,\,\kappa_\perp\right\rbrace.
\end{align}
The eigenvalues and band-velocitiesof $\mathcal H$ are
\begin{align}
\label{eqev}
\varepsilon_s({\bs k}) &= (-1)^s\,\epsilon_k\,, \quad
\epsilon_k = v_0\,\sqrt{\kappa_\perp^2+k_z^2}\,, \quad s\in\lbrace1,2\rbrace\,,\nn
%%%%%%%%%%%%%
{\bs v}^{(0,s)}(\bs{k})
&\equiv \nabla_{\bs{k}}\varepsilon_s(\bs{k})
= \frac{(-1)^s\,v_0^2}{\epsilon_k}\left\lbrace\kappa_\perp\cos\phi,\,
\kappa_\perp\sin\phi,\,k_z\right\rbrace ,
\end{align}
respectively.
The BC and OMM associated with band $s$ are \cite{xiao_review, sundaram99_wavepacket, graf-Nband, graf_thesis}
\begin{align}
\label{eqbcomm}
\bs\Omega_s({\bs k}) &=
\frac{(-1)^{s+1}\,v_0^3\,k_z}{2\,(k_0+\kappa_\perp)\,\epsilon_k^3}
\left\lbrace\kappa_\perp\cos\phi,\,\kappa_\perp\sin\phi,\,k_z\right\rbrace, \quad
%%%%%%%%%%%%%%%%%
{\bs m}({\bs k}) =
\frac{-\,e\,v_0^3\,k_z}{2\,(k_0+\kappa_\perp)\,\epsilon_k^2}
\left\lbrace\kappa_\perp\cos\phi,\,\kappa_\perp\sin\phi,\,k_z\right\rbrace.
\end{align}
The VNR carries a continuous ring of such dipole-like sources, one attached to every point of the nodal loop, as seen in Fig.~\ref{fig:bc-dip-mono}.
The OMM-induced energy shift is
\begin{align}
\varepsilon_m({\bs k}) =
\frac{e\,v_0^3\,k_z\left(B_x\,\kappa_\perp\cos\phi+B_y\,\kappa_\perp\sin\phi+B_z\,k_z\right)}
{2\,\epsilon_k^2\,(k_0+\kappa_\perp)}\,.
\end{align}
Since the OMM-induced parts are identical for both bands, we attach no $s$ index to them.

For performing the integrals, it is convenient to switch to toroidal coordinates,
\begin{align}
\label{eqtorus}
k_z = \kappa\sin\gamma\,, \quad \kappa_\perp = \kappa\cos\gamma\,, \quad \gamma\in[0,2\pi)\,.
\end{align}
In these coordinates, $k_0$ is the major radius (distance from the origin to a point on the node's circle), and $\kappa$ is the radius of the torus cross-section.

The linearised model of Eq.~\eqref{eqham} provides the minimal two-band realisation of a VNR. Although the two bands touch along an extended nodal line rather than at an isolated point, the ring remains topologically nontrivial through the vortex-like winding of the pseudospin texture around it. Away from the ring, the gapped bulk inherits a dipolar BC texture, rather than the monopolar BC familiar from Weyl semimetals or the $\pi$-quantised Zak phase of the conventional (non-vortex) nodal-ring. We now summarise this topological origin, identify the invariant that survives away from the node, and explain why the BC-dipole is the relevant quantity controlling the transport phenomena studied in this work.

Unlike a conventional $\mathcal{PT}$-symmetric nodal-ring, whose topology is captured by a single $\pi$-quantised Zak phase evaluated along a loop encircling the ring~\cite{schnyder_nodal,biao_nodal,zak_nodal}, the VNR topology is encoded in a family of Pontryagin (skyrmion) indices, one for each $k_z$ slice away from the ring plane~\cite{vortex-nrsm}. For fixed $k_z\neq0$, the pseudospin texture $\hat{\bs d}(\bs k)=\bs d(\bs k)/|\bs d(\bs k)|$ defines a smooth map of the $(k_x,k_y)$ plane into the Bloch sphere, classified by the integer-valued degree
\begin{align}
\label{eq-pontryagin}
Q(k_z) = \frac{1}{4\,\pi}\int dk_x\,dk_y\;\hat{\bs d}\cdot\left(\partial_{k_x}\hat{\bs d}\times\partial_{k_y}\hat{\bs d}\right).
\end{align}
Evaluating Eq.~\eqref{eq-pontryagin} on the linearised Hamiltonian of Eq.~\eqref{eqham} gives $Q(k_z)=\mathrm{sgn}(k_z)$, so that the pseudospin texture winds once around the Bloch sphere with opposite orientation on either side of the nodal plane.

This Pontryagin index differs fundamentally from the Zak phase that characterises the conventional nodal-ring. Whereas the Zak phase is a one-dimensional invariant defined along a path threading the ring, $Q(k_z)$ is a two-dimensional invariant defined on each momentum-space slice away from the ring. Its topological character therefore lies not in a quantised Berry phase accumulated along a loop, but in the winding of the pseudospin texture across the full transverse plane. Because $Q(k_z)$ changes sign discontinuously across the nodal plane, $Q(0^+)=-Q(0^-)$, the ring carries no net topological charge of the monopolar kind: the contributions from the two sides cancel exactly, consistent with the smooth and finite BC distribution found for the VNR~\cite{ips-dipole-vnr}. This cancellation can be verified explicitly from the BC of Eq.~\eqref{eqbcomm}. In the toroidal coordinates of Eq.~\eqref{eqtorus}, $\bs\Omega_s$ reduces to
\begin{align}
\label{eq-vnr-flux}
\bs\Omega_s = (-1)^{s+1}\,\frac{\sin\gamma}{2\,\kappa\,(k_0+\kappa\cos\gamma)}\,\hat{\bs n}\,,\quad
\hat{\bs n} = \left\lbrace\cos\gamma\cos\phi,\,\cos\gamma\sin\phi,\,\sin\gamma\right\rbrace\,,
\end{align}
with $\hat{\bs n}$ the local outward unit normal to a torus of cross-sectional radius $\kappa$ encircling the nodal-ring. Since the torus surface element is $d\bs S=\kappa\,(k_0+\kappa\cos\gamma)\,d\gamma\,d\phi\,\hat{\bs n}$, the net flux through any such torus is
\begin{align}
\oint\bs\Omega_s\cdot d\bs S = \frac{(-1)^{s+1}}{2}\int_0^{2\pi}d\phi\int_0^{2\pi}d\gamma\,\sin\gamma=0\,,
\end{align}
by the same odd-in-$k_z$ parity argument that recurs for the Hopf node in Sec.~\ref{sechopf}. The BC is therefore purely radial relative to the ring, and its sign, fixed by $\sin\gamma$, is positive on the $k_z>0$ side of the ring and negative on the $k_z<0$ side, vanishing identically in the nodal plane itself.

Despite carrying no net flux, the BC profile is highly nontrivial: the odd factor $\sin\gamma=k_z/\kappa$ partitions momentum space into two lobes of opposite sign separated by the plane $k_z=0$, with magnitude maximal away from the ring plane and vanishing on it. At each point along the ring, this pattern is that of a local BC-dipole in the transverse $(\kappa_\perp,k_z)$ plane, so the VNR does not carry a single, point-like BC-dipole, as the HSM does at its node, but a continuous ring of such dipole-like sources, one attached to every point of the nodal loop. This dipolar texture, inherited from the vortex-like winding of the pseudospin field, is the transport-relevant geometric remnant of the VNR topology. The OMM in Eq.~\eqref{eqbcomm} inherits the same $k_z$-odd angular dependence as the BC, so the two quantities are locked together in both magnitude and angular structure, exactly as for the HSM. The consequence for transport is that the parity of the relevant integrands under $k_z\to-k_z$ is controlled by this explicit factor of $k_z$, a feature we return to repeatedly when contrasting the VNR against the conventional nodal-ring in Sec.~\ref{seccompare}.

\subsection{Three-band Hopf semimetal (HSM)}
\label{sechopf}

%%%%%%%%%%%%%
\begin{figure*}[t!]
\subfigure{\includegraphics[width=0.27 \textwidth]{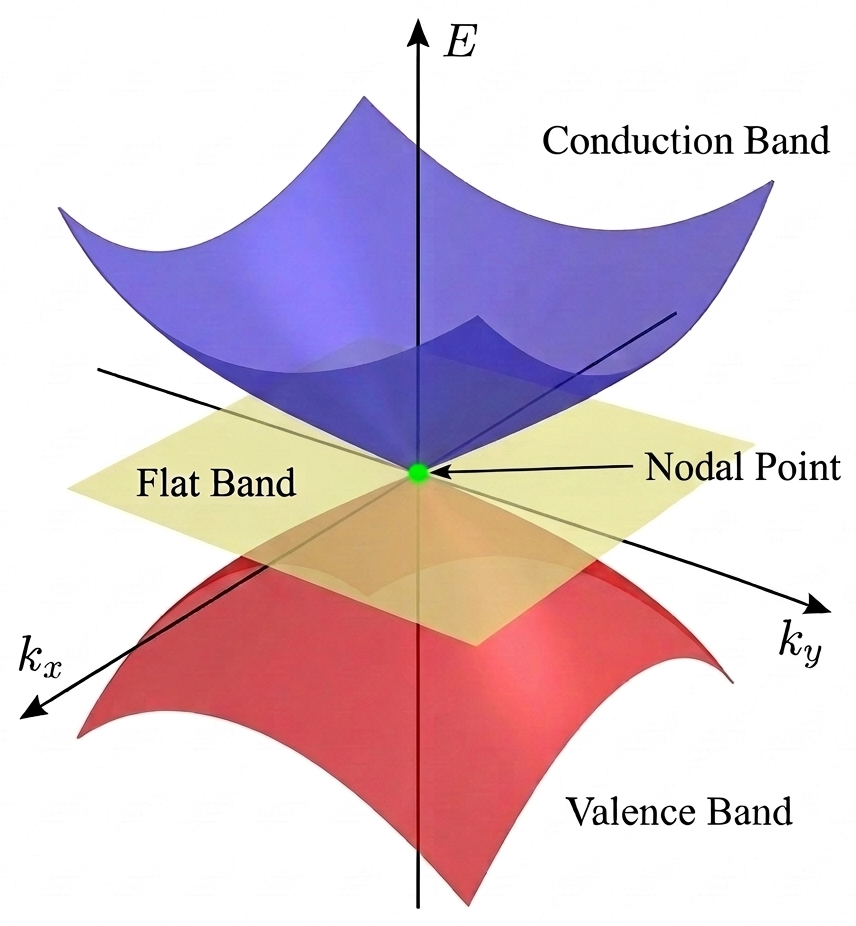}} \hspace{1.5 cm}
\subfigure{\includegraphics[width=0.6 \textwidth]{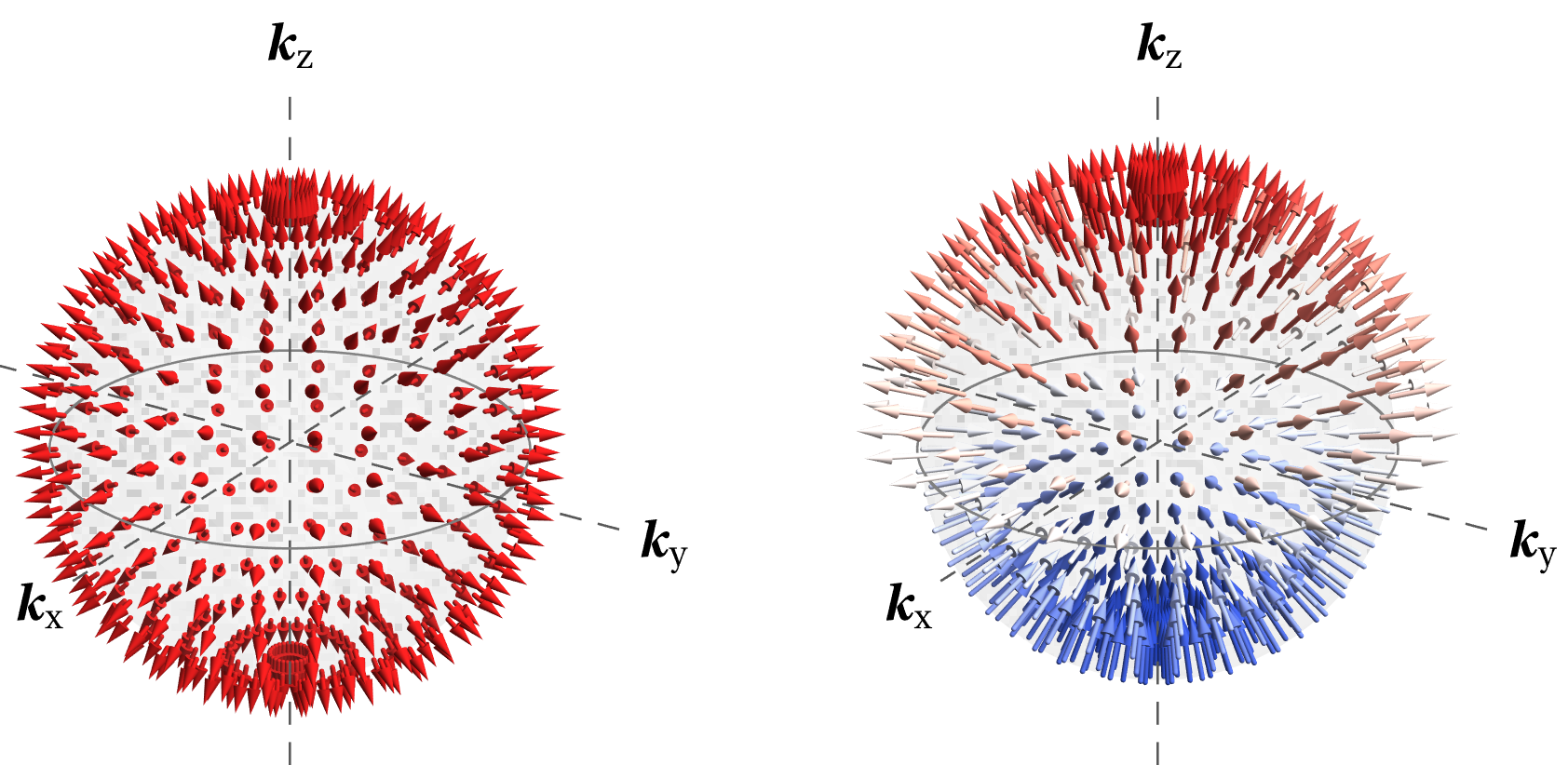}}
\caption{Pictorial depiction of some unique features of HSMs and TSMs:
(a) Dispersion $E$ against the $k_x k_y$-plane showing two bands crossing at the nodal point (highlighted in green), setting $k_z = 0$, and the flat-band (highlighted in yellow).
(b) Monopole-like BC field near the node of a TSM with $\mathcal C=\pm2$, exhibiting a net outward flux, and the dipolar BC field near the node of an HSM ($\mathcal C=0$), characterised by opposite field lines in the two hemispheres and vanishing total flux through the sphere.\label{fig:hopf}}
\end{figure*}

HSMs whose nodal points host BC-dipoles rather than monopoles were introduced in Ref.~\cite{graf-hopf}. Multifold Hopf semimetals (MMHSs) host BC-dipoles at nodes where linearly dispersing bands cross. We consider the simplest case, with a threefold-degenerate node, described by the effective continuum Hamiltonian~\cite{graf-hopf,hopf2,hopf3}
\begin{align}
\label{eqbcd2}
H_{h} = v_0\left(k_x\,\lambda_1+k_y\,\lambda_2+k_z\,\lambda_5\right),
\end{align}
where
\begin{align}
\lambda_1 = \begin{bmatrix}0&1&0\\1&0&0\\0&0&0\end{bmatrix}, \quad
\lambda_2 = \begin{bmatrix}0&-i&0\\i&0&0\\0&0&0\end{bmatrix}, \quad
\lambda_5 = \begin{bmatrix}0&0&-i\\0&0&0\\i&0&0
\end{bmatrix}.
\end{align}
The eigenvalues are
\begin{align}
\label{eqevhopf}
\varepsilon_s({\bs k}) = \begin{cases}
(-1)^s\,v_0\,k & \text{for } s\in\lbrace1,2\rbrace\\
0 & \text{for } s=0
\end{cases},
\end{align}
where $s=0$ is a nondispersive flat-band. A Hopf node hosts a flat-band at zero energy, pinned by the chiral symmetry $\mathcal{S} = \mathrm{diag}(1,-1,-1)$, which anticommutes with $H_h$. This forces the spectrum to be symmetric under $\varepsilon \to -\varepsilon$ and, since the Hilbert space is three-dimensional, at least one state must remain at zero energy.

The dispersion of HSM coincides with that of the isotropic pseudospin-1 triple-point semimetal (TSM) studied in Refs.~\cite{ips-spin1-ph, ips-internode}: two linearly-dispersing bands $\varepsilon_\pm = \pm \,v_0\,k$, together with a nondispersive flat-band at $\varepsilon_0=0$. The TSM is the threefold analogue of the pseudospin-$1/2$ WSM. It is therefore instructive to compare the quantum geometries of the two systems, since their band dispersions are identical. The corresponding bandstructure is shown in Fig.~\ref{fig:hopf}(a). For the Hopf system, we use spherical-polar coordinates,
\begin{align}
\label{eqtrs2}
k_x = k\,\sin\gamma\cos\phi\,,\quad
k_y = k\,\sin\gamma\sin\phi\,,\quad
k_z = k\,\cos\gamma\,,
\end{align}
with $\gamma\in[0,\pi]$, in contrast to the toroidal coordinates used for the VNR.

A single node of the TSM, with chirality $\chi$, is represented by~\cite{ips-spin1-ph,ips-internode}
\begin{align}
\label{eqhamspin1}
\mathcal{H}_{\rm TSM}(\mathbf  k) = v_0 \left(k_x \,\mathcal S_x + k_y \,\mathcal S_y 
+ k_z \,\mathcal S_z \right),
\end{align}
where $ \boldsymbol{\mathcal S} =  \lbrace {\mathcal S }_x,\, {\mathcal S }_y,\, {\mathcal S }_z \rbrace$ represents the angular-momentum vector operator in the spin-$ 1$ representation. We choose
\begin{align}
\mathcal{S}_x = \frac {1} {\sqrt{2}}
\begin{pmatrix}
0&1&0\\1&0&1\\0&1&0
\end{pmatrix} ,\quad
\mathcal{S}_y =\frac{1}{\sqrt{2}}
\begin{pmatrix}
0&- i &0\\ i &0&-  i \\
0&  i  &0
\end{pmatrix},\quad
%%%%%%%%
\mathcal{S}_z =
\begin{pmatrix}
1&0&0\\0&0&0\\0&0&-1
\end{pmatrix} .
\end{align}
As shown in Ref.~\cite{ips-spin1-ph}, the magnetoelectric response of the dispersive bands in the TSM closely parallels that of the WSM if the flat-band is ignored, differing only by simple numerical prefactors. It is interesting to note how far the analogy of the TSM with a Weyl node extends. Anticommutation of the Pauli matrices implies \begin{align}
\left(\bs k\cdot\bs\sigma\right)^2 = k^2\,\mathbb I_{2\times2}\,,
\end{align}
which is why the Weyl spectrum is exactly $\pm v_0 k$ with no flat-band, and why the two bands are particle-hole symmetric with equal and opposite BC. For the spin-$1$ matrices, we have instead
\begin{align}
\left(\bs k\cdot\bs S\right)^2 \neq k^2\,\mathbb I_{3\times3}\,.
\end{align}
This failure of anticommutation underlies the zero-energy flat-band noted above.

Using an orthonormal set of eigenvectors, the BC and OMM of the HSM are
\begin{align}
\label{eqbcommhopf}
{\bs\Omega}_s(\bs k) = \begin{cases}
\dfrac{k_z\,\bs k}{k^4} & \text{for } s=1,2 \\[6pt]
\dfrac{-2\,k_z\,\bs k}{k^4} & \text{for } s=0
\end{cases}
%%%%%%%%%%
\text{ and }
{\bs m}_s(\bs k) = \begin{cases}
\dfrac{(-1)^{s+1}\,e\,v_0\,k_z\,\bs k}{2\,k^3} & \text{ for } s=1,2 \\[6pt]
0 & \text{ for } s=0
\end{cases},
\end{align}
respectively.
The BC is symmetric for both dispersive bands, and dipole charges is carried by the flat-band as well.
The resulting BC texture for $s=2$ is illustrated in Figs.~\ref{fig:bc-dip-mono} and ~\ref{fig:hopf}(b). The dipolar singularity is characterised by oppositely-emanating field lines in the two hemispheres and vanishing net flux through any enclosing sphere.

Some discussions are in order. Both the HSM and the TSM are described in the form
$H=v_0\,\bs k\cdot\bs M$, with $\bs M=(M_x,M_y,M_z)$ a triplet of
$3\times 3$ matrices, yet the two triplets are not related by a change of
basis. This is the sole reason why one system hosts a BC-monopole and the
other a BC-dipole. The TSM uses the spin-$1$ angular-momentum operators,
\begin{align}
\bs M^{\rm TSM}=\left(\mathcal S_x,\mathcal S_y,\mathcal S_z\right),
\qquad
[\mathcal S_i,\mathcal S_j]=i\,\epsilon_{ijk}\,\mathcal S_k\,,
\label{eq-somat}
\end{align}
which close under the $\mathfrak{so}(3)$ algebra and furnish an irreducible
representation of $SO(3)$. In terms of the Gell-Mann matrices,
\begin{align}
\mathcal S_x=\frac{\lambda_1+\lambda_6}{\sqrt2}\,,\quad
\mathcal S_y=\frac{\lambda_2+\lambda_7}{\sqrt2}\,,\quad
\mathcal S_z=\frac{\lambda_3+\sqrt3\,\lambda_8}{2}\,,
\end{align}
so $\bs M^{\rm TSM}$ is built from six of the eight Gell-Mann generators
and rotates as an $SO(3)$ vector. The HSM, by contrast, uses only three
generators, $\bs M^{\rm H}=\left(\lambda_1,\lambda_2,\lambda_5\right)$,
whose commutators do not close on themselves:
\begin{align}
[\lambda_1,\lambda_2]=2\, i\,\lambda_3\,,\quad
[\lambda_2,\lambda_5]=i\,\lambda_7\,,\quad
[\lambda_1,\lambda_5]=-\,i\,\lambda_6\,.
\label{eq-commutators}
\end{align}
The appearance of the generators $\lambda_3$, $\lambda_6$, and
$\lambda_7$ on the right-hand sides means that the triplet
$\{\lambda_1,\lambda_2,\lambda_5\}$ generates the full $\mathfrak{su}(3)$
algebra and cannot be unitarily rotated into the spin-$1$ triplet of
Eq.~\eqref{eq-somat}. Equivalently, no $U\in U(3)$ exists with
$U\,\lambda_i\,U^\dagger=\mathcal S_i$ for $i=x,y,z$. The HSM triplet
therefore does not transform as an $SO(3)$ vector, and the only continuous
spatial symmetry it preserves is the $U(1)$ rotation about $\hat{\bs z}$
identified in Sec.~\ref{secsymmetry}. This algebraic distinction propagates directly into the eigenvectors and, hence, every quantity built from them (e.g., BC). Because $\bs M^{\rm TSM}$ rotates as an $SO(3)$ vector, the eigenvectors of $\mathcal H_{\rm TSM}$ are obtained from the
$\mathcal S_z$ eigenstates by the same rotation that maps $\bs {\hat z}$
onto $\bs{\hat k}$. Consequently, the pseudospin texture is the isotropic hedgehog shown
in Fig.~\ref{fig:hopf}(b), and the dispersive-band's BC is a radial field whose flux through any sphere enclosing the node is quantised and nonzero. The HSM triplet lacks this rotational covariance:
the $\hat{\bs z}$-axis is singled out by $\lambda_5$, the eigenvectors of
$H_h$ are not spin-coherent states, and the resulting BC carries an extra
factor of $k_z$. The dipolar texture BC shown in Fig.~\ref{fig:hopf}(b) is outward in the northern
hemisphere, inward in the southern one, and vanishing on the equator, making
the net flux through any enclosing sphere vanish identically.
In summary, it is the loss of $SO(3)$ covariance in
the HSM that converts the BC-monopole of the TSM into the BC-dipole of the Hopf node.

It is instructive to contrast the flat-band properties of the TSM and the HSM, since the roles of the BC and OMM are effectively interchanged. For any $N$-band Hamiltonian, the band projectors resolve the identity, $\sum_s P_s(\bs k)=\mathbb I $, at every $\bs k$. The curvature of the full $N$-dimensional bundle formed by all bands together is therefore that of a trivial flat bundle, which vanishes identically. Since the BC is additive across the orthogonal decomposition into individual one-dimensional eigenspaces, the curvatures of the separate bands must cancel pairwise in the sum, giving the general identity $\sum_s\bs\Omega_s =0$. In the TSM, this identity is realised through the spin-1 generators of the $SO(3)$ algebra, which enforce a particle-hole relation $\bs\Omega^{+1}_\chi=-\,\bs\Omega^{-1}_\chi$ between the dispersive bands, leaving the flat-band topologically inert with identically zero BC. Its OMM is instead doubled, $G_0=2$, relative to the dispersive bands: the OMM involves virtual interband transitions weighted by energy denominators, and since the flat-band sits symmetrically between $+\epsilon_k$ and $-\epsilon_k$, contributions from both dispersive bands add constructively. In the HSM, the two dispersive bands possess identical BC, $\bs\Omega_1=\bs\Omega_2 = k_z\,\bs k/k^4$, while the flat-band carries $\bs\Omega_0=-2\,k_z\,\bs k/k^4$, consistent with the same identity $\sum_s\bs\Omega_s =0$. The same interband coupling structure produces dispersive-band OMMs with opposite signs, $\bs m_s\propto(-1)^{s+1}$, whose contributions cancel exactly, leaving the flat-band with a vanishing OMM. 

The HSM of Eq.~\eqref{eqbcd2} is the gapless descendant of a 3d Hopf insulator~\cite{graf-hopf,hopf2,hopf3,hopf-ins-surface-states}, whose topology, unlike that of a Chern insulator, is classified not by $\mathcal C$ but by an integer Hopf invariant arising from the homotopy group $\pi_3(S^2)=\mathbb{Z}$. For a fully gapped two-band system, the occupied Bloch state defines a smooth map $\hat{\bs d}(\bs k):\mathbb T^3\to S^2$, classified by
\begin{align}
\label{eq-hopf_inv}
\mathcal H = - \, \frac{1}{4\,\pi^2}\int_{\rm BZ}d^3 k\,\bs A(\bs k)\cdot\bs\Omega(\bs k),
\end{align}
where $\bs A(\bs k)$ and $\bs\Omega(\bs k)=\nabla_{\bs k}\times\bs A(\bs k)$ are the Berry connection and BC of the occupied band. $\mathcal H$ equals the linking number of the preimages of two distinct points on the Bloch sphere. Tuning a mass parameter closes the bulk gap at a high-symmetry point of the BZ and produces the threefold band-touching of Eq.~\eqref{eqbcd2}, at which the eigenstate map becomes singular and $\mathcal H$ ceases to be a sharply defined invariant of the gapless phase. The band-touching point is instead protected by the chiral crystalline symmetries of the Hopf phase, which forbid all symmetry-allowed mass terms~\cite{graf-hopf,grushin-multifold}, so that $\mathcal C$ vanishes identically on every sphere surrounding it. What survives from the parent insulator is the geometric organisation away from the node: a dipolar BC texture with vanishing net flux but a nontrivial first angular moment. This dipolar texture is what the transport analysis below probes. It also reflects the larger eigenstate manifold of a three-band system: a normalised two-band Bloch spinor is parametrised by $\mathbb{CP}^1\simeq S^2$, whereas the three-component eigenstates of Eq.~\eqref{eqbcd2} parametrise the four-dimensional $\mathbb{CP}^2$, whose richer geometry supports a dipolar BC texture with no two-band analogue.

From Eq.~\eqref{eqbcommhopf}, the dispersing bands ($s=1,2$) carry $\bs\Omega_s(\bs k) \propto {k_z\,\bs k}/{k^4} = {\cos\gamma}/{k^2} \, \hat{\bs k}$,
where $\gamma$ is the polar angle of the spherical coordinates in Eq.~\eqref{eqtrs2}. The explicit factor of $k_z$ singles out the $\hat{\bs z}$ direction, reducing the rotational symmetry of the BC texture from the full $SO(3)$ symmetry of a monopole to the axial $C_{\infty v}$ symmetry characteristic of a dipole: the BC points outward in the northern hemisphere ($k_z>0$, $\cos\gamma>0$), inward in the southern hemisphere ($k_z<0$, $\cos\gamma<0$), and vanishes identically on the equatorial plane $k_z=0$. The net Berry flux through any sphere of radius $k$ surrounding the node consequently vanishes,
\begin{align}
\oint_{S^2}\bs\Omega_s\cdot d\bs S
= \int_0^{2\,\pi}d\phi\int_0^\pi d\gamma\,\sin\gamma\,\cos\gamma = 0\,,
\end{align}
so that $\mathcal C$ is zero on every enclosing surface. Unlike the Nielsen--Ninomiya cancellation~\cite{chiral_ABJ} between distinct Weyl nodes of opposite chirality, this vanishing is achieved at each Hopf node individually, through exact cancellation between the outward flux in the upper hemisphere and the inward flux in the lower hemisphere. The Hopf node is therefore neither a source nor a sink of BC flux in the monopolar sense. Its topological information is encoded in the angular distribution of the BC rather than in its net flux. The qualitative similarity between the two systems' transport response, analysed in the remainder of this paper, traces back to this shared dipolar structure of the BC field.

Since the OMM inherits the same $\cos\gamma$ dependence as the BC, the two quantities are rigidly locked in both magnitude and angular structure. Both are even under $\bs k\to-\bs k$, whereas the band velocity and the gradient of the OMM-induced energy shift are odd. Every integrand entering the internode charge-pumping vector $\bs{\mathcal I}^\chi$ of Sec.~\ref{secinternode} is therefore odd in $\bs k$ and vanishes upon integration. This parity structure underlies the absence of the internode-scattering-induced chiral-anomaly contribution.

%%%%%%%%%%%%%%%%%%%
\subsection{Symmetries of the model Hamiltonians}
\label{secsymmetry}

Since the magnetotransport responses derived below are strongly constrained by the symmetries of the underlying Hamiltonians, we identify here the exact spatial and discrete symmetries of the untilted VNR and HSM.

For the VNR, let us discuss the relevant symmetries:
%%%%%%%%%%%%%%%%%
\begin{enumerate}

\item The low-energy linearised  Hamiltonian $\mathcal{H}_\ell(\bs k)$ possesses a continuous $U(1)$ rotational symmetry about the $k_z$-axis, corresponding in real space to the continuous rotational subgroup $C_\infty$ about $\, \hat{\bs z}$. Under an azimuthal rotation $k_x+i\,k_y\to e^{i \, \varphi} \, (k_x+i\,k_y)$, $\mathcal{H}_\ell(\bs k)$ transforms covariantly as
\begin{align}
\mathcal{H}_\ell(\mathcal{R}_z \,\bs k) = U(\varphi)\,\mathcal{H}_\ell(\bs k)\,U^\dagger(\varphi)\,,
\quad U(\varphi) = e^{-i\,\varphi\,\sigma_z/2}\,.
\end{align}
This can be verified by noting that
\begin{align}
U\,\sigma_x\,U^\dagger &= \cos\varphi\,\sigma_x+\sin\varphi\,\sigma_y\,,\quad
U\,\sigma_y\,U^\dagger = -\sin\varphi\,\sigma_x+\cos\varphi\,\sigma_y\,,\quad
U\,\sigma_z\,U^\dagger = \sigma_z\,,
\end{align}
which precisely compensates the rotation of the in-plane components of $\bs d(\bs k)$, leaving $d_z=v_0\,\kappa_\perp$ invariant. The discrete rotation $C_{2z}$ is a subgroup of this continuous symmetry.

\item $\mathcal{H}_\ell(\bs k)$ is invariant under reflection through the $k_z=0$ plane:
\begin{align}
\mathcal{H}_\ell(k_x,k_y,-k_z) = U_\sigma\,\mathcal{H}_\ell(k_x,k_y,k_z)\,U_\sigma^\dagger\,,
\quad U_\sigma = \sigma_z\,,
\end{align}
since $U_\sigma$ leaves $d_z$ invariant while transforming $d_x\to-d_x$ and $d_y\to-d_y$, exactly compensating the sign reversal of $k_z$. This mirror symmetry exchanges the two sides of the nodal-ring plane while leaving the ring plane invariant. Consequently, any momentum-space quantity odd under $k_z\to-k_z$ integrates to zero over the full BZ (or over a closed Fermi surface), whereas even quantities are symmetry-allowed.
%%%%%%%%%%%%%%%%%%%%%%%%%%%%%'

\item The combination $\Theta = T\,M_y$ is an antiunitary symmetry, where $T$ is the time-reversal operator and $M_y$ denotes the mirror reflection $k_y\to-k_y$. This symmetry relates the two in-plane momentum directions and further constrains the allowed planar-Hall components. Time-reversal symmetry alone is absent, as can be seen from the sign change of the in-plane components of $\bs d(\bs k)$ under $k_z\to-k_z$, which cannot be compensated by a unitary operator in the absence of the accompanying mirror operation. Since $\bs\Omega_s$ and $\bs m$ inherit this same sign change, plain time reversal is broken precisely by the dipolar Berry geometry rather than by any external perturbation. Consequently, the ordinary Onsager--Casimir relation $\sigma_{ij}(\bs B)=\sigma_{ji}(-\bs B)$ does not hold for the VNR. Only its $\Theta$-covariant generalisation survives, worked out explicitly in Sec.~\ref{secfull}, and it is this weaker constraint that ultimately permits the dipolar BC to activate conductivity terms odd in $\bs B$, of the kind forbidden for a monopolar, time-reversal-symmetric semimetal.

\end{enumerate}

Let us now discuss the 3-band HSM:
\begin{enumerate}

\item The Hamiltonian $H_h$ possesses a continuous $U(1)$ rotational symmetry about the $k_z$-axis, corresponding in real space to the continuous rotational subgroup $C_\infty$ about $\hat{\bs z}$. Under an azimuthal rotation $k_x+i\,k_y\to e^{i \, \varphi}(k_x+i\,k_y)$, the Hamiltonian transforms covariantly as
\begin{align}
H_h(\mathcal{R}_z \,\bs k) = U(\varphi)\,H_h(\bs k)\,U^\dagger(\varphi)\,,
\quad U(\varphi) = \text{diag}(1,e^{i \, \varphi},1)\,.
\end{align}
This can be verified by noting that
\begin{align}
U\, \lambda_1 \,U^\dagger &= \cos\varphi\,\lambda_1+\sin\varphi\,\lambda_2\,,\quad
U\, \lambda_2 \,U^\dagger = -\sin\varphi\,\lambda_1+\cos\varphi\,\lambda_2\,,\quad
U\, \lambda_5 \,U^\dagger = \lambda_5\,,
\end{align}
which precisely compensates the rotation of $k_x$ and $k_y$, leaving $k_z\,\lambda_5$ invariant. Although the Hamiltonian possesses only the axial symmetry $U(1)$, its eigenvalues depend only on $k=|\mathbf{k}|$, yielding an accidental $O(3)$-symmetric spectrum. The reduction from $SO(3)$ to $U(1)$ therefore resides entirely in the eigenstates and the associated quantum geometry, rather than in the energy dispersion.
%%%%%%%%%

\item $H_h$ is nvariant under reflection through the $k_z=0$ plane:
\begin{align}
H_h(k_x,k_y,-k_z) = U_\sigma\,H_h(k_x,k_y,k_z)\,U_\sigma^\dagger\,,
\quad U_\sigma = \mathrm{diag}(1,1,-1)\,,
\end{align}
since $U_\sigma$ leaves $\lambda_1$ and $\lambda_2$ invariant while transforming $\lambda_5\to-\lambda_5$, exactly compensating the sign reversal of $k_z$. This mirror symmetry exchanges the northern and southern hemispheres of momentum space while leaving the equatorial plane invariant. Consequently, any momentum-space quantity odd under $k_z\to-k_z$ integrates to zero over the full BZ (or over a closed Fermi surface), whereas even quantities are symmetry-allowed.
%%%%%%%%%

\item $H_h$ anticommutes with the unitary and Hermitian operator,
\begin{align}
\mathcal S = \text{diag}(1,-1,-1)\,, \quad
\mathcal{S}\,H_h(\bs k)\,\mathcal{S}^\dagger = -H_h(\bs k)\,,
\end{align}
satisfying $\mathcal{S}^2=\mathbb{I}_{3\times3}$ and $\mathcal{S}=\mathcal{S}^\dagger$. This readily follows from
\begin{align}
\mathcal{S} \, \lambda_1 \,\mathcal{S}^\dagger = -\lambda_1\,,\quad
\mathcal{S} \, \lambda_2\,\mathcal{S}^\dagger = -\lambda_2\,,\quad
\mathcal{S} \, \lambda_5\,\mathcal{S}^\dagger = -\lambda_5\,.
\end{align}
The chiral symmetry forces the spectrum to occur in $\pm\varepsilon$ pairs, while leaving a zero-energy flat-band ($s=0$) pinned at $\varepsilon=0$, consistent with the spectrum in Eq.~\eqref{eqevhopf}. The existence of the flat-band is a general consequence of chiral symmetry in odd-dimensional Hilbert spaces: because eigenvalues occur in opposite-sign pairs, at least one state must remain at zero energy. Although this symmetry fixes the spectrum, it does not determine the quantum geometry of the bands. Consequently, both the HSM and the TSM possess symmetry-protected zero-energy flat-bands, yet their BC and OMM exhibit fundamentally different structures. As shown below, it is this difference in quantum geometry, rather than the existence of the flat-band itself, that leads to the contrasting transport responses of the two systems.
%%%%%%%%%%%%%%%%%%%%%%

\item Time-reversal symmetry is absent for this case, which can be understood from following discussion. For an antiunitary time-reversal operator $\mathcal{T}= U_{\mathcal T} \, \mathcal{K}$ (where $\mathcal K$ denotes complex conjugation), the condition $\mathcal{T} \,H_h(\bs k) \,\mathcal{T}^{-1}=H_h(-\bs k)$ requires
\begin{align}
U_{\mathcal T} \, H_h^*(\bs k) \, U_{\mathcal T}^\dagger = -\, H_h(\bs k)\,,
\end{align}
since $H_h(-\bs k)=-H_h(\bs k)$. Using $\lambda_1^*=\lambda_1$, $\lambda_2^*=-\lambda_2$, $\lambda_5^*=-\lambda_5$, and matching coefficients of $k_x$, $k_y$, $k_z$ separately, this translates into
\begin{align}
\label{eqtr}
U_{\mathcal T} \, \lambda_1 \, U_{\mathcal T}^\dagger = -\lambda_1\,,\quad
U_{\mathcal T} \, \lambda_2\,  U_{\mathcal T}^\dagger = \lambda_2\,,\quad
U_{\mathcal T} \, \lambda_5\, U_{\mathcal T}^\dagger = \lambda_5\,.
\end{align}
The last two conditions state that $U_{\mathcal T}$ commutes with both $\lambda_2$ and $\lambda_5$. These generators do not commute. Their commutator closes onto the remaining imaginary Gell-Mann matrix, $[\lambda_2,\lambda_5]=i\,\lambda_7$. Consequently, $\{\lambda_2,\lambda_5,\lambda_7\}$ span the defining spin-1 representation of $\mathfrak{so}(3)$ on $\mathbb{C}^3$, which is irreducible. By Schur's lemma, any operator commuting with an irreducible set of matrices must be proportional to the identity, $U_{\mathcal T}=e^{i\,\alpha} \, \mathbb{I}_{3\times3}$. A scalar multiple of the identity acts trivially under conjugation, giving $U_{\mathcal T} \, \lambda_1 \, U_{\mathcal T}^\dagger=\lambda_1\neq-\lambda_1$, in direct contradiction with the first condition of Eq.~\eqref{eqtr}. Hence no unitary $U_{\mathcal T}\in U(3)$ exists, and $\mathcal T$ is broken. As for the VNR, the residual antiunitary symmetry $\Theta$ identified in Sec.~\ref{secfull} survives this breaking and relates $\sigma_{ij}(\bs B)$ to its transpose at reversed $\bs B$ only up to the sign factors $\eta_i\eta_j$ of Eq.~\eqref{eq:theta_covariance}, rather than enforcing the plain relation $\sigma_{ij}(\bs B)=\sigma_{ji}(-\bs B)$. This weaker, $\Theta$-covariant constraint is what allows the HSM's dipolar BC and OMM to generate the odd-in-$\bs B$ conductivity terms established in Sec.~\ref{sechopfres}.
%%%%%%%%%%%%%%%%%%%%%%%%%

\item Particle-hole symmetry is absent, which is substantiated by the following discussion. If an antiunitary particle-hole operator $\mathcal C=U_{\mathcal C} \, \mathcal K$ were present, it would satisfy $\mathcal{C}\, H_h(\bs k) \, \mathcal{C}^{-1}=-H_h(-\bs k)=H_h(\bs k)$, requiring
\begin{align}
\label{eqph}
U_{\mathcal C}\,\lambda_1 \,U_{\mathcal C}^\dagger = \lambda_1\,,\quad
U_{\mathcal C}\,\lambda_2 \,U_{\mathcal C}^\dagger = -\lambda_2\,,\quad
U_{\mathcal C}\,\lambda_5 \,U_{\mathcal C}^\dagger = -\lambda_5\,.
\end{align}
The quickest way to rule this out is to invoke the chiral symmetry established above. If a valid $\mathcal C$ existed, the composite antiunitary operator $\mathcal T'\equiv\mathcal S\mathcal C$ would satisfy
\begin{align}
\mathcal T'H_h(\bs k)\mathcal T'^{-1}
= \mathcal S \,\big[- H_h(-\bs k)\big] \, \mathcal S^\dagger
= -\mathcal S \,H_h(-\bs k)\, \mathcal S^\dagger = H_h(-\bs k)\,,
\end{align}
i.e., it would be a legitimate time-reversal operator. Since the Hamiltonian admits no time-reversal symmetry, no such particle-hole operator can exist.

\end{enumerate}

Since we later contrast the VNR against its conventional counterpart in Sec.~\ref{seccompare}, we record here the corresponding symmetries of the NVNR. The minimal model of an NVNR, obtained by gapping a conventional $\mathcal{PT}$-symmetric nodal-ring with a mass term $\Delta$, is captured by~\cite{ips-nlsm-ph,ips-gnr-strain}
\begin{align}
\label{eqhamnvnr}
\mathcal{H}_{\rm N}(\bs k) = {\bs d}_{\rm N}(\bs k)\cdot\bs\sigma\,,\quad
{\bs d}_{\rm N}(\bs k) = \left\lbrace v_0\,\kappa_\perp,\,v_z\,k_z,\,\Delta\right\rbrace\,,
\end{align}
using the same cylindrical coordinates as Eq.~\eqref{eqtrs}. Unlike the VNR of Eq.~\eqref{eqham}, the pseudospin components $d_x$ and $d_y$ here carry no explicit dependence on the azimuthal angle $\phi$. The nodal-ring is encircled by a fixed, non-winding pseudospin texture rather than a vortex, and a finite $\Delta$ is required to generate any BC at all.

Let us now discuss the relevant symmetries of the NVNR:
\begin{enumerate}

\item The continuous axial rotational symmetry about the $k_z$-axis is realised trivially, $\mathcal{H}_{\rm N}(\mathcal R_z\,\bs k) = \mathcal H_{\rm N}(\bs k)$, since $\bs d_{\rm N}$ depends only on $\kappa_\perp$ and $k_z$ and carries no explicit $\phi$-dependence to compensate. This is simpler than the covariant $U(1)$ symmetry of the VNR, and is the direct Hamiltonian-level statement of the absence of any vortex winding in the pseudospin texture.

\item Time-reversal symmetry, $\mathcal T\,\mathcal H_{\rm N}(\bs k)\,\mathcal T^{-1}=\mathcal H_{\rm N}(-\bs k)$, holds with the simplest antiunitary operator $\mathcal T=\mathcal K$ (complex conjugation, $U_{\mathcal T}=\mathbb I_{2\times2}$). This follows since $d_x$ and $\Delta$ are even under $\bs k\to-\bs k$ while $d_y=v_z\,k_z$ is odd, and $\sigma_y^*=-\sigma_y$ supplies the compensating sign. This is the opposite of the VNR, for which plain time-reversal is explicitly broken.

\item No unitary mirror symmetry through the $k_z=0$ plane exists for the NVNR. Reproducing $\mathcal H_{\rm N}(k_x,k_y,-k_z)$ from $\mathcal H_{\rm N}(k_x,k_y,k_z)$ would require a unitary $U_\sigma$ satisfying $U_\sigma\,\sigma_x\,U_\sigma^\dagger=\sigma_x$, $U_\sigma\,\sigma_y\,U_\sigma^\dagger=-\sigma_y$, and $U_\sigma\,\sigma_z\,U_\sigma^\dagger=\sigma_z$. Commuting with both $\sigma_x$ and $\sigma_z$ forces $U_\sigma\propto\mathbb I_{2\times2}$, by the same Schur's-lemma argument used for the HSM above, which is incompatible with the required sign flip on $\sigma_y$. Since $\mathcal{H}_{\rm N}$ depends on $k_x$ and $k_y$ only through $\kappa_\perp$, the mirror relation is already captured by the antiunitary $\mathcal T=\mathcal K$ found above and does not survive as a separate, standalone unitary symmetry. This is the reverse of the VNR, which possesses a genuine unitary mirror $U_\sigma=\sigma_z$ but no standalone time-reversal symmetry.

\end{enumerate}

The VNR and the NVNR are therefore related by a symmetry exchange. The vortex winding that generates the VNR's BC-dipole is precisely what breaks time-reversal while preserving a unitary mirror, whereas gapping the conventional nodal-ring restores time-reversal at the cost of that mirror. Because plain time reversal is intact for the NVNR, its conductivity tensor obeys the ordinary Onsager--Casimir relation $\sigma_{ij}(\bs B)=\sigma_{ji}(-\bs B)$ without the weakening found above for the VNR and the HSM, so the odd-in-$\bs B$ diagonal terms permitted by their dipolar BC are symmetry-forbidden for the NVNR. This is the symmetry root of the qualitative contrast drawn when comparing the two systems' magnetotransport in Sec.~\ref{seccompare}.

%======================================================================
\section{Semiclassical Boltzmann formalism and planar-Hall set-ups}
\label{secmethod}
%======================================================================

We compute the linear magnetoelectric conductivity within the semiclassical Boltzmann framework, valid at weak magnetic fields, using a momentum-independent relaxation time $\tau$. This relaxation-time approximation (RTA) replaces the full scattering integral by a phenomenological rate $1/\tau$. The scattering lifetime drives the fermionic distribution towards a steady state, allowing all time dependence to be dropped when the external fields are static. The detailed derivation of the linearised Boltzmann equation and its solution can be found in our earlier works~\cite{ips-kush-review,ips_rahul_ph_strain,rahul-jpcm,ips-ruiz,ips-rsw-ph,ips-shreya}.

Throughout, we work in the weak-field regime $|\bs B|\ll\mu^2/(e\,v_0^2)$, which ensures that the inter-Landau-level spacing is negligible compared with other energy scales, justifying a continuous treatment of the dispersion. Equivalently, the Fermi momentum $\kappa_F$ satisfies $\kappa_F\,\ell_B\gg1$, where $\ell_B\equiv1/\sqrt{e\,|\bs B|}$ is the magnetic length. The phase-space factor $D_s$ is expanded as
\begin{align}
D_s = 1 - e\left(\bs B\cdot\bs\Omega_s\right)
+ e^2\left(\bs B\cdot\bs\Omega_s\right)^2
- e^3\left(\bs B\cdot\bs\Omega_s\right)^3
+ \mathcal{O}(|\bs B|^4)\,.
\label{Exp_D}
\end{align}
The same weak-field condition ensures that the OMM energy correction is small relative to the bare dispersion, $|\varepsilon_m(\bs k)|\ll|\varepsilon_s(\bs k)|$, since
\begin{align}
|\bs B\cdot\bs m|\equiv e\,|\varepsilon_s|\,|\bs B\cdot\bs\Omega_s|\ll|\varepsilon_s|\,.
\end{align}
This hierarchy permits a Taylor expansion of the derivative of the Fermi-Dirac distribution $f_0(\mathcal E_s,\mu,T)\equiv\left(1+e^{(\mathcal E_s-\mu)/T}\right)^{-1}$, where $\mu$ is the chemical potential and $T$ is the temperature. Retaining terms through cubic order in $|\bs B|$, we obtain
\begin{align}
f_{\rm prime}(\mathcal E_s) \equiv \frac{\partial f_0(\mathcal E_s)}{\partial\mathcal E_s}
= f_0'(\varepsilon_s) + \varepsilon_m\,f_0''(\varepsilon_s)
+ \frac{1}{2}(\varepsilon_m)^2 f_0'''(\varepsilon_s)
+ \frac{1}{6}(\varepsilon_m)^3 f_0''''(\varepsilon_s)
+ \mathcal{O}(|\bs B|^4)\,,
\label{Exp_f}
\end{align}
where a prime denotes differentiation of $f_0(u)$ with respect to its argument.
We work in the $T\to 0$ limit, in which $f_0'(\mathcal E_s)\to-\delta(\mathcal E_s-\mu)$. Results for $T>0$ can be obtained by using the relation~\cite{mermin}
\begin{align}
\label{eqsigmat}
\sigma^s_{ij}(T) = -\int_{-\infty}^\infty\sigma^s_{ij}(T=0)\, 
\partial_{\mathcal E_s} f_0(\mathcal E_s,\mu,T) \,.
\end{align}

%%%%%%%%%%%%%%%%%%
\subsection{Identifying distinct sources of net current}

%%%%%%%%%%%%%%%%%%%%%
\begin{figure}[t!]
\centering
\includegraphics[width= 1 \textwidth]{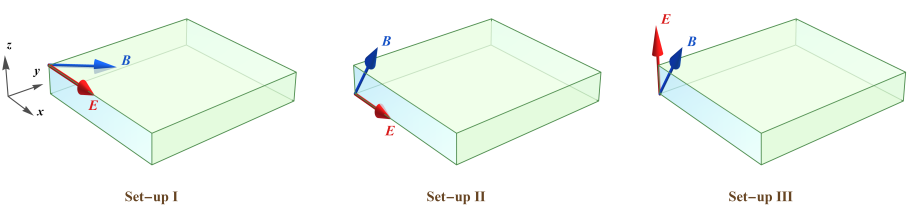}
\caption{\label{figsetups}
The three inequivalent planar-Hall configurations analysed in this work, shown relative to the $k_z=0$ plane and the Cartesian coordinates with unit vectors $( \hat{\bs x}, \, \hat{\bs y}, \, \hat{\bs z})$. Red and blue arrows denote the applied electric ($\bs E$) and magnetic ($\bs B$) fields, respectively. Set-up~I: $\bs E=E_x \, \hat{\bs x}$, $\bs B=B_x \, \hat{\bs x}+B_y \, \hat{\bs y}$. Set-up~II: $\bs E=E_x \, \hat{\bs x}$, $\bs B=B_x \, \hat{\bs x}+B_z \, \hat{\bs z}$. Set-up~III: $\bs E=E_z \, \hat{\bs z}$, $\bs B=B_x \, \hat{\bs x}+B_z \, \hat{\bs z}$. In each set-up, $\theta$ denotes the angle between $\bs B$ and the $x$-axis, so that $B_x=|\bs B|\,\cos\theta$ together with $B_y=|\bs B|\,\sin\theta$ in set-up~I, or $B_z=|\bs B|\,\sin\theta$ in set-ups~II and~III.}
\end{figure}
%%%%%%%%%%%%%%%%%%%%%%%%%%%%

The generic expression for the non-AH, non-LFO contribution to the conductivity, for the band $s$, is
\begin{align}
{\bar \sigma}^s_{ij} &= -\,e^2 \,\tau \int\frac{d^3\bs k}{(2\,\pi)^3}\,D_s
\left[(w_s)_i + e \, (\bs w_s\cdot\bs\Omega_s)B_i\right]
\left[(w_s)_j + e \, (\bs w_s\cdot\bs\Omega_s)B_j\right] f_0'(\mathcal E_s)\,.
\label{eq_elec}
\end{align}
For a given field configuration, we define the planar components of $\bar \sigma$ as those lying in the plane spanned by $\bs E$ and $\bs B$. These decompose into a component longitudinal to $\bs E$ and an in-plane transverse component, referred to as the longitudinal magnetoconductivity (LMC) and the planar-Hall conductivity (PHC), respectively.

An important qualitative difference from our earlier analyses of monopolar Weyl, multifold, and nodal-line semimetals~\cite{ips-rsw-ph,ips-spin1-ph,ips-nlsm-ph,ips-gnr-strain} concerns the parity of this field expansion. There, the BC and OMM are monopole-sourced, and $\bar\sigma_{ij}$ of Eq.~\eqref{eq_elec} was found to contain only even powers of $|\bs B|$, so that a quadratic-order expansion sufficed; the odd-in-$\bs B$ terms appeared exclusively in the out-of-plane AH and LFO-induced parts below. Here, both the VNR and the HSM host a BC-dipole rather than a monopole, and the resulting $\bs\Omega_s$ and $\bs m_s$ carry an additional angular factor ($k_z$ for the VNR, $\cos\gamma$ for the Hopf node) relative to their monopolar counterparts. This flips the $k_z$-parity of several transport integrands and, as we show explicitly in the results below, allows $\bar\sigma_{ij}$ itself to acquire odd-in-$\bs B$ terms, together with an out-of-plane response, both of which are symmetry-forbidden for a BC-monopole. The root cause is the broken time-reversal symmetry identified in Sec.~\ref{secsymmetry}: because $\bs\Omega_s$ and $\bs m_s$ do not flip sign under $\bs k\to-\bs k$ as they would for a monopole, the ordinary Onsager--Casimir relation $\sigma_{ij}(\bs B)=\sigma_{ji}(-\bs B)$ no longer holds, and only its generalised, $\Theta$-covariant form survives, derived below in Eq.~\eqref{eq:theta_covariance}. It is this weaker constraint, rather than the field expansion alone, that ultimately licenses the odd-in-$\bs B$ terms. For this reason we expand all three contributions, $\bar\sigma_{ij}$, $\sigma^{\rm{ah}}_{ij}$, and $\sigma^{\rm{lf}}_{ij}$, consistently up to cubic order in $|\bs B|$ throughout this work, rather than truncating $\bar\sigma_{ij}$ at quadratic order as was sufficient in the monopolar case.

The anomalous-Hall (AH) part, denoted by $\sigma^{\rm{ah}}_s$, contribute to out-of $\bs E \bs B$-plane response, which we will refer to as the out-of-plane conductivity. Expanding to order $|\bs B|^3$, the AH part reads
\begin{align}
\label{eq-ah}
(\sigma^{\rm{ah}}_s)_{ij} &= - \,e^2 \, \epsilon_{ijl}\int\frac{d^3\bs k}{(2\pi)^3}\,\Omega_s^l
\left[f_0(\varepsilon_s) + \varepsilon_m \,f_0'(\varepsilon_s)
+ \frac{1}{2} \,(\varepsilon_m)^2 \,f_0''(\varepsilon_s)
+ \frac{1}{6}\,(\varepsilon_m)^3 \, f_0'''(\varepsilon_s)
+ \order{|\bs B|^4}\right].
\end{align}
The first term is independent of $\bs B$ and survives for our systems. Additional nonzero terms appear only when the OMM part is correctly accounted for, showing the importance of the OMM contributions.

Last but not the least are the contributions from the action of the LFO \cite{ips-rsw-ph, ips_tilted_dirac, ips-spin1-ph, ips-nlsm-ph, ips-gnr-strain}
\begin{align}
\label{lf_cond1}
\left(\sigma_s^{\rm lf}\right)_{ij} &= -\,e^2 \, \tau
\int\frac{d^3\bs k}{(2\pi)^3}
\left[(w_s)_i + e(\bs w_s\cdot\bs\Omega_s)B_i\right]
f_0'(\mathcal E_s) \, \frac{\partial\mathcal Y_s}{\partial E_j}\,,\nn
\mathcal Y_s &= \sum_{n=1}^{\infty}\left(e\,\tau\,D_s\right)^n\check L^n\left[D_s\left\lbrace\bs w_s + e(\bs w_s\cdot\bs\Omega_s)\bs B\right\}\cdot\bs E\right], \quad
%%%%%%%%%%
\check L = (\bs w_s\times\bs B)\cdot\nabla_{\bs k}\,.
\end{align}
The symbol $\check L$ denotes the LFO. Each action of $\check L$ contributes one linear power of $|\bs B|$, while additional powers enter through $\bs u^{m}$ and the series expansions of $D_s$ and $f_0'(\mathcal E_s)$. We will find that LFO-induced part comprises both in-plane and out-of-plane components of conductivity.

While plotting some representative curves, we will use numerical values from Table~\ref{tab-params}.
%%%%%%%%%%%%%%%%%%%%%%%%
\begin{table}[t!]
\centering
\begin{tabular}{|c|c|c|}
\hline
 Parameter &     Natural Units  \\ \hline
$v_0$ from Ref.~\cite{phe_nlsm} & $0.0004$\\  
\hline
$v_z$ from Ref.~\cite{phe_nlsm} &  $0.00045$
\\  
\hline
$k_0$ from Refs.~\cite{phe_nlsm, pikulin-gauge-nlsm} &  $ 160 $  \text{eV} 
\\
\hline
$\tau$ from Ref.~\cite{yang1} &  $15.2  $ eV$^{-1}$  
\\  \hline
$ B $ from Ref.~\cite{phe_nlsm} &  $ 0 $ --- $ 100 $ eV$^2 $ \\ \hline
$\mu$ from Ref.~\cite{phe_nlsm} &  $0.04 $ eV  \\ \hline
\end{tabular}
\caption{\label{tab-params}The ranges of values for the various parameters, chosen to generate representative plots of the conductivity tensors, are tabulated here.}
\end{table}

%%%%%%%%%%%%%%%%%%%%%%%%%%%%%%%%%%%%%

\subsection{Conductivity tensor structure and reduction to three set-ups}
\label{secfull}

Here, we consider only the three planar-Hall set-ups shown in
Fig.~\ref{figsetups}, because they are sufficient, together with the
symmetry relations derived below, to determine eight of the nine
components of $\sigma_{ij}(\bs B)$ in each of its three physically
distinct contributions $\bar\sigma_{ij}$,
$\sigma^{\rm{ah}}_{ij}$, and $\sigma^{\rm{lf}}_{ij}$ [cf.\
Eqs.~\eqref{eq_elec}, \eqref{eq-ah}, and \eqref{lf_cond1}], for
$\bs B$ lying in the $xz$-plane ($B_y=0$). A natural question is then:
to what extent do these three set-ups, together with the symmetries of
Sec.~\ref{secsymmetry}, determine the full three-dimensional tensor
$\sigma_{ij}(B_x,B_y,B_z)$ for an arbitrary magnetic-field direction?
We now show that the full tensor is fixed by the combined data on the
$xy$- and $xz$-planes, up to a single undetermined component, which
resides in the $xz$-plane.

$\sigma_{ij}(\bs B)$ is a rank-2 tensor relating the current density to
the applied electric field via $j_i=\sigma_{ij}\, E_j$. Thus, in 3d,
$\bs\sigma$ can be represented as $3\times3$ matrix:
\begin{align}
\label{eq:full_tensor}
\bs{\sigma}(\bs{B})
= \begin{pmatrix}
\sigma_{xx} & \sigma_{xy} & \sigma_{xz} \\[3pt]
\sigma_{yx} & \sigma_{yy} & \sigma_{yz} \\[3pt]
\sigma_{zx} & \sigma_{zy} & \sigma_{zz}
\end{pmatrix} .
\end{align}
Each component $\sigma_{ij}$ is, in general, a nontrivial function of
$\bs{B}$. In a planar-Hall measurement, $\bs E$ is applied along a
fixed Cartesian axis and the resulting current $\bs j=\bs\sigma\cdot\bs E$
is measured along all three directions. Fixing $\bs E$ along a single
axis $\hat e_j$ selects the $j$-th column of the tensor:
$j_i=\sigma_{ij} \, E_j$ (no sum over $j$).

In the absence of rotational symmetry, six distinct planar-Hall
configurations can be constructed by choosing the electric field along
one Cartesian axis and allowing the magnetic field to lie in either of
the two coordinate planes containing that axis:
\begin{align}
(E_x,xy),\quad(E_x,xz),\quad(E_y,xy),\quad(E_y,yz),\quad(E_z,xz),\quad(E_z,yz)\,.
\end{align}
The continuous rotational symmetry $C_{\infty z}$ of the untilted
Hamiltonians identifies the pairs
\begin{align}
(E_x,xy)\sim(E_y,xy),\quad
(E_x,xz)\sim(E_y,yz),\quad
(E_z,xz)\sim(E_z,yz),
\end{align}
so that the six candidate geometries collapse into three inequivalent
classes. The three set-ups defined below then probe the following tensor
columns. Set-up~I probes column~1, yielding $\sigma_{xx}$,
$\sigma_{yx}$, $\sigma_{zx}$ as functions of $(B_x,B_y)$. Set-up~II
again probes column~1, but with $\bs B$ tilted out of the nodal plane,
giving $\sigma_{xx}$, $\sigma_{yx}$, $\sigma_{zx}$ as functions of
$(B_x,B_z)$. Set-up~III probes column~3, yielding $\sigma_{xz}$,
$\sigma_{yz}$, $\sigma_{zz}$. The three set-ups provide direct access
to columns~1 and~3 of the conductivity tensor, corresponding to six of
the nine components.

For $\mathcal{H}_\ell$, invariance under arbitrary rotations $R_\phi$
about the $\hat z$-axis implies the covariance relation
\begin{align}
\label{eq:covariance}
\sigma_{ij}(R_\phi \, \bs B) = (R_\phi)_{ik} \, (R_\phi)_{jl} \,\sigma_{kl}(\bs B)\,.
\end{align}
For $\phi=\pi/2$ ($\hat{\bs x}\to \, \hat{\bs y}$,
$\hat{\bs y}\to- \, \hat{\bs x}$, $\hat{\bs z}\to \, \hat{\bs z}$),
Eq.~\eqref{eq:covariance} determines column~2 from column~1 only when
the rotated field stays inside a plane already probed by the three
set-ups. This holds in the $xy$-plane: the field whose \emph{image}
under $R_{\pi/2}$ equals $(B_x,B_y,0)$ is the \emph{preimage}
\begin{align}
R_{\pi/2}^{-1}(B_x,B_y,0) = (B_y,-B_x,0)\,,
\end{align}
which remains within set-up~I's domain. Applying Eq.~\eqref{eq:covariance}
with $R_{\pi/2}$ and using $(R_{\pi/2})_{ik}(R_{\pi/2})_{jl}$ then gives
\begin{align}
\label{eq:rotation_relations}
\sigma_{yy}(B_x,B_y,0) &= \sigma_{xx}(B_y,-B_x,0)\,,\quad
\sigma_{xy}(B_x,B_y,0) = -\,\sigma_{yx}(B_y,-B_x,0)\,,\quad
\sigma_{zy}(B_x,B_y,0) = \sigma_{zx}(B_y,-B_x,0)\,.
\end{align}
For the first two relations, the components $\sigma_{xx}$ and
$\sigma_{yx}$ are even under $\bs B\to-\bs B$ in set-up~I, so the
preimage form $(B_y,-B_x,0)$ is equivalent to $(-B_y,B_x,0)$ and these
two entries may equivalently be written in the more symmetric form
quoted below. The third relation, by contrast, involves $\sigma_{zx}$,
which is \emph{odd} under $\bs B\to-\bs B$, so that
\begin{align}
\sigma_{zx}(B_y,-B_x,0) = -\,\sigma_{zx}(-B_y,B_x,0)\,.
\end{align}
Keeping the preimage argument $(B_y,-B_x,0)$ together with the positive
sign in Eq.~\eqref{eq:rotation_relations} is therefore the consistent
choice, and reproduces the same physical result for $\sigma_{zy}$ as the
more compact statement
\begin{align}
\label{eq:rotation_relations_compact}
\sigma_{yy}(B_x,B_y,0) &= \sigma_{xx}(-B_y,B_x,0)\,,\quad
\sigma_{xy}(B_x,B_y,0) = -\,\sigma_{yx}(-B_y,B_x,0)\,,\quad
\sigma_{zy}(B_x,B_y,0) = -\,\sigma_{zx}(-B_y,B_x,0)\,,
\end{align}
which we adopt in the main text. It fails in the $xz$-plane, since
$R_{\pi/2}(B_x,0,B_z)=(0,B_x,B_z)$ lies in the $yz$-plane, a
configuration none of the three set-ups computes.

For the VNR, this gap is closed instead by the antiunitary symmetry
$\Theta=T\,M_y$ of item~3 in Sec.~\ref{secsymmetry}. Under $M_y$ a
polar vector transforms as $(E_x,E_y,E_z)\to(E_x,-E_y,E_z)$ and an
axial vector with the opposite parity,
$(B_x,B_y,B_z)\to(-B_x,B_y,-B_z)$, so that composing with $T$ gives
$\Theta:\bs E\to(E_x,-E_y,E_z)$ and
$\Theta:\bs B\to(B_x,-B_y,B_z)$. Being antiunitary, $\Theta$ relates
$\sigma_{ij}$ to its transpose,
\begin{align}
\label{eq:theta_covariance}
\sigma_{ij}(\bs B) = \eta_i\,\eta_j\,\sigma_{ji}(B_x,-B_y,B_z)\,,\quad
\eta_x=\eta_z=1\,,\quad \eta_y=-1\,,
\end{align}
which reduces to the familiar $\sigma_{ij}(\bs B)=\sigma_{ji}(-\bs B)$
when $\eta_i=1$ for all $i$, as expected for plain time reversal. At
$B_y=0$ the right-hand side becomes a same-point relation,
\begin{align}
\label{eq:theta_xz}
\sigma_{xy}(B_x,0,B_z) = -\,\sigma_{yx}(B_x,0,B_z)\,,\quad
\sigma_{zy}(B_x,0,B_z) = -\,\sigma_{yz}(B_x,0,B_z),
\end{align}
recovering $\sigma_{xy}$ and $\sigma_{zy}$ from $\sigma_{yx}$ and
$\sigma_{yz}$, both already computed in set-ups~II and~III. Setting
$i=j=y$ in Eq.~\eqref{eq:theta_covariance} gives
$\sigma_{yy}(B_x,B_y,B_z)=\sigma_{yy}(B_x,-B_y,B_z)$, which is even
in $B_y$. At $B_y=0$, relevant for the $xz$-plane, this reduces to a
trivial identity and therefore imposes no constraint on
$\sigma_{yy}(B_x,0,B_z)$. Eight of the nine tensor components are thus
fixed for the VNR when $\bs B$ lies in the $xz$-plane: three from
column~1 (set-up~II), three from column~3 (set-up~III), and two more,
$\sigma_{xy}$ and $\sigma_{zy}$, from Eq.~\eqref{eq:theta_xz}. Only
$\sigma_{yy}(B_x,0,B_z)$ remains open.

The HSM possesses the same antiunitary symmetry, even though items~4
and~5 rule out plain time reversal and particle-hole symmetry. Since
$\lambda_1^*=\lambda_1$ and $\lambda_2^*=-\lambda_2$,
$\lambda_5^*=-\lambda_5$, one finds
$\mathcal U\,H_h^*(\bs k)\,\mathcal U^\dagger
=H_h(-k_x,k_y,-k_z)$ with $\mathcal U=\mathrm{diag}(1,-1,1)$.
This is the action of $\Theta$ on momentum, and
Eqs.~\eqref{eq:theta_covariance} and~\eqref{eq:theta_xz} therefore
hold for the HSM as well. Only $\sigma_{yy}(B_x,0,B_z)$ remains open
in the $xz$-plane for both systems. Set-ups~I and~II both probe
column~1, but belong to distinct symmetry classes: the $C_{\infty z}$
symmetry acts only within the $xy$-plane and cannot transform an
in-plane $\bs B$ into one containing an out-of-plane component.
Set-up~III is the only geometry that directly probes column~3, and no
rotation about the symmetry axis can transform an in-plane electric
field into an axial one.

Although the covariance relation~\eqref{eq:covariance} was motivated
above for the full tensor $\sigma_{ij}(\bs B)$, it in fact holds
separately for each of the three physically distinct contributions,
$\bar\sigma_{ij}$, $\sigma^{\rm{ah}}_{ij}$, and
$\sigma^{\rm{lf}}_{ij}$, given respectively by
Eqs.~\eqref{eq_elec}, \eqref{eq-ah}, and \eqref{lf_cond1}, since every
microscopic ingredient entering these three expressions, namely
$\bs w_s$, $\bs\Omega_s$, $\bs m_s$, $D_s$, $\mathcal E_s$ and
$\check L$, is itself covariant under $U(\varphi)$. We establish this
order by order in $|\bs B|$ in Appendix~\ref{app_nonAH}. The reduction
to three set-ups, together with
Eqs.~\eqref{eq:rotation_relations_compact} and~\eqref{eq:theta_xz},
therefore applies separately to each contribution rather than only to
their sum
$\sigma_{ij}=\bar\sigma_{ij}+\sigma^{\rm{ah}}_{ij}
+\sigma^{\rm{lf}}_{ij}$, and the components identified as unfixed
above remain unfixed for all three contributions alike.

The AH part obeys, in addition, a constraint that follows purely from
its algebraic structure rather than from the rotational symmetry.
Since $\sigma^{\rm{ah}}_{ij}$ is built from the Levi-Civita symbol
in Eq.~\eqref{eq-ah}, it is manifestly antisymmetric in $i,j$ for
every $\bs B$, so the diagonal components
$\sigma^{\rm{ah}}_{xx}$, $\sigma^{\rm{ah}}_{yy}$, and
$\sigma^{\rm{ah}}_{zz}$ vanish identically, independently of any
rotational argument. Combining this pointwise antisymmetry with the
covariance relation for $(i,j)=(y,x)$ in set-up~I produces a further,
nontrivial prediction: $\sigma^{\rm{ah}}_{yx}(\bs B)$ must be
invariant under a $\pi/2$ rotation of the in-plane field,
$B_x\to-B_y$, $B_y\to B_x$, and can therefore depend on
$(B_x,B_y)$ only through the rotationally invariant combination
$\bs B^2=B_x^2+B_y^2$. This is precisely the structure found
explicitly for $\sigma^{\rm{ah}}_{yx}$ in Eq.~\eqref{eqsigyxset1}
for the VNR, and in the analogous HSM result, which provides an
internal consistency check on the calculation quite independent of
the explicit evaluation of the momentum integrals.

We now turn to the full three-dimensional tensor. The key ingredient
is the continuous rotational symmetry $C_{\infty z}$ about the
$\hat z$-axis, which both the VNR and the HSM possess. Any magnetic
field $\bs B=(B_x,B_y,B_z)$ can be rotated about $\hat z$ into the
$xz$-plane. Defining
\begin{align}
B_\perp = \sqrt{B_x^2+B_y^2}\,,\quad
\theta_B=\arctan\!\left(\frac{B_y}{B_x}\right),
\end{align}
we have $\bs B=R_{\theta_B}(B_\perp,0,B_z)$, where $R_{\theta_B}$ is
a proper rotation about $\hat z$ by angle $\theta_B$. Substituting
this into the covariance relation of Eq.~\eqref{eq:covariance} gives
\begin{align}
\label{eq:full3d}
\sigma_{ij}(B_x,B_y,B_z)
= (R_{\theta_B})_{ik}\,(R_{\theta_B})_{jl}\,
\sigma_{kl}(B_\perp,0,B_z)\,.
\end{align}
Eq.~\eqref{eq:full3d} states that the full 3d tensor is completely
determined by its values on the $xz$-plane at $(B_\perp,0,B_z)$. It
is therefore enough to ask how many components of
$\sigma_{kl}(B_\perp,0,B_z)$ are fixed by the three set-ups and by
the symmetries of Sec.~\ref{secsymmetry}.

As established above, eight of the nine components of
$\sigma_{kl}(B_x,0,B_z)$ are determined in the $xz$-plane:
set-up~II gives $\sigma_{xx}(B_x,0,B_z)$,
$\sigma_{yx}(B_x,0,B_z)$, and $\sigma_{zx}(B_x,0,B_z)$;
set-up~III gives $\sigma_{xz}(B_x,0,B_z)$,
$\sigma_{yz}(B_x,0,B_z)$, and $\sigma_{zz}(B_x,0,B_z)$;
the antiunitary symmetry $\Theta$ of Eq.~\eqref{eq:theta_xz} gives
$\sigma_{xy}(B_x,0,B_z)=-\,\sigma_{yx}(B_x,0,B_z)$ and
$\sigma_{zy}(B_x,0,B_z)=-\,\sigma_{yz}(B_x,0,B_z)$.
The single remaining component is $\sigma_{yy}(B_x,0,B_z)$. It is not
obtainable from the three set-ups by symmetry alone: the rotational
covariance relation relates it to $\sigma_{xx}(0,-B_x,B_z)$, an
$(E_x,\bs B\in yz)$ configuration outside the set of geometries
probed by the three set-ups, and the antiunitary symmetry $\Theta$
gives only $\sigma_{yy}(B_x,B_y,B_z)=\sigma_{yy}(B_x,-B_y,B_z)$,
which is a trivial identity at $B_y=0$. The AH part contributes
nothing to $\sigma_{yy}$ for any $\bs B$, since
$\sigma^{\rm{ah}}_{ij}$ is antisymmetric in $i,j$, so the missing
piece resides in $\bar\sigma_{yy}$ and $\sigma^{\rm{lf}}_{yy}$.
Neither of these is physically undetermined by the model, however.
Eq.~\eqref{eq_elec} for $\bar\sigma_{ij}$ carries no $\bs E$-dependence
at all, being built only from $D_s$, $\bs w_s$, and $\bs\Omega_s$,
already known for both models at every $\bs k$; setting $i=j=y$ in
that same integral fixes $\bar\sigma_{yy}(B_x,0,B_z)$ directly,
without appeal to symmetry or to a new field configuration.
Likewise, $\check L=(\bs w_s\times\bs B)\cdot\nabla_{\bs k}$ in
Eq.~\eqref{lf_cond1} acts only on $\bs k$, so $\mathcal Y_s$ is an
exactly linear functional of $\bs E$ with a $\bs k$-dependent
coefficient vector; its $y$-component,
$\partial\mathcal Y_s/\partial E_y$, is fixed by the same recursive
procedure already used for the other two columns. Both components
are therefore determined by the model; what is missing is only their
explicit closed form, which requires further momentum integration
rather than new physics.

It is instructive to compare this situation with the $xy$-plane,
where the data from set-up~I is supplemented by the rotational
covariance relation of Eq.~\eqref{eq:rotation_relations_compact}.
There, we obtain $\sigma_{yy}$, $\sigma_{xy}$, and $\sigma_{zy}$
directly from the corresponding components in column~1 via
\begin{align}
\sigma_{yy}(B_x,B_y,0) &= \sigma_{xx}(B_x,B_y,0)\big|_{B_x\leftrightarrow B_y}\,,\nn
\sigma_{xy}(B_x,B_y,0) &= -\,\sigma_{yx}(B_x,B_y,0)\big|_{B_x\leftrightarrow B_y}\,,\nn
\sigma_{zy}(B_x,B_y,0) &= -\,\sigma_{zx}(B_x,B_y,0)\big|_{B_x\leftrightarrow B_y}\,.
\end{align}
Moreover, at $B_z=0$, set-up~III reduces to the $xy$-plane and
provides $\sigma_{xz}(B_x,0,0)$, $\sigma_{yz}(B_x,0,0)$, and
$\sigma_{zz}(B_x,0,0)$. Rotating these to an arbitrary in-plane
field using Eq.~\eqref{eq:full3d} then determines the full column~3
in the $xy$-plane. Thus, in the $xy$-plane, all nine components of
the conductivity tensor are determined by combining set-up~I with
set-up~III at $B_z=0$, together with the rotational covariance
relation. The single component whose closed form is still missing,
$\sigma_{yy}(B_x,0,B_z)$, is therefore specific to the $xz$-plane,
and does not arise from any deficiency of the $xy$-plane data.

Obtaining the explicit expression for $\sigma_{yy}(B_x,0,B_z)$ is
most directly organised as a fourth planar-Hall set-up, $(E_y,xz)$
with $\bs E=E_y\,\boldsymbol{\hat y}$ and
$\bs B=B_x\,\boldsymbol{\hat x}+B_z\,\boldsymbol{\hat z}$, which
evaluates Eqs.~\eqref{eq_elec}--\eqref{lf_cond1} with $i=j=y$. This
is a bookkeeping choice rather than a theoretical necessity: as noted
above, both $\bar\sigma_{ij}$ and $\mathcal Y_s$ are defined for
arbitrary $i,j$ and $\bs B$ without reference to any particular
set-up, so the calculation proceeds by the same methods already used
for column~1 and column~3. We leave this explicit evaluation for
future work.

Combining this counting with Eq.~\eqref{eq:full3d}, we conclude that
the full three-dimensional tensor $\sigma_{ij}(B_x,B_y,B_z)$ is
determined by the combined data on the $xy$- and $xz$-planes,
together with the symmetries of Sec.~\ref{secsymmetry}, up to a
single undetermined function of two variables,
$\sigma_{yy}(B_\perp,0,B_z)$, whose closed form has not been
evaluated here even though it is fixed by the same microscopic
formulas as every other component. In this sense, the three set-ups
are ``almost complete'': they fix the explicit form of the full
tensor everywhere except along the $yy$-channel of the $xz$-plane
slice, which is then propagated to the rest of momentum space by
Eq.~\eqref{eq:full3d}. No relation to an already-computed component
exists in the $xz$-plane for this channel, since the required
rotation $R_{\pi/2}$ maps $(B_x,0,B_z)$ into the $yz$-plane, which
none of the three set-ups probes; its explicit form would instead
follow from direct evaluation of Eqs.~\eqref{eq_elec}
and~\eqref{lf_cond1} at $i=j=y$.

%======================================================================
\section{Magnetoelectric conductivity of vortex nodal-ring}
\label{secvnr}
%======================================================================

We expand $\bar\sigma_{ij}$, $\sigma^{\rm{ah}}_{ij}$, and $\sigma^{\rm{lf}}_{ij}$ consistently up to cubic order in $\bs B$, retaining odd as well as even powers, as anticipated in Secs.~\ref{secsymmetry} and~\ref{secmethod}. The model breaks time-reversal symmetry through its dipolar BC and OMM, which are $\mathcal T$-odd quantities fixed by the band structure. They do not reverse with $\bs B$ in a measurement. Onsager reciprocity~\cite{onsager1,onsager2,onsager3} must therefore be applied with the combined operation that reverses $\bs B$, $\bs\Omega_s$, and $\bs m_s$ simultaneously. In its generalised form, the relation
\begin{align}
\sigma_{ij}(\bs B,\bs\Omega_s,\bs m_s)=\sigma_{ji}(-\bs B,-\bs\Omega_s,-\bs m_s)\,,
\end{align}
requires the diagonal components to be even under the simultaneous reversal of $\bs B$, $\bs\Omega_s$, and $\bs m_s$. It does not require them to be even in $\bs B$ alone, since an odd power of $\bs B$ can be compensated by an odd total power of the dipolar $\bs\Omega_s$ and $\bs m_s$.

The mirror symmetry of Sec.~\ref{secsymmetry} decides which of these odd powers can actually appear. It maps the axial vector $\bs B$ to $(-B_x,-B_y,B_z)$ and leaves the diagonal components unchanged, which gives
\begin{align}
\sigma_{ii}(B_x,B_y,B_z)=\sigma_{ii}(-B_x,-B_y,B_z)\,.
\end{align}
A diagonal component can therefore contain only even total powers of the in-plane field components $B_x$ and $B_y$, and $B_z$ is the sole field component that can carry an odd power. At the level of the integrand, the extra factor of $k_z$ in the dipolar BC and OMM makes $\Omega_s^z$ and $m_s^z$ even in $k_z$, while $\Omega_s^{x,y}$ and $m_s^{x,y}$ are odd. Each power of $B_z$ then multiplies a factor that is even in $k_z$ and survives the integration. Each power of $B_x$ or $B_y$ multiplies an odd factor and survives only in pairs. For a BC-monopole, the isotropic texture $\bs\Omega_s\propto\hat{\bs k}$ makes $\Omega_s^z$ odd in $k_z$ and forces the relevant angular average to vanish. Only even powers of $\bs B$ then appear in practice. The dipolar factor reduces this spherical symmetry to $C_{\infty z}$ and removes the accidental cancellation.

Odd powers of $B_z$ are thus permitted in the diagonal components of set-ups~II and~III, but not in those of set-up~I, where $\bs B$ lies in the $xy$-plane. They are indeed generated in the individual BC- and OMM-dependent sub-parts of $\bar\sigma_{xx}$ and $\sigma^{\rm{lf}}_{xx}$ in set-up~II, and of $\bar\sigma_{zz}$ and $\sigma^{\rm{lf}}_{zz}$ in set-up~III. An off-diagonal component is instead related to its transpose and is not constrained on its own. The dipolar factor activates odd-in-$\bs B$ terms in these components as well, namely in $\bar\sigma_{zx}$, $\sigma^{\rm{ah}}_{zx}$, and $\sigma^{\rm{lf}}_{zx}$ in set-up~I, in $\bar\sigma_{zx}$ and $\sigma^{\rm{lf}}_{zx}$ in set-up~II, and in $\bar\sigma_{xz}$ and $\sigma^{\rm{lf}}_{xz}$ in set-up~III. All of these appear below. To our knowledge, this out-of-plane and odd-in-$\bs B$ in-plane conductivity, arising from the non-AH contribution alone and excluding the LFO-induced parts, constitutes the first such observation in the magnetotransport literature on topological semimetals. To write the results compactly, we introduce the variable
\begin{align}
k_\mu \equiv \sqrt{k_0^2-\mu^2/v_0^2}\,.
\end{align}
Throughout this section, we decompose
\begin{align}
\label{eqbarsigdec}
\bar\sigma_{ij} = \sigma^{\text{d}}_{ij} + \sigma^{\text{bc}}_{ij} + \sigma^{\text{m}}_{ij} + \sigma^{\text{conc}}_{ij},
\end{align}
corresponding to the Drude term, the BC-only term, the OMM-only term, and the BC-OMM concurrent contribution. The AH and LFO parts are presented separately.

%%%%%%%%%%%%%%%%%%%%%%%%%%%%%%%%%%%%%%%%s
\subsection{Set-up I: 
\texorpdfstring{$\bs{E}=E_x\, {\bs{\hat x}}$}{E}, 
\texorpdfstring{$ \bs{B}=B_x \,{\bs{\hat x}} + B_y \,{\bs{\hat y}} $}{B}, and
\texorpdfstring{$ \bs{B}^2 = B_x^2 + B_y^2 $}{BB}}
\label{secset1}

The non-AH part comprises
%%%%%%%%%%%%%%%%%%%%%%%%
\begin{align}
\label{eqsigxxset1}
\sigma^{\rm d}_{xx} &= \frac{\tau \, e^2 \, k_0 \,\mu }{8\,\pi}\,,
\end{align}
%%%%%%%%%%%%%%%
\begin{align}
\sigma^{\rm bc}_{xx} &= \frac{ - \,  \tau \, e^4 \, v_0^4 \,}{512\,\pi\,\mu^7}
\Big[B_x^2  \left(-24\,k_0^5\,v_0^4+
24\,k_0^4\,k_\mu\,v_0^4+76\,k_0^3\,v_0^2 \,\mu^2
-64\,k_0^2\,k_\mu\,v_0^2 \,\mu^2-93\,k_0\,\mu^4+64\,k_\mu\,\mu^4\right)\nn
& \hspace{2.25 cm} 
+B_y^2\,k_0\left(-8\,k_0^4\,v_0^4+8\,k_0^3\,k_\mu\,v_0^4+4\,k_0^2\,v_0^2 \,
\mu^2+\mu^4\right)\Big],\nn
%%%%%%%%%%%%%%%%%
\sigma^{\rm m}_{xx} &= \frac{\tau \,e^4}{512\,\pi\,\mu^7\,k_\mu^4}
\big[B_x^2 \,
\{-120\,k_0^9\,v_0^8+152\,k_0^7\,v_0^6\,\mu^2+121\,k_0^5\,v_0^4\,\mu^4
-218\,k_0^3\,v_0^2\,\mu^6+65\,k_0\,\mu^8\nn
& \hspace{2.25 cm}
+ k_\mu(120\,k_0^8\,v_0^8-92\,k_0^6\,v_0^6\,\mu^2
-152\,k_0^4\,v_0^4\,\mu^4+144\,k_0^2\,v_0^2\,\mu^6-24\,\mu^8)\}\nn
&\hspace{2.25 cm} +B_y^2 \,
\{-40\,k_0^9\,v_0^8-120\,k_0^7\,v_0^6\,\mu^2+403\,k_0^5\,v_0^4\,\mu^4
-286\,k_0^3\,v_0^2\,\mu^6+43\,k_0\,\mu^8\nn
& \hspace{2.25 cm} +k_\mu(40\,k_0^8\,v_0^8+140\,k_0^6\,v_0^6\,\mu^2-328\,k_0^4\,v_0^4\,\mu^4+144\,k_0^2\,v_0^2\,\mu^6-8\,\mu^8)\}\Big],\nn
\sigma^{\rm conc}_{xx} &= \frac{\tau e^4 v_0^2}{256\,\pi\,\mu^7\,k_\mu^2}
\Big[B_x^2\{24\,k_0^7\,v_0^6-81\,k_0^3\,v_0^2\,\mu^4+57\,k_0\,\mu^6
+k_\mu(-24\,k_0^6\,v_0^6-12\,k_0^4\,v_0^4 \,\mu^2+72\,k_0^2\,v_0^2\,\mu^4-32\,\mu^6)\}\nn
& \hspace{2.25 cm} + k_0\,B_y^2
\,\{8\,k_0^6\,v_0^6-64\,k_0^4\,v_0^4 \,\mu^2+69\,k_0^2\,v_0^2\,\mu^4-13\,\mu^6
+ k_\mu(-8\,k_0^5\,v_0^6+60\,k_0^3\,v_0^4 \,\mu^2-40\,k_0\,v_0^2\,\mu^4)\}\Big],
\end{align}
%%%%%%%%%%%%%%%%%%%%%
\begin{align}
\label{eqsigyxset1}
\sigma^{\rm bc}_{yx} &= \frac{\tau \, e^4 \, v_0^4 \, B_x \,B_y} {256\,\pi\,\mu^7}
\left[8\,k_0^5\,v_0^4-36\,k_0^3\,v_0^2 \,\mu^2
+ 47\,k_0\,\mu^4-8 \,k_\mu\,(k_0^2\,v_0^2-2\,\mu^2)^2\right],\nn
%%%%%%%%%%%%%%%%%
\sigma^{\rm m}_{yx} &= \frac{\tau \,e^4\, B_x\, B_y} {256\,\pi\,\mu^7 \,k_\mu^4}
\big[-\,40\,k_0^9\,v_0^8+136\,k_0^7\,v_0^6\,\mu^2-141\,k_0^5\,v_0^4\,\mu^4
+34\,k_0^3\,v_0^2\,\mu^6+11\,k_0\,\mu^8\nn
&\quad+4\,k_\mu(10\,k_0^8\,v_0^8-29\,k_0^6\,v_0^6\,\mu^2
+22\,k_0^4\,v_0^4\,\mu^4-2\,\mu^8)\big],\nn
%%%%%%%%%%%%%%
\sigma^{\rm conc}_{yx} &= -\,\frac{\tau \,e^4\, v_0^2\, B_x\, B_y}
{128\,\pi\,\mu^7\,k_\mu^2}
\big[-8\, k_0^7\,v_0^6-32\,k_0^5\,v_0^4 \,\mu^2+75 \,k_0^3\,v_0^2\,\mu^4-35\,k_0\,\mu^6
% \nn &\quad
+4\,k_\mu(2\,k_0^6\,v_0^6+9\,k_0^4\,v_0^4 \,\mu^2-14\,k_0^2\,v_0^2\,\mu^4+4\,\mu^6)\big],
\end{align}
%%%%%%%%%%%%%%%%
\begin{align}
\label{eqsigzxset1}
\sigma^{\rm bc}_{zx} &= \frac{\tau\,e^3\,v_0\,B_x} {2048\,\pi}
\big[-224+\frac{ {\boldsymbol B}^2\,e^2\,v_0^4}{\mu^{10}} \,
\{336\,k_0^6\,v_0^6-680\,k_0^4\,v_0^4 \,\mu^2+342\,k_0^2\,v_0^2\,\mu^4-21\,\mu^6
% \nn & \quad
+k_\mu \,(v-336\,k_0^5\,v_0^6-128\,k_0\,v_0^2\,\mu^4)\}\big],\nn
%%%%%%%%%%%%%%%%%%%%%%%%%%
\sigma^{\rm m}_{zx} &= \frac{-\,3 \,\tau \,e^3\, B_x}
{2048\,\pi\,v_0^5\,\mu^{10} \,k_\mu^6}
\big[-32\,\mu^{10}\, v_0^6\, k_\mu^6
+ v_0^6 \,k_\mu^5\,(2688\,k_0^6\,v_0^6-1400\,k_0^4\,v_0^4 \,\mu^2
+100\,k_0^2\,v_0^2\,\mu^4+3\,\mu^6)\}
\nn& \quad
+ {\boldsymbol B}^2\,e^2 \,v_0^4 \,k_\mu\{k_0^3\, v_0^4(-2688\,k_0^8\,v_0^8
+8120\,k_0^6\,v_0^6\,\mu^2-8640\,k_0^4\,v_0^4\,\mu^4+3712\,k_0^2\,v_0^2\,\mu^6-512\,\mu^8)
\big],\nn
%%%%%%%%%%%%%%%%%%%%%%%%%%%%%%%%%%
\sigma^{\rm conc}_{zx} &= \frac{\tau\, e^5 v_0 \,B_x\, 
{\boldsymbol B}^2}{1024\,\pi\,k_\mu^4\,\mu^{10}}
\big[1512\,k_0^{10} v_0^{10}-3504\,k_0^8\,v_0^8\,\mu^2+
2313\,k_0^6\,v_0^6\,\mu^4-155\,k_0^4\,v_0^4\,\mu^6
 \nn &\quad
-173\,k_0^2\,v_0^2\,\mu^8+7\,\mu^{10}+k_0\,k_\mu\,v_0^2
\, (-1512\,k_0^8\,v_0^8+2748\,k_0^6\,v_0^6\,\mu^2
% \nn & \quad
-1128\,k_0^4\,v_0^4\,\mu^4-160\,k_0^2\,v_0^2\,\mu^6+64\,\mu^8)\big]\,.
\end{align}

%%%%%%%%%%%%%%%%%%%%%%%%%%%%%
\begin{figure}[t!]
\centering
\subfigure{\includegraphics[width=0.75 \textwidth]{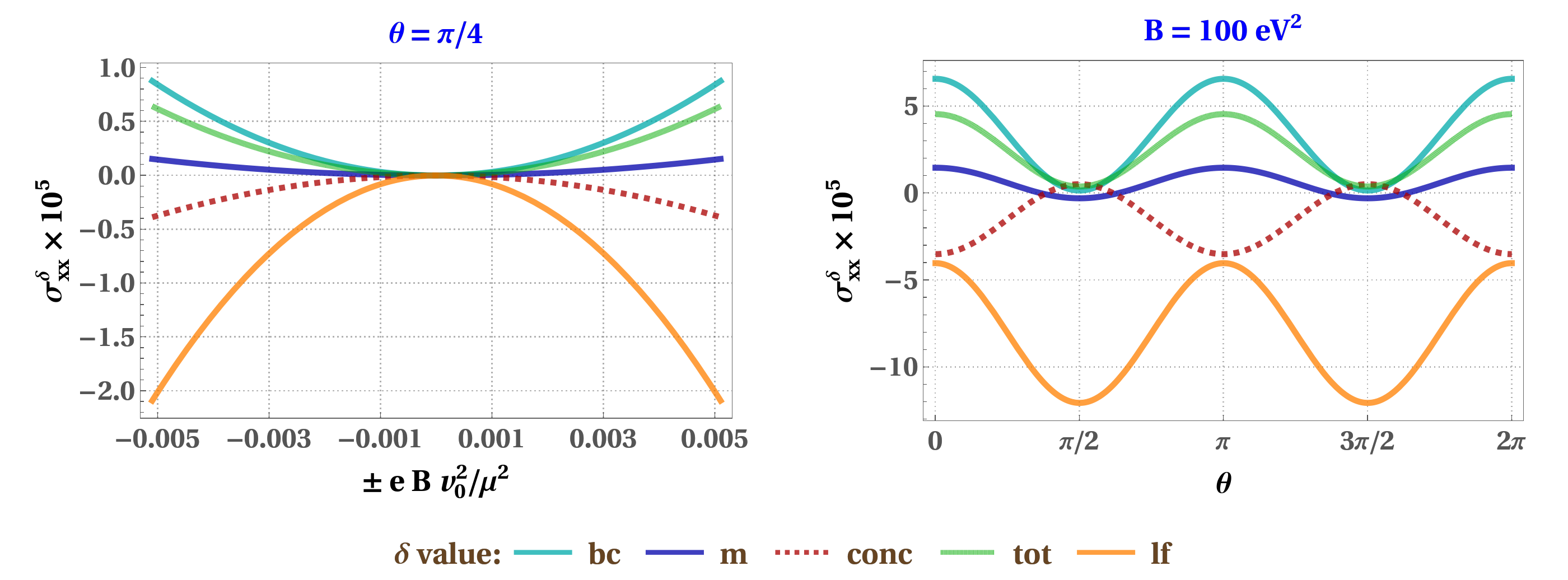}}
\subfigure{\includegraphics[width=0.75 \textwidth]{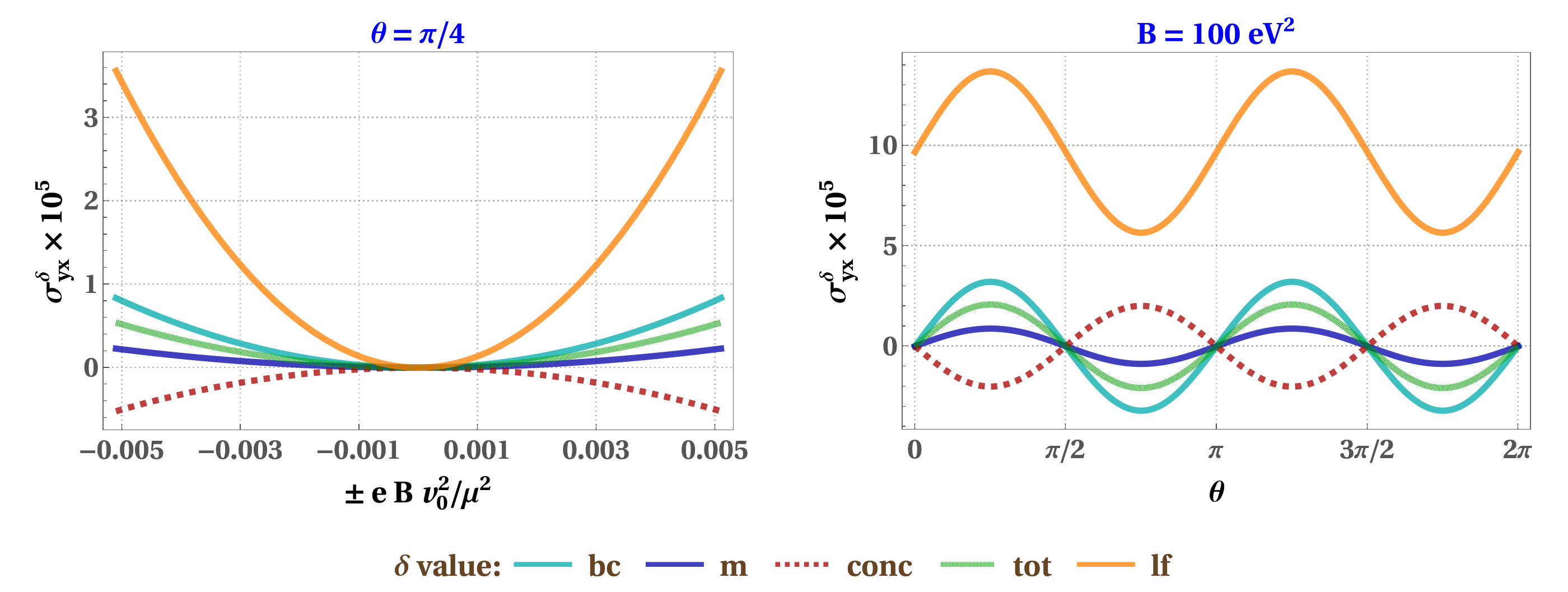}}
\subfigure{\includegraphics[width=0.75 \textwidth]{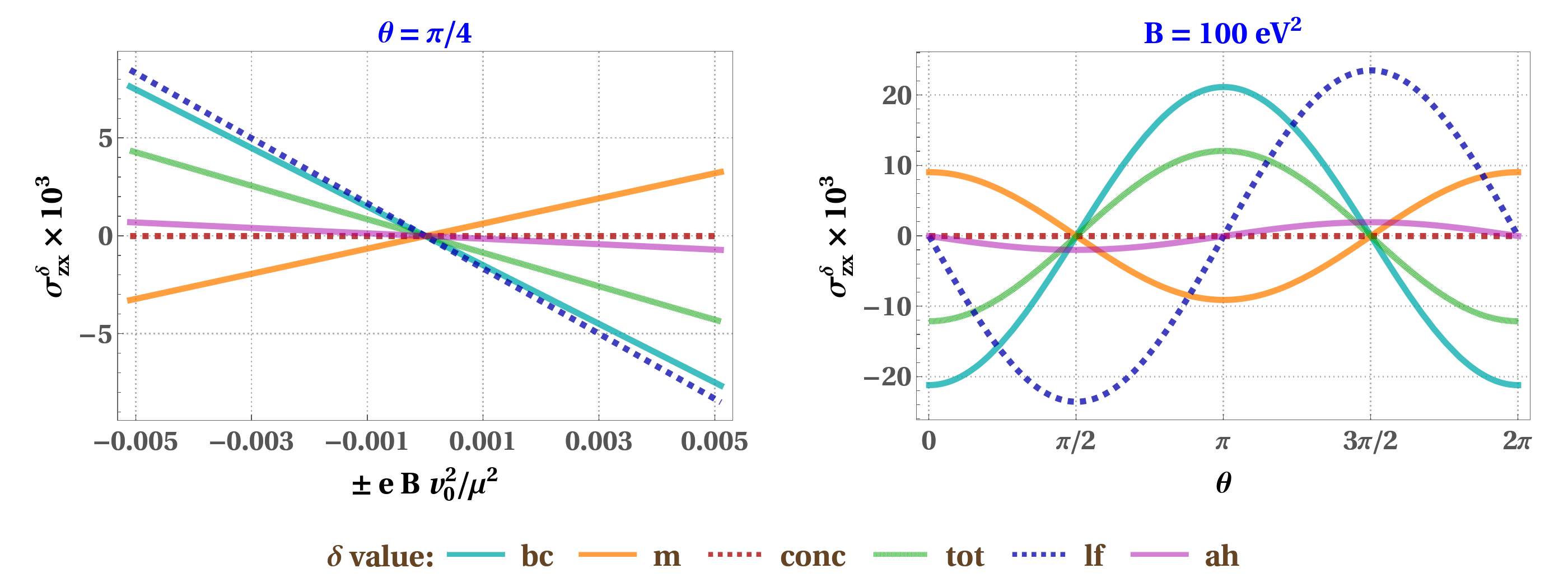}}
\caption{Set-up I for VNR: Behaviour of conductivity curves (in eV) as functions of ${e\, B \,v_0^2} / {\mu^2}$ and $\theta$. Here we have omitted showing the $\sigma_{yx}^{\rm ah}$ curve as its order of magnitude is highly incompatible with the remaining curves.
\label{figset1vnr}}
\end{figure}
%%%%%%%%%%%%%%%%

The AH part comprises
\begin{align}
\label{eqahvnr1}
\sigma^{\rm{ah}}_{zx} &= \frac{-\,e^3\, k_0 \,v_0^2 \,B_y}
{1024\, k_\mu^5 \,\pi\,\mu^{10}}\,
\Big[32 \,k_\mu^5 \,\mu^6( \mu^2-2 \,k_0 \,k_\mu\, v_0^2 + 2\, k_\mu^2\, v_0^2)
+9 \,{\boldsymbol B}^2 \, e^2\{-k_\mu^5 \,v_0^4\,
(336\, k_\mu^4\, v_0^4+532\, k_\mu^2 \, v_0^2 \,\mu^2+201 \, \mu^4)
\nn &\hspace{2 cm}
+ k_0 \,(336\, k_\mu^8\, v_0^8 + 364 \, k_\mu^6 \, v_0^6 \,\mu^2
+61\, k_\mu^4 \,v_0^4\,\mu^4-6\, k_\mu^2\, v_0^2\,\mu^6+\mu^8)\}\Big],\nn
\sigma^{\rm{ah}}_{yx} &= \frac{e^2}{256\,\pi\,v_0}
\, \Big[-32\,\mu
+\frac{3 \,{\boldsymbol B}^2 \, e^2 \,v_0^3} {k_\mu\,\mu^7}
\{-40 \,k_0^5 \,v_0^5 + 44 \, k_0^3 \,v_0^3 \,\mu^2-8 \,k_0\, v_0\,\mu^4
+ k_\mu \,(40\, k_0^4 \,v_0^5-24\, k_0^2\, v_0^3\,\mu^2+v_0 \,\mu^4)\}
\Big].
\end{align}
We note the appearance of an in-plane conductivity arising from the AH part.

The total response from the LFO is
\begin{align}
\sigma^{\rm{lf}}_{zx} =
\sigma^{\rm{lf}, 2}_{zx} + \sigma^{\rm{lf}, 3}_{zx} + \sigma^{\rm{lf}, 4}_{zx}\,,
\end{align}
%%%%%%%%%%%%%%%%%%%%
\begin{align}
  \sigma^{\rm{lf}, 2}_{zx}&= \frac{e^3\,B_y }{512 \left(k_0-k_\mu \right)^9 
   k_\mu^3 \, \pi \, v_0^6 \, \mu^8}
\big[-64 \, k_0  \left( k_0-k_{\mu } \right)^9  k_\mu^3 \,
v_0^8 \, \mu^8  \,{\bs B}^2\,  e^2\big\{
36864 \, k_0^{13}  \left(k_0-k_\mu \right)
 \left(9 \, k_0^3-2 \, k_0 \, k_\mu^2+3 \, k_\mu^3 \right)
\, v_0^{16}\nn
& \hspace{2.5 cm} \qquad \qquad\quad
-1024 \, k_0^{11} \left(1531 \, k_0^4-1369 \, k_0^3 \, k_\mu
-242 \, k_0^2 \, k_\mu^2+563 \, k_0 \, k_\mu^3-303 \, k_\mu^4\right) v_0^{14} \, \mu^2\nn
& \hspace{2.5 cm}\qquad \qquad\quad
+256 \, k_0^9
\left(
11853 \, k_0^4
-9277 \, k_0^3 \, k_\mu
-1286 \, k_0^2 \, k_\mu^2
+2731 \, k_0 \, k_\mu^3
-1269 \, k_\mu^4\right) v_0^{12} \, \mu^4 \nn
& \hspace{2.5 cm}\qquad \qquad\quad
-256 \, k_0^7\left(12023 \, k_0^4
-7988 \, k_0^3 \, k_\mu-868 \, k_0^2 \, k_\mu^2+1617 \, k_0 \, k_\mu^3
-609 \, k_\mu^4\right) v_0^{10} \, \mu^6
\nn & \hspace{2.5 cm}\qquad \qquad\quad
+96 \, k_0^5
\left(18101 \, k_0^4-9765 \, k_0^3 \, k_\mu
-860 \, k_0^2 \, k_\mu^2+1288 \, k_0 \, k_\mu^3-350 \, k_\mu^4\right) v_0^8 \, \mu^8\nn
& \hspace{2.5 cm}\qquad \qquad\quad
+4 \, k_0^3\left(-133773 \, k_0^4+54811 \, k_0^3 \, k_\mu+4342 \, k_0^2 \, k_\mu^2-4325 \, k_0 \, k_\mu^3+657 \, k_\mu^4
\right) v_0^6 \, \mu^{10}
\nn & \hspace{2.5 cm}\qquad \qquad\quad
+k_0\left(82017 \, k_0^4-23009 \, k_0^3 \, k_\mu
-1982 \, k_0^2 \, k_\mu^2+903 \, k_0 \, k_\mu^3-33 \, k_\mu^4\right)v_0^4 \, \mu^{12}
\nn & \hspace{2.5 cm}\qquad \qquad\quad
+ \left(-5210 \, k_0^3
+865 \, k_0^2 \, k_\mu
+86 \, k_0 \, k_\mu^2-8 \, k_\mu^3\right)
 v_0^2 \, \mu^{14}+ \left(105 \, k_0-8 \, k_\mu \right)\mu^{16}\big\}\big] \, , \nn
%%%%%%%%%%%%%%%%%%
%%%%%%%%%%%%%%%%%%%%%%%%%%%%%%%%%%
 \sigma^{\rm{lf}, 3}_{zx}&= \frac{e^5 \, v_0^3 \, B_x \,{\bs B}^2}
 {64 \, k_\mu^2 \, \pi \, \mu^6}
\big[2 \, k_0^3 \left(k_0 - k_\mu \right)
\left(k_0^2 - 7 \, k_\mu^2 \right)v_0^6
+k_0\left(-7 \, k_0^3+6 \, k_0^2 \, k_\mu+3 \, k_0 \, k_\mu^2+4 \, k_\mu^3
\right) v_0^4 \, \mu^2
% \nn & \hspace{2.5 cm} \qquad 
+ \left(2 \, k_0^2+7 \, k_\mu^2\right)v_0^2 \, \mu^4+3 \, \mu^6 \big] \, , \nn
%%%%%%%%%%%%%%%%%%%%
 \sigma^{\rm{lf}, 4}_{zx}&=
 \frac{3 \, e^5 \, k_0^2\,  v_0^8  \left(k_0-k_\mu \right) B_y \,{\bs B}^2}
 {16 \, \pi \, \mu^4}\,,
\end{align}
%%%%%%%%%%%%%%%%%%%%ss
\begin{align}
\sigma^{\text{lf}}_{xx}&=\frac{
\tau^3 \, e^4 \,v_0^6 \,  k_0^2 \left(-k_0+k_\mu \right)
\left(B_x^2+3 \, B_y^2 \right)
}{16 \, \pi \, \mu^3}, \, \nn
%%%%
\sigma^{\text{lf}}_{yx}&=   \frac{\tau^2 \, e^4 \, v_0 }{64 \, \pi \, \mu^5}
\Big[ {\bs B}^2 \Big\{
\left(k_0-k_\mu \right)^2\left(3 \, k_0^2
+6 \, k_0 \, k_\mu+7 \, k_\mu^2
\right) v_0^6
+ \frac{ 8 \, k_0^5 \left(-k_0 + k_\mu \right) v_0^6
+4 \, k_0^4 \, v_0^4 \, \mu^2+k_0^2 \, v_0^2 \, \mu^4+3 \, \mu^6}
{k_\mu^2} \Big\}\nn& \hspace{ 2 cm}
+8 \,\tau \, k_0^2\,v_0^5  \,\mu^2  \left(k_0-k_\mu \right)  B_x \, B_y  \Big].
 \end{align}
See also Appendix~\ref{appset1vnr} for the intermediate steps.

%%%%%%%%%%%%%%%%%%%%%%%%%%%%%
\begin{figure}[t!]
\centering
\subfigure{\includegraphics[width=0.75 \textwidth]{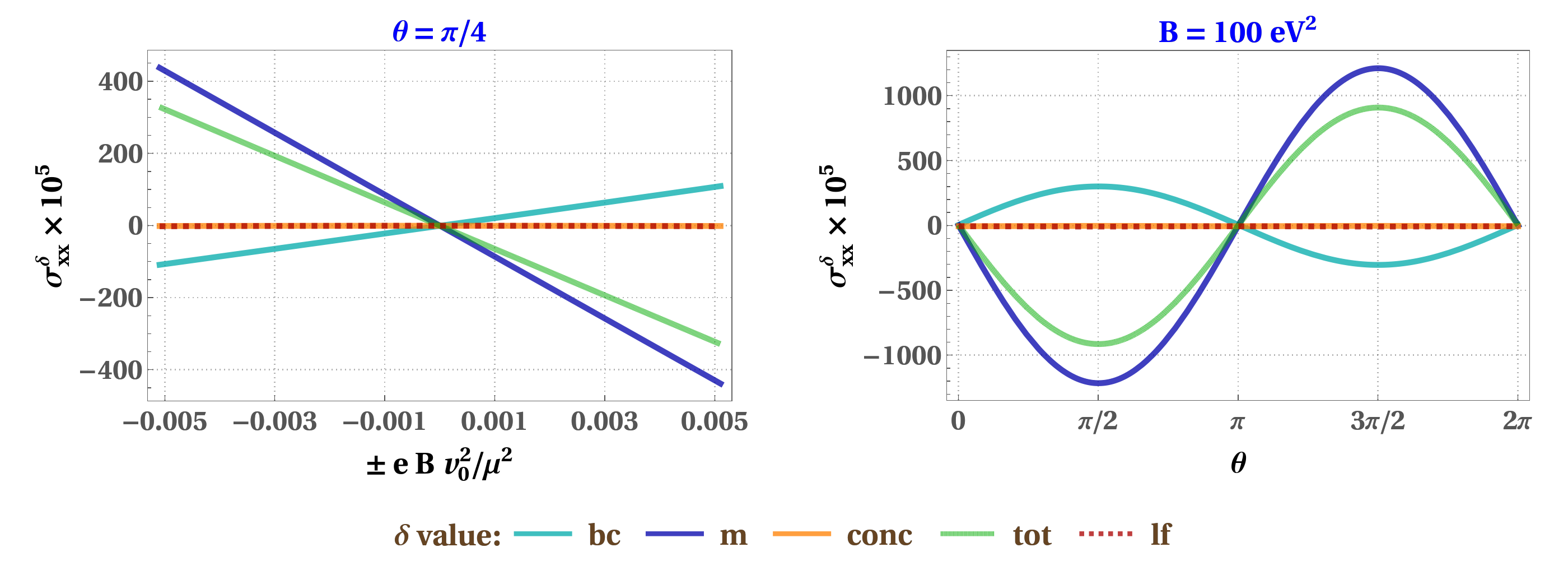}}
\subfigure{\includegraphics[width=0.75 \textwidth]{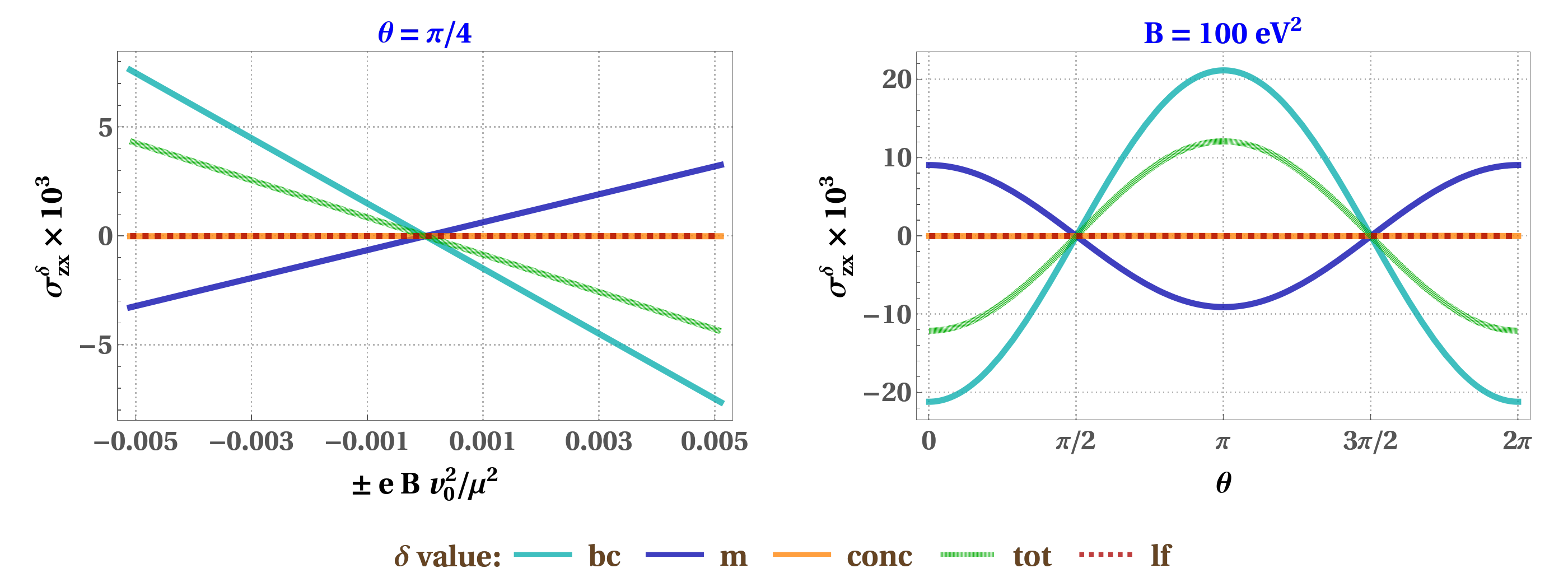}}
\subfigure{\includegraphics[width=0.75 \textwidth]{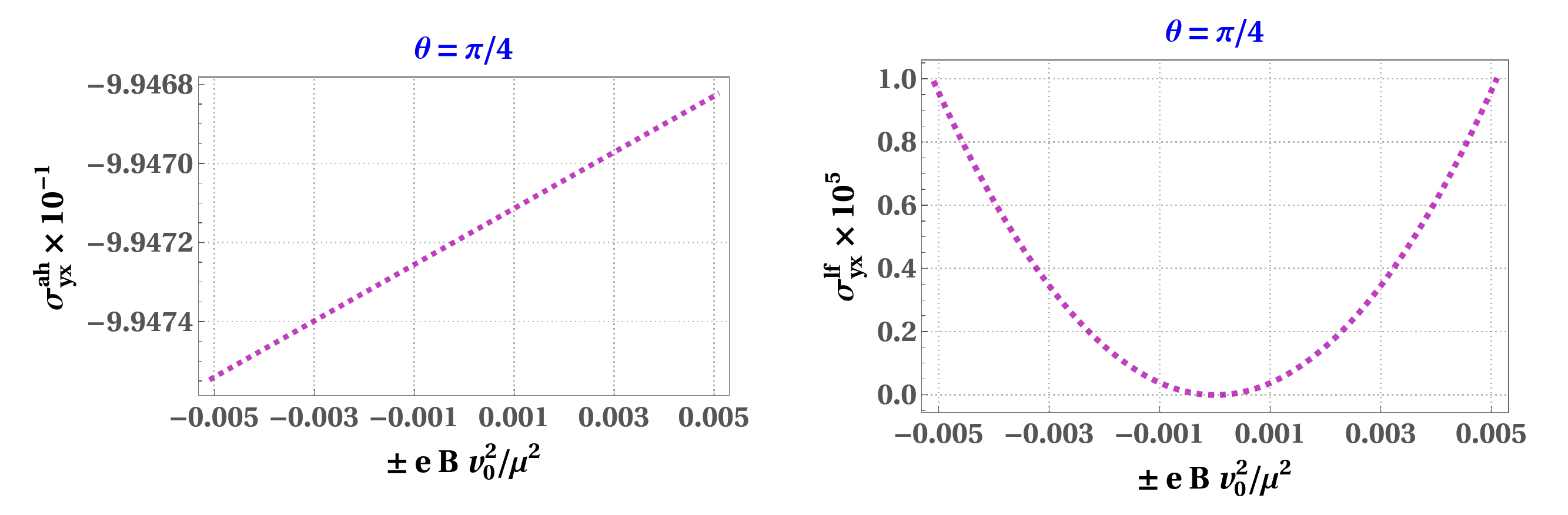}}
\caption{Set-up II for VNR: Behaviour of the conductivity curves (in eV) as functions of ${e\, B \,v_0^2} / {\mu^2}$ and $\theta$. For $\sigma_{yx}$, we have omitted showing the $\theta$-dependence, and both panels show the field-strength dependence of the AH and LFO parts instead.\label{figset2vnr}}
\end{figure}
%%%%%%%%%%%%%%%%

The parity and angular structure of the response can be read directly off the analytic
expressions. The longitudinal component $\bar\sigma_{xx}$ is even under
$\bs B\to-\bs B$: every sub-part is quadratic in the field, with independent
coefficients for $B_x^2$ and $B_y^2$. The planar-Hall $\bar\sigma_{yx}$ is
proportional to $B_x\,B_y$, vanishing when $\bs B$ is aligned with either in-plane
axis and exhibiting the characteristic $\sin(2\theta)$ profile. The out-of-plane
$\bar\sigma_{zx}$, in contrast, is odd in $\bs B$, comprising a linear term in
$B_x$ and a cubic term $\propto B_x\bs B^2$. Its BC-only and OMM-only linear
coefficients, $-7\tau\, e^3\,v_0/(64\,\pi)$ and $3\tau \,e^3\,v_0/(64\,\pi)$, are independent
of $k_0$ and $\mu$ and partially cancel to leave a total
$-\tau\, e^3\, v_0\,\,B_x/(16\,\pi)$. The concurrent contribution begins only at cubic
order, and all cubic terms are controlled by the weak-field parameter
$e\,B\,v_0^2/\mu^2$, so the out-of-plane response remains nearly linear in the field.

The AH and LFO contributions exhibit a similar structure. The in-plane
anomalous-Hall conductivity
$\sigma^{\rm{ah}}_{yx}=-e^2\,\mu/(8\,\pi\, v_0)+\mathcal{O}(\bs B^2)$ forms a plateau
that depends on the field only through $\bs B^2$ and carries no $\theta$-dependence
at this order. The out-of-plane $\sigma^{\rm{ah}}_{zx}$ is linear in $B_y$ at
leading order and therefore odd in $B_y$, vanishing at $\theta=0$ and $\pi$. For
the LFO part, $\sigma^{\rm{lf}}_{xx}\propto B_x^2+3\,B_y^2$ has a coefficient that
is negative for all parameters, reflecting $k_\mu<k_0$. The planar-Hall
$\sigma^{\rm{lf}}_{yx}$ contains an isotropic $\bs B^2$ term at order $\tau^2$ and
a $B_xB_y$ term at order $\tau^3$, the latter with a positive coefficient. At low
fields the out-of-plane $\sigma^{\rm{lf}}_{zx}$ is dominated by
$-\tau^2\,e^3\,k_0\,v_0^2\,B_y/(8\,pi)$, which is odd in $B_y$. Since the LFO terms carry
higher powers of $\tau$ than the BC-only and OMM-only contributions, their
relative weight grows with increasing $\tau$.

The numerical results for some representative parameters are
displayed in Fig.~\ref{figset1vnr}, with the field-strength dependence at
$\theta=\pi/4$ in the left panels and the angular dependence at fixed $|\bs B|$
in the right panels. The $B_x^2$ coefficients of $\bar\sigma_{xx}$ are positive
for the BC-only and OMM-only parts and negative for the concurrent part, while
the corresponding $B_y^2$ coefficients are considerably smaller, with the
OMM-only and concurrent contributions changing sign relative to their $B_x^2$
partners. This sign reversal produces the small dip of the OMM-only curve and the
small rise of the concurrent curve near $\theta=\pi/2$ --- an asymmetry between
the two in-plane field orientations that has no analogue in an isotropic
monopolar semimetal. The LFO curve is negative, roughly three times larger in
magnitude near $\theta=\pi/2$ than near $\theta=0$, and exceeds the total non-AH
curve by a factor of about three. In $\sigma_{yx}$, the non-AH curves follow the
$\sin(2\theta)$ profile, whereas the LFO curve remains positive and oscillates
about the offset set by its $\bs B^2$ term, peaking near $\theta=\pi/4$ and
reaching a minimum near $3\pi/4$. In $\sigma_{zx}$, the non-AH curves follow
$-\cos\theta$ with the BC-only and OMM-only contributions in antiphase, while the
LFO and AH curves follow $-\sin\theta$, vanishing at $\theta=0$ and $\pi$.

%%%%%%%%%%%%%%%%%%%%%%%%%%%%%%%%%%%%%%%%%%%%%%%%%%%%%%%%%%%%%%%
\subsection{Set-up II: \texorpdfstring{$\bs{E}=E_x\,  {\bs{\hat x}}$}{E-field},
\texorpdfstring{$ \bs{B}= B_x \, {\bs{\hat x}} + B_z\, \bs{\hat z} $}{B-field}, and
\texorpdfstring{$ \bs{B}^2 = B_x^2 + B_z^2 $}{BB}}
\label{secset2}

For the non-AH part, we have
\begin{align}
\sigma^{\rm d}_{xx} & = \frac{\tau \, e^2 \, k_0 \,  \mu }{8 \, \pi} \,,
\end{align}
%%%
\begin{align}
\label{eqsigxxset2}
\sigma^{\rm bc}_{xx} & =
\frac{\tau \, e^3 \,v_0 }{2048 \,\pi\, \mu^{10}}\big[ 32 \,B_z \,\mu^{10}
- 4 \,e \,v_0^3 \,\mu^3\, \big\{B_x^2 \left(
k_0 \left(-24 \,k_0^4\, v_0^4 + 76 \,k_0^2 \,v_0^2 \,\mu^2 - 93\, \mu^4 \right)
+ 8 \,k_\mu \left(3 \,k_0^4\, v_0^4 - 8 \,k_0^2\, v_0^2\, \mu^2 + 8 \,\mu^4 \right)
\right)\nn & \hspace{ 2.5 cm}
%%%%%%%%%%%%%%%%
+ 4 \,B_z^2 \,k_0 \left(
8 \,k_0^4 \,v_0^4 - 12\, k_0^2 \,v_0^2 \,\mu^2 + 3 \,\mu^4
+ k_\mu\left(-8 \,k_0^3 \,v_0^4 + 8\, k_0\, v_0^2 \,\mu^2\right)
\right) \big\}
%%%%%%%%%%%%%%%%%%
\nn & \hspace{ 2.5 cm}
+\, B_z\, e^2 \,v_0^4 \big\{
4 \,B_z^2 \Big(
-112 \,k_0^6 \,v_0^6 + 200\, k_0^4 \,v_0^4\, \mu^2 - 90 \,k_0^2 \,v_0^2 \,\mu^4 + 5 \,\mu^6
\nn & \hspace{2.5 cm}
%%%%%%%%%%%%%%%%%%%%%%%%%%%%%%%%%%%
+ 16 \,k_0 \,k_\mu \,v_0^2 \left(7 \,k_0^4 \,v_0^4 - 9 \,k_0^2 \,v_0^2 \,\mu^2 + 2 \,\mu^4 \right)
\Big) + B_x^2 \Big(
1008 \,k_0^6 \,v_0^6 - 2360 \,k_0^4\, v_0^4 \,\mu^2 + \nn & \hspace{ 2.5 cm}
%%%%%%%%%%%%%%%%%%%%%%%%%%%%%%%%%%
+ 1698\, k_0^2\, v_0^2 \,\mu^4 - 279 \,\mu^6- 16 \,k_0\, k_\mu\, v_0^2 
\left(63 \,k_0^4\, v_0^4 - 116 \,k_0^2 \,v_0^2 \,\mu^2 + 56 \,\mu^4 \right)
\Big)\big\}\big]\, , 
\end{align}
%%%%
\begin{align}
\sigma^{\rm m}_{xx} & =\frac{ -\,\tau \, e^3 \,\,\mu^{-10}}
{2048\,\pi\,v_0^5\,k_\mu^6}\,\Big[
4\,B_z\,k_\mu^2\,v_0^2\,\Big\{
8\,k_\mu^4\,v_0^4\,\mu^8 
\left( 16\,k_0\,(k_\mu-k_0)\,v_0^2 + 13\,\mu^2 \right)
-4\,e\,B_z\,k_\mu^2\,v_0^5\,\mu^3 \, \big\{ 40\,k_0^7\,v_0^6
+16\,k_0^5\,v_0^4\,\mu^2 \nn & \hspace{2.5 cm}
%%%%%%%%%%%%%%%%%%%%%%%%
-71\,k_0^3\,v_0^2\,\mu^4+15\,k_0\,\mu^6 +\,k_\mu \left(
-40\,k_0^6\,v_0^6-36\,k_0^4\,v_0^4\,\mu^2
+48\,k_0^2\,v_0^2\,\mu^4+4\,\mu^6\right)\big\}
\nn  &\hspace{ 2.5 cm}
%%%%%%%%%%%%%%%%%%%%%%%%%%%%%%
+\,B_z^2\,e^2\,v_0^4 \big\{-2688\,k_0^{10}\,v_0^{10}
+6216\,k_0^8\,v_0^8\,\mu^2-4468\,k_0^6\,v_0^6\,\mu^4
+1095\,k_0^4\,v_0^4\,\mu^6
-210\,k_0^2\,v_0^2\,\mu^8
%%%%%%%%%%%%%%%%
\nn & \hspace{2.5 cm}
+55\,\mu^{10} +\,k_\mu\left(2688\,k_0^9\,v_0^{10}-4872\,k_0^7\,v_0^8\,\mu^2+2368\,k_0^5\,v_0^6\,\mu^4
-352\,k_0^3\,v_0^4\,\mu^6+128\,k_0\,v_0^2\,\mu^8\right)\big\}\Big\}
\nn &\hspace{ 2.5 cm}
+\,B_x^2\,e\,v_0^3\,\Big\{-4\,k_\mu^2\,v_0^2\,\mu^3\big\{-120\,k_0^9\,v_0^8
+152\,k_0^7\,v_0^6\,\mu^2+121\,k_0^5\,v_0^4\,\mu^4-218\,k_0^3\,v_0^2\,\mu^6+65\,k_0\,\mu^8
\nn& \hspace{2.5 cm}
%%%%%%%%%%%%%%%%%%%%%
+\,k_\mu\left(120\,k_0^8\,v_0^8-92\,k_0^6\,v_0^6\,\mu^2-152\,k_0^4\,v_0^4\,\mu^4
+144\,k_0^2\,v_0^2\,\mu^6-24\,\mu^8\right)\big\}
\nn & \hspace{ 2.5 cm}
+\,B_z\,e\,v_0 \big\{k_\mu^6\,v_0^6
\left(24192\,k_0^6\,v_0^6
-5880\,k_0^4\,v_0^4\,\mu^2
-700\,k_0^2\,v_0^2\,\mu^4
+27\,\mu^6 \right)+\,k_\mu\,v_0^2
\big(-24192\,k_0^{11}\,v_0^{10}\nn
&\hspace{ 2.5 cm}
+66360\,k_0^9\,v_0^8\,\mu^2
-59360\,k_0^7\,v_0^6\,\mu^4
+16808\,k_0^5\,v_0^4\,\mu^6
+480\,k_0^3\,v_0^2\,\mu^8
-192\,k_0\,\mu^{10} \big)\big\}\Big\}\Big],
\end{align}
%%%%%------------------------------
%%%%%%%%%%%%%%%%%%%%%%%%%%%%%%%%%%%%%%%%%%%
\begin{align}
\sigma^{\rm conc}_{xx} & =
\frac{\tau \, e^4}{1024\,k_\mu^4\,\pi\,\mu^{10}}
\Big[ 4\,B_z^2\,k_\mu^2\,v_0^2 \Big\{
-4\,k_\mu^2\,v_0^2\,\mu^3
\left\{ 8\,k_0^5\,v_0^4
- 32\,k_0^3\,v_0^2\,\mu^2
+ 21\,k_0\,\mu^4 +\,k_\mu\left(-8\,k_0^4\,v_0^4
+ 28\,k_0^2\,v_0^2\,\mu^2
- 8\,\mu^4\right)\right\}
\nn &\hspace{2.5 cm}
+\,B_z\,e\,v_0\big\{
-504\,k_0^8\,v_0^8+ 1944\,k_0^6\,v_0^6\,\mu^2
- 2255\,k_0^4\,v_0^4\,\mu^4+ 880\,k_0^2\,v_0^2\,\mu^6
- 65\,\mu^8 \nn &\hspace{2.5 cm}
%%%%%%%%%%%%%%%%%%%%
+\,k_\mu \left(
504\,k_0^7\,v_0^8- 1692\,k_0^5\,v_0^6\,\mu^2
+ 1472\,k_0^3\,v_0^4\,\mu^4- 324\,k_0\,v_0^2\,\mu^6 \right)\big\}\Big\}
\nn& \hspace{2.5cm}
+\,B_x^2 \, \Big\{4\,k_\mu^2\,v_0^2\,\mu^3
\big\{
24\,k_0^7\,v_0^6
- 81\,k_0^3\,v_0^2\,\mu^4
+ 57\,k_0\,\mu^6 \nn &\hspace{2.5cm}
+\,k_\mu \left(
-24\,k_0^6\,v_0^6- 12\,k_0^4\,v_0^4\,\mu^2
+ 72\,k_0^2\,v_0^2\,\mu^4- 32\,\mu^6\right) \big\} \nn & \hspace{2.5cm}
+3\,B_z\,e\,v_0\big\{k_\mu^4\,v_0^4
\left(1512\,k_0^6\,v_0^6
- 1120\,k_0^4\,v_0^4\,\mu^2
+ 113\,k_0^2\,v_0^2\,\mu^4
+ 15\,\mu^6 \right) \nn &\hspace{ 2.5 cm}
%%%%%%%%%
+ k_\mu \left(-1512\,k_0^9\,v_0^{10}+ 3388\,k_0^7\,v_0^8\,\mu^2
- 2360\,k_0^5\,v_0^6\,\mu^4
+ 480\,k_0^3\,v_0^4\,\mu^6
+ 16\,k_0\,v_0^2\,\mu^8 \right) \big\}\Big\} \Big],
\end{align}
%%%%%%%%%%%%%%%%%%%%%%%%%
\begin{align}
\label{eqsigzxset2}
\sigma^{\rm bc}_{zx} & =
\frac{\tau\, e^3\, v_0\, B_x}{2048\,\pi\,\mu^{10}}
\big[B_x^2\, e^2\, v_0^4
\big\{336\,k_0^6\, v_0^6- 680\,k_0^4\, v_0^4\, \mu^2+ 342\,k_0^2\, v_0^2\, \mu^4
- 21\,\mu^6  + k_\mu \left(-336\,k_0^5\, v_0^6+ 512\,k_0^3\, v_0^4\, \mu^2
- 128\,k_0\, v_0^2\, \mu^4\right)\big\}
\nn& \hspace{2.5cm}- 4 \big\{
%%%%
56\,\mu^{10}
+ 8\,B_z\, e\, k_0\, v_0^3\, \mu^3
\left(
8\,k_0^4\, v_0^4- 16\,k_0^2\, v_0^2\, \mu^2
+ 5\,\mu^4+ k_\mu \left(
-8\,k_0^3\, v_0^4+ 12\,k_0\, v_0^2\, \mu^2
\right) \right)
\nn& \hspace{2.5cm}
+ 3\,B_z^2\, e^2\, v_0^4\big(112\,k_0^6\, v_0^6- 200\,k_0^4\, v_0^4\, \mu^2
+ 90\,k_0^2\, v_0^2\, \mu^4
- 5\,\mu^6
\nn & \hspace{2.5cm}
+ k_\mu \left(-112\,k_0^5\, v_0^6+ 144\,k_0^3\, v_0^4\, \mu^2
- 32\,k_0\, v_0^2\, \mu^4\right)\big)\big\}\big], 
\end{align}
%%%%%
\begin{align}
\sigma^{\rm m}_{zx} & = \frac{\tau \, e^3 \, B_x}{2048 \, k_\mu^5 \, (k_0 + k_\mu)^2 \, \pi \, v_0^5 \, \mu^6}\big[
 4 \, k_\mu^2 \, v_0^2\big\{
24 \, k_\mu^3  \left(k_0 + k_\mu \right)^2 \, v_0^4 \, \mu^6
- 8 \, B_z \, e \, k_\mu^2 \, v_0^3 \, \mu^3
\,\big(10 \, k_0^4 \, v_0^4
- 6 \, k_0^2 \, v_0^2 \, \mu^2+ 4 \, \mu^4 
\nn& \hspace{2.5cm}
+ k_\mu\,   k_0 \, v_0^2\left(-10 \, k_0^2 \, v_0^2+  \mu^2
\right) \big)
+ B_z^2 \, e^2 \, v_0^4\big(-2016 \, k_0^7 \, v_0^6
+ 3766 \, k_0^5 \, v_0^4 \, \mu^2- 2060 \, k_0^3 \, v_0^2 \, \mu^4
\nn& \hspace{2.5cm}
+ 358 \, k_0 \, \mu^6+ k_\mu \left( 2016 \, k_0^6 \, v_0^6
- 2758 \, k_0^4 \, v_0^4 \, \mu^2
+ 931 \, k_0^2 \, v_0^2 \, \mu^4- 93 \, \mu^6\right)\big)\big\}\nn& \hspace{2.5cm}
+ 3 \, B_x^2 \, e^2 \, v_0^4 \,\big\{672 \, k_0^9 \, v_0^8
- 1694 \, k_0^7 \, v_0^6 \, \mu^2
+ 1354 \, k_0^5 \, v_0^4 \, \mu^4
- 330 \, k_0^3 \, v_0^2 \, \mu^6+ 6 \, k_0 \, \mu^8
\nn& \hspace{2.5cm}
+ k_\mu \left(
-672 \, k_0^8 \, v_0^8
+ 1358 \, k_0^6 \, v_0^6 \, \mu^2
- 761 \, k_0^4 \, v_0^4 \, \mu^4
+ 88 \, k_0^2 \, v_0^2 \, \mu^6
+ 3 \, \mu^8\right)\big\} \big]\, , 
\end{align}
%%%%%%%%%%%%%%%%%%%%%%
\begin{align}
\sigma^{\rm conc}_{zx} & =\frac{- \, \tau \, e^4 \, B_x}
{1024 \, k_\mu^3  \left(k_0 + k_\mu \right)^2  \pi \, v_0^2 \, \mu^6}
\Big[ 4 \, B_z
\big\{
8 \, \left(k_0 - k_\mu \right) \, k_\mu^3 
\left(3 \, k_0^2 + 3 \, k_0 \, k_\mu + 2 \, k_\mu^2 \right)
\, v_0^6 \, \mu^3  + B_z \, e \, v_0^3\big(- 378 \, k_0^7 \, v_0^6+ 978 \, k_0^5 \, v_0^4 \, \mu^2
\nn & \hspace{2.5cm}
- 782 \, k_0^3 \, v_0^2 \, \mu^4
+ 174 \, k_0 \, \mu^6+ k_\mu \left(378 \, k_0^6 \, v_0^6- 789 \, k_0^4\,  v_0^4 \, \mu^2
+ 446\,  k_0^2 \, v_0^2 \, \mu^4
- 51 \, \mu^6 \right)\big)
\big\}\nn
& \hspace{2.5cm}
+ B_x^2 \, e \, v_0^3
\big\{
378 \, k_0^7 \, v_0^6
- 498 \, k_0^5 \, v_0^4 \, \mu^2
+ 58 \, k_0^3 \, v_0^2 \, \mu^4
+ 50 \, k_0 \, \mu^6
+ k_\mu \big(
- 378 \, k_0^6 \, v_0^6
\nn& \hspace{2.5cm}
+ 309 \, k_0^4 \, v_0^4 \, \mu^2 + 52 \, k_0^2 \, v_0^2 \, \mu^4
- 7 \, \mu^6\big)\big\}\Big]\,.
\end{align}
%%%

The AH part reads
\begin{align}
 \sigma^{\rm ah}_{yx} &= \frac{ - \, \mu^{-10}}  {256 \, \pi \, v_0  \, k_\mu^3}
\Big[2 \, e^2 \, k_\mu^2
\Big\{16 \, k_\mu \, \mu^{11}
+ 8 \, B_z \, e \, k_\mu \, v_0^3 \, \mu^6
\left(2 \, k_0^3 \, v_0^2- 2 \, k_0^2 \, k_\mu \, v_0^2- 3 \, k_0 \, \mu^2+ 2 \, k_\mu \, \mu^2
\right)\nn & \hspace{ 2.5 cm}
+ 15 \, B_z^2 \, e^2 \, k_\mu \, v_0^4 \, \mu^3\left(
8 \, k_0^4 \, v_0^4
- 8 \, k_0^3 \, k_\mu \, v_0^4
- 8 \, k_0^2 \, v_0^2 \, \mu^2
+ 4 \, k_0 \, k_\mu \, v_0^2 \, \mu^2
+ \mu^4\right)
%%%
\nn & \hspace{ 2.5 cm}
- 7 \, B_z^3 \, e^3 \, v_0^5\big(
144 \, k_0^6 \, v_0^6
- 144 \, k_0^5 \, k_\mu \, v_0^6- 212 \, k_0^4 \, v_0^4 \, \mu^2
+ 140 \, k_0^3 \, k_\mu \, v_0^4 \, \mu^2
+ 77 \, k_0^2 \, v_0^2 \, \mu^4
\nn & \hspace{ 2.5 cm}
- 25 \, k_0 \, k_\mu \, v_0^2 \, \mu^4- 4 \, \mu^6\big)\Big\}
+ 3 \, B_x^2 \, e^4 \, v_0^3\Big\{k_\mu^2 \, v_0 \, \mu^3\big(
40 \, k_0^5 \, v_0^4- 40 \, k_0^4 \, k_\mu \, v_0^4- 44 \, k_0^3 \, v_0^2 \, \mu^2
%%%%%%%%%%%%%
\nn & \hspace{ 2.5 cm}
+ 24 \, k_0^2 \, k_\mu \, v_0^2 \, \mu^2
+ 8 \, k_0 \, \mu^4- k_\mu \, \mu^4\big)
+ B_z \, e \big(
1008 \, k_0^8 \, v_0^8- 1008 \, k_0^7 \, k_\mu \, v_0^8- 2212 \, k_0^6 \, v_0^6 \, \mu^2
\nn & \hspace{ 2.5 cm}
+ 1708 \, k_0^5 \, k_\mu \, v_0^6 \, \mu^2
+ 1503 \, k_0^4 \, v_0^4 \, \mu^4
- 775 \, k_0^3 \, k_\mu \, v_0^4 \, \mu^4
- 312 \, k_0^2 \, v_0^2 \, \mu^6
+ 75 \, k_0 \, k_\mu \, v_0^2 \, \mu^6
% \nn & \hspace{ 2.5 cm} 
+ 8 \, \mu^8\big)\Big\}\Big].
\end{align}

The LFO-induced contributions have been derived term-by-term in Appendix~\ref{appset2vnr}. Let us define
\begin{align}
\sigma^{\rm{lf}}_{yx}=\tau^2 \, \sigma^{\rm{lf},2}_{yx}+\tau^4\, \sigma^{\rm{lf},4}_{yx}\,,
\end{align}
where the superscripts $2$ and $4$ represent the coefficients of $\tau^2$ and $\tau^4$ accompanying the expressions, respectively.
The combined explicit forms are provided below:
%%%%%%%%%%%%%%%%%%%%%%%%%
\begin{align}
\sigma^{\rm{lf}, 2}_{yx} & =\frac{e^4 \, v_0}{128 \, k_\mu^4 \, \pi \, \mu^8}\Big[
B_x^2 \,\Big\{
2 \, k_\mu^2 \, \mu^3
\left\{
-8 \, k_0 \, \left( k_0 - k_\mu \right)^2
\left( k_0 + k_\mu \right)
\left( k_0^2 + k_\mu^2 \right) v_0^6
+ 4 \left( k_0^4 - k_\mu^4 \right) v_0^4 \, \mu^2
+ \left( k_0^2 + 3 \, k_\mu^2 \right) v_0^2 \, \mu^4
+ 3 \, \mu^6\right\}\nn & \hspace{2.5 cm}
+ B_z \, e \, v_0 \big\{
- 72 \, k_0^6 \left( k_0 - k_\mu \right)
\left( 7 \, k_0^2 - 3 \, k_\mu^2 \right) v_0^8+ 4 \, k_0^4\left( 248 \, k_0^3- 185 \, k_0^2 \, k_\mu
- 120 \, k_0 \, k_\mu^2+ 93 \, k_\mu^3 \right) v_0^6 \, \mu^2
\nn & \hspace{2.5 cm}
+ k_0^2 \left( -439 \, k_0^3+ 132 \, k_0^2 \, k_\mu+ 343 \, k_0 \, k_\mu^2
- 184 \, k_\mu^3 \right) v_0^4 \, \mu^4
\nn & \hspace{2.5 cm}
+ \left( -82 \, k_0^3
+ 88 \, k_0^2 \, k_\mu
- 79 \, k_0 \, k_\mu^2
+ 24 \, k_\mu^3 \right) v_0^2 \, \mu^6
+ 33 \, k_0 \, \mu^8\big\}\Big\}\nn & \hspace{2.5 cm}
+ 2 \, B_z^2 \Big\{2 \, k_\mu^2 \, \mu^3
\big\{ 8 \, k_0^3 \left( k_0 - k_\mu \right)^2
\left( k_0 + k_\mu \right) v_0^6+ 4 \, k_0^2
\left( 5 \, k_0^2 - 6 \, k_0 \, k_\mu + k_\mu^2 \right) v_0^4 \, \mu^2
\nn & \hspace{2.5 cm}
%%%%%%%%%%%%%%%%%%%%%%%%%%%%%%%
+ \left( -37 \, k_0^2
+ 24 \, k_0 \, k_\mu
+ k_\mu^2 \right) v_0^2 \, \mu^4
+ 9 \, \mu^6
\big\}+ 3 \, B_z \, e \, v_0
\big\{
8 \, k_0^3 \left( k_0 - k_\mu \right)
\left( 9 \, k_0^5 - 2 \, k_0^3 \, k_\mu^2 + 3 \, k_\mu^5 \right) v_0^8
\nn & \hspace{2.5 cm}
- 4 \, k_0^2 \left( 42 \, k_0^5- 33 \, k_0^4 \, k_\mu
- 20 \, k_0^3 \, k_\mu^2
+ 18 \, k_0^2 \, k_\mu^3
+ 3 \, k_\mu^5 \right) v_0^6 \, \mu^2 \nn & \hspace{2.5 cm}
+ k_0
\left( 125 \, k_0^4
- 68 \, k_0^3 \, k_\mu
- 94 \, k_0^2 \, k_\mu^2
+ 60 \, k_0 \, k_\mu^3
- 3 \, k_\mu^4 \right) v_0^4 \, \mu^4
\nn & \hspace{2.5 cm}
- 2 \left( k_0 - k_\mu \right)
\left( 17 \, k_0^2
+ 11 \, k_0 \, k_\mu
- 4 \, k_\mu^2 \right) v_0^2 \, \mu^6
+ \left( 5 \, k_0 - 2 \, k_\mu \right) \mu^8
\big\}\Big], \nn
%%%%%%%%%%%%%%%%%
\sigma_{yx}^{\rm{lf}, 4}&=\frac{B_z \, e^5 \, k_0 \, v_0^2}{8 \, k_\mu^4 \, \pi \, \mu^4}\,\big[
B_x^2 \, k_\mu^2 \, v_0^2 \left\{8 \, k_0^3
\left(-k_0+k_\mu
\right)v_0^4
+
k_0\left(
9 \, k_0-5 \, k_\mu
\right)
v_0^2 \mu^2
-
\mu^4
\right\}
\nn
& \qquad\qquad\qquad
+2 \, B_z^2
\left\{
4 \, k_0^5
\left(k_0-k_\mu\right)v_0^6+ k_0^3 \left(
-7 \, k_0+5 \, k_\mu\right)v_0^4 \mu^2+2 \, k_0^2 \, v_0^2 \mu^4
+\mu^6\right\}\Big\}\big]\, ,
\end{align}
%%%%%%%%%%%%%%%%%%%%%%%%%%%%%%
\begin{align}
\sigma_{xx}^{\rm{lf}}
&=
\frac{
\tau^3 \,e^4 \, v_0  
}{
64 \, k_\mu^4 \, \pi \, \mu^6
}
\Big[
B_z^2
\Big\{
8 \, k_0 \, k_\mu^3 \, v_0^3 \, \mu^3
\left(
-2 \, k_0^3\,  v_0^2
+2 \, k_0^2 \, k_\mu\,  v_0^2
+k_\mu\, \mu^2
\right)
+
B_z \, e
\big\{
-120 \, k_0^8 \, v_0^8
+120 \, k_0^7 \, k_\mu \, v_0^8
+96 \, k_0\,  k_\mu \, v_0^2 \, \mu^6
\nn
& \qquad \qquad \qquad 
+3 \, k_\mu^2\,  v_0^2 \, \mu^6
+23 \, \mu^8
+
12 \, k_0^6
\left(
10 \, k_\mu^2 \, v_0^8
+9 \, v_0^6 \, \mu^2
\right)
-
24 \, k_0^5
\left(
5 \, k_\mu^3\,  v_0^8
+2 \, k_\mu \, v_0^6\,  \mu^2
\right)
+
k_0^4
\big(
-156 \, k_\mu^2 \, v_0^6 \, \mu^2
\nn
& \qquad \qquad \qquad 
+167 \, v_0^4 \, \mu^4
\big)
+
16 \, k_0^3
\left(
6 \, k_\mu^3 \, v_0^6 \, \mu^2
-11 \, k_\mu \, v_0^4 \, \mu^4
\right)
+
k_0^2
\left(
33 \, k_\mu^2 \, v_0^4\,  \mu^4
-178 \, v_0^2\,  \mu^6
\right)
\big\}
\Big\}
\nn
& \qquad\qquad\qquad
+
B_x^2
\Big\{
4 \, k_0^2\,  k_\mu^3 \, v_0^3 \, \mu^3
\left(
k_0^2\,  v_0^2
-k_0 \, k_\mu \, v_0^2
-\mu^2
\right)
+
B_z \, e
\big\{
122 \, k_0^8\,  v_0^8
-122 \, k_0^7\,  k_\mu \, v_0^8
+13 \, k_\mu^2\,  v_0^2\,  \mu^6
+6 \, \mu^8
\nn
& \qquad \qquad \qquad 
+
4 \, k_0^3 \, k_\mu \, v_0^4\,  \mu^2
\left(
-41 \, k_\mu^2\,  v_0^2
+\mu^2
\right)
-
3 \, k_0^6
\left(
50 \, k_\mu^2 \, v_0^8
+69 \, v_0^6 \, \mu^2
\right)
+
2 \, k_0^5
\left(
75 \, k_\mu^3\, v_0^8
+73 \, k_\mu\,  v_0^6\,  \mu^2
\right)
\nn
& \qquad \qquad \qquad 
+
k_0^4
\left(
239 \, k_\mu^2 \, v_0^6 \, \mu^2
+54 \, v_0^4 \, \mu^4
\right)
+
k_0^2
\left(
-102 \, k_\mu^2 \, v_0^4 \, \mu^4
+25 \, v_0^2\,  \mu^6
\right)
+
2 \, k_0
\left(
19 \, k_\mu^3\,  v_0^4\,  \mu^4
-5 \, k_\mu \, v_0^2\,  \mu^6
\right)
\big\}
\Big\}
\Big], \nn
%%%%
\sigma_{zx}^{\rm{lf}}
&=
\frac{
\tau^3 \,  e^5 \,B_x
}{
64 \, k_\mu^4 \, \pi \, \mu^6
}
\big[
B_x^2 \, k_\mu \, v_0^5
\big\{
-2 \, k_0^7\,  v_0^4
+3 \, k_0^6 \, k_\mu\,  v_0^4
+8 \, k_0^5\,  v_0^2 \, \mu^2
+
4 \, k_0 \, k_\mu^4\,  v_0^2 \, \mu^2
+
4 \, k_\mu^3 \, \mu^4
-
3 \, k_0^4
\left(
5 \, k_\mu^3 \, v_0^4
+4 \, k_\mu \, v_0^2\,  \mu^2
\right)
\nn
& \qquad \qquad \qquad 
+
2 \, k_0^3
\left(
7 \, k_\mu^4 \, v_0^4
-3 \, \mu^4
\right)
+
k_0^2
\left(
7 \, k_\mu^3\,  v_0^2 \, \mu^2
+9 \, k_\mu\,  \mu^4\right)
\big\}-B_z^2
\big\{
40 \, k_0^8 \, v_0^9
-40 \, k_0^7 \, k_\mu \, v_0^9
+3 \, k_\mu^2 \, v_0^3 \, \mu^6
\nn & \qquad \qquad \qquad
+11 \, v_0 \, \mu^8
-
4 \, k_0^6
\left(
26 \, k_\mu^2 \, v_0^9
+29 \, v_0^7 \, \mu^2
\right)+8 \, k_0^5\left(13 \, k_\mu^3\,  v_0^9+12 \, k_\mu \, v_0^7 \, \mu^2\right)
+3 \, k_0^4\left(52 \, k_\mu^2 \, v_0^7\,  \mu^2+41 \, v_0^5 \, \mu^4\right)\nn
& \qquad \qquad \qquad
-8 \, k_0^3\left(13 \, k_\mu^3 \, v_0^7 \, \mu^2
+10 \, k_\mu\,  v_0^5\,  \mu^4\right)-k_0^2\left(
55 \, k_\mu^2 \, v_0^5 \, \mu^4
+58 \, v_0^3 \, \mu^6
\right)+16 \, k_0\left(k_\mu^3 \, v_0^5\,  \mu^4+2 \, k_\mu \, v_0^3 \, \mu^6\right)\big\}\big]\,.
\end{align}

The linear-in-$B_z$ term in $\bar\sigma_{xx}$ is the sole odd contribution at leading
order and dominates the even quadratic terms in the weak-field regime. Its presence
is permitted because the sign change of $B_z$ is compensated by that of the dipolar
$\bs\Omega_s$ and $\bs m_s$, and it has no analogue for a BC-monopole. In
Eq.~\eqref{eqsigxxset2}, the BC-only linear-in-$B_z$ coefficient reduces to the
parameter-independent value $\tau e^3v_0/(64\pi)$ --- positive and equal in
magnitude to the corresponding coefficient in set-up~I. The OMM-only and concurrent
linear coefficients, in contrast, retain an explicit dependence on $k_0$, $k_\mu$,
and $\mu$ inherited from the toroidal Fermi surface, as they involve the
OMM-corrected dispersion rather than the pure BC integral. The transverse
$\bar\sigma_{zx}$ carries an overall factor of $B_x$; at $B_z=0$ this set-up
coincides with set-up~I, and its linear term reduces to the
$-\tau e^3v_0B_x/(16\pi)$ of Eq.~\eqref{eqsigzxset1}. Terms containing $B_z$
modify it only when the field is tilted out of the $xy$-plane. Through the symmetry
relation of Eq.~\eqref{eq:theta_covariance} at $B_y=0$, $\bar\sigma_{zx}$ here is
identical to $\bar\sigma_{xz}$ of set-up~III. For the AH part, the $B^0$ term is
the same plateau $-e^2\mu/(8\pi v_0)$ as in set-up~I, and the linear-in-$B_z$ term
arises because $B_z$ couples directly to the out-of-plane component of the BC.

For the LFO-induced part, the out-of-plane field $B_z$ activates the out-of-plane
responses of $\bs\Omega_s$ and $\bs m_s$, resulting in a rich structure. At
$n=1$ ($\propto\tau^2$), only $\sigma^{\rm{lf}}_{yx}$ is generated, with its
kinetic-only part vanishing upon integration; the BC, OMM, and concurrent sub-parts
carry the field dependence $\{B_x^2,B_z^2,B_zB_x^2,B_z^3\}$
(Table~\ref{tab:lf_setup2}), which is even in $\bs B$ at leading order. At
$n=2$ ($\propto\tau^3$), the longitudinal component
$\sigma^{\rm lf,h}_{xx}\propto B_x^2+\alpha B_z^2$ reflects the standard LMC/PHC
anisotropy of a 3d semimetal, while the in-plane transverse
$\sigma^{\rm{lf}}_{zx}$ receives contributions $\propto\{B_x^3,B_xB_z^2\}$. At
$n=3$ ($\propto\tau^4$), only the kinetic part
$\sigma^{\rm lf,h}_{yx}\propto\{B_zB_x^2,B_z^3\}$ survives.

The numerical results for the parameters of Table~\ref{tab-params} are shown in
Fig.~\ref{figset2vnr}, with the field-strength dependence at $\theta=\pi/4$ in the
left panels and the angular dependence at fixed $|\bs B|$ in the right panels
($B_x=|\bs B|\cos\theta$, $B_z=|\bs B|\sin\theta$). In the left panels, the non-AH
$\bar\sigma_{xx}$ is odd and nearly linear in $B_z$, with an OMM-only contribution
several times larger than the BC-only part and of opposite slope, consistent with
the coefficients of Eq.~\eqref{eqsigxxset2}. The concurrent and LFO curves are
invisible on this scale, as expected from their quadratic or higher-order origin.
The transverse $\bar\sigma_{zx}$ is linear in $B_x$, again with BC-only and
OMM-only curves of opposite slope. The AH curve $\sigma^{\rm{ah}}_{yx}$ sits on
its plateau with only a weak tilt from the linear-in-$B_z$ term, while the LFO
curve $\sigma^{\rm{lf}}_{yx}$ is even and quadratic. In the right panels, the
linear terms produce a $\sin\theta$ profile for $\bar\sigma_{xx}$, vanishing at
$\theta=0$ and $\pi$ where $B_z=0$, and a $\cos\theta$ profile for $\sigma_{zx}$,
vanishing at $\theta=\pi/2$ and $3\pi/2$ where $B_x=0$. The BC-only and OMM-only
curves are in antiphase in both cases. For $\sigma_{yx}$, the $\theta$-dependence
is omitted, and both panels display the AH and LFO curves against the field
strength.

%%%%%%%%%%%%%%%%%%%%%%%%%%%%%%%%%
\subsection{Set-up III: 
\texorpdfstring{$\bs{E}=E_z\,\bs{\hat{z}}$}{E} and 
\texorpdfstring{$\bs{B}= B_x\, \bs{\hat{x}}+B_z\,\bs{\hat{z}}$}{B}, and
\texorpdfstring{$\bs{B}^2 = B_x^2 + B_z^2$}{BB}}
\label{secset3}

%%%%%%%%%%%%%%%%%%%%%%%%%%%%%
\begin{figure}[t!]
\centering
\subfigure{\includegraphics[width=0.75 \textwidth]{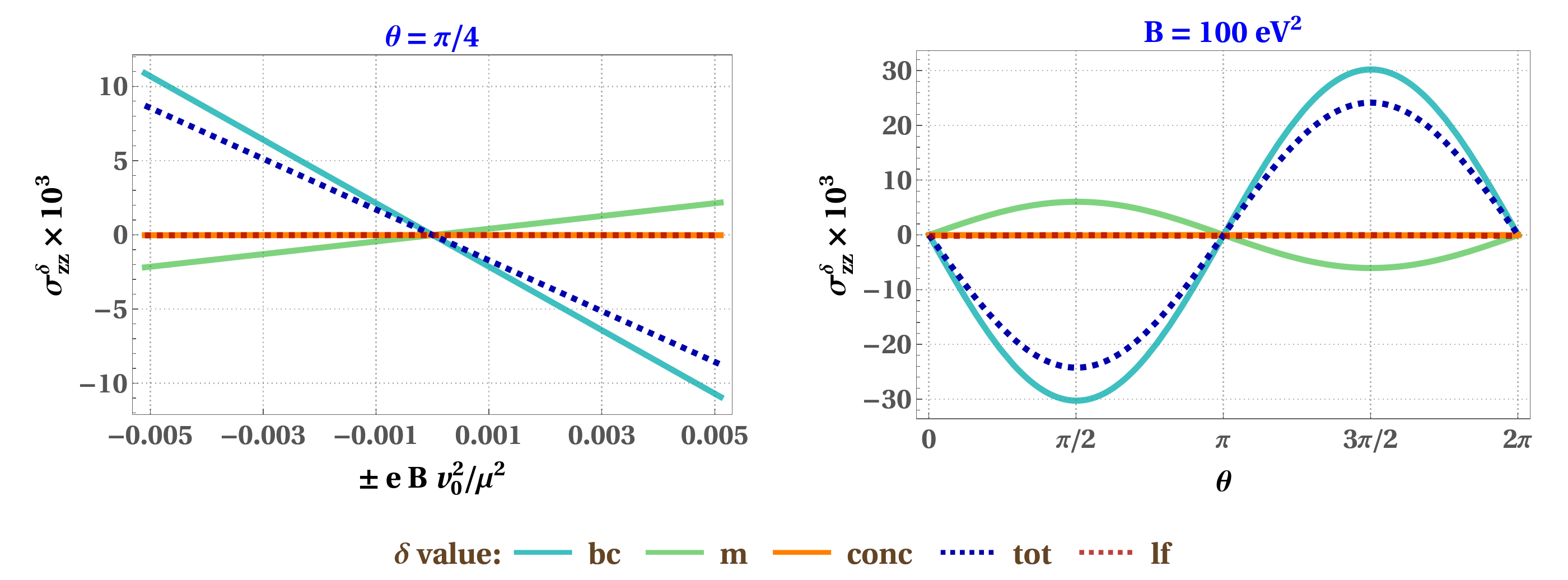}}
\subfigure{\includegraphics[width=0.75 \textwidth]{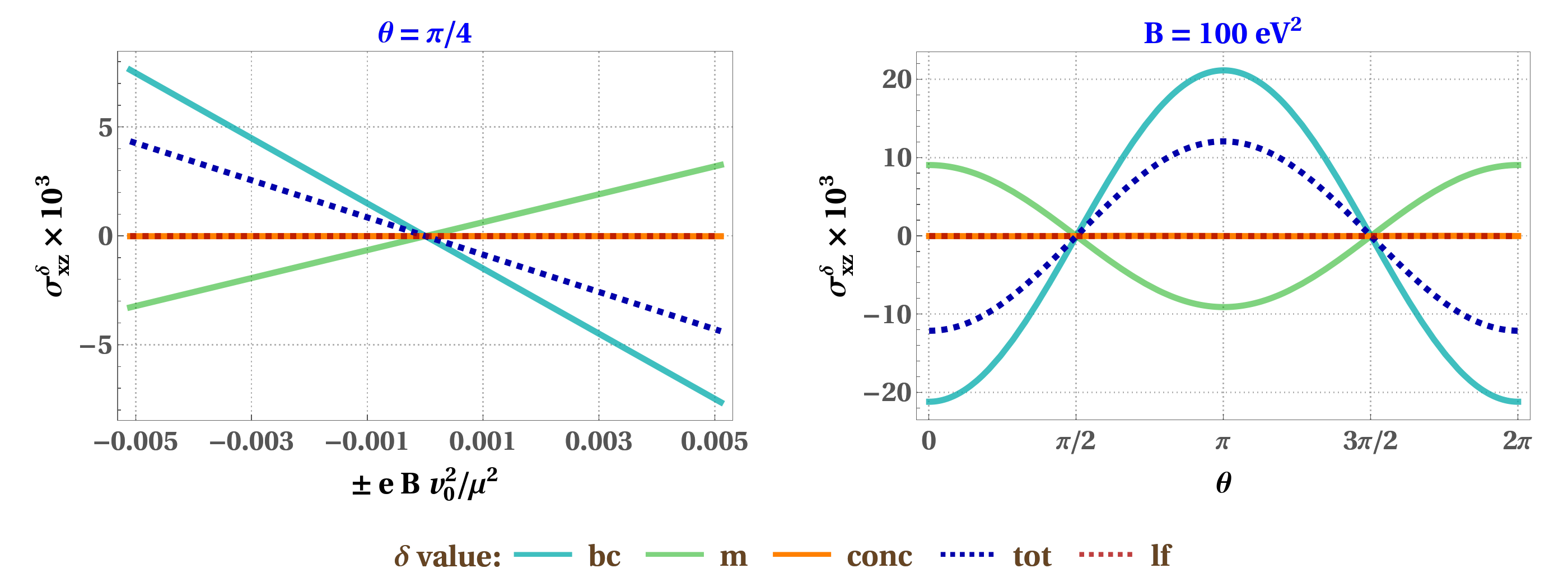}}
\subfigure{\includegraphics[width=0.75 \textwidth]{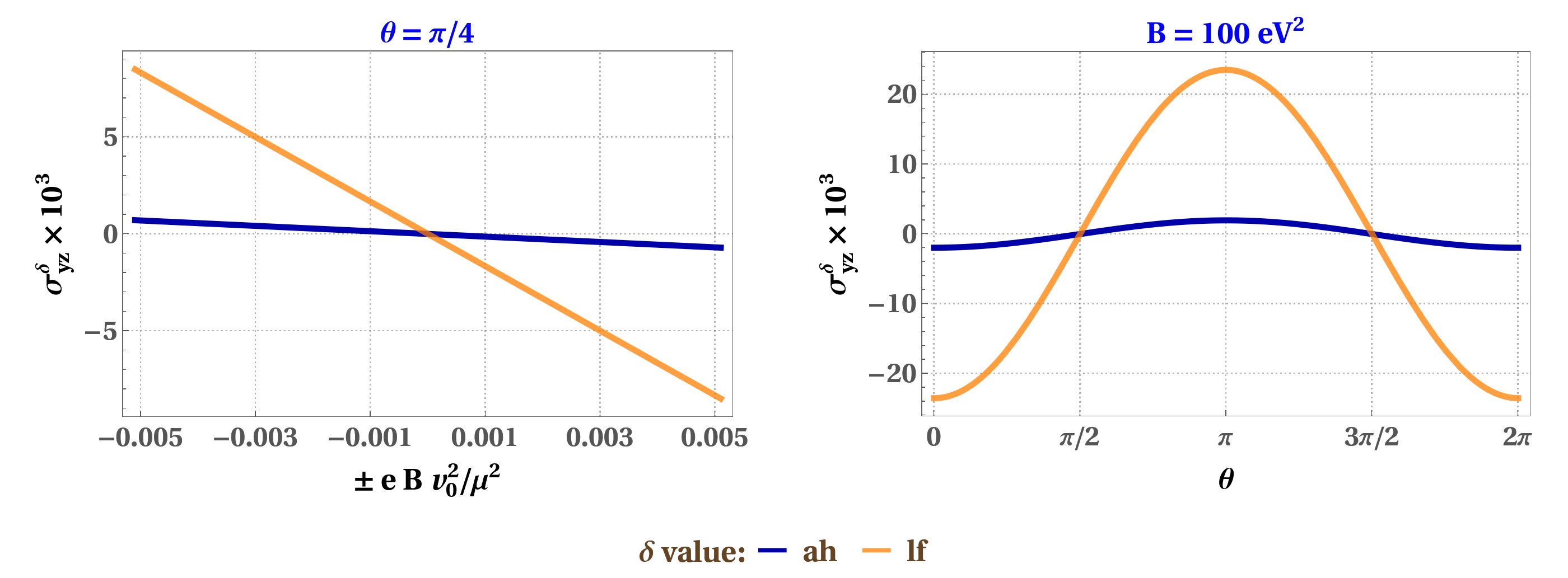}}
\caption{Set-up III for VNR: Behaviour of the conductivity curves (in eV) as functions of ${e\, B \,v_0^2} / {\mu^2}$ and $\theta$.\label{figset3vnr}}
\end{figure}
%%%%%%%%%%%%%%%%%%%%%%%%%%%%%%%%

The non-AH part of the response has the following nonzero components:
%%%%%%%%%%%%%%%%%%%%%%%%%%
\begin{align}
\label{eqsigzzset3}
\sigma^{\rm d}_{zz} & = \frac{\tau \, e^2 \, k_0 \, \mu }{4 \, \pi}  \,,
\end{align}
%%%%%%%%%%%%%%%%%%%%%%%
\begin{align}
\sigma^{\rm bc}_{zz} & =
\frac{\tau \, v_0 \,e^3 }{512\,\pi\,\mu^{10}}\big[-80\,B_z\,\mu^{10}
-4 \,e \,k_0 \,v_0^3 \,\mu^3\big\{2 \,B_z^2 \left(-8 \,k_0^4 \,v_0^4
+ 8 \,k_0^3 \,k_\mu \,v_0^4+ 4 \,k_0^2 \,v_0^2 \,\mu^2+ \mu^4\right)
\nn &\qquad\qquad\qquad
+ B_x^2 \left(8 \,k_0^4 \,v_0^4- 8\, k_0 \,k_\mu^3\, v_0^4- 12\, k_0^2 \,v_0^2 \,\mu^2+ 3\, \mu^4\right)\big\}
\nn&\quad\qquad\qquad\quad
+ B_z \,e^2 \,v_0^4\big\{2 \,B_z^2 \big(
112 \,k_0^6 \,v_0^6- 120 \,k_0^4 \,v_0^4 \,\mu^2+ 18 \,k_0^2 \,v_0^2\, \mu^4+ \mu^6
+ k_\mu \left(-112 \,k_0^5 \,v_0^6+ 64\, k_0^3 \,v_0^4\, \mu^2\right)
\big)
\nn&\quad\qquad\qquad\quad
+ B_x^2 \left(-336 \,k_0^6 \,v_0^6
+ 440 \,k_0^4 \,v_0^4 \,\mu^2- 126\, k_0^2 \,v_0^2 \,\mu^4
+ 3 \,\mu^6+ 16 \,k_0\, k_\mu\, v_0^2
\left(21 \,k_0^4 \,v_0^4
- 17 \,k_0^2\, v_0^2 \,\mu^2+ 2\, \mu^4\right)\right)\big\}\big]\, ,
\end{align}
%%%%%%%%%%%%%%%%%%%%%%%
%%%%%%%%%%%%%%%%%%%%%
\begin{align}
\sigma^{\rm m}_{zz} & =\frac{\tau \,e^3}
{512\,k_\mu^4\,\pi\,v_0^3\,\mu^{10}} \big[
2\,B_z\,k_\mu^2\,v_0^2 \big\{
8\,k_\mu^2\,v_0^2\,\mu^{10}- 4\,B_z\,e\,k_0\,k_\mu^2\,v_0^5\,\mu^3
\,\big(40\,k_0^4\,v_0^4
- 56\,k_0^2\,v_0^2\,\mu^2+ 13\,\mu^4\nn
%%%%%-
& \hspace{2.5 cm}+ 4\,k_0\,k_\mu\,v_0^2 \left(-10\,k_0^2\,v_0^2 + 9\,\mu^2 \right)
\big)+ B_z^2\,e^2\,v_0^4
\big(
-2688\,k_0^8\,v_0^8
+ 6328\,k_0^6\,v_0^6\,\mu^2
- 4780\,k_0^4\,v_0^4\,\mu^4\nn
%%%%%%%
& \hspace{2.5 cm}+ 1177\,k_0^2\,v_0^2\,\mu^6- 37\,\mu^8+ k_\mu\left( 2688\,k_0^7\,v_0^8
- 4984\,k_0^5\,v_0^6\,\mu^2
+ 2624\,k_0^3\,v_0^4\,\mu^4- 320\,k_0\,v_0^2\,\mu^6\right)\big)
%%%%%%%
\big\}\nn
& \hspace{2.5 cm}+ B_x^2\,e\,v_0^3
%%%%%%%%
\Big\{
-4\,k_0\,k_\mu^2\,v_0^2\,\mu^3
\big(
-40\,k_0^6\,v_0^6+ 80\,k_0^4\,v_0^4\,\mu^2- 41\,k_0^2\,v_0^2\,\mu^4
+ \mu^6+ k_0\,k_\mu\,v_0^2
\big( 40\,k_0^4\,v_0^4 \nn
%%%%%%
& \qquad  \qquad \qquad \qquad
- 60\,k_0^2\,v_0^2\,\mu^2
+ 16\,\mu^4 \big)\big)
+B_z\,e\,v_0\big(k_\mu^4\,v_0^4\left(8064\,k_0^6\,v_0^6
- 8120\,k_0^4\,v_0^4\,\mu^2+ 1500\,k_0^2\,v_0^2\,\mu^4- 9\,\mu^6\right)\nn
%%%%%%%%
& \hspace{2.5 cm} + k_\mu
\left(-8064\,k_0^9\,v_0^{10}+ 20216\,k_0^7\,v_0^8\,\mu^2- 16704\,k_0^5\,v_0^6\,\mu^4
+ 4800\,k_0^3\,v_0^4\,\mu^6+ 256\,k_0\,v_0^2\,\mu^8\right)\big)
\Big\}\big]\, , \end{align}
%%%%%%%%%%%%%%%%%%%%%%%
\begin{align}
\sigma^{\rm conc}_{zz} & =\frac{\tau\,e^4}{256\,k_\mu^4\,\pi\,\mu^{10}}
\big[
-2\,B_z^2\,k_\mu^2\,v_0^2
%%%%%
%%%%
\big\{
-4\,k_\mu^2\,v_0^2\,\mu^3
\left(
k_\mu
\left(-8\,k_0^4\,v_0^4 + 28\,k_0^2\,v_0^2\,\mu^2 - 4\,\mu^4
\right)
+ k_0
\left(8\,k_0^4\,v_0^4 - 32\,k_0^2\,v_0^2\,\mu^2 + 17\,\mu^4 \right)
\right)
\nn
%%%%
& \qquad   \qquad \qquad \quad
- 3\,B_z\,e\,v_0
\big(
168\,k_0^8\,v_0^8 - 648\,k_0^6\,v_0^6\,\mu^2 + 737\,k_0^4\,v_0^4\,\mu^4
- 274\,k_0^2\,v_0^2\,\mu^6 + 17\,\mu^8
\nn
%%%%
& \qquad  \qquad \qquad  \quad 
+ k_\mu \left( -168\,k_0^7\,v_0^8 + 564\,k_0^5\,v_0^6\,\mu^2
- 476\,k_0^3\,v_0^4\,\mu^4 + 96\,k_0\,v_0^2\,\mu^6 \right)\big)\big\}\nn
& \qquad  \qquad \qquad \quad
+ B_x^2\, \big\{-4\,k_0\,k_\mu^4\,v_0^4\,\mu^3 \left(
8\,k_0^4\,v_0^4 - 8\,k_0^2\,v_0^2\,\mu^2 + \mu^4
+ k_\mu\left(-8\,k_0^3\,v_0^4 + 4\,k_0\,v_0^2\,\mu^2 \right)\right) \nn
%%%%%-
& \qquad  \qquad \qquad \quad
+ B_z\,e\,v_0 \big(
- k_\mu^4\,v_0^4
\left( 1512\,k_0^6\,v_0^6 - 2400\,k_0^4\,v_0^4\,\mu^2 + 765\,k_0^2\,v_0^2\,\mu^4 - 23\,\mu^6 \right)
\nn
%%%%%
& \qquad  \qquad \qquad \quad
+ k_0\,k_\mu\,v_0^2 \left(1512\,k_0^8\,v_0^8 - 4668\,k_0^6\,v_0^6\,\mu^2 + 4932\,k_0^4\,v_0^4\,\mu^4
- 1976\,k_0^2\,v_0^2\,\mu^6 + 208\,\mu^8 \right)\big)\big\}\big]\, .
\end{align}
%%%%%
\begin{align}
\label{eqsigxzset3}
\sigma^{\rm bc}_{xz} & =\frac{\tau \, e^3 \, v_0 \, B_x}
{2048 \, \pi \, \mu^{10}}\big[- 4\big\{56 \, \mu^{10}+ 8 \, B_z \, e \, k_0 \, v_0^3 \, \mu^3
\left(8 \, k_0^4 \, v_0^4- 16 \, k_0^2 \, v_0^2 \, \mu^2+ 5 \, \mu^4
+ k_\mu \left(- 8 \, k_0^3 \, v_0^4+ 12 \, k_0 \, v_0^2 \, \mu^2\right)\right)\nn
&\qquad\qquad\qquad
+ 3 \, B_z^2 \, e^2 \, v_0^4\big(112 \, k_0^6 \, v_0^6
- 200 \, k_0^4 \, v_0^4 \, \mu^2+ 90 \, k_0^2 \, v_0^2 \, \mu^4- 5 \, \mu^6 \nn
&\qquad\qquad\qquad
+ k_\mu \left(
- 112 \, k_0^5 \, v_0^6
+ 144 \, k_0^3 \, v_0^4 \, \mu^2
- 32 \, k_0 \, v_0^2 \, \mu^4
\right)
\big)
\big\}
\nn
&\qquad\qquad\qquad
+ B_x^2 \, e^2 \, v_0^4
\big\{
336 \, k_0^6 \, v_0^6
- 680 \, k_0^4 \, v_0^4 \, \mu^2
+ 342 \, k_0^2 \, v_0^2 \, \mu^4
- 21 \, \mu^6\nn
&\qquad\qquad\qquad
+ k_\mu \left(
- 336 \, k_0^5 \, v_0^6
+ 512 \, k_0^3 \, v_0^4 \, \mu^2
- 128 \, k_0 \, v_0^2 \, \mu^4\right)
\big\}\big],
\end{align}
%%%%%%%%%%%%%%%
\begin{align}
\sigma^{\rm m}_{xz} & =\frac{\tau \, e^3 \, B_x}{2048 \, k_\mu^6 \, \pi \, v_0 \, \mu^{10}}
\big[
- 3 \, B_x^2 \, e^2
\big\{
2688 \, k_0^{12} \, v_0^{12}
- 9464 \, k_0^{10} \, v_0^{10} \, \mu^2
+ 12364 \, k_0^8 \, v_0^8 \, \mu^4
- 7185 \, k_0^6 \, v_0^6 \, \mu^6
+ 1691 \, k_0^4 \, v_0^4 \, \mu^8
\nn&\qquad\qquad\qquad\qquad
- 91 \, k_0^2 \, v_0^2 \, \mu^{10}- 3 \, \mu^{12}+ k_\mu\big(- 2688 \, k_0^{11} \, v_0^{12}
+ 8120 \, k_0^9 \, v_0^{10} \, \mu^2- 8640 \, k_0^7 \, v_0^8 \, \mu^4
+ 3712 \, k_0^5 \, v_0^6 \, \mu^6
\nn&\qquad\qquad\qquad\qquad
- 512 \, k_0^3 \, v_0^4 \, \mu^8\big)\big\}
+ 4 \, k_\mu^2 \, v_0^2\big\{
24 \, k_\mu^4 \, \mu^{10}
+ 8 \, B_z \, e \, k_\mu^2 \, v_0 \, \mu^3
\big(
40 \, k_0^7 \, v_0^6
- 64 \, k_0^5 \, v_0^4 \, \mu^2
+ 33 \, k_0^3 \, v_0^2 \, \mu^4\nn &\qquad\qquad\qquad\qquad
- 9 \, k_0 \, \mu^6
+ k_\mu \left(- 40 \, k_0^6 \, v_0^6+ 44 \, k_0^4 \, v_0^4 \, \mu^2- 16 \, k_0^2 \, v_0^2 \, \mu^4
+ 4 \, \mu^6\right)\big)
\nn &\qquad\qquad\qquad\qquad
+ B_z^2 \, e^2\big(8064 \, k_0^{10} \, v_0^{10}
- 23128 \, k_0^8 \, v_0^8 \, \mu^2+ 23804 \, k_0^6 \, v_0^6 \, \mu^4- 10573 \, k_0^4 \, v_0^4 \, \mu^6
+ 1926 \, k_0^2 \, v_0^2 \, \mu^8
\nn&\qquad\qquad\qquad\qquad
- 93 \, \mu^{10}
+ k_\mu
\left(- 8064 \, k_0^9 \, v_0^{10}+ 19096 \, k_0^7 \, v_0^8 \, \mu^2
- 15264 \, k_0^5 \, v_0^6 \, \mu^4+ 4824 \, k_0^3 \, v_0^4 \, \mu^6- 544 \, k_0 \, v_0^2 \, \mu^8
\right)\big)\big\}\big], 
\end{align}
%%%%
\begin{align}
\sigma^{\rm conc}_{xz} & =\frac{\tau \, e^4 \, B_x}
{1024 \, k_\mu^4 \, \pi \, \mu^{10}}\big[B_x^2 \, e \, v_0
\big(1512 \, k_0^{10} \, v_0^{10}- 3504 \, k_0^8 \, v_0^8 \, \mu^2+ 2313 \, k_0^6 \, v_0^6 \, \mu^4
- 155 \, k_0^4 \, v_0^4 \, \mu^6- 173 \, k_0^2 \, v_0^2 \, \mu^8
+ 7 \, \mu^{10}
\nn &\qquad\qquad\qquad\qquad
+ k_\mu
\left(- 1512 \, k_0^9 \, v_0^{10}+ 2748 \, k_0^7 \, v_0^8 \, \mu^2- 1128 \, k_0^5 \, v_0^6 \, \mu^4
- 160 \, k_0^3 \, v_0^4 \, \mu^6+ 64 \, k_0 \, v_0^2 \, \mu^8\right)\big)\nn
&\qquad\qquad\qquad\qquad
+ 4 \, B_z \, k_\mu \, v_0^3 \big\{8 \, k_\mu^3 \, v_0 \, \mu^3
\big(- 8 \, k_0^5 \, v_0^4+ 8 \, k_0^3 \, v_0^2 \, \mu^2
- 3 \, k_0 \, \mu^4+ k_\mu \left( 8 \, k_0^4 \, v_0^4- 4 \, k_0^2 \, v_0^2 \, \mu^2+ 2 \, \mu^4
\right)\big)\nn &\qquad\qquad\qquad\qquad
+ B_z \, e\, \big(
1512 \, k_0^9 \, v_0^8- 4668 \, k_0^7 \, v_0^6 \, \mu^2+ 5012 \, k_0^5 \, v_0^4 \, \mu^4
- 2124 \, k_0^3 \, v_0^2 \, \mu^6+ 276 \, k_0 \, \mu^8
\nn&\qquad\qquad\qquad\qquad
+ k_\mu^3 \left(- 1512 \, k_0^6 \, v_0^8
+ 2400 \, k_0^4 \, v_0^6 \, \mu^2- 845 \, k_0^2 \, v_0^4 \, \mu^4+ 51 \, v_0^2 \, \mu^6
\right)\big)\big\}\big]\, ,
\end{align}
%%%%%%%%%%%%%%%%%%%%%%%%

For the AH conductivity, we get 
\begin{align}
 \sigma^{\rm{ah}}_{yz} &=
-\frac{v_0^2}
{1024 \, \pi \, \mu^{10} \, k_\mu^5}
\Big[-4 \, B_x \, e^3 \, k_\mu^2
\Big\{
8 \, k_0 \, k_\mu^2 \, \mu^6
\left(
2 \, k_0^3 \, v_0^2
- 2 \, k_0^2 \, k_\mu \, v_0^2
- 2 \, k_0 \, \mu^2
+ k_\mu \, \mu^2
\right)
- 6 \, B_z \, e \, k_\mu^2 \, v_0 \, \mu^3
\big\{
-40 \, k_0^5 \, v_0^4
\nn
& \hspace{2.5 cm}\quad
+ 44 \, k_0^3 \, v_0^2 \, \mu^2
- 8 \, k_0 \, \mu^4
+ k_\mu
\left(
40 \, k_0^4 \, v_0^4
- 24 \, k_0^2 \, v_0^2 \, \mu^2
+ \mu^4
\right)
\big\}
+ 3 \, B_z^2 \, e^2
\big\{
1008 \, k_0^8 \, v_0^8
\nn
& \hspace{2.5 cm} \quad
- 2212 \, k_0^6 \, v_0^6 \, \mu^2
+ 1503 \, k_0^4 \, v_0^4 \, \mu^4
- 312 \, k_0^2 \, v_0^2 \, \mu^6
+ 8 \, \mu^8
+ k_\mu
\big(
-1008 \, k_0^7 \, v_0^8
+ 1708 \, k_0^5 \, v_0^6 \, \mu^2\nn
& \hspace{2.5 cm} \quad
- 775 \, k_0^3 \, v_0^4 \, \mu^4+ 75 \, k_0 \, v_0^2 \, \mu^6
\big)\big\}\Big\}+ 9 \, B_x^3 \, e^5 \, k_0 \,\Big\{
336 \, k_0^9 \, v_0^8- 980 \, k_0^7 \, v_0^6 \, \mu^2+ 985 \, k_0^5 \, v_0^4 \, \mu^4\nn
& \hspace{2.5 cm} \quad
- 380 \, k_0^3 \, v_0^2 \, \mu^6+ 40 \, k_0 \, \mu^8
%%%
- k_\mu^5 \, v_0^4\left(336 \, k_0^4 \, v_0^4+ 5 \, \mu^4- 140 \, k_0^2 \, v_0^2 \, \mu^2
\right)\Big\}\Big].
\end{align}

The sources of the contributions from the LFO are detailed in Appendix~\ref{appset3vnr}. Here, we show the final forms in the following:
%%%%%%%%%%%%%%
\begin{align}
\sigma_{yz}^{\rm{lf}} = \tau^2 \, \sigma_{yz}^{\rm{lf}, 2}+ \tau^4 \, \sigma_{yz}^{\rm{lf}, 4}\,,
\end{align}
 where
\begin{align} 
\sigma_{yz}^{\rm{lf}, 2}&=\frac{e^3 \, v_0 \, B_x }{512 \, k_\mu^3 \, \pi \, \mu^8}
\Big[B_x^2 \, e^2 \, v_0\, \Big\{
648 \, k_0^8 \, v_0^8-648 \, k_0^7 \, k_\mu\,  v_0^8
-4 \, k_0^6\left(90 \, k_\mu^2\,  v_0^8 +401 \, v_0^6\,  \mu^2\right)
+40 \, k_0^5\left(9 \, k_\mu^3 \, v_0^8+32 \, k_\mu \, v_0^6 \, \mu^2\right)
\nn & \qquad \qquad \qquad
+
8 \, k_0^4
\left(
62 \, k_\mu^2 \, v_0^6 \, \mu^2
+
153 \, v_0^4 \, \mu^4
\right)
-
k_0^3
\left(
316 \, k_\mu^3 \, v_0^6\,  \mu^2
+
665 \, k_\mu \, v_0^4 \, \mu^4
\right)
-
8 \, k_0^2
\left(
20 \, k_\mu^2 \, v_0^4 \, \mu^4
+
31 \, v_0^2\,  \mu^6
\right)
\nn
& \qquad \qquad \qquad 
+
k_0
\left(
47 \, k_\mu^3 \, v_0^4\,  \mu^4
+
33 \, k_\mu \, v_0^2 \, \mu^6
\right)
+
8
\left(
k_\mu^2 \, v_0^2 \, \mu^6
+
\mu^8
\right)
\Big\}
-
4
\Bigg\{
16  \, k_0 \, k_\mu^3 \, v_0 \, \mu^8
-
4 \, B_z \, e \, k_\mu \, \mu^3
\Big\{
8 \, k_0^6 \, v_0^6
\nn
& \qquad \qquad \qquad 
-8 \, k_0^5\,  k_\mu\,  v_0^6
-4 \, k_0 \, k_\mu \, v_0^2 \, \mu^4
+
k_\mu^2 \, v_0^2 \, \mu^4
-3 \, \mu^6
+
8 \, k_0^3 \, k_\mu\,  v_0^4
\left(
k_\mu^2 \, v_0^2
+
\mu^2
\right)
-
4 \, k_0^4
\left(
2 \, k_\mu^2 \, v_0^6
+
3 \, v_0^4 \, \mu^2
\right)
\nn
& \qquad \qquad \qquad
+
k_0^2
\left(
4 \, k_\mu^2 \, v_0^4 \, \mu^2
+
7 \, v_0^2 \, \mu^4
\right)
\Big\}
+
B_z^2 \, e^2 \, v_0^3
\Big\{
648 \, k_0^8 \, v_0^6
-648 \, k_0^7\,  k_\mu \, v_0^6
+
16 \, k_\mu^2 \, \mu^6
\nn
& \qquad \qquad \qquad 
-
4 \, k_0^6
\left(
90 \, k_\mu^2\,  v_0^6
+
373 \, v_0^4 \, \mu^2
\right)
+
8 \, k_0^5
\left(
45 \, k_\mu^3\,  v_0^6
+
146 \, k_\mu\,  v_0^4 \, \mu^2
\right)
+
4 \, k_0^4
\left(
170 \, k_\mu^2 \, v_0^4 \, \mu^2
+
273 \, v_0^2 \, \mu^4
\right)
\nn
& \qquad \qquad \qquad 
-
k_0^3
\left(
500 \, k_\mu^3 \, v_0^4 \, \mu^2
+
589 \, k_\mu \, v_0^2 \, \mu^4
\right)
%%%%%
-
64 \, k_0^2
\left(
5 \, k_\mu^2\,  v_0^2 \, \mu^4
+
4 \, \mu^6
\right)
+
23 \, k_0
\left(
5 \, k_\mu^3\,  v_0^2\,  \mu^4
+
3 \, k_\mu\,  \mu^6
\right)
\Big\}
\Bigg\}
\Big] \, ,
\nn 
\sigma_{yz}^{\rm{lf}, 4}
&=
\frac{
  e^5 \, k_0 \, v_0^6 \,B_x 
}{
16 \, k_\mu \, \pi \, \mu^4
}
\left[
3 \, B_x^2 \, k_0
\left(
-k_0^2 \, v_0^2
+
k_0\,  k_\mu \, v_0^2
+
\mu^2
\right)
+
2 \, B_z^2
\left(
2 \, k_0^3\,  v_0^2
-
2 \, k_0^2 \, k_\mu \, v_0^2
-
k_\mu \mu^2
\right)
\right],
\end{align}
%%%%%%%%%%%%%%%%%%%%%
\begin{align}
\sigma_{xz}^{\rm{lf}}
&=
\frac{
 \tau^3\, e^5 \, v_0^3 \,B_x 
}{
64 \, k_\mu^3 \, \pi \, \mu^6
}
\big[
B_x^2
\big\{
\left(
-2 \, k_0^3 \, v_0^2
+
2 \, k_0^2 \, k_\mu \, v_0^2
+
k_\mu\,  \mu^2
\right)
\left(
k_0^4\,  v_0^4
-
4 \, k_0^2 \, v_0^2\,  \mu^2
+
3 \, \mu^4
\right)
\nn
& \qquad \qquad \qquad
-
k_\mu^2\,  v_0^2
\left(
2 \, k_0^5 \, v_0^4
-
2 \, k_0^4\,  k_\mu \, v_0^4
-
8 \, k_0^3 \, v_0^2 \, \mu^2
+
7 \, k_0^2\,  k_\mu \, v_0^2 \, \mu^2
+
6 \, k_0 \, \mu^4
-
6 \, k_\mu\,  \mu^4
\right)
\big\}
\nn
& \qquad\qquad\qquad
+
B_z^2
\big\{
40 \, k_0^7 \, v_0^6
-
96 \, k_0^5 \, v_0^4 \, \mu^2
+
80 \, k_0^3 \, v_0^2 \, \mu^4
-
32 \, k_0 \, \mu^6
+
k_\mu
\big(
-40 \, k_0^6 \, v_0^6
+
76 \, k_0^4 \, v_0^4 \, \mu^2
\nn
& \qquad \qquad \qquad
-
47 \, k_0^2\,  v_0^2 \, \mu^4
+
11 \, \mu^6
\big)
-
k_\mu^2 \, \, v_0^2
\left(
40 \, k_0^5\,  v_0^4
-
32 \, k_0^3 \, v_0^2\,  \mu^2
+k_\mu
\left(
-40 \, k_0^4\,  v_0^4
+12 \, k_0^2\,  v_0^2 \, \mu^2
+\mu^4\right)\right)\big\}\big],
\end{align} 
\begin{align}
\sigma_{zz}^{\rm{lf}}
&=
-\frac{\tau^3 \,  e^4\,  k_0\,  v_0^4}{8 \, \pi \, \mu^{6}\, k_\mu}\,
B_x^2
\left[
k_\mu\, \mu^5
+
B_z \, e\,  v_0^3
\left\{
2 \, k_0^2 \left( k_0-k_{\mu } \right)
\left(2\, k_0 + 3k_\mu \right)v_0^2
+
\left(-2\, k_0^2-4\, k_0\, k_\mu + k_\mu^2 \right)\mu^2
\right\}
\right].
\end{align}

Appendix~\ref{app_GNR} presents the corresponding NVNR results, against which the characteristics above may be compared. $\bar\sigma_{xz}$ of this set-up coincides
with $\bar\sigma_{zx}$ of set-up~II through the symmetry relation of
Eq.~\eqref{eq:theta_covariance} at $B_y=0$, and therefore carries the same overall
factor of $B_x$, reducing at $B_z=0$ to the linear term
$-\,\tau\, e^3\,v_0\, B_x/(16\,\pi)$ of Eq.~\eqref{eqsigzxset1}. The longitudinal
$\bar\sigma_{zz}$ contains the Drude term together with an odd term linear in
$B_z$, again permitted by the compensation of the sign change of $B_z$ by the
dipolar $\bs\Omega_s$ and $\bs m_s$. In Eq.~\eqref{eqsigzzset3}, the BC-only
linear-in-$B_z$ coefficient reduces to the parameter-independent value
$-\,5\,\tau \,e^3\,v_0/(32\, \pi)$, which is negative and exactly ten times larger in
magnitude than the corresponding coefficient $\tau \,e^3\,v_0/(64\,\pi)$ for
$\bar\sigma_{xx}$ in set-up~II. Consequently, the BC-only slope of
$\bar\sigma_{zz}$ dominates over its OMM-only counterpart, in contrast to set-up~II
where the OMM-only slope is larger. The OMM-only and concurrent linear
coefficients again depend on $k_0$, $k_\mu$, and $\mu$ through the toroidal Fermi
surface. In the AH part, $\sigma^{\rm{ah}}_{yz}$ carries an overall factor of
$B_x$. In the LFO part, the term $ \sigma^{\rm lf,h}_{yz}=-\,\tau^2\,e^3 \,k_0\,v_0^2 \,B_x
/(8 \, \pi)$ of Appendix~\ref{app-lf-vnr} is linear in $B_x$, whereas the BC- and OMM-dependent
sub-parts of $\sigma^{\rm{lf}}_{zz}$ and $\sigma^{\rm{lf}}_{xz}$ begin at cubic
order and carry $\tau^3$.

The numerical results of Fig.~\ref{figset3vnr} for the parameters of
Table~\ref{tab-params} --- field-strength dependence at $\theta=\pi/4$ in the left
panels, angular dependence at fixed $|\bs B|$ in the right panels --- confirm these predictions.
In the left panels, $\bar\sigma_{zz}$ is odd and nearly linear in $B_z$, with the
BC-only part larger than the OMM-only part and of opposite slope, as anticipated
from the coefficient ratio above. The BC-only, OMM-only, and total curves of
$\bar\sigma_{xz}$ coincide with those of $\bar\sigma_{zx}$ in
Fig.~\ref{figset2vnr}, as required by the symmetry relation. The concurrent and
LFO contributions to $\bar\sigma_{zz}$ and $\sigma_{xz}$ are negligible on this
scale. The out-of-plane $\sigma_{yz}$ has only AH and LFO parts, both odd and
nearly linear in the field, with the LFO curve roughly an order of magnitude
larger than the AH curve. In the right panels, $\bar\sigma_{zz}$ follows a
$\sin\theta$ profile, vanishing at $\theta=0$ and $\pi$ where $B_z=0$, while the
curves $\bar\sigma_{xz}$, $\sigma^{\rm{ah}}_{yz}$, and $\sigma^{\rm{lf}}_{yz}$
follow a $\cos\theta$ profile, vanishing at $\theta=\pi/2$ and $3\,\pi/2$ where
$B_x=0$.

%%%%%%%%%%%%%%%%%%%%%%%%
\begin{table*}[t!]
\centering
\caption{\label{tab-nonAH-ah}
Nonzero components of $\bar\sigma$ and $\sigma^{\textrm{ah}}$ across the three set-ups of the VNR and the Hopf Berry dipole.
$B$ represents the magnitude of the magnetic field in the respective set-up.}
\begin{ruledtabular}
\begin{tabular}{lcccc}
 & & & \multicolumn{2}{c}{Out-of-plane transverse} \\
 \cline{4-5}
 & Longitudinal & In-plane transverse & Non-AH & AH \\
\colrule
Set-up~I
 & $\bar\sigma_{xx} \propto B^0 , \,B_x^2,\, B_y^2$
 & $\bar\sigma_{yx} \propto B_x\, B_y$
 & $\bar\sigma_{zx} \propto B_x,\, B_x\, B^2$
 & \makecell[c]{$\sigma^{\textrm{ah}}_{zx} \propto B_y,\, B_y^3,\, B_y\, B_x^2$ \\[2pt]
    $\sigma^{\textrm{ah}}_{yx} \propto B^0,\, B^2$} \\[16pt]
Set-up~II
 & \makecell[c]{$\bar\sigma_{xx} \propto B_z,\, B_x^2,\, B_z^2,$ \\
    $B_z\, B_x^2,\, B_z^3$}
 & $\bar\sigma_{zx} \propto B_x,\, B_x\, B_z,\, B_x^3,\, B_x\, B_z^2$
 & $\cdots$
 & \makecell[c]{$\sigma^{\textrm{ah}}_{yx} \propto B^0,\, B_z,\, B_z^2,$ \\
    $B_z^3,\, B_x^2,\, B_x^2\, B_z$} \\[16pt]
Set-up~III
 & \makecell[c]{$\bar\sigma_{zz} \propto B_z,\, B_x^2,\, B_z^2,$ \\
    $B_z\, B_x^2,\, B_z^3$}
 & $\bar\sigma_{xz} \propto B_x,\, B_x\, B_z,\, B_x^3,\, B_x\, B_z^2$
 & $\cdots$
 & \makecell[c]{$\sigma^{\textrm{ah}}_{yz} \propto B_x,\, B_x\, B_z,$ \\
    $B_x^3,\, B_x\, B_z^2$} \\
\end{tabular}
\end{ruledtabular}
\end{table*}

\begin{table}[ht!]
\caption{\label{tab:lf_setup1}
$B_i$-dependence in the LFO-induced terms for set-up~I of the VNR and the Hopf Berry dipole, organised by the values of $n$ and physical origins.}
\begin{ruledtabular}
\begin{tabular}{lcccc}
 & $\sigma^{\textrm{lf, h}}$ & $\sigma^{\textrm{lf, bc}}$ & $\sigma^{\textrm{lf, m}}$ & $\sigma^{\textrm{lf, conc}}$ \\
\colrule
$n=1$, $\sigma_{zx}$
 & $B_y$
 & $B_y\,(B_x^2 + B_y^2)$
 & $B_y\,(B_x^2 + B_y^2)$
 & $B_y\,(B_x^2 + B_y^2)$ \\[4pt]
$n=1$, $\sigma_{yx}$
 & $0$
 & $B_x^2 + B_y^2$
 & $B_x^2 + B_y^2$
 & $0$ \\[4pt]
$n=2$, $\sigma_{xx}$
 & $B_x^2 + 3\,B_y^2$
 & $0$ & $0$ & $0$ \\[4pt]
$n=2$, $\sigma_{yx}$
 & $B_x\,B_y$
 & $0$ & $0$ & $0$ \\[4pt]
$n=2$, $\sigma_{zx}$
 & $0$
 & $B_x\,(B_x^2 + B_y^2)$
 & $B_x\,(B_x^2 + B_y^2)$
 & $0$ \\[4pt]
$n=3$, $\sigma_{zx}$
 & $B_y\,(B_x^2 + B_y^2)$
 & $0$ & $0$ & $0$ \\
\end{tabular}
\end{ruledtabular}
\end{table}

\begin{table}[ht!]
\caption{\label{tab:lf_setup2}
$B_i$-dependence in the LFO-induced terms for set-up~II of the VNR and the Hopf Berry dipole, organised by the values of $n$ and physical origins.}
\begin{ruledtabular}
\begin{tabular}{lcccc}
 & $\sigma^{\textrm{lf, h}}$ & $\sigma^{\textrm{lf, bc}}$ & $\sigma^{\textrm{lf, m}}$ & $\sigma^{\textrm{lf, conc}}$ \\
\colrule
$n=1$, $\sigma_{yx}$
 & $0$
 & \makecell[c]{$B_x^2, B_z^2,$  $B_z\,B_x^2, B_z^3$}
 & \makecell[c]{$B_x^2, B_z^2,$ $B_z\,B_x^2, B_z^3$}
 & $B_z\,B_x^2, B_z^3$ \\[4pt]
$n=2$, $\sigma_{xx}$
 & $B_x^2, B_z^2$
 & $B_z\,B_x^2, B_z^3$
 & $B_z\,B_x^2, B_z^3$
 & $0$ \\[4pt]
$n=2$, $\sigma_{zx}$
 & $0$
 & $B_x^3, B_x\,B_z^2$
 & $B_x^3, B_x\,B_z^2$
 & $0$ \\[4pt]
$n=3$, $\sigma_{yx}$
 & $B_z\,B_x^2, B_z^3$
 & $0$ & $0$ & $0$ \\
\end{tabular}
\end{ruledtabular}
\end{table}

\begin{table}[ht!]
\caption{\label{tab:lf_setup3}
$B_i$-dependence in the LFO-induced terms for set-up~III of the VNR and the Hopf Berry dipole, organised by the values of $n$ and physical origins.}
\begin{ruledtabular}
\begin{tabular}{lcccc}
 & $\sigma^{\textrm{lf, h}}$ & $\sigma^{\textrm{lf, bc}}$ & $\sigma^{\textrm{lf, m}}$ & $\sigma^{\textrm{lf, conc}}$ \\
\colrule
$n=1$, $\sigma_{yz}$
 & $B_x$
 & \makecell[c]{$B_x\,B_z, B_x^3,$   $B_x\,B_z^2$}
 & \makecell[c]{$B_x\,B_z, B_x^3,$   $B_x\,B_z^2$}
 & $B_x^3, B_x\,B_z^2$ \\[4pt]
$n=2$, $\sigma_{zz}$
 & $B_x^2$
 & $B_z\,B_x^2$
 & $B_z\,B_x^2$
 & $0$ \\[4pt]
$n=2$, $\sigma_{xz}$
 & $0$
 & $B_x^3, B_x\,B_z^2$
 & $B_x^3, B_x\,B_z^2$
 & $0$ \\[4pt]
$n=3$, $\sigma_{yz}$
 & $B_x^3, B_x\,B_z^2$
 & $0$ & $0$ & $0$ \\
\end{tabular}
\end{ruledtabular}
\end{table}

%======================================================================
\section{Magnetoelectric conductivity of the Hopf semimetal}
\label{sechopfres}
%======================================================================

Similar to the treatment of the VNR case, we expand $\bar\sigma_{ij}$, $\sigma^{\rm{ah}}_{ij}$, and $\sigma^{\rm{lf}}_{ij}$ consistently up to cubic order in $\bs B$, retaining odd as well as even powers, as anticipated in Secs.~\ref{secsymmetry} and~\ref{secmethod}. The HSM breaks time-reversal symmetry, since its dipolar BC satisfies $\bs\Omega_s(-\bs k)=\bs\Omega_s(\bs k)$ rather than $-\bs\Omega_s(\bs k)$, as shown in Appendix~\ref{sec:AHeval}. As in Sec.~\ref{secvnr}, Onsager reciprocity~\cite{onsager1,onsager2,onsager3} must therefore be applied with the combined operation that reverses $\bs B$, $\bs\Omega_s$, and $\bs m_s$ simultaneously. In its generalised form, the relation
\begin{align}
\sigma_{ij}(\bs B,\bs\Omega_s,\bs m_s)=\sigma_{ji}(-\bs B,-\bs\Omega_s,-\bs m_s)\,,
\end{align}
requires the diagonal components to be even under the simultaneous reversal of $\bs B$, $\bs\Omega_s$, and $\bs m_s$. It does not require them to be even in $\bs B$ alone, since an odd power of $\bs B$ can be compensated by an odd total power of the dipolar $\bs\Omega_s$ and $\bs m_s$.

The mirror symmetry of Sec.~\ref{secsymmetry} restricts the diagonal components exactly as in Sec.~\ref{secvnr}. They contain only even total powers of the in-plane field components $B_x$ and $B_y$, and $B_z$ is the sole field component that can carry an odd power. Here, the extra factor $\cos\gamma$ in the dipolar $\bs\Omega_s$ and $\bs m_s$ makes $\Omega_s^z$ and $m_s^z$ even in $k_z$, while $\Omega_s^{x,y}$ and $m_s^{x,y}$ are odd. Each power of $B_z$ then multiplies a factor that is even in $k_z$ and survives the integration. For a BC-monopole, the isotropic texture $\bs\Omega_s\propto\hat{\bs k}$ makes $\Omega_s^z$ odd in $k_z$ and forces the relevant angular average to vanish, leaving only even powers of $\bs B$. The Hopf node's dipolar factor reduces this spherical symmetry to $C_{\infty z}$ and removes the accidental cancellation.

Odd powers of $B_z$ are thus permitted in the diagonal components of set-ups~II and~III, but not in those of set-up~I, where $\bs B$ lies in the $xy$-plane. They are indeed generated in the individual BC- and OMM-dependent sub-parts of $\bar\sigma_{xx}$ and $\sigma^{\rm{lf}}_{xx}$ in set-up~II, and of $\bar\sigma_{zz}$ and $\sigma^{\rm{lf}}_{zz}$ in set-up~III. The dipolar factor also activates odd-in-$\bs B$ terms in the off-diagonal components $\bar\sigma_{zx}$, $\sigma^{\rm{ah}}_{zx}$, and $\sigma^{\rm{lf}}_{zx}$ in set-up~I, in $\bar\sigma_{zx}$ and $\sigma^{\rm{lf}}_{zx}$ in set-up~II, and in $\bar\sigma_{xz}$ and $\sigma^{\rm{lf}}_{xz}$ in set-up~III. Thus same phenomenon recurs here in a three-band system, whose dipolar BC and OMM originate from an $SU(3)$ node algebra rather than from the toroidal Fermi-surface topology of the VNR. This confirms that it is a generic consequence of the dipolar structure of a BC field and not of a specific microscopic realisation. Notation-wise, we follow the convention outlined in Eq.~\eqref{eqbarsigdec}.

%%%%%%%%%%%%%%%%%%
\subsection{Set-up I: 
\texorpdfstring{$\bs{E}=E_x\, {\bs{\hat x}}$}{E}, 
\texorpdfstring{$ \bs{B}=B_x \,{\bs{\hat x}} + B_y \,{\bs{\hat y}} $}{B}, and
\texorpdfstring{$ \bs{B}^2 = B_x^2 + B_y^2 $}{BB}}
\label{secset1-hopf}

The non-AH part comprises
\begin{align}
\label{eqhfI-nonah}
\sigma^{\rm d}_{xx} &= \frac{\tau\,e^2\,\mu^2}{6\,\pi^2\,v_0}\,,\quad
\sigma^{\rm bc}_{xx} = \frac{\tau\,e^4\,v_0^3
\, (24\,B_x^2+B_y^2)}{210 \,\pi^2\,\mu^2}\,,\quad
\sigma^{\rm m}_{xx} = \frac{\tau\,e^4\,v_0^3
\,(16\,B_x^2+3\,B_y^2)}{840 \,\pi^2\,\mu^2}\,,\quad
%%%%%%%%%%%%%
\sigma^{\rm conc}_{xx} = - \, \frac{\tau\,e^4\,v_0^3
\,(29\,B_x^2+5\,B_y^2)}{420 \,\pi^2\,\mu^2}\,,\nn
%%%%%%%%%%%%%%%%%
\sigma^{\rm bc}_{yx} &= \frac{23\,\tau\,e^4\,v_0^3\,B_x\,B_y}{210 \,\pi^2\,\mu^2}\,,\quad
\sigma^{\rm m}_{yx} = \frac{13\,\tau\,e^4\,v_0^3\,B_x\,B_y}{840 \,\pi^2\,\mu^2}\,,\quad
\sigma^{\rm conc}_{yx} = - \, \frac{2\,\tau\,e^4\,v_0^3\,B_x\,B_y}{35\,\pi^2\,\mu^2}\,,\nn
\sigma^{\rm bc}_{zx} &= - \, \frac{\tau\,e^3\,v_0\,B_x
\,({\boldsymbol B}^2 \, e^2\,v_0^4+14\,\mu^4)}{105\,\pi^2\,\mu^4}\,,\quad
%%%%%%%%%%%%%%
\sigma^{\rm m}_{zx} = \frac{\tau\,e^3\,v_0\,B_x
\,(-5 {\boldsymbol B}^2 \, e^2\,v_0^4+84\,\mu^4)}{1680 \,\pi^2\,\mu^4}\,,\quad
\sigma^{\rm conc}_{zx} = \frac{\tau\,e^5\,v_0^5\,B_x {\boldsymbol B}^2}{280 \,\pi^2\,\mu^4}\,.
\end{align}

%%%%%%%%%%%%%%%%%%%%%%%%%%%%%
\begin{figure}[t!]
\centering
\subfigure{\includegraphics[width=0.75 \textwidth]{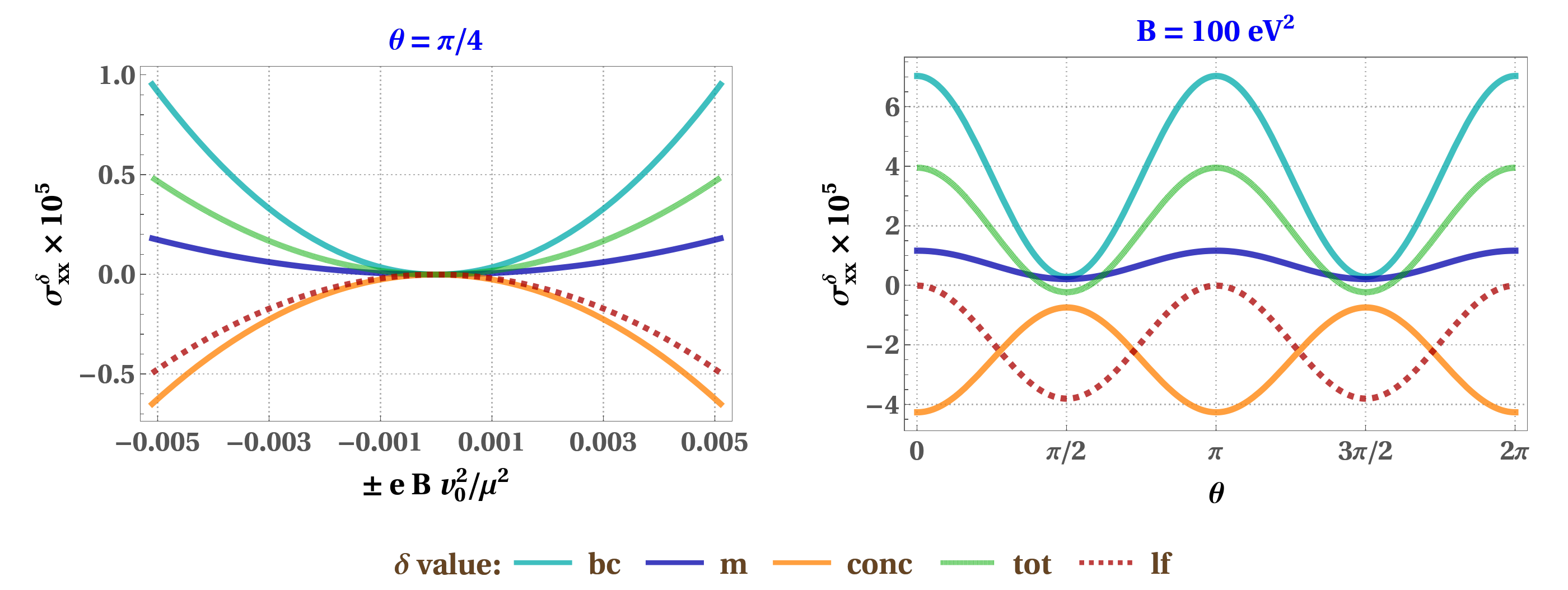}}
\subfigure{\includegraphics[width=0.75 \textwidth]{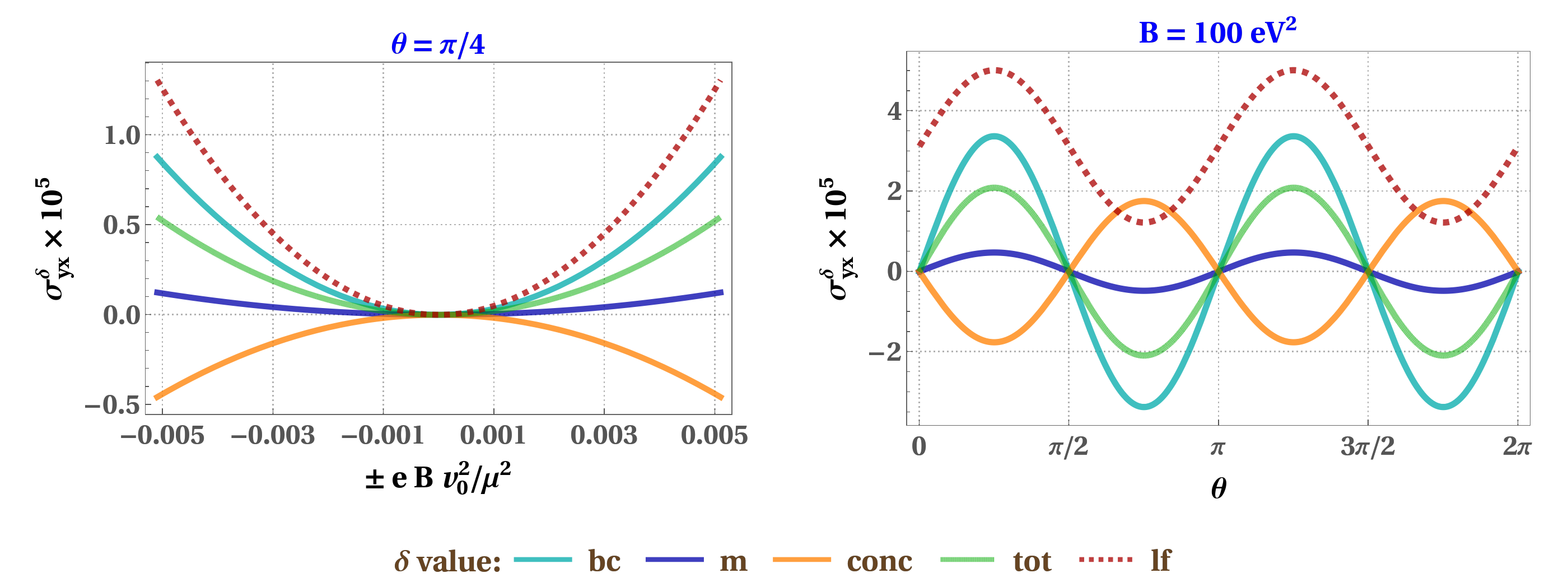}}
\subfigure{\includegraphics[width=0.75 \textwidth]{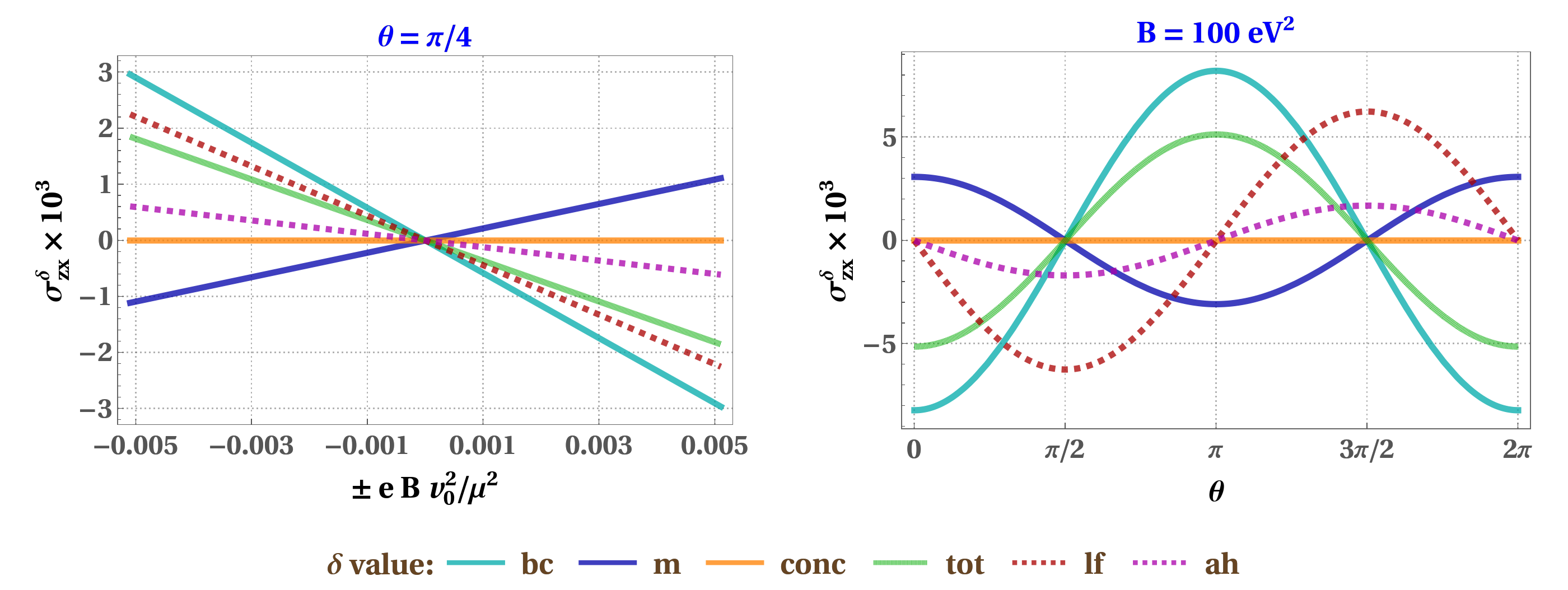}}
\caption{Set-up I for HSM: Behaviour of conductivity curves (in eV) as functions of ${e\, B \,v_0^2} / {\mu^2}$ and $\theta$.
\label{figset1hf}}
\end{figure}
%%%%%%%%%%%%%%%%

The AH part comprises
\begin{align}
\label{eqhfI-ah}
\sigma^{\rm{ah}}_{yx} &= \frac{e^2}{840 \,\pi^2\,v_0}
\left(-140\,\mu+\frac{3 \, {\boldsymbol B}^2 \, e^2\,v_0^4}{\mu^3}\right ), \quad
\sigma^{\rm{ah}}_{zx} = - \, \frac{e^3\,v_0\,B_y
\left( {\boldsymbol B}^2 \, e^2\,v_0^4+14\,\mu^4 \right )}
{840 \,\pi^2\,\mu^5}\,.
\end{align}

Let us now consider the LFO-induced part. The out-of-plane conductivity is
\begin{align}
& \sigma^{\rm{lf}}_{zx} = 
\sigma^{\rm{lf}, 2}_{zx}+\sigma^{\rm{lf}, 3}_{zx}
+\sigma^{\rm{lf}, 4}_{zx}\,,\nn &
%%%%%%%%%%%%%%%%%%%%%%%%%%%%%%%%%
\sigma^{\rm{lf}, 2}_{zx} = - \, \frac{\tau^2 \,e^3\,v_0\,B_y}{840 \,\pi^2\,\mu^3}
\left (23\, {\boldsymbol B}^2 \, e^2\,v_0^4 + 140\,\mu^4\right),\quad
%%%%%%%%%%%%%%%%%
\sigma^{\rm{lf}, 3}_{zx} = \frac{\tau^3 \,e^5\,v_0^5\,B_x \,{\boldsymbol B}^2}
{12\,\pi^2\,\mu^2}\,,\quad
%%%%%%%%%%%%%%%%
\sigma^{\rm{lf}, 4}_{zx} = \frac{\tau^4 \,e^5\,v_0^5\,B_y {\boldsymbol B}^2}
{6\,\pi^2\,\mu}\,.
\end{align}
The in-plane response is
\begin{align}
\label{eqhfI-lfin}
\sigma^{\rm{lf}}_{xx} &= - \, \frac{\tau^3\, e^4\,v_0^3\,B_y^2}{6\,\pi^2}\,,\quad
\sigma^{\rm{lf}}_{yx} = \frac{\tau^2 \, e^4\,v_0^3}{12\,\pi^2\,\mu}
\left( {\bs B}^2+ 2\, \tau\, \mu \,B_x\,B_y\right ).
\end{align}
The sources of the above terms are explained in Appendix~\ref{app-lf-hopf1}.

In $\bar\sigma_{xx}$, the ratios of the $B_x^2$ to $B_y^2$ coefficients are
$24:1$ for the BC-only part, $16:3$ for the OMM-only part, and $29:5$ for the
concurrent part, with positive, positive, and negative signs for all parameters.
Unlike the VNR, no sub-part changes sign between the $B_x^2$ and $B_y^2$ channels.
The three planar-Hall coefficients are proportional to $B_xB_y$ with the same sign
pattern. In the out-of-plane $\bar\sigma_{zx}$, the linear coefficients of the
BC-only and OMM-only parts are $-14\tau e^3v_0/(105\pi^2)$ and
$\tau e^3v_0/(20\pi^2)$, standing in the parameter-independent ratio $-8/3$ and
leaving a total of $-\tau e^3v_0B_x/(12\pi^2)$, while the concurrent part appears
only at cubic order. The corresponding ratio for the VNR is $-7/3$, so the partial
cancellation of the BC-only and OMM-only pieces is a common feature of both
systems.

For the AH part, Eq.~\eqref{eqhfI-ah} gives
$\sigma^{\rm{ah}}_{yx}=-e^2\mu/(6\pi^2v_0)+e^4v_0^3\bs B^2/(280\pi^2\mu^3)$,
which depends on the field only through $\bs B^2$ and is therefore exactly
independent of $\theta$. The out-of-plane $\sigma^{\rm{ah}}_{zx}$ is odd in
$B_y$, with linear coefficient $-e^3v_0/(60\pi^2\mu)$. For the LFO part,
Eq.~\eqref{eqhfI-lfin} shows that
$\sigma^{\rm{lf}}_{xx}=-\tau^3e^4v_0^3B_y^2/(6\pi^2)$ depends on $B_y^2$ alone and
vanishes for $\bs B\parallel\bs E$. The planar-Hall $\sigma^{\rm{lf}}_{yx}$ is
proportional to $\bs B^2+2\tau\mu B_xB_y$, that is, to
$1+\tau\mu\sin(2\theta)$, and remains positive at all angles for $\tau\mu<1$. The
term $\sigma^{\rm{lf}}_{zx}\simeq-\tau^2e^3v_0\mu B_y/(6\pi^2)$ is odd
in $B_y$ and exceeds the linear AH term by a factor of $10\tau^2\mu^2$.

The numerical results for the parameters of Table~\ref{tab-params} are displayed
in Fig.~\ref{figset1hf} (field-strength dependence at $\theta=\pi/4$ in the left
panels, angular dependence at fixed $|\bs B|$ in the right panels). The BC-only and
OMM-only curves of $\sigma_{xx}$ and $\sigma_{yx}$ are positive, while the
concurrent curves are negative. The $\sigma_{xx}$ curves peak at $\theta=0$ and
$\pi$ and are weakest near $\theta=\pi/2$ because of the coefficient ratios above,
whereas the LFO curve vanishes at $\theta=0$ and $\pi$ and is largest in magnitude
at $\theta=\pi/2$. The non-AH $\sigma_{yx}$ curves vanish at $\theta=0$, $\pi/2$,
$\pi$, and $3\pi/2$ and are extremal near $\theta=\pi/4$, following the
$\sin(2\theta)$ profile, while the LFO curve remains positive and is modulated by
the factor $1+\tau\mu\sin(2\theta)$. In $\sigma_{zx}$, the BC-only curve is
negative and the OMM-only curve positive, and the two partly cancel. The BC-only
and total curves follow $-\cos\theta$, the OMM-only curve follows $+\cos\theta$,
and the LFO and AH curves follow $-\sin\theta$.

The parity structure of the dipolar response is fixed by the analytic expressions.
Every sub-part of $\sigma_{xx}$ and $\sigma_{yx}$ carries only even powers of
$\bs B$, so both are symmetric under $\bs B\to-\bs B$. The out-of-plane
$\sigma_{zx}$ is odd in $\bs B$, as every sub-part carries an overall factor of
$B_x$ or $B_y$; consequently, the field-strength curves of Fig.~\ref{figset1hf}
pass through the origin. This odd-in-$\bs B$ term is permitted by the generalised
Onsager relation used in Secs.~\ref{secvnr} and~\ref{sechopfres}, since the sign
change of $\bs B$ is compensated by that of the dipolar $\bs\Omega_s$ and
$\bs m_s$. In this set-up it appears only in the off-diagonal column of the
tensor, while the diagonal $\sigma_{xx}$ and the in-plane transverse $\sigma_{yx}$
remain even.

%%%%%%%%%%%%%%%%%%%%%%%%%%%%%%%%%%%%%%%%%%%%%%%%%%%%%%%%%%%%%%%
\subsection{Set-up II: \texorpdfstring{$\bs{E}=E_x\,  {\bs{\hat x}}$}{E-field},
\texorpdfstring{$ \bs{B}= B_x \, {\bs{\hat x}} + B_z\, \bs{\hat z} $}{B-field}, and
\texorpdfstring{$ \bs{B}^2 = B_x^2 + B_z^2 $}{BB}}
\label{secset2-hopf}

%%%%%%%%%%%%%%%%%%%%%%%%%%%%%
\begin{figure}[t!]
\centering
\subfigure{\includegraphics[width=0.75 \textwidth]{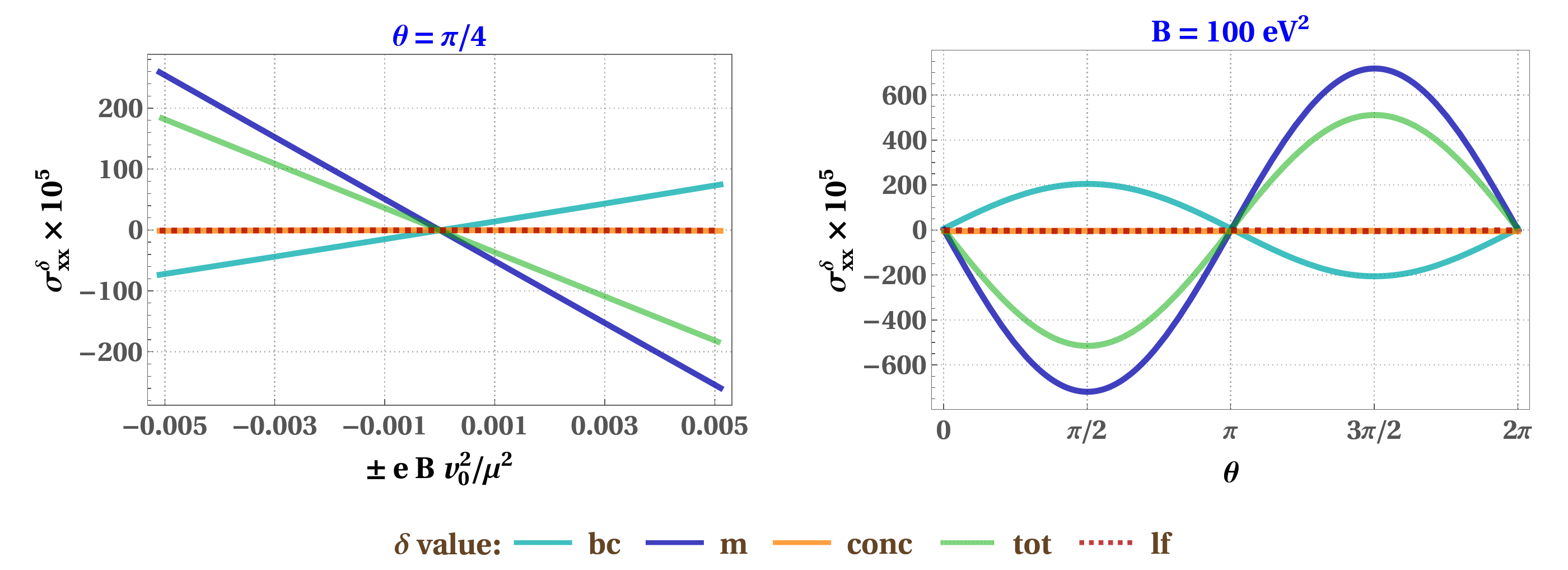}}
\subfigure{\includegraphics[width=0.75 \textwidth]{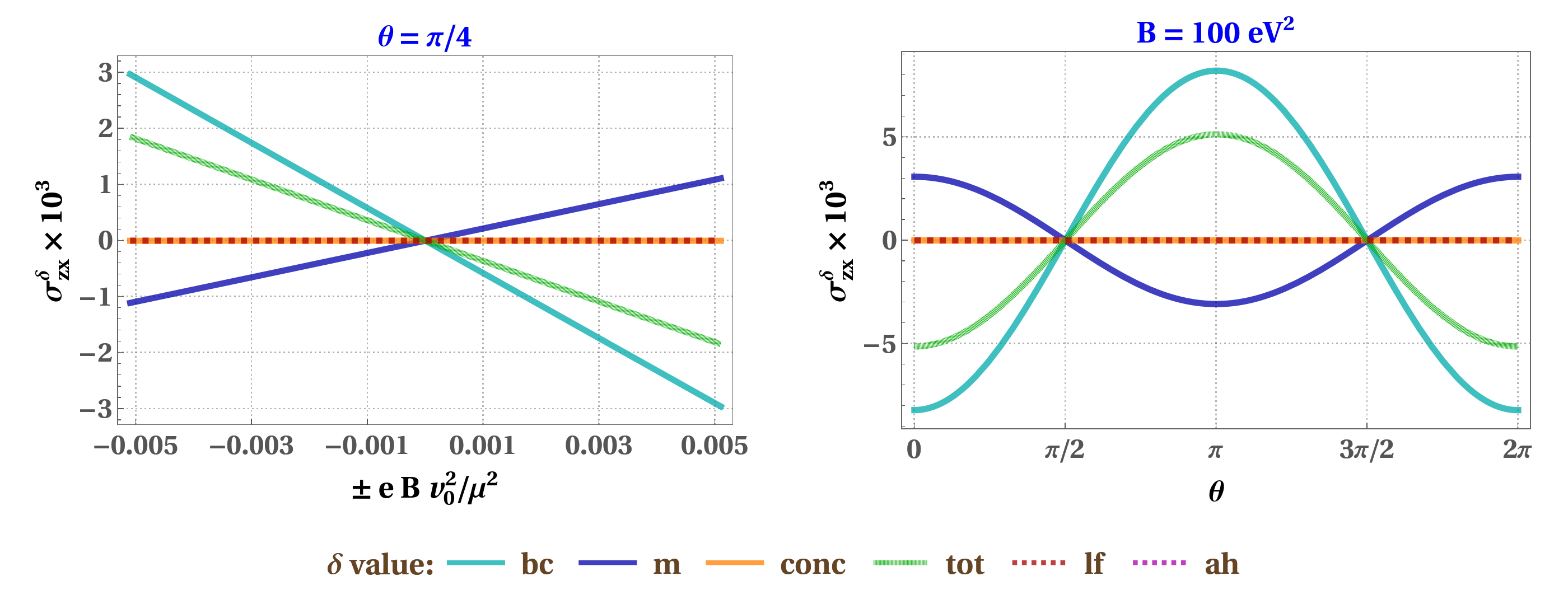}}
\subfigure{\includegraphics[width=0.75 \textwidth]{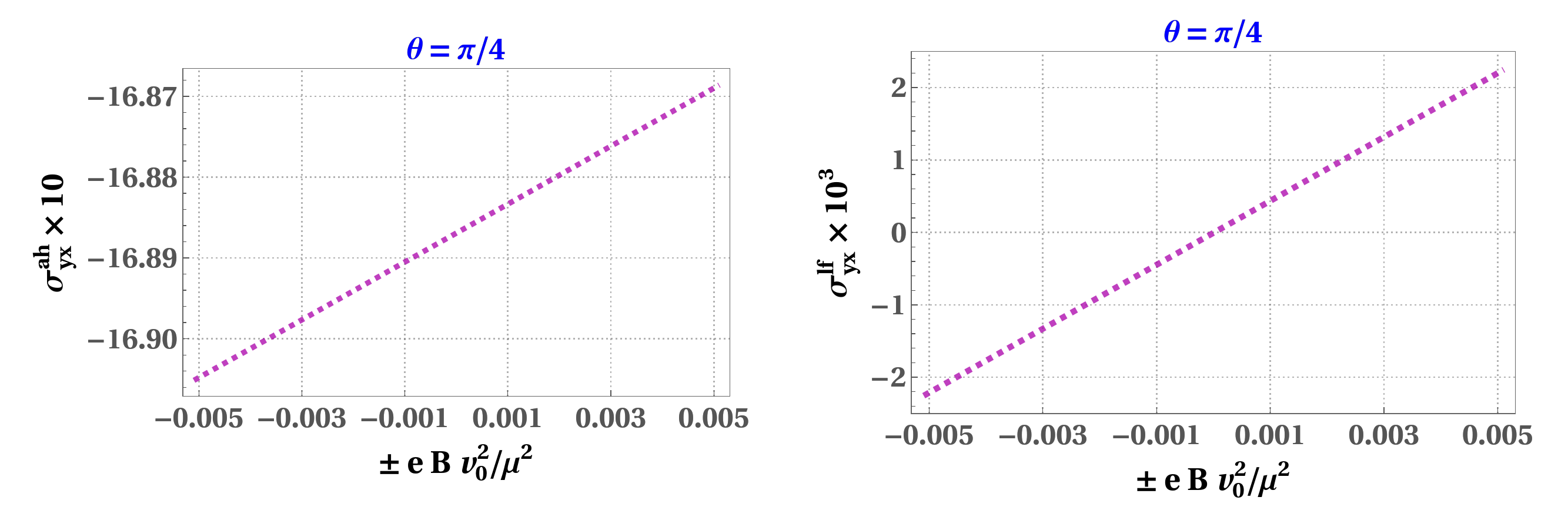}}
\caption{Set-up II for HSM: Behaviour of the conductivity curves (in eV) as functions of ${e\, B \,v_0^2} / {\mu^2}$ and $\theta$. For the out-of-plane components, we have omitted showing the $\theta$-dependence.\label{figset2hf}}
\end{figure}
%%%%%%%%%%%%%%%%%%%%%%%%%%%%%%%%

The non-AH part is given by
\begin{align}
\sigma^{{\rm d}}_{xx}& =\frac{\tau\, e^2\,\mu^2  }{6\,\pi^2\, v_0} ,\quad
%%%%%%%%%%%%%%%%%%
\sigma^{\rm bc}_{xx}=\frac{\tau\,
e^3 \,v_0 }{630 \, \pi^2 \, \mu^4}
\left[ B_z \left( 36 \, B_x^2+5 \, B_z^2\right)e^2\, v_0^4
+9\left(8 \, B_x^2+B_z^2\right)e \, v_0^2 \, \mu^2
+
21 \, B_z \, \mu^4\right] , \nn
\sigma^{\rm m}_{xx}&=\frac{\tau \,
e^3 \, v_0 }{5040 \, \pi^2 \, \mu^4}
\left[B_z\left(63 \, B_x^2+50 \, B_z^2\right)e^2\, v_0^4+
6 \left(16 \, B_x^2+9 \, B_z^2\right)e \, v_0^2 \,\mu^2
-588 \, B_z \, \mu^4\right], \nn
\sigma^{\rm conc}_{xx}&=
-\frac{ \tau\, e^4 \,v_0^3}{840 \,\pi^2 \,\mu^4} \left[
B_z  \left(21\, B_x^2 + 20 \,B_z^2 \right)e \, v_0^2
+ 2  \left(29 \, B_x^2 + 15\, B_z^2 \right)\mu^2\right] , \nn
%%%%%%%%%%%%%%%%%
\sigma^{\rm bc}_{zx}&= \frac{\tau \, e^3\,v_0\, B_x  }{
210 \, \pi^2 \, \mu^4}
\left[\left(-2 \, B_x^2 + 5\, B_z^2 \right) \, e^2\, v_0^4
+13\, B_z \, e\, v_0^2\, \mu^2
-28 \, \mu^4 \right], \nn
%%%%%%%%%%%%%%%%%%%%%%%
\sigma^{\rm m}_{zx} & = \frac{\tau  \, e^3 \,v_0\, B_x }{1680 \, \pi^2 \, \mu^4}
\left[-\left(5 \, B_x^2
+4 \, B_z^2\right)e^2\, v_0^4
+8 \, B_z \, e \,v_0^2\, \mu^2
+84 \, \mu^4\right], \quad
%%%%%%%%%%%%%%%%%
\sigma^{\rm conc}_{zx} = 
\frac{\tau\, e^4 \,v_0^3\,B_x  }{280 \, \pi^2 \, \mu^4}
\left[B_x^2 \, e\, v_0^2-6 \, B_z \, \mu^2\right].
\end{align}

In this geometry, $B_z$ couples to the dipole axis and the AH response takes the form of
\begin{align}
\label{eqhfII-ah}
\sigma^{\rm{ah}}_{yx} &= \frac{e^2}{2520 \,\pi^2\,v_0\,\mu^5}
\Big[\,5\,B_z \left( 3\,B_x^2 + 7\,B_z^2 \right) e^3\,v_0^6
+ 9 \left( B_x^2 + 5\,B_z^2 \right) e^2\,v_0^4 \,\mu^2 
+ 126\,B_z \,e\,v_0^2\,\mu^4 - 420\,\mu^6 \Big]\,.
\end{align}

The steps to obtain the LFO-induced contributions are detailed in Appendix~\ref{app-lf-hopf2}. The final forms are shown below:
%%%%%%%%%%%
\begin{align}
\sigma^{\text{lf}}_{yx} & =
\sigma_{yx}^{\rm{lf}, 2}+\sigma_{yx}^{\rm{lf}, 4} ,\nn
%%%%%%%%%%%%%%%%%%%
\sigma_{yx}^{\rm{lf}, 2}&=\frac{\tau^2\, e^3\, v_0 }{840 \, \pi^2 \, \mu^3}
\left[B_z\left(13 \, B_x^2-18 \, B_z^2\right)e^2\, v_0^4
+70\left(B_x^2-B_z^2\right)e \, v_0^2\,\mu^2
+140 \, B_z \, \mu^4\right] ,\quad
%%%%%%%%%%%%%%%%%%%%%%%%%%%%%%%
\sigma_{yx}^{\rm{lf}, 4} = -\frac{\tau^4  \, e^5\, v_0^5}{6 \, \pi^2 \, \mu}
\,B_z\,{\bs B}^2 \,,
\end{align}
%%%%%%%%%%%%%%%%%%%%%%%%%%%%%%%%%%%%%%%%%%
\begin{align}
    \sigma^{\rm{lf}}_{xx} &=\frac{ \tau^3\, e^4 \, v_0^3 \, B_z }{60 \, \pi^2 \, \mu^2}\left[
-8 \, B_x^2 \, e\, v_0^2+5 \, B_z\left(B_z \, e \, v_0^2-2 \, \mu^2\right)
\right] , \quad
%%%%%%%%%%%%%%%%%%%%%%%%%
    \sigma^{\text{lf}}_{zx} 
=\frac{\tau^3  \, e^4 \, v_0^3\, B_x
}{60 \, \pi^2 \, \mu^2}\left[\left( 5 \, B_x^2 - 8 \, B_z^2\right)\, e\, v_0^2
+10 \, B_z\mu^2\right].
\end{align}

For the AH part, the term $126B_zev_0^2\mu^4$ in Eq.~\eqref{eqhfII-ah}
contributes a linear tilt $e^3v_0B_z/(20\pi^2\mu)$, which arises because $B_z$
couples directly to the dipole axis through $\Omega^z_s$. In contrast to
set-up~I, $\sigma^{\rm{ah}}_{yx}$ is therefore no longer a function of
$\bs B^2$ alone and acquires a $\theta$-dependence. In the LFO sector,
$\sigma^{\rm{lf}}_{yx}$ contains the linear term
$\tau^2e^3v_0\mu B_z/(6\pi^2)$; the BC-only and OMM-only parts of
$\sigma^{\rm{lf}}_{xx}$ carry even terms in $\{B_x^2,B_z^2\}$ at
$\mathcal{O}(B^2)$, together with odd cubic terms of the form
$B_z\{B_x^2,B_z^2\}$; and the out-of-plane component $\sigma^{\rm{lf}}_{zx}$
contains a bilinear $B_xB_z$ at $\mathcal{O}(B^2)$, together with the
cubic structures $B_x^3$ and $B_xB_z^2$ inherited from the BC and OMM dipoles.
For the non-AH part, the diagonal $\bar\sigma_{xx}$ contains, alongside the even
terms $\{B_x^2,B_z^2\}$, the odd-in-$B_z$ terms $\{B_z,B_zB_x^2,B_z^3\}$
inherited from the dipolar $\bs\Omega_s$ and $\bs m_s$. $\bar\sigma_{zx}$ coincides with $\bar\sigma_{xz}$ of set-up~III through the symmetry relation of Eq.~\eqref{eq:theta_covariance} at $B_y=0$ and is therefore
given by Eq.~\eqref{eqhfIII-nonah-xz}, with an overall factor of $B_x$. The numerical results for the parameters of Table~\ref{tab-params}, shown in
Fig.~\ref{figset2hf}, confirm this behaviour. In the left panels, the non-AH
$\bar\sigma_{xx}$ is odd and nearly linear in $B_z$, with an OMM-only part
several times larger than the BC-only part and of opposite slope. The transverse
$\bar\sigma_{zx}$ is linear in $B_x$, again with BC-only and OMM-only curves of
opposite slope. The right panels show the $\sin\theta$ profile of $\bar\sigma_{xx}$,
vanishing at $\theta=0$ and $\pi$ where $B_z=0$, and the $\cos\theta$ profile of
$\bar\sigma_{zx}$, vanishing at $\theta=\pi/2$ and $3\pi/2$ where $B_x=0$. For
$\sigma_{yx}$, the $\theta$-dependence is omitted, and both panels display the AH
and LFO curves against the field strength. The AH curve is the plateau with its
small linear tilt, and the LFO curve is linear in $B_z$ at low fields.

The odd-in-$\bs B$ behaviour in the diagonal components does not violate Onsager
reciprocity. The diagonal of the full tensor is constrained to be even under the
joint reversal $(\bs B,\bs\Omega_s,\bs m_s)\to(-\bs B,-\bs\Omega_s,-\bs m_s)$,
and each odd-in-$B_z$ term in $\sigma_{xx}$ is accompanied by an odd total power
of the dipolar $\bs\Omega_s$ and $\bs m_s$.

%%%%%%%%%%%%%%%%%%%%%%%%%%%%%

\subsection{Set-up III: 
\texorpdfstring{$\bs{E}=E_z\,\bs{\hat{z}}$}{E} and 
\texorpdfstring{$\bs{B}= B_x\, \bs{\hat{x}}+B_z\,\bs{\hat{z}}$}{B}, and
\texorpdfstring{$\bs{B}^2 = B_x^2 + B_z^2$}{BB}}
\label{secset3-hopf}

%%%%%%%%%%%%%%%%%%%%%%%%%%%%%
\begin{figure}[t!]
\centering
\subfigure{\includegraphics[width=0.75 \textwidth]{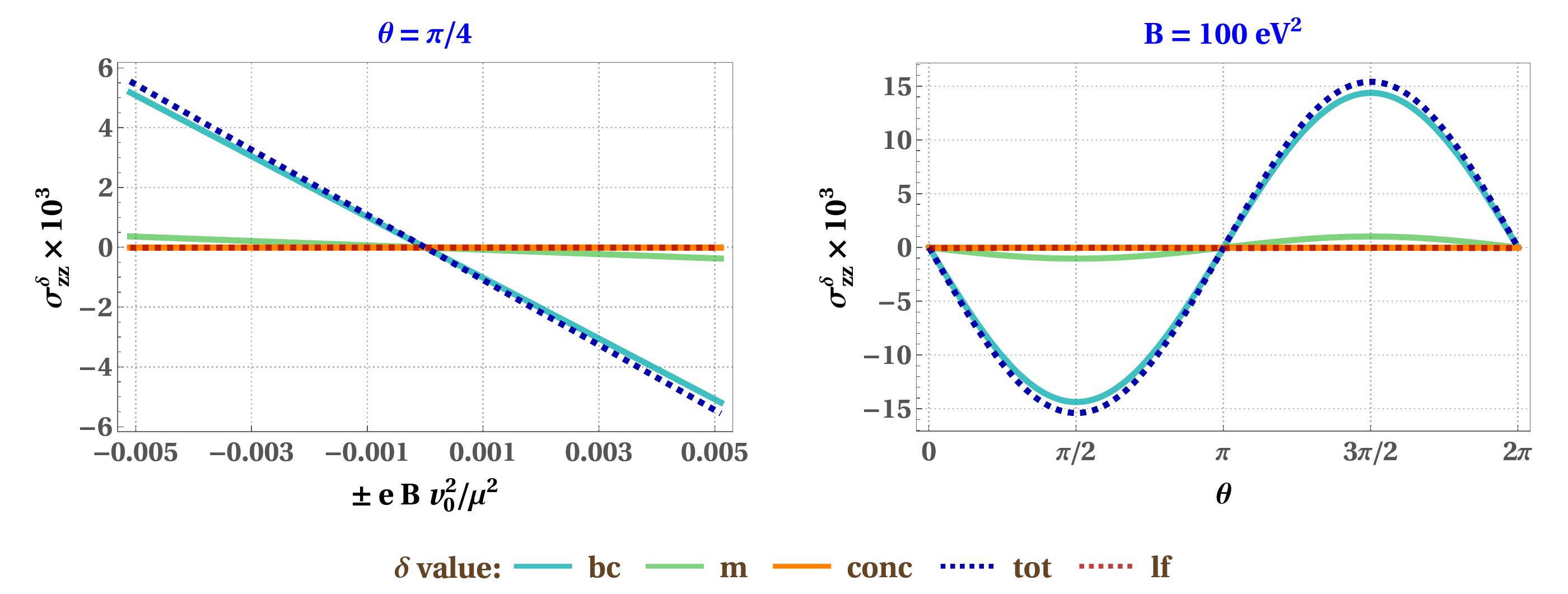}}
\subfigure{\includegraphics[width=0.75 \textwidth]{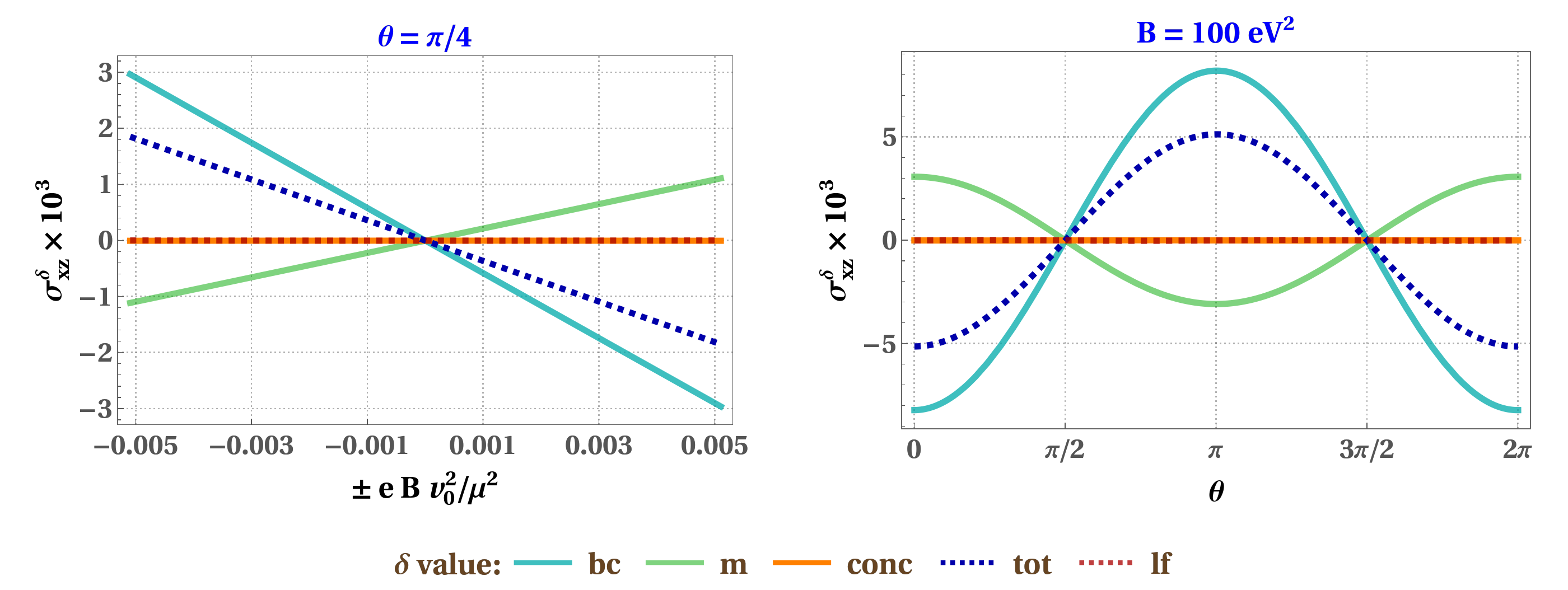}}
\subfigure{\includegraphics[width=0.75 \textwidth]{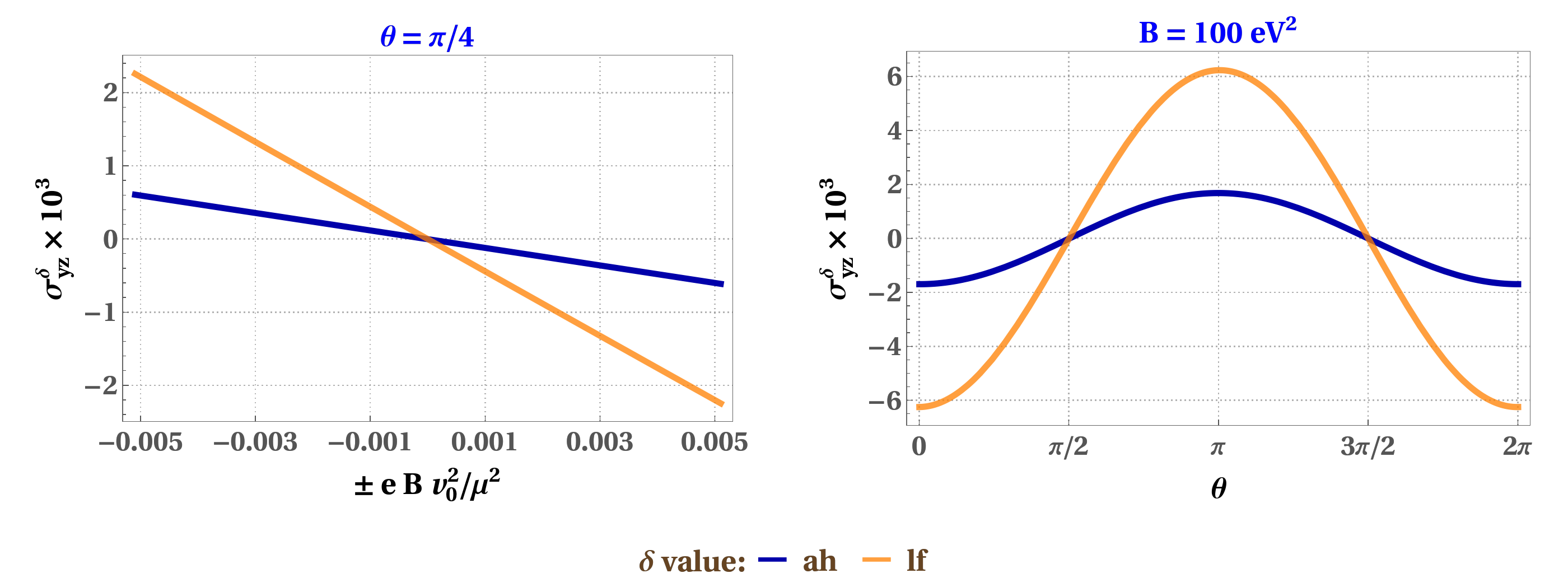}}
\caption{Set-up III for HSM: Behaviour of the conductivity curves (in eV) as functions of ${e\, B \,v_0^2} / {\mu^2}$ and $\theta$.\label{figset3hf}}
\end{figure}
%%%%%%%%%%%%%%%%%%%%%%%%%%%%%%%%

For the non-AH part, we have
\begin{align}
\label{eqhfIII-nonah}
& \sigma^{\rm d}_{zz} = \frac{\tau \,e^2 \,\mu^2}{6\, \pi^2\,v_0}\,,\quad
%%%%%%%%%%%%%%%%%%
\sigma^{\rm bc}_{zz} = \frac{\tau \,e^3 \,v_0}{630 \, \pi^2\,\mu^4}
\Big[ B_z \left( -3\,B_x^2 + 8\,B_z^2 \right) e^2\,v_0^4
+ 3 \left( 3\,B_x^2 + 8\,B_z^2 \right) e\,v_0^2\,\mu^2
- 147\,B_z\,\mu^4 \Big]\,,\nn &
%%%%%%%%%%%%%%%%%%
\sigma^{\rm m}_{zz} = \frac{\tau \,e^3 \,v_0}{5040 \, \pi^2\,\mu^4}
\Big[ B_z \left( 33\,B_x^2 + 26\,B_z^2 \right) e^2\,v_0^4
+ 6 \left( 16\,B_x^2 + 17\,B_z^2 \right) e\,v_0^2\,\mu^2
- 84\,B_z\,\mu^4 \Big]\,,\nonumber \\
%%%%%%%%%%%%%%%%%%
& \sigma^{\rm conc}_{zz} = \frac{\tau \,e^4 \,v_0^3}{840 \, \pi^2\,\mu^4}
\Big[ 5\,B_z \left( 3\,B_x^2 + 2\,B_z^2 \right) e\,v_0^2
- 2 \left( B_x^2 + 5\,B_z^2 \right) \mu^2 \Big],
\end{align}
%%%%%%%%%%%%%%%%%%%%%%%%
\begin{align}
\label{eqhfIII-nonah-xz}
& \sigma^{\rm bc}_{xz} = \frac{\tau \,e^3 \,v_0\,B_x}{210 \, \pi^2\,\mu^4}
\Big[ \left( -2\,B_x^2 + 5\,B_z^2 \right) e^2\,v_0^4
+ 13\,B_z \,e\,v_0^2\,\mu^2 - 28\,\mu^4 \Big],\nn &
%%%%%%%%%%%%%%%%%
\sigma^{\rm m}_{xz} = \frac{ -\, \tau \,e^3 \,v_0\,B_x}
{1680 \, \pi^2\,\mu^4}
\Big[ \left( 5\,B_x^2 + 4\,B_z^2 \right) e^2\,v_0^4
+ 8\,B_z \,e\,v_0^2\,\mu^2 + 84\,\mu^4 \Big]\,,\quad
%%%%%%%%%%%%%%%%
\sigma^{\rm conc}_{xz} 
= \frac{\tau \,e^4 \,v_0^3\,B_x}{280 \, \pi^2\,\mu^4}
\left( B_x^2\,e\,v_0^2 - 6\,B_z\,\mu^2 \right).
\end{align}
By Eq.~\eqref{eq:theta_covariance}, at $B_y=0$, this coincides with $\bar\sigma_{zx}$ of set-up~II as expected. The out-of-plane transverse component $\bar\sigma_{yz}$ vanishes in the non-AH response.

The AH part is
\begin{align}
 \sigma^{\rm{ah}}_{yz} &= 
 \frac{ -\,e^3\,v_0 \, B_x \left[\left(B_x^2\,+\,5\,B_z^2\right) e^2\,v_0^4 
 + 6\,B_z\,e\,v_0^2\, \mu^2\,+\,14\,\mu^4\right]}{840\,\pi^2\,\mu^5} \,.
\end{align}
The LFO-induced parts add up to
\begin{align}
\sigma^{\text{lf}}_{yz} &
=
\sigma_{yz}^{\rm{lf}, 2}+\sigma_{yz}^{\rm{lf}, 4} ,\nn
\sigma_{yz}^{\rm{lf}, 2}&=\frac{\tau^2\, e^3 \,v_0\,
B_x  }{840 \, \pi^2 \, \mu^3}
\left[ \left( -23 \, B_x^2 
+8 \, B_z^2 \right)\, e^2\, v_0^4
+140 \, B_z \, e \, v_0^2 \, \mu^2
-140 \, \mu^4\right] \, , \quad 
%%%%%%%%%%%%%%%%%%%%%%%%%%%%%
\sigma_{yz}^{\rm{lf}, 4}=\frac{\tau^4 \,e^5 \,v_0^5\, B_x\,{\bs B}^2}
{6 \, \pi^2 \, \mu}\,,
\end{align}
%%%%%%%%%%%%%%%%%%%%%%%
\begin{align}
  \sigma^{\text{lf}}_{xz} &
=  \frac{ \tau^3\, e^4 \, v_0^3 \, B_x 
}{60 \, \pi^2 \, \mu^2}
\left[3 \, B_x^2 \, e \, v_0^2
+10 \, B_z\left(- B_z \, e\, v_0^2 +\mu^2\right)\right] \, , \quad 
  \sigma^{\text{lf}}_{zz}=\frac{ \tau^3\, e^4 \, v_0^3\,
B_x^2 }{60 \, \pi^2 \, \mu^2}
\left[13 \, B_z \, e\, v_0^2
-10 \, \mu^2\right]
\end{align}
%=%%%%
The term-by-term derivations are shown in Appendix~\ref{app-lf-hopf3}.

In $\bar\sigma_{zz}$, every sub-part contains, alongside the even terms
$\{B_x^2,B_z^2\}$, the odd-in-$B_z$ terms $\{B_z,B_z\,B_x^2,B_z^3\}$ inherited from
the dipolar $\bs\Omega_s$ and $\bs m_s$. The transverse $\bar\sigma_{xz}$ carries
an overall factor of $B_x$; at $B_z=0$ its linear coefficients reduce to the total
$-\tau e^3v_0B_x/(12\pi^2)$ of set-up~I, and by Eq.~\eqref{eq:theta_covariance}
it coincides with $\bar\sigma_{zx}$ of set-up~II. $\sigma^{\rm{ah}}_{yz}$ carries an overall factor of $B_x$ with linear coefficient $-e^3\,v_0/(60\,\pi^2\,\mu)$. For the kinematic
$n=1$ term in the LFO-induced part, $-\tau^2 \,e^3\,v_0\,\mu\, B_x/(6\,\pi^2)$ is linear in $B_x$; the $n=1$ BC- and
OMM-dependent parts combine into a $B_xB_z$ bilinear at order $B^2$; the $n=1$
BC-, OMM-, and concurrent parts combine at order $B^3$ into a single
$B_x \,(23\,B_x^2-8\,B_z^2)$ structure; and the $n=3$ term contributes a
$\tau^4$ piece $\propto B_x\,(B_x^2+B_z^2)$. The LFO-induced $\sigma^{\rm{lf}}_{yz}$
is therefore dominated at low fields by the linear-in-$B_x$ term, which
exceeds the linear AH term by a factor of $10\,\tau^2\,\mu^2$, as in set-up~I.
The curves shown in
Fig.~\ref{figset3hf} (field-strength dependence at $\theta=\pi/4$ in the left
panels, angular dependence at fixed $|\bs B|$ in the right panels with
$B_x=|\bs B|\cos\theta$ and $B_z=|\bs B|\sin\theta$), confirm these predictions.
In the left panels, $\bar\sigma_{zz}$ is odd and nearly linear in $B_z$, with the
BC-only part dominating the OMM-only part, as expected from the coefficients of
Eq.~\eqref{eqhfIII-nonah}. The curve $\bar\sigma_{xz}$ is linear in $B_x$, with
BC-only and OMM-only curves of opposite slope. The AH and LFO curves of
$\sigma_{yz}$ are linear in $B_x$, with a ratio of
$10\tau^2\mu^2\simeq3.7$ for these parameters. In the right panels,
$\bar\sigma_{zz}$ follows a $\sin\theta$ profile, vanishing at $\theta=0$ and
$\pi$ where $B_z=0$, while the curves $\bar\sigma_{xz}$,
$\sigma^{\rm{ah}}_{yz}$, and $\sigma^{\rm{lf}}_{yz}$ follow a $\cos\theta$
profile, vanishing at $\theta=\pi/2$ and $3\pi/2$ where $B_x=0$.

Comparing the three set-ups, the pattern is transparent from the analytic
expressions. In set-up~I, where the field lies in the equatorial plane, only the
off-diagonal $\sigma_{zx}$ carries odd powers of $\bs B$, while $\sigma_{xx}$ and
$\sigma_{yx}$ are even. Once $B_z\neq0$ in set-ups~II and~III, the out-of-plane
field component couples to the dipole axis $\hat{\bs z}$. The odd-in-$\bs B$ terms
then propagate into the individual BC- and OMM-dependent sub-parts of the
diagonal $\sigma_{xx}$ (set-up~II) and $\sigma_{zz}$ (set-up~III), always with
the compensating odd total power of the dipolar $\bs\Omega_s$ and $\bs m_s$
required by the generalised Onsager relation. No such propagation occurs for a BC-monopole. The isotropic texture $\bs\Omega_s\propto\hat{\bs k}$ forces every BC-
and OMM-dependent integrand to be even in $\bs B$, and the corresponding
monopolar $\sigma_{xx}$ and $\sigma_{zz}$ would be symmetric in $\bs B$ in all
three set-ups. The odd curves of $\sigma_{xx}$ and $\sigma_{zz}$ in
Figs.~\ref{figset2hf} and~\ref{figset3hf} illustrate this contrast.

%%%%%%%%%%%%%%%%%%%%%%
\subsection{AH conductivity from the flat-band}

There exists an AH part of the conductivity from the flat-band, which we explain in this subsection. Within the semiclassical Boltzmann formalism, every contribution to $\bar\sigma_{ij}$ and
$\sigma^{\rm{lf}}_{ij}$ derives from $-\partial_\varepsilon f(\varepsilon)$ and is
therefore controlled entirely by the partially-filled dispersive band with index
$s=2$. The intrinsic AH conductivity, however, is fundamentally a
Fermi-sea quantity,
\begin{align}
\sigma^{\rm{ah}}_{xy} = -\,e^2 \sum_s \int\!\frac{d^3k}{(2\pi)^3}\,
f(\varepsilon_s)\,\Omega_{s,z}(\bs k)\,,
\end{align}
summed over all bands weighted by their occupation, not restricted to
states at the Fermi level. This distinction was emphasised for the HSM by
Graf and Pi\'echon~\cite{graf-hopf}, who showed that the resulting AH
conductivity is an odd function of $\mu$ with a jump-discontinuity,
precisely at the node, of size equal to the flat-band's own BC-dipole charge.
Since the flat-band's own OMM vanishes identically, $\bs m_0=0$ (Sec.~\ref{sechopf}), it acquires no field-dependent correction of its own; the flat-band contributes to $\sigma^{\rm{ah}}_{ij}$ only through the $\bs B$-independent sea term evaluated below, and the field-dependent structure of $\sigma^{\rm{ah}}_{yz}$ discussed later in this section originates entirely from the dispersive bands' nonzero $\bs m_s = (-1)^{s+1}\,e\,v_0\,k\,\bs\Omega_s / 2 $.\footnote{This relation does not apply to $s=0$ since the flat-band has zero OMM but nonzero BC.}

For $\mu>0$, as assumed throughout this work, the two remaining bands are
never both empty. The lower dispersive branch, $\varepsilon_1=-v_0 k\leq0<\mu$
for every $\bs k$, and the flat-band, $\varepsilon_0=0<\mu$, are fully
occupied at every momentum, unlike $s=2$, which is occupied only for
$k<k_F\equiv\mu/v_0$. Using $\Omega_{s,z}(\bs k)=k_z^2/k^4$ for $s=1,2$ and
$\Omega_{0,z}(\bs k)=-2\,k_z^2/k^4$, the three-band sum rule
$\sum_s\Omega_{s,z}=0$ makes the combined contribution of all three bands
vanish identically wherever all three are simultaneously filled, that is, for
$k<k_F$. The only surviving region is $k>k_F$, where $s=2$ has emptied out
but $s=0,1$ remain filled:
\begin{align}
\sum_s f(\varepsilon_s)\,\Omega_{s,z}(\bs k) =
-\,\frac{k_z^2}{k^4}\,\Theta(k-k_F)\,.
\end{align}
The transverse components $\sigma^{\rm{ah}}_{xz}$ and $\sigma^{\rm{ah}}_{yz}$ receive no analogous sea contribution. Since $\bs\Omega_s(\bs k)$ is purely radial for every band, its in-plane components carry an explicit $\cos\phi$ or $\sin\phi$ factor, and the azimuthal integral of either vanishes identically, independently of the $k_F$-cutoff structure above. Hence $\sigma^{\rm{ah}}_{xz}\big|_{\rm sea}=\sigma^{\rm{ah}}_{yz}\big|_{\rm sea}=0$, and this flat-band mechanism is confined entirely to the $xy$ plane.
Carrying out the angular integral and regularising the radial integral with a
momentum cutoff $k_\Lambda$,\footnote{The continuum model is only a low-energy
description near the node, so, as in Ref.~\cite{graf-hopf}, this piece is not
meaningful beyond the range of validity of the linearisation.} one finds
\begin{align}
\sigma^{\rm{ah}}_{xy}\big|_{\rm sea} = \frac{e^2}{6\,\pi^2}
\left(k_\Lambda - k_F\right).
\end{align}
The $k_F$-dependent piece here produces, up to the antisymmetry
$\sigma^{\rm{ah}}_{yx}=- \,\sigma^{\rm{ah}}_{xy}$, a plateau term
$-\,e^2 \,\mu/(6\,\pi^2 v_0)$. That
plateau is the standard Karplus--Luttinger contribution of the single active
Fermi surface, $s=2$, treated in isolation. This term is nonzero whenever a band is fully occupied over
an extended region of the BZ, but it is cutoff-dependent in the present
continuum model and only becomes universal and finite in a genuine lattice
completion, such as the multiband Hopf insulator lattice model of
Ref.~\cite{graf-hopf}. It therefore stands outside the scope of the
field-dependent magnetoelectric response analysed in the rest of this paper,
but it is a necessary ingredient for any comparison with a zero-field part $\sigma_{xy}$.

%%%%%%%%%%%%%%%%%
\subsection{Internode-scattering contribution}
\label{secinternode}

\begin{figure*}[t!]
\subfigure{\includegraphics[width=0.6 \textwidth]{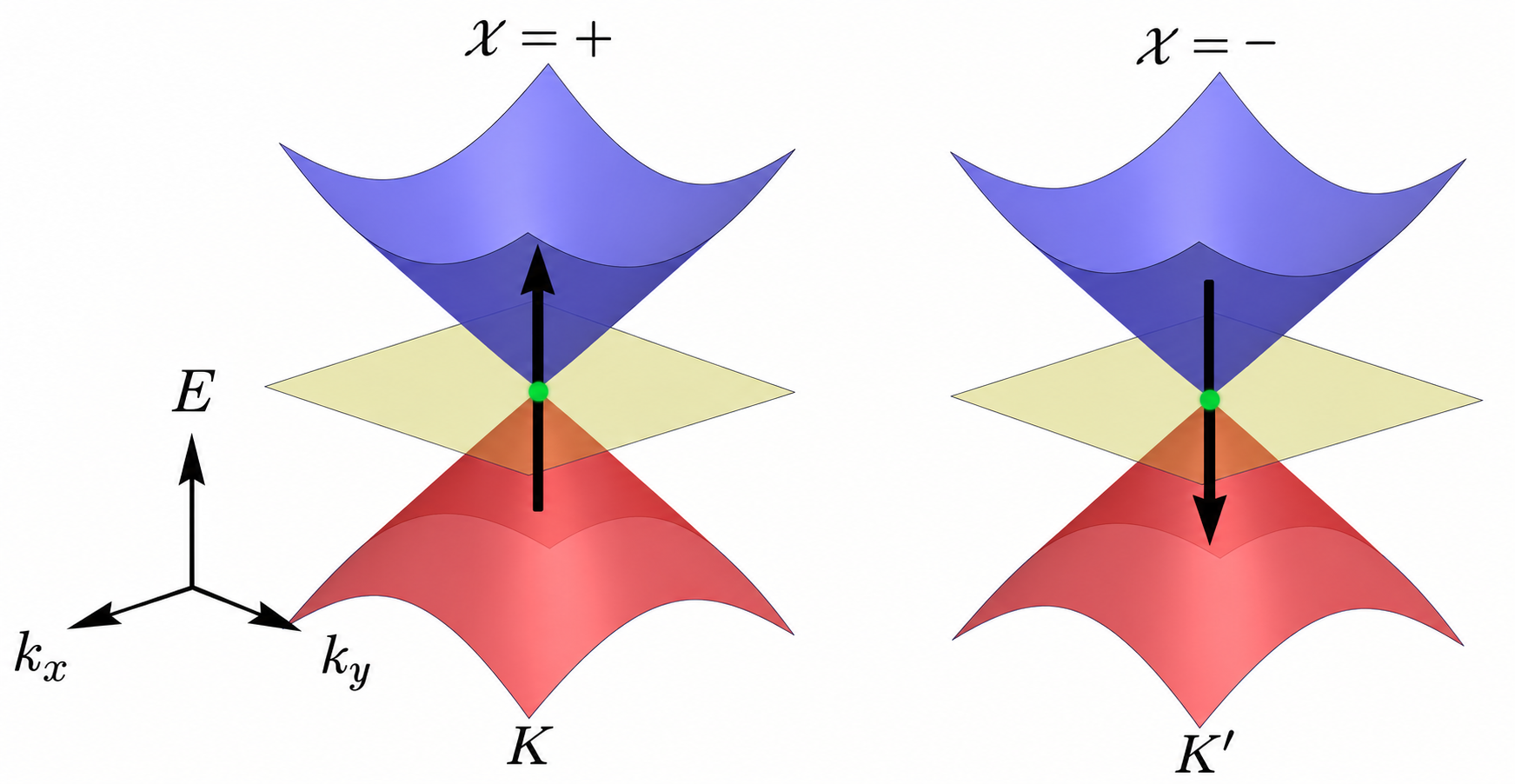}} 
\caption{\label{figfs2} Dispersion of a valley-HSM in the $k_xk_y$-plane at $k_z=0$. $K$ ($\chi=+1$) and $K'$ ($\chi=-1$) denote the two valleys, each of which hosts two linearly-dispersing bands and a flat-band (yellow) meeting at a threefold-degenerate node (green). The black arrows schematically indicate the BC-dipole orientations $\bs\xi_\chi=\chi\,\hat{\bs z}$, which lie along the $k_z$-axis.}
\end{figure*}

A contribution from internode scattering requires at least two nodes. We therefore consider the valley-HSM of Ref.~\cite{graf-hopf}, which hosts a pair of conjugate Hopf nodes labelled by $\chi=\pm1$, as depicted in Fig.~\ref{figfs2}.

%%%%%%%%%%%%%%%%%%%%%%

The geometric object that distinguishes the two nodes is the orientation of the BC-dipole axis. As established in Ref.~\cite{graf-hopf}, the vectors $\bs\xi_\chi$ along the dipole axes satisfy $\bs\xi_\chi=-\,\bs\xi_{-\chi}$, which makes the BC-profile of one node the negative of the other. Equivalently, the Berry dipole points up in one valley and down in the other. We take the node with $\chi=+1$ as the reference, with $\bs\xi_{+1}=\hat{\bs z}$ and $\bs\Omega_s^{\chi}=\chi\,\bs\Omega_s$, where $\bs\Omega_s$ is given in Eq.~\eqref{eqbcommhopf}. The conjugate node then has $\bs\xi_{-1}=-\,\hat{\bs z}$, and the BC and the OMM, which inherits the orientation flip [cf.\ Eq.~\eqref{eqbcommhopf}], satisfy
\begin{align}
\bs\Omega_s^{\chi}=-\,\bs\Omega_s^{-\chi}\,,\qquad
\bs m_s^{\chi}=-\,\bs m_s^{-\chi}\,,
\end{align}
while the dispersions and the band velocities are identical at the two nodes. The internode conductivity follows from the generic treatment of Ref.~\cite{ips-internode}. The formalism, the definition of the charge-pumping vector $\bs{\mathcal I}^\chi$ that fixes the shift $\delta\mu_\chi$ of the local chemical potentials, and the explicit momentum integrations are collected in Appendix~\ref{app-boltz-internode}. Carrying out the integrations gives
%%%%%%%%%%%%%%%%%%%
\begin{align}
\label{eq-internode_zero}
\bs{\mathcal I}^\chi=\bs 0\,,\qquad
\left(\sigma_s^{\chi,\rm inter}\right)_{ij} = 0\,,
\end{align}
for each node separately.

The vanishing is a direct consequence of the dipolar geometry. On the Fermi sphere of a given node, the BC profile $\bs\Omega_s\propto k_z\,\bs k/k^4$ carries an extra factor of $\cos\gamma=\hat{\bs d}\cdot\hat{\bs k}$ relative to a monopole, and the OMM, hence also the OMM-induced energy shift $\varepsilon_m^\chi=-\,\bs m_s^\chi\cdot\bs B$, inherits the same angular dependence. Both $\bs\Omega_s^\chi$ and $\varepsilon_m^\chi$ are therefore even under $\bs k\to-\bs k$, whereas the band velocity $\bs v_s^\chi$ and the OMM-induced velocity $\bs u_m^\chi=\nabla_{\bs k}\varepsilon_m^\chi$ are odd. Every integrand of $\bs{\mathcal I}^\chi$ contains one of the odd factors $\bs v_s^\chi$, $\bs\Omega_s^\chi\cdot\bs v_s^\chi$, $\bs u_m^\chi$, or $\bs\Omega_s^\chi\cdot\bs u_m^\chi$, multiplied by even functions, which include the derivatives of the Fermi function. Each integrand is thus odd in $\bs k$, and the integral over a single Fermi sphere vanishes identically. Since the argument applies to the unexpanded $\bs{\mathcal I}^\chi$, the result holds to all orders in $\bs B$. In particular, the BC term of $\bs{\mathcal I}^\chi$ measures the Berry flux through the Fermi surface, which is zero for a BC-dipole. The density of states of node $\chi$ does acquire a correction linear in $B_z$ with the sign $\chi$, as shown in Appendix~\ref{app-internode}. It only multiplies $\bs{\mathcal I}^\chi$, however, and therefore generates no internode current. The vanishing is thus a property of each node and does not rely on a cancellation between the two nodes.

The contrast with a BC-monopole pair is instructive. The two nodes of a Weyl pair carry opposite monopole charges $\mathcal C_\chi=\mp 1$, and their BC takes the form $\bs\Omega_s^\chi\propto\chi\,\hat{\bs k}/k^2$, which is odd under $\bs k\to-\bs k$. Then $\bs\Omega_s^\chi\cdot\bs v_s^\chi\propto\chi/k^2$ is even on the Fermi sphere, the angular integral $\int_0^\pi\sin\gamma\,d\gamma$ is nonvanishing, and $\bs{\mathcal I}^\chi$ is finite and proportional to $\chi\,\bs B$. The internode conductivity contains two such factors, the sign $\chi$ drops out, and the contributions of the two nodes add, which yields the familiar $\sigma_{ij}^{\rm inter}\propto B_i\,B_j$ form. The BC changes sign between the two nodes in both the monopole and the dipole case. The decisive difference lies in the parity of the BC under $\bs k\to-\bs k$, which is odd for a monopole and even for a dipole. It determines whether the Berry flux through the Fermi surface, and with it the pumping of charge between the nodes, is nonzero.

Physically, the chiral anomaly requires a net imbalance of chiral charge to be pumped between the nodes at a rate $\propto\bs E\cdot\bs B$. In BC-monopole systems this pumping is sustained because the BC flux through any surface enclosing a node is nonzero, and the node is a genuine source or sink of BC field lines carrying a quantised $\mathcal C$. In the HSM, however, the node carries a BC-dipole, and the net flux through any enclosing surface vanishes by construction, since the monopole charge at each node is identically zero. Consequently, the local chemical potentials are not shifted ($\delta\mu_\chi\propto\bs{\mathcal I}^\chi\cdot\bs E=0$), which reflects the absence of a monopole source or sink of BC at the Hopf node. The internode-scattering-induced negative longitudinal magnetoresistance, ubiquitous in Weyl and multifold semimetals~\cite{son13_chiral,ips-internode}, is therefore absent in HSMs.

A related remark concerns the rest of this section, which refers to a single node. Every dipole-activated channel, which contains an odd number of factors of $\bs\Omega_s$ or $\bs m_s$, changes sign under $\chi\to-\chi$. In a valley-HSM, whose two valleys are related as described above and share the same chemical potential and velocity, such channels cancel between the two nodes. This is in line with the cancellation of the AH conductivity and of the geometric magnetocurrent of the two valleys noted in Ref.~\cite{graf-hopf}. These channels survive in a topological HSM with a single Hopf crossing, which is the case covered by the single-node results of this section.

%%%%%%%%%%%%%%%
\begin{table}[t!]
\caption{\label{tab-nonAH_AH-TSM}
$B_i$-dependence in the non-anomalous-Hall and anomalous-Hall
contributions to the magnetoelectric conductivity of the TSM in set-up~I, organised by physical origin.}
\begin{ruledtabular}
\begin{tabular}{lccccc}
 & $\sigma^{\rm d}$
 & $\sigma^{\rm bc}$
 & $\sigma^{\rm m}$
 & $\sigma^{\rm conc}$
 & $\sigma^{\rm{ah}}$ \\
\colrule
$\sigma_{xx}$
 & $B^{0}$
 & $B_x^2,  B_y^2$
 & $B_x^2$
 & $B_x^2,  B_y^2$
 & $-$ \\[4pt]
$\sigma_{yx}$
 & $0$
 & $B_x\,B_y$
 & $B_x\,B_y$
 & $B_x\,B_y$
 & $0$ \\[4pt]
$\sigma_{zx}$
 & $0$
 & $0$
 & $0$
 & $0$
 & $B_y,  B_y\, {\boldsymbol B}^2$ \\
\end{tabular}
\end{ruledtabular}
\end{table}

\begin{table}[t!]
\caption{\label{tab-lf-TSM}
$B_i$-dependence in the Lorentz-force-induced conductivity
components of the TSM in set-up~I,
organised by the values of $n$ and physical origins.
``$0$'' denotes a symmetry-forbidden vanishing of the
angular integrand.}
\begin{ruledtabular}
\begin{tabular}{lcccc}
 & $\sigma^{\text{lf, h}}$ & $\sigma^{\text{lf, bc}}$
 & $\sigma^{\text{lf, m}}$ & $\sigma^{\text{lf, conc}}$ \\
\colrule
$n=1$, $\sigma_{zx}$
 & $B_y$
 & $B_y\, {\boldsymbol B}^2$
 & $B_y\, {\boldsymbol B}^2$
 & $B_y\, {\boldsymbol B}^2$ \\[4pt]
$n=2$, $\sigma_{xx}$
 & $B_y^2$
 & $0$ & $0$ & $0$ \\[4pt]
$n=2$, $\sigma_{yx}$
 & $B_x\,B_y$
 & $0$ & $0$ & $0$ \\[4pt]
$n=3$, $\sigma_{zx}$
 & $B_y\, {\boldsymbol B}^2$
 & $0$ & $0$ & $0$ \\
\end{tabular}
\end{ruledtabular}
\end{table}

%======================================================================
\section{Comparison and discussion}
\label{seccompare}
%======================================================================

In this section, we will compare the characterisitcs of magnetoelectric response of the two
dipolar systems, the VNR and the HSM, and then contrast them with the NVNR~\cite{ips-nlsm-ph,ips-gnr-strain} and the TSM~\cite{ips-spin1-ph}, respectively. The comparison is organised according to the three physically distinct contributions: $\bar\sigma_{ij}$, $\sigma^{\rm{ah}}_{ij}$, and $\sigma^{\rm{lf}}_{ij}$.  

%%%%%%%%%%%%%%%%%%%%%%%%
\subsection{VNR versus HSM}
\label{secvnr-vs-hsm}

We begin by comparing the VNR and the HSM directly, before contrasting each against its monopolar counterpart in the following two subsections. The two systems realise the same qualitative object, a BC-dipole, through unrelated microscopic mechanisms: the VNR dipole is a Fermi-surface-topology effect tied to the toroidal geometry of a two-band nodal line, while the HSM dipole is a node-algebra effect tied to the $SU(3)$ structure of a threefold-degenerate point. A further distinction lies in how the dipole is distributed in momentum space: the HSM carries a single, point-like BC-dipole fixed at its node, whereas the VNR carries a continuous ring of such dipole-like sources, one attached to every point of the nodal loop. This difference in origin and distribution leaves a clear imprint on the transport response.

For the non-AH part, both systems activate the out-of-plane component
$\bar\sigma_{zx}$ already in set-up~I, a response forbidden for a BC-monopole. In
the VNR, this term is controlled by the major and minor radii of the torus, while
in the HSM it is controlled instead by the single length scale $\mu/v_0$. In both
systems the term is odd and nearly linear in $\bs B$, since its cubic corrections
are controlled by the weak-field parameter. The linear coefficients of the BC-only
and OMM-only parts have opposite signs and stand in the parameter-independent
ratios $-7/3$ for the VNR and $-8/3$ for the HSM. They therefore partly cancel in
both systems, and the concurrent part starts at cubic order
[Eqs.~\eqref{eqsigzxset1} and~\eqref{eqhfI-nonah}]. A further distinction appears
in $\sigma_{xx}$ itself: the VNR's $B_x^2$ and $B_y^2$ channels carry opposite
relative signs for the OMM-only and concurrent contributions, an anisotropy
inherited from the toroidal geometry, whereas the HSM keeps the same sign in both
channels for every contribution, reflecting the higher, $C_{\infty v}$-compatible
symmetry of the simple radial-times-$\cos\gamma$ texture.

Set-ups~II and~III add a shared feature. Each system possesses an antiunitary
symmetry that acts on momentum as $\bs k \to (-k_x, k_y, -k_z)$, namely
$\Theta = T \, M_y$ for the VNR and its counterpart for the HSM
(Sec.~\ref{secsymmetry}). Both therefore obey $\sigma_{xz} = \sigma_{zx}$ at
$B_y = 0$ through Eq.~\eqref{eq:theta_covariance}. The same symmetry leaves the
diagonal components unrestricted in $B_x$ and $B_z$. Both systems accordingly host
odd-in-$\bs B$ longitudinal responses, namely $\bar\sigma_{xx}$ in set-up~II and
$\bar\sigma_{zz}$ in set-up~III for the VNR, and $\bar\sigma_{xx}$ in set-up~II
for the HSM. The linear dependence on $B_z$ gives them a $\sin\theta$ profile,
where the OMM-only part dominates in set-up~II and the BC-only part in set-up~III
for the parameters of Table~\ref{tab-params}.

For the AH part, both systems possess a nonzero $\sigma^{\rm{ah}}_{yx}$ plateau at
$\bs B = 0$ [Eqs.~\eqref{eqahvnr1} and~\eqref{eqhfI-ah}]. A finite AH response
therefore survives even without an applied field in either case. In set-up~II, the
leading field dependence of the VNR plateau is a tilt linear in $B_z$. The plateau
values are $-e^2 \, \mu/(8 \, \pi \, v_0)$ for the VNR and
$-e^2 \, \mu/(6 \, \pi^2 \, v_0)$ for the HSM.

For the LFO-induced part, both systems generate dipole-activated terms that are
symmetry-forbidden for a BC-monopole, but the internode-scattering discussion of
Sec.~\ref{secinternode} applies only to the HSM. The VNR is a single continuous
nodal-ring rather than a pair of point nodes, and the vanishing of the
chiral-anomaly-like internode response found there has no direct analogue in the VNR. On the shared side,
both systems generate odd-in-$\bs B$ LFO contributions in the diagonal components
$\sigma^{\rm lf}_{xx}$ in set-up~II and $\sigma^{\rm lf}_{zz}$ in set-up~III,
sourced by the BC-only and OMM-only parts, in line with the generalised Onsager
relation. The LFO terms carry higher powers of $\tau$ than the BC-only and
OMM-only terms, and for the parameters of Table~\ref{tab-params} they exceed the
total non-AH curves of the VNR several times over in $\sigma_{xx}$ and
$\sigma_{yx}$.

The two Hamiltonians illustrate how the BC-dipole is the geometric origin of the
unconventional magnetotransport in this class of semimetals, in a form largely
independent of the number of bands and of the microscopic mechanism generating the
dipolar $\bs\Omega_s(\bs k)$. The VNR provides the two-band,
Fermi-surface-topology-driven realisation: the dipole is a toroidal-Fermi-surface
effect, no gap or extra band is required, and its most striking response is the
appearance of $\bar\sigma_{zx}$ in set-up~I. The HSM provides the three-band,
node-algebra-driven realisation: the dipole is an $SU(3)$ feature of the node
itself and suppresses the chiral-anomaly-like internode response. In both cases the
OMM plays a nontrivial role that cannot be dropped: several
$\sigma^{\rm{ah}}$ and $\sigma^{\rm lf}$ contributions in set-ups~II and~III
arise only when the OMM-induced energy shift $\varepsilon_m(\bs k)$ and the OMM
velocity correction $\bs u_m = \nabla_{\bs k} \varepsilon_m$ are consistently
retained. We have cross-checked our results against the symmetries of the
respective Hamiltonians, namely $C_{2z}$, $M_z$, and $\Theta = T \, M_y$ for the
VNR, and $C_{\infty v}$ combined with the chiral symmetry $\mathcal S$ and an
antiunitary symmetry acting on momentum as $\bs k \to (-k_x, k_y, -k_z)$ for the
Hopf node.

%%%%%%%%%%%%%%%%%%%%%%%%%%%%%%%%%%%%
\subsection{VNR versus NVNR}
\label{secvnr-vs-nvnr}

As established in Sec.~\ref{secsymmetry}, the VNR and the NVNR realise a
complementary pair of discrete symmetries: the VNR possesses a unitary mirror
$M_z = \sigma_z$ but breaks time-reversal, while gapping the conventional nodal
ring restores time-reversal at the cost of that mirror. This exchange underlies the
contrast developed below. Under the VNR's mirror, the out-of-plane component of an
axial vector such as $\bs\Omega_s$ is even under $k_z \to -k_z$, while its in-plane
components are odd. This is exactly the pattern realised by
$\bs\Omega_s^{\text{VNR}}$ below: an odd in-plane part and an even, locally
nonzero, $\Omega^z_s \propto k_z^2$. The NVNR possesses no such mirror. Its
time-reversal symmetry instead forces the global antisymmetry
$\bs\Omega_s(-\bs k) = -\bs\Omega_s(\bs k)$, consistent with the purely in-plane,
$\Delta$-sourced texture found below, whose vanishing $z$-component removes every
out-of-plane and $k_z$-odd-in-plane signature identified in this subsection.

The conductivity tensor of the NVNR is summarised in Appendix~\ref{app_GNR}, and
the corresponding nonzero components and LFO field dependencies are collected in
Tables~\ref{tab-gnr_nonAH_ah}--\ref{tab-gnr-lf3}. Throughout, two geometric
quantities control the entire contrast:
\begin{align}
\label{eqn-bc-gnr-vnr}
\bs\Omega_s^{\text{NVNR}},\,\bs m_s^{\text{NVNR}} &\propto \Delta \, \{k_y, -k_x, 0\}
\Longrightarrow \Omega^z_s = m^z_s = 0\,,\nn
\bs\Omega_s^{\text{VNR}},\,\bs m_s^{\text{VNR}} &\propto k_z \, \{\kappa_\perp \cos\phi, \kappa_\perp \sin\phi, k_z\}
\Longrightarrow \Omega^z_s,\, m^z_s \neq 0\,.
\end{align}
The VNR's $\bs\Omega_s$ and $\bs m$ therefore carry an explicit $k_z$ factor, while
the NVNR's carry an overall factor of the mass gap $\Delta$ and vanish identically
along $\hat{\bs z}$. Consequently, the NVNR contribution to $\bar\sigma$ or
$\sigma^{\rm{ah}}$ originating from $\bs\Omega_s$ or $\bs m_s$ scales as
$\Delta$, whereas the corresponding VNR contribution is $\Delta$-independent and
depends on $k_z/(k_0 + \kappa_\perp)$.

The most immediate qualitative difference is the nonvanishing out-of-plane non-AH
response $\bar\sigma_{zx}$ of the VNR in set-up~I, which is absent for the NVNR.
More broadly, the NVNR non-AH response collapses to a much smaller set of
components: set-up~I retains only the in-plane $\bar\sigma_{xx}$ and
$\bar\sigma_{yx}$; set-up~II only $\bar\sigma_{xx}$ (Drude plus $B_x^2$
corrections); and set-up~III only $\bar\sigma_{zz}$ (Drude plus $B_x^2$). The
in-plane transverse parts $\bar\sigma_{zx}$ (set-up~II) and $\bar\sigma_{xz}$
(set-up~III), as well as all out-of-plane non-AH components, vanish identically,
because $B_z$ cannot project onto the in-plane BC when $\Omega^z_s = 0$. The VNR,
by contrast, yields a nonzero in-plane transverse conductivity in all three
set-ups, and set-up~I additionally produces an out-of-plane response. The reason is
that the three-dimensional character of the VNR's BC and OMM allows $B_z$ to couple
to $\Omega^z_s$ and $m^z$, generating a nonvanishing result even when $\bs B$ has
an out-of-plane component.

The role of the $k_z$-odd structure is most transparent in the non-AH response. In
set-up~I ($B_z = 0$), both $\bs B \cdot \bs\Omega_s$ and
$\varepsilon_m = -\bs B \cdot \bs m_s$ collapse to their in-plane projections in
both systems, but the VNR retains a residual contribution through
$w_z \Omega^z_s$ inside $\bs w_s \cdot \bs\Omega_s$, since
$\Omega^z_s \propto k_z$ is intrinsically nonzero. This term is absent in the
NVNR, where $\Omega^z_s = m^z_s = 0$ throughout the BZ. In set-ups~II and~III,
$B_z \neq 0$ activates the products $B_z \, \Omega^z_s$ and $B_z \, m^z_s$,
together with $B_z$-driven corrections to $\varepsilon_m$; every such term is
annihilated in the NVNR and generically nonzero in the VNR.

Turning to individual components, in set-up~I the VNR yields three nonzero non-AH
components, $\bar\sigma_{xx}$, $\bar\sigma_{yx}$, and $\bar\sigma_{zx}$, whereas
the NVNR yields only the first two. The longitudinal $\bar\sigma_{xx}$ contains
the Drude term together with an $\mathcal{O}(|\bs B|^2)$ BC- and OMM-induced
response of the form $B_x^2 + \alpha_i \, B_y^2$, whose anisotropy coefficient
$\alpha_i$ ($i \in \{\text{bc}, \rm m, \text{conc}\}$) is fixed by the angular
structure of the in-plane $\bs\Omega_s$ and $\bs m_s$. For the NVNR all three
sub-parts share $\alpha_i = 3$. For the VNR the coefficients differ between
sub-parts on account of the distinct $\phi$- and $\kappa_\perp$-dependence of
$\bs\Omega_s$ and $\bs m_s$ themselves. The planar-Hall component $\bar\sigma_{yx}$, with
$\sigma^{\rm d}_{yx} = 0$, receives only $B_x \, B_y$-dependent contributions from
the geometric parts --- the standard $\sin(2\theta)$ planar-Hall signature, present
in both systems and differing only in system-specific prefactors. The
distinguishing feature of set-up~I is the out-of-plane $\bar\sigma_{zx}$: in the
VNR it acquires a linear-in-$B_x$ term together with a cubic term
$\propto B_x \, {\boldsymbol B}^2$, whereas it vanishes identically in the NVNR.
Its presence or absence is dictated entirely by the $k_z$-parity of the BC and OMM
prefactors, so an odd-in-$\bs B$ out-of-plane conductivity in set-up~I constitutes
a clean experimental fingerprint of the vortex pseudospin texture that the NVNR
cannot mimic. Its linear coefficient $-\tau \, e^3 \, v_0/(16 \, \pi)$ gives a
$\cos\theta$ profile that changes sign as $\bs B$ rotates through $\hat{\bs y}$.

A second fingerprint of the VNR appears in the diagonal components once
$B_z \neq 0$. The VNR's $\bar\sigma_{xx}$ in set-up~II and $\bar\sigma_{zz}$ in
set-up~III each contain a term linear in $B_z$, together with odd cubic terms,
whereas the NVNR retains only the Drude term and even corrections in $B_x^2$. The
linear term requires the couplings $B_z \, \Omega^z_s$ and $B_z \, m^z_s$, which
vanish identically for the NVNR. Symmetry does not forbid it in the VNR, since the
antiunitary symmetry $\Theta$ constrains a diagonal component only to be even in
$B_y$ and leaves its dependence on $B_x$ and $B_z$ unrestricted. This is
consistent with the generalised Onsager relation of Sec.~\ref{secvnr}, in which the
sign change of $\bs B$ is compensated by that of the dipolar $\bs\Omega_s$ and
$\bs m_s$. The linear dependence on $B_z$ gives it a $\sin\theta$ profile that
vanishes where $B_z = 0$. For the parameters of Table~\ref{tab-params}, the
OMM-only part dominates in set-up~II and the BC-only part in set-up~III. The
transverse curves $\bar\sigma_{zx}$ in set-up~II and $\bar\sigma_{xz}$ in
set-up~III coincide, as required by Eq.~\eqref{eq:theta_covariance} at $B_y = 0$.

The AH contribution is $\tau$-independent and determined entirely by the intrinsic
Bloch geometry through $\bs\Omega_s$ and $\bs m_s$. In the zero-field limit, the
NVNR's $\bs\Omega_s^{\text{NVNR}} \propto \Delta \, \{k_y, -k_x, 0\}/\varepsilon_s^3$
renders all relevant integrals vanishing: $\sigma^{\rm{ah}}_{xx}$ is excluded by
the Levi-Civita symbol, $\sigma^{\rm{ah}}_{yx}$ vanishes because
$\Omega^z_s = 0$, and for $\sigma^{\rm{ah}}_{zx}$ the in-plane vortex texture is
annihilated by the angular integration. The NVNR therefore possesses no intrinsic
AH response at $\bs B = 0$. The VNR is qualitatively different: its smoke-ring
texture endows each $k_z \neq 0$ layer with a nontrivial planar
$\mathcal{C}(k_z) = +1$ on both sides of the nodal plane~\cite{vortex-nrsm}, so
that $\sigma^{\rm{ah},0}_{yx}\big|_{\bs B = 0}$ is nonvanishing and, in a lattice
regularisation, saturates at the maximal value $(e^2/h)(1/a)$, with $a$ the lattice
constant~\cite{vortex-nrsm}. This appears in our low-energy calculation as the
$B^0$-dependent term $-e^2 \, \mu/(8 \, \pi \, v_0)$ of
$\sigma^{\rm{ah}}_{yx}$ in Eq.~\eqref{eqahvnr1} for set-ups~I (in-plane) and~II
(out-of-plane), where it is the nearly constant curve of Fig.~\ref{figset2vnr},
with no counterpart in the NVNR.

Set-up~I is the only geometry in which the AH contribution produces an in-plane
response. Since $B_z = 0$, only the in-plane components of $\bs\Omega_s$ and
$\bs m$ enter Eq.~\eqref{eq-ah}, and both systems retain a nonzero
$\sigma^{\rm{ah}}_{zx}$ driven by $\bs E \times \bs\Omega_s$-type terms. For the
NVNR,
\begin{align}
\sigma^{\text{ah,NVNR}}_{zx} &= - \, \frac{e^3 \, v_z \, v_0 \, k_0 \, \Delta^2 \, B_y}
{16 \, \pi \, \mu^4}
\left(1 + \frac{9 \, e^2 \, v_z^2 \, v_0^2 \, \Delta^2 \, {\boldsymbol B}^2}
{4 \, \mu^6}\right), \quad
\sigma^{\text{ah,NVNR}}_{yx} = 0\,,
\end{align}
while for the VNR $\sigma^{\rm{ah}}_{zx}$ has the same field dependence
$\{B_y, B_y^3, B_y B_x^2\}$ but different prefactors, and
$\sigma^{\rm{ah}}_{yx}$ is additionally nonvanishing with $B^0$ and $B^2$
contributions --- the $B^0$ term being the intrinsic maximal AH signal identified
above, and the $B^2$ term originating in the $\varepsilon_m$-expansion of the Fermi
function in Eq.~\eqref{eq-ah}.

Set-up~II maximises the VNR--NVNR contrast. In the NVNR,
$\Omega^z_s = m^z_s = 0$ annihilates every AH term generated by $B_z$ after the
$\phi$-integration, so the AH response is entirely blind to the out-of-plane field
component and $\sigma^{\text{ah,NVNR}}_{yx} = 0$. In the VNR, the terms
$B_z \, \Omega^z_s$ and $B_z \, m^z_s$ survive angular integration and generate
$\sigma^{\rm{ah}}_{yx}$ with the field dependence
$\{B^0, B_z, B_x^2, B_z^2, B_z^3, B_z \, B_x^2\}$. In set-up~III,
$\sigma^{\rm{ah}}_{yz}$ probes the in-plane components of $\bs\Omega_s$ via
$(\bs E \times \bs\Omega_s)_y \propto E_z \, \Omega^x_s$, and both systems retain a
nonvanishing signal. For the VNR, $\sigma^{\text{ah,VNR}}_{yz}$ carries the
$B$-dependence $\{B_x, B_x \, B_z, B_x \, B_z^2, B_x^3\}$, allowed by
$\Omega^z_s \neq 0$, whereas the counterpart $\sigma^{\rm{ah}}_{xz}$ vanishes in
the NVNR.

The LFO contribution originates in the iterative action of $\check L$, with the
iteration index $n$ generating contributions scaling as $\tau^{n+1}$. We decompose
$\sigma^{\rm{lf}}_{ij} = \sigma^{\rm lf,h}_{ij} + \sigma^{\rm lf,bc}_{ij}
+ \sigma^{\rm lf,m}_{ij} + \sigma^{\rm lf,conc}_{ij}$, where the subscript ``h''
denotes the purely non-topological parts and the remaining three are the BC-only,
OMM-only, and concurrent contributions:
\begin{enumerate}

\item In set-up~I at $n = 1$ ($\propto \tau^2$), the LFO generates the out-of-plane
$\sigma^{\rm{lf}}_{zx}$ and the in-plane $\sigma^{\rm{lf}}_{yx}$. The purely
part $\sigma^{\rm lf,h}_{zx} \propto B_y$ is the ordinary three-dimensional Hall response (an $\hat{\bs x}$-drive deflected by $B_y \, \hat{\bs y}$ into a
$\hat{\bs z}$-current), present in both systems. The
$(\text{bc}, \text{m}, \text{conc})$ sub-parts of $\sigma^{\rm{lf}}_{zx}$ acquire
cubic terms $\propto B_y \, {\boldsymbol B}^2$ in both systems. The VNR
additionally receives a qualitatively new in-plane response
$\sigma^{\rm lf,bc}_{yx}, \sigma^{\rm lf,m}_{yx} \propto {\boldsymbol B}^2$,
absent in the NVNR; its persistence is a direct manifestation of the BC texture
inherited from the smoke-ring winding of the pseudospin $\bs d(\bs k)$. At $n = 2$ ($\propto \tau^3$), three components ($xx$, $yx$,
$zx$) are generated. The longitudinal
$\sigma^{\rm lf,h}_{xx} \propto (B_x^2 + 3 \, B_y^2)$ is the standard positive LMC
of a 3d semimetal, and the planar-Hall
$\sigma^{\rm lf,h}_{yx} \propto B_x \, B_y$ is common to both systems. The
$(\text{bc}, \text{m})$ sub-parts of $\sigma^{\rm{lf}}_{zx}$ carry the cubic
structure $B_x \, {\boldsymbol B}^2$, finite in the VNR and absent in the NVNR. At
$n = 3$ ($\propto \tau^4$), only the purely geometric part is absent and
$\sigma^{\rm lf,h}_{zx} \propto B_y \, {\boldsymbol B}^2$ survives, common to both systems.
The isotropic ${\boldsymbol B}^2$ term of $\sigma^{\rm lf}_{yx}$ prevents it from vanishing at $\theta = 0$, and the
$B_x^2 + 3 \, B_y^2$ dependence of $\sigma^{\rm lf}_{xx}$ makes it three times
larger at $\theta = \pi/2$ than at $\theta = 0$.

\item In set-up~II, the out-of-plane field $B_z$ activates the out-of-plane
responses of $\bs\Omega_s$ and $\bs m_s$, and the LFO response becomes markedly
richer in the VNR. At $n = 1$, only $\sigma^{\rm{lf}}_{yx}$ is generated, and its
$h$-part vanishes upon integration. The
$(\text{bc}, \text{m}, \text{conc})$ sub-parts carry the field dependence
$\{B_x^2, B_z^2, B_z \, B_x^2, B_z^3\}$ in the VNR, whereas only the
$B_z \, B_x^2$ term survives in the NVNR. At $n = 2$,
$\sigma^{\rm lf,h}_{xx} \propto B_x^2 + \alpha \, B_z^2$ is the standard LMC/PHC
anisotropy of a 3d semimetal, common to both systems. The BC-only and OMM-only
sub-parts of $\sigma^{\rm lf}_{xx}$ additionally carry odd-in-$B_z$ terms of the
form $B_z \, \{B_x^2, B_z^2\}$ in the VNR, which vanish in the NVNR. The in-plane
transverse $\sigma^{\rm{lf}}_{zx}$ displays the sharpest VNR/NVNR contrast: the VNR
receives a contribution $\propto \{B_x^3, B_x \,B_z^2\}$ that is structurally
forbidden in the NVNR. At $n = 3$, only
$\sigma^{\rm lf,h}_{yx} \propto \{B_z \, B_x^2, B_z^3\}$ survives, common to both systems.

\item In set-up~III at $n = 1$, only $\sigma^{\rm{lf}}_{yz}$ is generated. The
part $\sigma^{\rm lf,h}_{yz} \propto B_x$ is the ordinary three-dimensional Hall response, present in both systems. In the VNR the
$(\text{bc}, \text{m}, \text{conc})$ sub-parts carry the structures
$\{B_x \, B_z, B_x^3, B_x \, B_z^2\}$, all finite; in the NVNR the source term
$(\bs{\mathcal N}_{1,2})_y$ vanishes identically, so the mixed $B_x \, B_z$
structure is absent altogether, the surviving geometric contributions
$\sigma^{\text{lf,bc}}_{yz} \propto \Delta^2 \, B_x^3$ and
$\sigma^{\text{lf,m}}_{yz} \propto \Delta^2 \, \{B_x^3, B_x \, B_z^2\}$ are
uniformly suppressed by $\Delta^2$ and vanish at gap closure, and
$\sigma^{\text{lf,conc}}_{yz} = 0$. The $B_x \, B_z$ term is therefore unique to
the VNR in this geometry. At $n = 2$, $\sigma^{\rm lf,h}_{zz} \propto B_x^2$ is the
perpendicular-field LMC of a 3d metal, common to both systems; a
$\propto B_z^2$ contribution is absent in both because
$B_z \, \hat{\bs z} \parallel \bs E$. The BC and OMM sub-parts of
$\sigma^{\rm{lf}}_{zz}$ carry the cubic structure $B_z \, B_x^2$ in the VNR,
whereas in the NVNR the source term $\bs{\mathcal N}_{2,3}$ vanishes identically,
leaving only
$\sigma^{\rm lf,h}_{zz} \propto B_x^2 \, (\mu^2 - \Delta^2)$. The in-plane
transverse $\sigma^{\rm{lf}}_{xz}$ at $n = 2$ is the set-up~III counterpart of
$\sigma^{\rm{lf}}_{zx}$ in set-up~II:
$\sigma^{\rm{lf}}_{xz} \propto \{B_x^3, B_x \,B_z^2\}$ for the VNR, and vanishes
identically for the NVNR. At $n = 3$ ($\propto \tau^4$), only
$\sigma^{\rm lf,h}_{yz} \propto \{B_x^3, B_x\, B_z^2\}$ survives, common to both systems and
$\Delta$-independent in the NVNR.

\end{enumerate}

%%%%%%%%%%%%%%%%%%%%%%%%
\subsection{HSM versus TSM}
\label{secHopf-vs-TSM}

For the pseudospin-1 node, the BC and OMM are radial monopoles,
\begin{align}
\label{eqOmegaSpin1}
\bs{\Omega}^{s-1}_s &\propto - \, \chi \, s \, \frac{\hat{\bs k}}{v_0^2 \, k^2}\,,\quad
\bs m^{s-1}_s \propto \frac{\hat{\bs k}}{k}\,,\quad
\mathcal C = 2 \, \chi\,,\quad
(\bs v \cdot \bs\Omega)^{\rm TSM} = - \, \frac{\chi \, v_0}{k^2}\,,
\end{align}
purely radial and isotropic on the Fermi sphere, with
$(\bs v \cdot \bs\Omega)$ a scalar on the Fermi sphere. For the Hopf BC-dipole
node, the same radial field is modulated by an extra odd factor along the dipole
axis:
\begin{align}
\bs{\Omega}^{\rm dip}_s &\propto \frac{\cos\gamma}{k^2} \, \hat{\bs k}
= \cos\gamma \times (\text{radial field})\,,\quad
\bs m^{\rm dip}_s \propto \frac{\cos\gamma}{k} \, \hat{\bs k}\,,\quad
\mathcal C = 0\,,\quad
(\bs v \cdot \bs\Omega)^{\rm H} = s \, v_0 \, \frac{k_z}{k^3}\,,
\end{align}
with $\gamma$ the polar angle ($k_z = k \cos\gamma$). The explicit $k_z$ factor in
the BC and OMM textures singles out the $\hat{\bs z}$ direction, lowering their
symmetry from $SO(3)$ to $C_{\infty v}$. Correspondingly,
$(\bs v \cdot \bs\Omega)^{\rm H} \propto k_z/k^3$ is odd under $k_z \to -k_z$:
rotations about $\hat{\bs z}$ remain symmetries, but rotations that tilt
$\hat{\bs z}$ do not leave the BC invariant. The three field configurations
therefore cannot be mapped into one another by a symmetry operation of the system,
and the corresponding conductivity tensors are not related by a simple coordinate
rotation --- the three set-ups probe genuinely independent projections of the BC-dipole.

Here, we can explicitly compare the qualitative features of conductivity only for
set-up~I, because Ref.~\cite{ips-spin1-ph} deals with only that particular $\bs E \bs B$-configuration. Since the HSM and the TSM possess identical band dispersions, the
Drude contribution is unchanged, which follows directly from Eq.~(22) of
Ref.~\cite{ips-spin1-ph}, specialised to the isotropic spin-1 case. The flat-band
contributes in neither system because its group-velocity vanishes identically. Any
qualitative differences therefore originate from the BC- and OMM-dependent terms,
whose quantum geometry differs.

In set-up~I, both systems exhibit a quadratic-in-$B$ longitudinal
magnetoconductivity with $\sigma_{xx}$ containing terms proportional to $B_x^2$,
$B_y^2$, and $B_x \, B_y$. However, the relative weights of the $B_x^2$ and
$B_y^2$ terms, together with the signs and magnitudes of the BC, OMM, and
concurrent contributions, differ markedly. For the TSM, Eqs.~(22)--(23) of
Ref.~\cite{ips-spin1-ph} give
$\sigma^{\rm bc}_{xx} \propto (8 \, B_x^2 + B_y^2)$,
$\sigma^{\rm m}_{xx} \propto B_x^2$, and
$\sigma^{\rm conc}_{xx} \propto -(3 \, B_x^2 + B_y^2)$. For the Hopf node,
$\sigma^{\rm bc}_{xx} \propto (24 \, B_x^2 + B_y^2)$,
$\sigma^{\rm m}_{xx} \propto (16 \, B_x^2 + 3 \, B_y^2)$, and
$\sigma^{\rm conc}_{xx} \propto -(29 \, B_x^2 + 5 \, B_y^2)$. These differences
are not merely numerical, despite the identical band dispersions: they originate
from the dipolar angular modulation of the BC and OMM in the HSM,
$\bs\Omega, \bs m \propto \cos\gamma \, \hat{\bs k}$, which reshapes the angular
integrals entering the conductivity and redistributes the relative weights of the
$B_x^2$ and $B_y^2$ terms. The enhancement is most pronounced for the $B_x^2$
terms.

The origin of this amplification for the BC-only part, whose integrand is
quadratic in $\bs B$, is
\begin{align}
\label{eq:M2xx_bc}
\mathcal M^{(2),\rm bc}_{xx} = e^2 \, \left[(\bs\Omega \cdot \bs B)^2 \, v_x^2 - 2 \, (\bs\Omega \cdot \bs B)(\bs v \cdot \bs\Omega) \, v_x \, B_x + (\bs v \cdot \bs\Omega)^2 \, B_x^2\right].
\end{align}
For $\bs B$ in the equatorial ($xy$) plane, only the first structure generates a
$B_y^2$ contribution; the second and third are pure $B_x^2$. Their angular
integrands scale as $\sin^5\gamma$, $\sin^3\gamma$, and $\sin\gamma$, respectively.
The dipolar modulation $\bs\Omega, \bs m \propto \cos\gamma \, \hat{\bs k}$ inserts
an additional $\cos^2\gamma$ into each, and the corresponding $\gamma$-dependent
ratios are
\begin{align}
\label{eq:cos2moments}
\frac{\langle \sin^5\gamma \, \cos^2\gamma \rangle}{\langle \sin^5\gamma \rangle} = \frac{1}{7}\,,\quad
\frac{\langle \sin^3\gamma \, \cos^2\gamma \rangle}{\langle \sin^3\gamma \rangle} = \frac{1}{5}\,,\quad
\frac{\langle \sin\gamma \, \cos^2\gamma \rangle}{\langle \sin\gamma \rangle} = \frac{1}{3}\,.
\end{align}
The dipolar suppression is therefore strongest for the equator-weighted tensor,
which alone feeds $B_y^2$, and weakest for the component
$(\bs v \cdot \bs\Omega)^2 \, B_x^2$, which contributes solely to $B_x^2$.
Consequently, in going from the monopolar TSM to the HSM, $B_y^2$ is suppressed by
a factor $1/7$ while $B_x^2$ retains contributions from the less-suppressed
structures, lifting the intrinsic ratio $B_x^2 : B_y^2 = 8 : 1$ to $24 : 1$ in the
BC-only response. The analogous analysis of the OMM and mixed BC-OMM tensor
structures underlies the corresponding $16 : 3$ and $29 : 5$ ratios in
$\bar\sigma^{\rm m}_{xx}$ and $\bar\sigma^{\rm conc}_{xx}$.

The planar-Hall conductivity is proportional only to $B_x \, B_y$ in both systems,
a signature of the standard planar-Hall response. In both the TSM and the HSM, the
BC and OMM contributions enter with the same positive sign, whereas the concurrent
contribution is negative. The overall structure of $\sigma_{yx}$ is therefore
identical in the two systems, differing only in the numerical coefficients
multiplying the individual contributions.

The out-of-plane non-AH response $\sigma_{zx}$ is qualitatively different from that
of the monopolar systems: it vanishes identically for the TSM but is nonzero for
the Hopf node. This distinction follows entirely from the $k_z$-parity of the
integrand. In set-up~I, $B_z = 0$, and the relevant $k_z$-parities are as follows.
For the TSM, $v_0 \, k_z/k$ (odd), $v_0 \, k_x/k$ (even),
$(\bs v \cdot \bs\Omega^{\rm TSM}) = -\chi \, v_0/k^2$ (even),
$(\bs B \cdot \bs\Omega^{\rm TSM}) = -\chi \, (B_x \, k_x + B_y \, k_y)/k^3$ (even),
and $\varepsilon^{\rm m}_{\rm TSM} = \chi \, e \, v_0 \, (B_x \, k_x + B_y \, k_y)/k^2$
(even). For the Hopf node, the velocity components are the same, but
$(\bs v \cdot \bs\Omega^{\rm H}) = s \, v_0 \, k_z/k^3$ (odd),
$(\bs B \cdot \bs\Omega^{\rm H}) = k_z \, (B_x \, k_x + B_y \, k_y)/k^4$ (odd), and
$\varepsilon^{\rm m}_{\rm H} = (-1)^{s+1} \, e \, k_z \, (B_x \, k_x + B_y \, k_y)/k^3$
(odd). Every BC- and OMM-dependent quantity that is even in $k_z$ for the TSM is
therefore odd in $k_z$ for the Hopf node, owing to the explicit $k_z$ factor in
$\bs\Omega^{\rm H}$ and $\bs m^{\rm H}$. Physically, the $z$-component of the
velocity, $v_0 \, k_z/k$, contributes one odd factor of $k_z$ to every
$\sigma_{zx}$ integrand. In the TSM, all BC- and OMM-dependent quantities are even
under $k_z \to -k_z$, so the overall integrand remains odd and vanishes upon
angular integration. In the HSM, the dipolar BC and OMM introduce an additional
factor of $k_z$, rendering these quantities odd; this parity change converts the
overall integrand from odd to even, yielding a finite angular average and hence a
nonzero $\sigma_{zx}$.

Set-ups~II and~III have no independent counterparts in the TSM, since they are
related to set-up~I by a global rotation $R \in SO(3)$: the TSM response in these
set-ups follows from Eqs.~(22)--(26) of Ref.~\cite{ips-spin1-ph} by the
substitution $B_y \to B_z$ in set-up~II and $B_x \to B_z$ in set-up~III, with all
coefficients unchanged. By contrast, the HSM supports several symmetry-allowed
structures absent in the TSM, including terms linear in $B_z$ and $B_x$, bilinear
terms proportional to $B_x \, B_z$, and cubic terms proportional to
$B_x^2 \, B_z$, $B_z^2 \, B_x$, $B_x^3$, and $B_z^3$. Among these, the odd powers
of $\bs B$ in the diagonal components are of particular interest. The HSM acquires
a term linear in $B_z$ in $\bar\sigma_{xx}$ in set-up~II, together with odd terms
in $\sigma^{\rm lf}_{xx}$ in set-up~II and in $\sigma^{\rm lf}_{zz}$ in set-up~III,
none of which exists for the TSM. For the TSM, the extra spherical symmetry of the
isotropic texture forces the relevant angular averages to vanish and leaves only
even powers of $\bs B$. For the HSM, the generalised Onsager relation of
Sec.~\ref{sechopfres} permits these terms, since the sign change of $\bs B$ is
compensated by that of the dipolar $\bs\Omega_s$ and $\bs m_s$.

For the AH part, the mirror-symmetry analysis of $\bs\Omega^l$ inside
Eq.~\eqref{eq-ah} determines which components of $\sigma^{\rm{ah}}_{ij}$ can be
nonzero in a given set-up. For the $SO(3)$-symmetric TSM, the in-plane AH response
$\sigma^{\rm{ah}}_{yx}$ vanishes in set-up~I because
$\langle \Omega_z \rangle_{\rm filled} = 0$ by the $z$-mirror symmetry of the
hedgehog texture, whereas the out-of-plane $\sigma^{\rm{ah}}_{zx}$ is finite
because the OMM-induced energy shift proportional to $B_y$ generates a nonzero
angular average of $\Omega_y$. For the $C_{\infty v}$-symmetric Hopf node, the BC-dipole breaks the $z$-mirror symmetry of the monopolar texture, so
$\langle \Omega_z \rangle_{\rm filled}$ is no longer constrained to vanish and both
$\sigma^{\rm{ah}}_{yx}$ and $\sigma^{\rm{ah}}_{zx}$ can be nonzero. In
set-up~I, the in-plane AH response of the dispersive TSM bands vanishes by the
$k_z$-parity of the monopolar BC, $\sigma^{\rm{ah}}_{yx}\big|_{\rm TSM} = 0$,
whereas the out-of-plane AH response acquires a term linear in $B_y$ and a
cubic-in-$\bs B$ correction,
\begin{align}
\label{eq-tsm-ah}
\sigma^{\rm{ah}}_{zx}\big|_{\rm TSM} = \frac{e^3 \, v_0 \, B_y}{120 \, \pi^2 \, \mu^5}
\left[-3 \, {\boldsymbol B}^2 \, e^2 \, v_0^4 - 10 \, \mu^4\right].
\end{align}
For the Hopf node, $\Omega^{\rm H}_z = k_z^2/k^4$ is even in $k_z$, so its
Fermi-sea integral is nonzero and produces the $\bs B$-independent contribution
$\sigma^{\rm{ah}}_{yx} = -e^2 \, \mu/(6 \, \pi^2 \, v_0)$ of
Eq.~\eqref{eqhfI-ah}. This is possible despite $\mathcal C = 0$ because
$\mathcal C$ is determined by the BC flux through a closed surface enclosing the
node (which vanishes), whereas the AH conductivity involves the volume integral of
$\Omega_z$ (which does not). For the out-of-plane component,
$\Omega^{\rm H}_y = k_z \, k_y/k^4$ is odd in $k_z$, and only the OMM-corrected
Fermi-surface terms survive. In the TSM, therefore, the AH response is confined to
the out-of-plane conductivity, whose leading contribution is linear in the magnetic
field, while the in-plane AH conductivity vanishes by the $k_z$-parity of the
monopolar BC. In the HSM, the even $k_z$-parity of $\Omega^{\rm H}_z$ permits a
finite in-plane AH conductivity, and the out-of-plane response begins only at
linear and cubic order in $\bs B$. For set-ups~II and~III, $B_z$ couples to the
dipole axis and the AH response acquires entirely new terms.

For the LFO-induced response, the qualitative differences between the TSM and the
HSM originate entirely from the dipolar angular modulation of the BC and OMM.
Although the two systems share the same band dispersion, the additional
$\cos\gamma$ factor changes the parity of the transport integrands, activating
additional nonzero responses. For the TSM, the Lorentz-force expansion together
with the BC and OMM symmetry leads to a strict correspondence between
magnetic-field order and current direction: out-of-plane currents carry even powers
of $\tau$ and odd powers of $B$, while in-plane currents carry odd powers of $\tau$
and even powers of $B$, with the quantum-geometric corrections feeding only the
out-of-plane response. Each topological term in $\bs{\mathcal N}_{n,\delta}$ enters
through the scalars $\bs\Omega_s \cdot \bs B$,
$\varepsilon_m = -\bs B \cdot \bs m_s$, or the vector
$\bs u_m = \nabla_{\bs k} \varepsilon_m$; upon performing the Fermi-surface angular
integral $\int d\Omega_{\hat{\bs k}}$, only integrands even under $k_z \to -k_z$
survive. For the radial monopole these terms are even, and the surviving
combinations land in the out-of-plane parts only, while the in-plane
$\bs{\mathcal N}_{1,2}$ integrand and the out-of-plane $bc/m$ part of
$\bs{\mathcal N}_{2,3}$ are odd and integrate to zero. For the dipole, every BC- or
OMM-dependent term carries one extra factor of $\cos\gamma$, which changes the
parity of the transport integrands and activates LFO contributions forbidden in the
monopolar case:
\begin{align}
\text{TSM:}\quad \text{BC/OMM} &\longrightarrow \text{out-of-plane at } n = 1 \nn
\text{Hopf:}\quad \text{BC/OMM} &\longrightarrow \begin{cases} \text{in-plane at } n = 1 \ (\bs{\mathcal N}_{1,2}) \\ \text{out-of-plane at } n = 2 \ (\bs{\mathcal N}_{2,3}) \end{cases}.
\end{align}

At $n = 1$ of set-up~I, the geometry-independent part $\sigma^{\rm lf,h}_{zx}$
takes the same form in the two systems with an identical prefactor, as expected
since it does not involve $\bs\Omega$ or $\bs m$ and depends only on the common
dispersion. The BC-only and OMM-only contributions to the in-plane transverse
$\sigma^{\rm lf}_{yx}$ are nonzero for the Hopf node but vanish identically for the
TSM: the additional $k_z$ factor carried by the dipolar BC and OMM textures
reverses the parity, rendering the corresponding integrands even and allowing
finite angular averages. The dipole therefore generates a quadratic-in-$\bs B$
in-plane planar-Hall Lorentz-force current that is symmetry-forbidden in the
monopolar case. The cubic-in-$\bs B$ out-of-plane components
$\sigma^{\rm lf,bc}_{zx}$ and $\sigma^{\rm lf,m}_{zx}$ are proportional to
$B_y \, {\boldsymbol B}^2$ in both systems; the concurrent contribution is positive
and partially compensates the negative BC- and OMM-induced corrections, reducing
the net suppression of the out-of-plane Lorentz-force conductivity.

At $n = 2$, the entire LFO-induced response is sourced by the geometry-independent
part for the TSM, while the BC-only and OMM-only contributions vanish --- the
direct counterpart of the $n = 1$ analysis. For the Hopf node, the
geometry-independent contributions again coincide with the TSM, and the qualitative
difference arises entirely from the BC- and OMM-dependent parts: the out-of-plane
contributions to $\sigma^{\rm lf}_{zx}$ vanish identically for the TSM but become
finite for the HSM, scaling as $B_x \, {\boldsymbol B}^2$. The parity change
renders the relevant angular averages even under $k_z \to -k_z$, activating the
out-of-plane Lorentz-force response forbidden in the monopolar case. The BC- and
OMM-induced contributions carry opposite signs, with the BC term dominating in
magnitude, so the net quantum-geometric correction remains positive and is governed
primarily by the BC-dipole.

At $n = 3$, the TSM response contributes only to the out-of-plane
$\sigma^{\rm lf}_{zx}$ through the geometry-independent component
\cite{ips-spin1-ph}, and the HSM exhibits exactly the same contribution:
\begin{align}
\label{eq-tsm-lf-n3}
\sigma^{\rm lf,h}_{zx}\big|_{\rm TSM} = \sigma^{\rm lf,h}_{zx}\big|_{\rm H} = \frac{\tau^4 \, e^5 \, v_0^5 \, B_y \, {\boldsymbol B}^2}{6 \, \pi^2 \, \mu}.
\end{align}
This is expected, since the integrand only carries a $(\bs v \times \bs B)$-
dependent term, common to the two systems through their identical dispersion.

In set-ups~II and~III, the TSM introduces no qualitatively new LFO-induced
response, since these are related to set-up~I by a global $SO(3)$ rotation, and all
conductivity tensors follow directly from those of set-up~I by the appropriate
rotation of the magnetic-field components. The HSM, lacking this rotational
equivalence because its BC and OMM textures have only $C_{\infty v}$ symmetry,
exhibits genuinely new LFO-induced responses with no monopolar analogue. From
Tables~\ref{tab:lf_setup1}--\ref{tab:lf_setup3}, in set-up~II at $n = 1$,
$\sigma^{\rm lf,bc}_{yx}$ acquires a term linear in $B_z$, in addition to the
$B_z \, B_x^2$ and $B_z^3$ terms; at $n = 2$, $\sigma^{\rm lf,bc}_{xx}$ contains
$B_z \, B_x^2$ and $B_z^3$ terms, and $\sigma^{\rm lf,bc}_{zx}$ contains $B_x^3$
and $B_x \, B_z^2$ terms; at $n = 3$, $\sigma^{\rm lf,h}_{yx}$ contains
$B_z \, B_x^2$ and $B_z^3$ terms. In set-up~III at $n = 1$,
$\sigma^{\rm lf,bc}_{yz}$ contains $B_x \, B_z$, $B_x^3$, and $B_x \, B_z^2$ terms;
at $n = 2$, $\sigma^{\rm lf,bc}_{zz}$ contains a $B_x^2 \, B_z$ term, and
$\sigma^{\rm lf,bc}_{xz}$ contains $B_x^3$ and $B_x \, B_z^2$ terms. Three new
families of responses therefore emerge: a finite low-field in-plane LF transverse
response $\sigma^{\rm lf,bc,m}_{yx} \propto {\boldsymbol B}^2$ in set-up~I,
forbidden in any isotropic monopolar nodal-point semimetal; a finite out-of-plane
$n = 2$ LFO-induced response
$\sigma^{\rm lf,bc,m}_{zx} \propto B_x \, {\boldsymbol B}^2$ from BC and OMM,
likewise forbidden in the monopolar case; and the appearance of $B_z$-parity-odd
LF-induced polynomials in set-ups~II and~III with no monopolar analogue. Each is
predictable on symmetry grounds alone from the $C_{\infty v}$ reduction of the
dipolar BC and OMM.

%%%%%%%%%%%%%%%%%%%%%%
\begin{table*}[ht!]
\centering
\caption{\label{tab-gnr_nonAH_ah}
Nonzero components of $\bar\sigma$ and $\sigma^{\rm{ah}}$ for the NVNR across the three
set-ups, kept up to $\mathcal{O}(|\bs B|^3)$.}
\begin{ruledtabular}
\begin{tabular}{lcccc}
 & & & \multicolumn{2}{c}{Out-of-plane transverse} \\
 \cline{4-5}
 & Longitudinal & In-plane transverse & Non-AH & AH \\
\colrule
Set-up~I
 & $\bar\sigma_{xx}\propto B^{0},  B_x^2,  B_y^2$
 & $\bar\sigma_{yx}\propto B_x\,B_y$
 & $\bar\sigma_{zx}=0$
 & $\sigma^{\rm{ah}}_{zx}\propto B_y,  B_y\, {\boldsymbol B}^2$;\quad
   $\sigma^{\rm{ah}}_{yx}=0$ \\[8pt]
Set-up~II
 & $\bar\sigma_{xx}\propto B^{0},  B_x^2$
 & $\bar\sigma_{zx}=0$
 & $\cdots$
 & $\sigma^{\rm{ah}}_{yx}=0$ \\[8pt]
Set-up~III
 & $\bar\sigma_{zz}\propto B^{0},  B_x^2$
 & $\bar\sigma_{xz}=0$
 & $\cdots$
 & $\sigma^{\rm{ah}}_{yz}\propto B_x,  B_x^3$ \\
\end{tabular}
\end{ruledtabular}
\end{table*}

\begin{table}[ht!]
\centering
\caption{\label{tab-gnr-lf}
$B_i$-dependence in the LFO-induced conductivity of the NVNR in
set-up~I, organised by the values of $n$
and physical origins.}
\begin{ruledtabular}
\begin{tabular}{lcccc}
 & $\sigma^{\text{lf, h}}$ & $\sigma^{\text{lf, bc}}$
 & $\sigma^{\text{lf, m}}$ & $\sigma^{\text{lf, conc}}$ \\
\colrule
$n=1$, $\sigma_{zx}$
 & $B_y$
 & $B_y\, {\boldsymbol B}^2$
 & $B_y\, {\boldsymbol B}^2$
 & $B_y\, {\boldsymbol B}^2$ \\[4pt]
$n=1$, $\sigma_{yx}$
 & $0$ & $0$ & $0$ & $0$ \\[4pt]
$n=2$, $\sigma_{xx}$
 & $B_x^2,  B_y^2$
 & $0$ & $0$ & $0$ \\[4pt]
$n=2$, $\sigma_{yx}$
 & $B_x\,B_y$
 & $0$ & $0$ & $0$ \\[4pt]
$n=2$, $\sigma_{zx}$
 & $0$ & $0$ & $0$ & $0$ \\[4pt]
$n=3$, $\sigma_{zx}$
 & $B_y\, {\boldsymbol B}^2$
 & $0$ & $0$ & $0$ \\
\end{tabular}
\end{ruledtabular}
\end{table}

%%%%%%%%%%%%%%%%
\begin{table}[ht!]
\centering
\caption{\label{tab-gnr-lf2}
$B_i$-dependence in the LFO-induced conductivity of the NVNR in
set-up~II, organised by the values of $n$
and physical origins. The absence of $\Omega^{z}$ in the NVNR nulls a
large family of terms that are finite in the VNR.}
\begin{ruledtabular}
\begin{tabular}{lcccc}
 & $\sigma^{\text{lf, h}}$ & $\sigma^{\text{lf, bc}}$
 & $\sigma^{\text{lf, m}}$ & $\sigma^{\text{lf, conc}}$ \\
\colrule
$n=1$, $\sigma_{yx}$
 & $0$
 & $B_z\,B_x^2$
 & $B_z\,B_x^2$
 & $0$ \\[4pt]
$n=2$, $\sigma_{xx}$
 & $B_x^2,  B_z^2$
 & $0$ & $0$ & $0$ \\[4pt]
$n=2$, $\sigma_{zx}$
 & $0$ & $0$ & $0$ & $0$ \\[4pt]
$n=3$, $\sigma_{yx}$
 & $B_z\,B_x^2,  B_z^3$
 & $0$ & $0$ & $0$ \\
\end{tabular}
\end{ruledtabular}
\end{table}

%%%%%%%%%%%%%%%%%%%%%%%
\begin{table}[ht!]
\centering
\caption{\label{tab-gnr-lf3}
$B_i$-dependence in the LFO-induced conductivity of the NVNR in
set-up~III, organised by the values of $n$
and physical origins.}
\begin{ruledtabular}
\begin{tabular}{lcccc}
 & $\sigma^{\text{lf, h}}$ & $\sigma^{\text{lf, bc}}$
 & $\sigma^{\text{lf, m}}$ & $\sigma^{\text{lf, conc}}$ \\
\colrule
$n=1$, $\sigma_{yz}$
 & $B_x$
 & $B_x^3,  B_x\,B_z^2$
 & $B_x^3,  B_x\,B_z^2$
 & $0$ \\[4pt]
$n=2$, $\sigma_{zz}$
 & $B_x^2$
 & $0$ & $0$ & $0$ \\[4pt]
$n=2$, $\sigma_{xz}$
 & $0$ & $0$ & $0$ & $0$ \\[4pt]
$n=3$, $\sigma_{yz}$
 & $B_x^3,  B_x\,B_z^2$
 & $0$ & $0$ & $0$ \\
\end{tabular}
\end{ruledtabular}
\end{table}

%======================================================================
%======================================================================
\section{Summary and future directions}
\label{secsum}
%======================================================================

We have carried out a complete analytical computation of the linear magnetoelectric conductivity tensor $\bs\sigma$ for two distinct representatives of a broader class of semimetals whose quantum geometry is organised by a BC-dipole rather than by a BC-monopole. The two representatives are complementary. The VNR~\cite{vortex-nrsm} has a smoke-ring pseudospin texture, which generates a nonsingular $\bs\Omega_s(\bs k)$ and $\bs m(\bs k)$ whose dipolar character is inherited from the toroidal topology of the Fermi surface, realised as a continuous ring of local BC-dipoles rather than a single point dipole. The three-band HSM~\cite{graf-hopf,hopf2,hopf3}, $H_h $, instead has a single, point-like dipolar $\bs\Omega_s(\bs k)\propto\cos\gamma$ that is an algebraic property of $H_h $ at the node, together with an additional zero-energy flat-band pinned by the chiral symmetry $\mathcal S=\mathrm{diag}(1,-1,-1)$. In both models we organised the response by the same three inequivalent planar-Hall configurations. Working within the RTA to the semiclassical Boltzmann formalism, we computed on equal footing the Drude, BC-only, OMM-only, concurrent, AH, and LFO-induced responses, expanding the integrands consistently to $\mathcal{O}(|\bs B|^3)$ and decomposing them according to the order $n\in\{1,2,3\}$ of the LFO.

For the VNR, the results are organised using $k_\mu $ to compactify the expressions. The tensor structure of $\bs\sigma$ is genuinely richer than that of the $\mathcal{PT}$-broken NVNR analysed earlier~\cite{ips-nlsm-ph,ips-gnr-strain}. In set-up~I, $\bar\sigma$ acquires a nonvanishing out-of-plane response $\sigma^{\rm bc}_{zx},\sigma^{\rm m}_{zx},\sigma^{\rm conc}_{zx}\propto B_x$ and $B_x \,{\boldsymbol B}^2$, which was forbidden for the NVNR both by symmetry and by the vanishing of $\Omega^z_s$. The intrinsic AH conductivity $\sigma^{\rm{ah}}_{yx}$ survives at zeroth order in $\bs B$, and the LFO-induced conductivity generates transverse and in-plane responses, including out-of-plane $\sigma^{\rm lf}_{zx}$ scaling as $\tau^2\,B_y \,{\boldsymbol B}^2/\mu^3$, longitudinal Hall components scaling as $\tau^3\,B_y^2$, and a purely $\tau^4$-driven cubic-in-$\bs B$ contribution at $n=3$ in all three set-ups. Odd-in-$\bs B$ terms of this kind, forbidden for a BC-monopole, recur throughout every set-up.

For the HSM, the three set-ups reveal a complementary set of dipolar signatures. First, the internode-scattering contribution vanishes identically to all orders in $\bs B$, $(\sigma_s^{\chi,\rm inter})_{ij}=0$. This is a direct consequence of the extra $\cos\gamma$ factor carried by the dipolar $\bs\Omega_s$ and $\bs m_s$, which makes both even in momentum and the charge-pumping vector $\bs{\mathcal I}^\chi$ vanish at each node, irrespective of the oppositely oriented BC-dipoles of the two conjugate nodes. The negative longitudinal magnetoresistance ubiquitous in Weyl and multifold monopole semimetals due to internode scattering is therefore absent in the HSM, providing the sharpest qualitative discriminant between BC-monopole and BC-dipole systems. Second, the reduction from $O(3)$ (spectrum) to $C_{\infty v}$ (eigenstates) activates the same kind of monopole-forbidden channels identified for the VNR above, here an out-of-plane non-AH response in set-up~I, an in-plane planar-Hall AH component $\sigma^{\rm{ah}}_{yx}\propto {\boldsymbol B}^2$ at $\mathcal{O}(|\bs B|^2)$ in set-up~I, and odd-in-$B_z$ terms that are linear, bilinear, and cubic in set-ups~II and~III, with no counterpart in the TSM. The LFO likewise induces such forbidden responses. These single-node channels are odd in the dipole orientation. They cancel between the two nodes of a valley-HSM and survive in a topological HSM with a single Hopf crossing. Remarkably, the same set of tensor components is activated in both the VNR and the HSM despite their different microscopic origins: whenever the reversal of $\bs B$ is tied to that of a dipolar $\bs\Omega_s$ and $\bs m_s$ rather than to an isotropic monopole texture, the generalised Onsager relation admits terms that are symmetry-forbidden in the monopolar case. This activation is therefore a generic fingerprint of a dipolar BC field rather than a feature specific to either model.

The odd-in-$\bs B$ terms of the diagonal components obey a simple selection rule, which holds for both systems. The mirror symmetry through the $k_z=0$ plane maps $\bs B$ to $(-B_x,-B_y,B_z)$ and leaves the diagonal components unchanged. These components therefore contain only even total powers of the in-plane field, and $B_z$ is the sole field component that can carry an odd power. The extra factor of $k_z$ in the dipolar BC and OMM makes $\Omega_s^z$ and $m_s^z$ even in $k_z$, and the odd powers of $B_z$ survive the integration. For a BC-monopole, $\Omega_s^z$ is odd in $k_z$ and the same terms cancel. Such terms appear only in set-ups~II and~III, namely in $\bar\sigma_{xx}$ and $\sigma^{\rm lf}_{xx}$ in set-up~II and in $\bar\sigma_{zz}$ and $\sigma^{\rm lf}_{zz}$ in set-up~III. They are absent from the diagonal components of set-up~I, where $B_z=0$. The longitudinal conductivity thus acquires a contribution that is odd under the reversal of the field component along the dipole axis, whereas it remains even under the joint reversal of the in-plane components. This direction dependence provides a further experimentally accessible discriminant between BC-dipole and BC-monopole systems.

Several avenues open up as natural extensions. A first concrete direction is the systematic derivation of nonlinear response for both systems in the same three set-ups. Ref.~\cite{ips-dipole-vnr} computed the linear longitudinal magnetoconductivity exactly for collinear fields along the anisotropy axis, together with the second-order electrochemical response within the RTA. In monopolar nodal-point semimetals, the intrinsic, disorder-independent second-order response driven by the Berry-curvature dipole (BCD)~\cite{bcd-inti} vanishes identically in the isotropic limit, because the hedgehog BC texture is spherically symmetric on the Fermi sphere. A finite BCD requires tilt or crystalline anisotropy~\cite{roy2022_nonlinear, ma2019_nlhe, kang2019_nlhe}. The same holds for the VNR and the HSM. Although their BC fields are intrinsically dipolar, the corresponding coefficients vanish for an untilted dispersion by the $k_z$-parity of the integrands and become finite only once a tilt is introduced~\cite{ips-dipole-vnr}. Computing the tilt-dependent coefficients in the toroidal coordinates of the VNR and the spherical coordinates of the Hopf node, in all three planar-Hall set-ups, would extend the collinear-field analysis of Ref.~\cite{ips-dipole-vnr} to the full second-order conductivity tensor. A closely related and experimentally cleaner direction is the finite-frequency response: the optical Kerr and Faraday rotations, the a.c. planar-Hall response, and the circular photogalvanic and shift-current response-features~\cite{juan_quantized, grushin-multifold, nonlin-photo, yadav23_magneto}.

A second direction concerns the thermoelectric and thermal analogues of the responses derived here. Applying the Mott and Wiedemann--Franz relations in the presence of the OMM will yield closed forms for the Peltier effect and the thermal Hall conductivity in each of the three set-ups. The finite $\Omega^z_s$ may generically produce a nonzero anomalous Nernst signal even in configurations where the ordinary Nernst effect is forbidden by symmetry. A third direction is to relax the two assumptions on which the present analysis rests: the momentum-independent $\tau$ and the RTA. Beyond the RTA \cite{timm, ips-exact-spin1, ips-exact-kwn, ips-exact-rsw}, a full linearised collision integral will include intra- and inter-nodal-loop processes for the VNR and interband scatterings for the HSM. Beyond the linearised Hamiltonian, a lattice-regularised study of the parent VNR Hamiltonian~\cite{vortex-nrsm} and of the lattice Hopf model~\cite{graf-hopf} would let one test the robustness of the intrinsic AH conductivity against ultraviolet regularisation, interband mixing away from the node, and finite temperature.  A fourth direction is to add strain, which would introduce pseudo-electromagnetic fields, opening the way to a systematic study of strain-induced response in both systems~\cite{ips-sanskar, ips-gnr-strain}. Electron-electron interactions may renormalise the geometric quantities entering the Boltzmann formalism. Identifying whether such interaction-driven modifications can be distinguished from disorder-driven ones is an interesting direction to follow.
A fifth idea of investigation is to resort to Kubo formula which can access conductivity for strong-field regime where semiclassical Boltzmann formalism will become inapplicable \cite{yadav23_magneto, dirac-qtm, ips-sanskar-2dgnr}.

Finally, connecting the abstract analysis presented here to real materials is the essential next step for both systems. Identifying candidate compounds that host smoke-ring pseudospin textures, for which the two-orbital lattice model of Ref.~\cite{vortex-nrsm} offers a natural starting point, as well as candidate three-band systems that realise a Hopf node in a controllable regime, and probing them through the discriminants proposed here, would provide a physical rather than a purely theoretical organising principle for topological magnetotransport.

%======================================================================
\begin{acknowledgments}
This research, leading to the results reported, has received funding from the Council of Science \& Technology (CST), U.P.
\end{acknowledgments}
%======================================================================%======================================================================

\appendix

%======================================================================
\section{Non-AH contribution to the magnetoelectric conductivity}
\label{app_nonAH}
%======================================================================

Here we give the detailed expansion of the non-AH, non-LFO contribution $\bar\sigma_s$ of Eq.~\eqref{eq_elec} up to order $|\bs{B}|^3$, and establish the covariance property of the three physically distinct contributions used in Sec.~\ref{secmethod}. We adopt the velocity notation of Ref.~\cite{ips-rsw-ph} throughout, adapting it to the present nodal-line system where the OMM-induced parts are band-independent.

The OMM generates a Zeeman-like energy shift $\varepsilon_m(\bs{k})$, which is linear in $|\bs{B}|$ and band-independent. We define the OMM-corrected energy and the corresponding modified Bloch velocity as
\begin{align}
\label{eq_omm_energy_vel}
\mathcal{E}_s \left(\bs{k} \right) = \varepsilon_s(\bs{k}) + \varepsilon_m(\bs{k})
\quad \text{and} \quad
\bs{w}_s(\bs{k}) = \bs{v}_s(\bs{k}) + \bs{u}_m(\bs{k}) \,,
\end{align}
where $\bs{v}_s = \nabla_{\bs{k}} \varepsilon_s$ is the bare Bloch velocity and $\bs{u}_m = \nabla_{\bs{k}} \varepsilon_m$ is the OMM-induced velocity correction. Although $\varepsilon_m$ is band-independent, the gradient $\bs{u}_m$ carries a band index because it depends on the momentum-space structure. We define
\begin{align}
\bs{W}_s = e \left ( \bs{w}_s \cdot \bs{\Omega}_s \right ) \bs{B} \,.
\end{align}
Following Ref.~\cite{ips-rsw-ph}, we further decompose $\bs{W}_s$ as
\begin{align}
(W_s)_i = (V_s)_i + (u_m)_i \,,\quad
(V_s)_i = e \sum_q (v_s)_q \, (\Omega_s)_q \, B_i \,,\quad
(u_m)_i = e \sum_q (u_m)_q \, (\Omega_s)_q \, B_i \,,
\end{align}
where $(V_s)_i$ is linear in $|\bs{B}|$ (arising from $\bs{v}_s$) and $(u_m)_i$ is quadratic in $|\bs{B}|$ (since $\bs{u}_m$ is itself linear in $|\bs{B}|$). The full velocity entering the conductivity expression is thus
\begin{align}
(w_s)_i + (W_s)_i = (v_s)_i + (u_m)_i + (V_s)_i + (u_m)_i \,,
\end{align}
with order in $|\bs{B}|$: $0, \, 1, \, 1, \, 2$, respectively.

The phase-space factor is
\begin{align}
D_s = \frac{1}{1 + e \left ( \bs{\Omega}_s \cdot \bs{B} \right )}
= 1 - e \left ( \bs{\Omega}_s \cdot \bs{B} \right )
+ e^2 \left ( \bs{\Omega}_s \cdot \bs{B} \right )^2
- e^3 \left ( \bs{\Omega}_s \cdot \bs{B} \right )^3
+ \mathcal{O}(|\bs{B}|^4) \,.
\end{align}
The Fermi-Dirac derivative is expanded about the unshifted dispersion $\varepsilon_s$ as
\begin{align}
\label{eq_fdexp}
f_0^\prime (\mathcal{E}_s) &= f_0^\prime (\varepsilon_s)
+ \varepsilon_m \, f_0^{\prime \prime} (\varepsilon_s)
+ \frac{\left(\varepsilon_m\right)^2}{2} \, f_0^{\prime \prime \prime} (\varepsilon_s)
+ \frac{\left(\varepsilon_m\right)^3}{3!} \, f_0^{\prime \prime \prime \prime} (\varepsilon_s)
+ \mathcal{O}(|\bs{B}|^4) \,.
\end{align}
The non-AH contribution (excluding the LFO-induced part) is given by
\begin{align}
\label{eq_sigmabar_master}
\left(\bar\sigma_s \right)_{i j}
= - \,e^2 \, \tau
\int \frac{ d^3 \bs{k}}{(2\,\pi)^3 } \, D_s \,
\left[  (w_s)_i + (W_s)_i \right ]
\left [ (w_s)_j + (W_s)_j \right] \, f^\prime_0 (\mathcal{E}_s)  \,.
\end{align}
We decompose this as
\begin{align}
\label{eq_decomp}
{\bar\sigma}_s  =
\sigma^{\text{d}}_s + \sigma^{\text{bc}}_s
+ \sigma^{\text{m}}_s
+ \sigma^{\text{conc}}_s\,,
\end{align}
where the Drude part is independent of $\bs{B}$, the BC-only part survives when the OMM is set to zero (i.e., $\varepsilon_m \to 0$, $\bs{u}_m \to \bs{0}$), the third part vanishes if the OMM is ignored, and the fourth part vanishes if either the BC or the OMM is ignored.

The Drude part is independent of $\bs{B}$ and remains unchanged:
\begin{align}
\left ( \sigma^{\text{d}}_s \right )_{i j} =
- \, \frac{e^2 \, \tau }{(2\,\pi)^3}
\int d^3 \bs{k} \, (v_s)_i \, (v_s)_{j} \,
f^\prime_0  (\varepsilon_s ) \,.
\end{align}

Setting $\varepsilon_m = 0$ and $\bs{u}_m = \bs{0}$ eliminates all OMM contributions. The remaining terms involve $\bs{v}_s$ and $\bs{V}_s$ only (since $\bs{u}_m$ also vanishes). The BC-only part is
\begin{align}
(\sigma^{\text{bc}}_s)_{i j} &=
- \, \frac{e^2 \, \tau }{(2\,\pi)^3}
\int d^3 \bs{k} \,  \mathcal{M}_{i j}   \,  f^\prime_0  (\varepsilon_s )\,,
\end{align}
where
\begin{align}
\mathcal{M}_{ij} = \left [ D_s \left \lbrace \frac{(v_s)_i (v_s)_j}{2} + (v_s)_i (V_s)_j + \frac{1}{2} (V_s)_i (V_s)_j \right \rbrace + i \leftrightarrow j \right ] - (v_s)_i (v_s)_j \,.
\end{align}
Expanding in powers of $|\bs{B}|$:
\begin{align}
\mathcal{M}_{i j} = \mathcal{M}_{i j}^{(1)} + \mathcal{M}_{i j}^{(2)} + \mathcal{M}_{i j}^{(3)} + \mathcal{O}(|\bs{B}|^4) \,.
\end{align}
\begin{align}
\label{eq_M}
\mathcal{M}_{i j}^{(1)} & = \left [
- \frac{e}{2} \left( \bs{\Omega}_s \cdot \bs{B}\right) (v_s)_i \, (v_s)_j
+ e \left ( \bs{v}_s \cdot \bs{\Omega}_s \right ) (v_s)_i \, B_j
\right ] + i \leftrightarrow j  \,, \nn
\mathcal{M}_{i j}^{(2)} & =
\Bigg [
\frac{e^2}{2} \left( \bs{\Omega}_s \cdot \bs{B}\right)^2  (v_s)_i \, (v_s)_j
- e^2 \left( \bs{\Omega}_s \cdot \bs{B}\right) \left ( \bs{v}_s \cdot \bs{\Omega}_s \right ) (v_s)_i \, B_j
+ \frac{e^2}{2} \left ( \bs{v}_s \cdot \bs{\Omega}_s \right )^2 B_i \, B_j  \Bigg ] + i \leftrightarrow j \,, \nn
\mathcal{M}_{i j}^{(3)} & =
\Bigg [
- \frac{e^3}{2}  \left (\bs{\Omega}_s \cdot \bs{B} \right )^3 (v_s)_i \, (v_s)_j
+ e^3 \left (\bs{\Omega}_s \cdot \bs{B} \right )^2 \left (\bs{v}_s \cdot \bs{\Omega}_s  \right ) (v_s)_i \, B_j
- \frac{e^3 }{2}\left (\bs{\Omega}_s \cdot \bs{B} \right ) \left (\bs{v}_s \cdot \bs{\Omega}_s  \right )^2 B_i \, B_j  \Bigg ] + i \leftrightarrow j \,.
\end{align}
The sign alternation $\{+,-,+,-,\ldots \}$ at successive orders follows from the geometric series expansion of $D_s$, and the three structures at each order correspond to the $D_s$ acting on $v_i v_j$, the cross term $v_i V_j$ modulated by $D_s$, and the $V_i V_j$ term, respectively.

The OMM shifts the dispersion by $\varepsilon_m(\bs{k})$. Let us define
\begin{align}
(u_m)_i = \partial_{k_i} \varepsilon_m \,,
\end{align}
which is the OMM-induced velocity correction, linear in $|\bs{B}|$. The OMM contributions enter through two responses: the velocity correction $\bs{u}_m$ (and the resulting $\bs{u}_m$), which modifies the total velocity, and the Taylor expansion of $f_0'(\mathcal{E}_s)$ about $\varepsilon_s$, which introduces powers of $\varepsilon_m$. We organise the OMM-dependent terms by separating them into purely OMM (containing $\varepsilon_m$ and $\bs{u}_m$ but no BC factors) and concurrent (involving both BC factors $\bs{\Omega}_s$ and OMM quantities).

The OMM-only terms arise from the terms involving $\bs{v}_s$ and $\bs{u}_m$ only (no $\bs{V}_s$, $\bs{u}_m$, and with $D_s$ set to unity), multiplied by the appropriate derivatives of $f_0$. The OMM-only part of the conductivity takes the form
\begin{align}
\label{eq_omm_general}
\left(\sigma^{m}_s\right)^{\text{OMM}}_{ij} = -e^2\,\tau \int \frac{d^3 \bs{k}}{(2\,\pi)^3} \left [ \mathcal{I}^{(1)}_{ij} + \mathcal{I}^{(2)}_{ij} + \mathcal{I}^{(3)}_{ij} \right ] + i \leftrightarrow j \,,
\end{align}
where the superscript denotes the order in $|\bs{B}|$. The integrands are, at linear order,
\begin{align}
\label{eq_omm_lin}
\mathcal{I}^{(1)}_{ij} =
(v_s)_i (u_m)_j  \, f_0'(\varepsilon_s)
+ \frac{(v_s)_i (v_s)_j}{2} \, \varepsilon_m \, f_0''(\varepsilon_s)  \,,
\end{align}
at quadratic order,
\begin{align}
\label{eq_omm_quad}
\mathcal{I}^{(2)}_{ij} =
\frac{(u_m)_i (u_m)_j}{2} \, f_0'(\varepsilon_s)
+  (v_s)_i (u_m)_j \, \varepsilon_m \,  f_0''(\varepsilon_s)
+ \frac{(v_s)_i (v_s)_j}{2}  \frac{\left(\varepsilon_m\right)^2 }{2} \, f_0'''(\varepsilon_s) \,,
\end{align}
and at cubic order,
\begin{align}
\label{eq_omm_cub}
\mathcal{I}^{(3)}_{ij} =
\frac{(u_m)_i (u_m)_j}{2}  \, \varepsilon_m \, f_0''(\varepsilon_s)
+ \frac{\left(\varepsilon_m\right)^2}{2!} \,  (v_s)_i (u_m)_j \, f_0'''(\varepsilon_s)
+ \frac{\left(\varepsilon_m\right)^3}{3!}\frac{(v_s)_i (v_s)_j}{2}  \, f_0''''(\varepsilon_s) \,.
\end{align}
Note the appearance of the fourth derivative $f_0''''$ for the first time at cubic order.

We can now express the OMM-only part compactly as
\begin{align}
\label{eq_sigma_OMM_master}
\left(\sigma^{m}_s\right)^{\text{OMM}}_{ij} = -e^2\,\tau \int \frac{d^3 \bs{k}}{(2\,\pi)^3} \left [
\mathcal{S}_{ij} \, f_0'(\varepsilon_s)
+ \mathcal{P}_{ij} \, f_0''(\varepsilon_s)
+ \frac{1}{2} \, \mathcal{T}_{ij} \, f_0'''(\varepsilon_s)
+ \frac{1}{6} \, \mathcal{V}_{ij} \, f_0''''(\varepsilon_s)
\right ] \,,
\end{align}
where, on keeping terms up to cubic order in $|\bs{B}|$,
\begin{align}
\label{eq_OMM_symbols}
\mathcal{S}_{ij} &= \left [
(v_s)_i \, (u_m)_j
+ \frac{(u_m)_i \, (u_m)_j}{2}
\right ] + i \leftrightarrow j \,,
\nn
\mathcal{P}_{ij} &= \left [
\frac{(v_s)_i \, (v_s)_j}{2} \, \varepsilon_m
+ (v_s)_i \, (u_m)_j \, \varepsilon_m
+ \frac{(u_m)_i \, (u_m)_j}{2} \, \varepsilon_m
\right ] + i \leftrightarrow j \,,
\nn
\mathcal{T}_{ij} &= \left [
\frac{(v_s)_i \, (v_s)_j}{2} \left(\varepsilon_m\right)^2
+ (v_s)_i \, (u_m)_j \left(\varepsilon_m\right)^2
\right ] + i \leftrightarrow j \,,
\nn
\mathcal{V}_{ij} &= \left [
\frac{(v_s)_i \, (v_s)_j}{2} \left(\varepsilon_m\right)^3
\right ] + i \leftrightarrow j \,.
\end{align}

The concurrent terms involve at least one factor of $\bs{\Omega}_s$ (from $D_s$, $\bs{V}_s$, or $\bs{u}_m$) together with at least one OMM quantity ($\varepsilon_m$ or $\bs{u}_m$). We can express the relevant part of the conductivity as
\begin{align}
\label{eq_sigma_conc_master}
\left(\sigma^{m}_s\right)^{\text{conc}}_{ij} = -e^2 \, \tau \int \frac{d^3\,  \bs{k}}{(2\,\pi)^3} \left [
\tilde{\mathcal{S}}_{ij} \, f_0'(\varepsilon_s)
+ \tilde{\mathcal{P}}_{ij} \, f_0''(\varepsilon_s)
+ \frac{1}{2} \, \tilde{\mathcal{T}}_{ij} \, f_0'''(\varepsilon_s)
\right ] \,,
\end{align}
where, on keeping terms up to cubic order in $|\bs{B}|$,
\begin{align}
\label{eq_conc_symbols}
\tilde{\mathcal{S}}_{ij} &= \Bigg [
f_0'(\varepsilon_s) \bigg \lbrace
e \left(\bs{u}_m \cdot \bs{\Omega}_s \right) (v_s)_i \, B_j
- e \left(\bs{\Omega}_s \cdot \bs{B} \right) (v_s)_i \, (u_m)_j
+ e \left(\bs{v}_s \cdot \bs{\Omega}_s \right) (u_m)_i \, B_j
\bigg \rbrace
\nn & \quad
+ e^2 \left(\bs{\Omega}_s \cdot \bs{B} \right)^2 (v_s)_i \, (u_m)_j
+ e \left(\bs{u}_m \cdot \bs{\Omega}_s \right) (u_m)_i \, B_j
- e^2 \left(\bs{\Omega}_s \cdot \bs{B} \right) \left(\bs{u}_m \cdot \bs{\Omega}_s \right) (v_s)_i \, B_j
\nn & \quad
- e^2 \left(\bs{\Omega}_s \cdot \bs{B} \right) \left(\bs{v}_s \cdot \bs{\Omega}_s \right) (u_m)_i \, B_j
+ e^2 \left(\bs{v}_s \cdot \bs{\Omega}_s \right) \left(\bs{u}_m \cdot \bs{\Omega}_s \right) B_i \, B_j
\nn & \quad
- \frac{e}{2} \left(\bs{\Omega}_s \cdot \bs{B} \right) (u_m)_i \, (u_m)_j
\Bigg ] + i \leftrightarrow j \,,
\nn
\tilde{\mathcal{P}}_{ij} &= \Bigg [
\varepsilon_m \bigg \lbrace
e \left(\bs{u}_m \cdot \bs{\Omega}_s \right) (v_s)_i \, B_j
- e \left(\bs{\Omega}_s \cdot \bs{B} \right) (v_s)_i \, (u_m)_j
- e^2 \left(\bs{\Omega}_s \cdot \bs{B} \right) \left(\bs{v}_s \cdot \bs{\Omega}_s \right) (v_s)_i \, B_j
\nn & \quad
+ e \left(\bs{u}_m \cdot \bs{\Omega}_s \right) (v_s)_i \, B_j
+ \frac{e^2}{2} \left(\bs{\Omega}_s \cdot \bs{B} \right)^2 (v_s)_i \, (v_s)_j
+ \frac{e^2}{2} \left(\bs{v}_s \cdot \bs{\Omega}_s \right)^2 B_i \, B_j
\bigg \rbrace
\nn & \quad
+ \varepsilon_m \bigg \lbrace
- e \left(\bs{\Omega}_s \cdot \bs{B} \right) \frac{(v_s)_i \, (v_s)_j}{2}
+ e \left(\bs{v}_s \cdot \bs{\Omega}_s \right) (v_s)_i \, B_j
\bigg \rbrace
\Bigg ] + i \leftrightarrow j \,,
\nn
\tilde{\mathcal{T}}_{ij} &= \left [
\left(\varepsilon_m \right)^2 \bigg \lbrace
e \left(\bs{v}_s \cdot \bs{\Omega}_s \right) (v_s)_i \, B_j
- \frac{e}{2} \left(\bs{\Omega}_s \cdot \bs{B} \right) (v_s)_i \, (v_s)_j
\bigg \rbrace
\right ] + i \leftrightarrow j \,.
\end{align}
Here, $\tilde{\mathcal{S}}_{ij}$ pairs with $f_0'$ and contains terms at $\mathcal{O}(|\bs{B}|^2)$ and $\mathcal{O}(|\bs{B}|^3)$, $\tilde{\mathcal{P}}_{ij}$ pairs with $f_0''$ and contains terms at $\mathcal{O}(|\bs{B}|^2)$ and $\mathcal{O}(|\bs{B}|^3)$, and $\tilde{\mathcal{T}}_{ij}$ pairs with $f_0'''$ and appears only at $\mathcal{O}(|\bs{B}|^3)$. We note that there is no concurrent term pairing with $f_0''''$, in contrast to the OMM-only response.

The quadratic-only concurrent terms can be identified as:
\begin{align}
\tilde{\mathcal{S}}_{ij} \big \rvert_{\mathcal{O}(|\bs{B}|^2)} & = \left [
e \left(\bs{u}_m \cdot \bs{\Omega}_s \right) (v_s)_i \, B_j
- e \left(\bs{\Omega}_s \cdot \bs{B} \right) (v_s)_i \, (u_m)_j
+ e \left(\bs{v}_s \cdot \bs{\Omega}_s \right) (u_m)_i \, B_j
\right ] + i \leftrightarrow j \,, \nn
\tilde{\mathcal{P}}_{ij} \big \rvert_{\mathcal{O}(|\bs{B}|^2)} & = \varepsilon_m \left [
- e \left(\bs{\Omega}_s \cdot \bs{B} \right) \frac{(v_s)_i \, (v_s)_j}{2}
+ e \left(\bs{v}_s \cdot \bs{\Omega}_s \right) (v_s)_i \, B_j
\right ] + i \leftrightarrow j  \,.
\end{align}
These are analogous to the $\mathcal{Q}_{ij}$ and $\mathcal{R}_{ij}$ of Ref.~\cite{ips-rsw-ph}, adapted to the present system.

%%%%%%%%%%%%%%%%%%%%%%%%%%%%%%%%%%%%%
\subsection{Covariance of the three physically distinct contributions}
\label{secAcovariance}

We now show that the covariance relation~\eqref{eq:covariance}, motivated in Sec.~\ref{secmethod} for the full tensor $\sigma_{ij}(\bs B)$, holds separately for each of $\bar\sigma_{ij}$, $\sigma^{\rm{ah}}_{ij}$, and $\sigma^{\rm{lf}}_{ij}$, given respectively by Eqs.~\eqref{eq_elec}, \eqref{eq-ah}, and \eqref{lf_cond1}. This follows because every microscopic ingredient entering these three expressions is itself covariant under $U(\varphi)$. The transport velocity $\bs w_s$ is the $\bs k$-gradient of the rotation-invariant dispersion $\varepsilon_s(\bs k)$, and therefore rotates as an ordinary vector under $\mathcal R_z$. Although $\bs\Omega_s$ and $\bs m_s$ are pseudovectors, a proper rotation about $\hat{\bs z}$ acts on a pseudovector in exactly the same way as on an ordinary vector. Consequently, $\bs\Omega_s$ and $\bs m_s$ inherit the same covariance as $\bs w_s$. The applied field $\bs B$, being a lab-frame vector, transforms in the same way. Consequently, the phase-space factor $D_s=1-e\,(\bs B\cdot\bs\Omega_s)$ and the field-corrected dispersion $\mathcal E_s$ are rotational scalars, since both are contractions of two covariant vectors. The LFO, $\check L=(\bs w_s\times\bs B)\cdot\nabla_{\bs k}$, is covariant for the same reason: $\bs w_s\times\bs B$ transforms as a vector under a proper rotation, and $\nabla_{\bs k}$ inherits the covariance of $\bs k$ established through $\mathcal H_\ell(\mathcal R_z\,\bs k)$ and $H_h(\mathcal R_z\,\bs k)$ in Sec.~\ref{secsymmetry}. Since $\mathcal Y_s$ is linear in $\bs E$ and is built by repeated action of the covariant operator $\check L$ on the covariant vector $D_s\{\bs w_s+e(\bs w_s\cdot\bs\Omega_s)\bs B\}$, its coefficient $\partial\mathcal Y_s/\partial E_j$ remains a covariant vector at every order $n$ in the series defining $\mathcal Y_s$. The $\bs k$-integration is invariant under the relabelling $\bs k\to\mathcal R_z\,\bs k$. Every ingredient entering $\bar\sigma_{ij}$, $\sigma^{\rm{ah}}_{ij}$, and $\sigma^{\rm{lf}}_{ij}$ therefore satisfies the covariance relation~\eqref{eq:covariance} order by order in $|\bs B|$, since $\Theta$ acts on the microscopic ingredients of the Boltzmann formalism with the same covariance established above for $U(\varphi)$.

%======================================================================
\section{Evaluation of the AH conductivity}
\label{sec:AHeval}
%======================================================================

The absence of time-reversal symmetry is directly reflected in the BC texture. For the dispersive bands, $\bs\Omega_s(-\bs k)=\bs\Omega_s(\bs k)$, since the combination $k_z\,\bs k$ is even under $\bs k\rightarrow-\bs k$. By contrast, time-reversal symmetry requires the BC to satisfy $\bs\Omega_s(-\bs k)=-\bs\Omega_s(\bs k)$, which is incompatible with the dipolar BC field of the HSM. The absence of time-reversal symmetry has an immediate transport consequence. The intrinsic AH conductivity contains the BC integral
\begin{align}
\sigma^{\rm{ah}}_{yx}\propto \int d^3{\bs k} \,\Omega_z  \left(- \, \frac{\partial f_0}{\partial\varepsilon}\right),
\end{align}
which is nonzero only because $\Omega_z(\bs k)$ is even under $\bs k\rightarrow-\bs k$. Thus, the intrinsic AH response presented in the main text is a direct manifestation of the broken time-reversal symmetry encoded in the BC-dipole. It is precisely this $\bs k$-even parity of the dipolar BC, incompatible with the standard time-reversal requirement $\bs\Omega_s(-\bs k)=-\bs\Omega_s(\bs k)$, that underlies the generalised Onsager relation
$\sigma_{ij}(\bs B,\bs\Omega_s,\bs m_s)=\sigma_{ji}(-\bs B,-\bs\Omega_s,-\bs m_s)$
used in Secs.~\ref{secvnr} and~\ref{sechopfres}: the sign change of $\bs B$ in an odd-in-$\bs B$ term is compensated by that of the dipolar $\bs\Omega_s$ and $\bs m_s$, so that individual contributions to the diagonal components can carry odd powers of $\bs B$.

In this section we lay out the procedure used to evaluate the AH component of the conductivity tensor, starting from the general expression of Eq.~\eqref{eq-ah}, and reducing it to a sequence of angular integrals through the use of the standard delta-function identities valid in the $T\to 0$ limit. We focus on the conduction band (dropping the index $s$) and work with the linearised dispersion
\begin{align}
\varepsilon(\bs{k})=v_0\,|\bs{k}|\,, \qquad
\bs{\Omega}(\bs{k})=-\,\frac{k_z\, \, \hat{\bs{k}}}{k^3}\,,\qquad
\bs{m}(\bs{k})=-\,\frac{e\,v_0\,k_z\, \, \hat{\bs{k}}}{2\,k^2}\,.
\end{align}
Let us consider the example of set-up I, for which
\begin{align}
\varepsilon_m(\bs{k})=\frac{e\,v_0\,\sin 2\,\gamma\,(B_x\cos\phi+B_y\sin\phi)}{4\,k}\,.
\end{align}
At zero temperature, $f_0(\varepsilon-\mu)\to\Theta(\mu-\varepsilon)$, and its derivatives are concentrated at the Fermi surface:
\begin{align}
f_0^{\prime}(\varepsilon-\mu)=-\,\delta(\varepsilon-\mu)\,,\quad
f_0^{\prime\prime}(\varepsilon-\mu)=-\,\delta^{\prime}(\varepsilon-\mu)\,,\quad
f_0^{\prime\prime\prime}(\varepsilon-\mu)=-\,\delta^{\prime\prime}(\varepsilon-\mu)\,.
\end{align}
Combined with the standard property $\int   d\varepsilon\,F(\varepsilon)\,\delta^{(n-1)}(\varepsilon-\mu)=(-1)^{n-1}F^{(n-1)}(\mu)$, this gives the working identity
\begin{align}
\label{eqIBP}
\int d\varepsilon\,F(\varepsilon)\,f_0^{(n)}(\varepsilon-\mu)
=(-1)^{n}\,F^{(n-1)}(\mu)\,,\qquad n\ge 1\,.
\end{align}
The $n=0$ term, by contrast, projects onto the entire Fermi sea:
\begin{align}
\label{eqFsea}
\int d\varepsilon\,F(\varepsilon)\,f_0(\varepsilon-\mu)
=\int_{0}^\mu d\varepsilon\,F(\varepsilon)\,.
\end{align}
To bring Eq.~\eqref{eq-ah} into a form where Eqs.~\eqref{eqIBP} and \eqref{eqFsea} can be applied, we pass to spherical momentum coordinates,
\begin{align}
k_x=k\sin\gamma\cos\phi\,,\quad k_y=k\sin\gamma\sin\phi\,,\quad k_z=k\cos\gamma\,,\qquad
d^3\bs{k}=k^2\sin\gamma\,dk\,d\gamma\,d\phi\,,
\end{align}
and trade the radial coordinate for the energy through
\begin{align}
\label{eqchange}
k=\frac{\varepsilon}{v_0}\,,\qquad dk=\frac{d\varepsilon}{v_0}\,.
\end{align}
The Jacobian factor $1/v_0$ generated by Eq.~\eqref{eqchange} is pulled out once and for all in front of the energy integral. The AH conductivity then takes the schematic form
\begin{align}
(\sigma^{\rm{ah}})_{ij}
&=-\,\frac{e^2\,\epsilon_{ijl}}{(2\,\pi)^3\,v_0}\,\sum_{n=0}^{3} \int d\varepsilon \int_{0}^{\pi}   d\gamma \int_{0}^{2\,\pi} d\phi\,
\mathcal{I}^{(n)}(\varepsilon,\gamma,\phi)\,f_0^{(n)}(\varepsilon-\mu)\,,
\end{align}
where, with $k$ replaced by $\varepsilon/v_0$ throughout,
\begin{align}
\mathcal{I}^{(0)}&=k^2\,\sin\gamma\,\Omega^l\,,\qquad
\mathcal{I}^{(1)}=k^2\, \sin\gamma\,\Omega^l\,\varepsilon_m\,,\nn
\mathcal{I}^{(2)}&=\tfrac{1}{2}\,k^2\, \sin\gamma\,\Omega^l\,\left(\varepsilon_m\right)^{2}\,,\qquad
\mathcal{I}^{(3)}=\tfrac{1}{6}\,k^2\, \sin\gamma\,\Omega^l\,\left(\varepsilon_m\right)^{3}\,.
\end{align}

\section{Current from the LFO-induced part}
\label{appLF}
%======================================================================

In this appendix, we deal with the Lorentz-force-operator (LFO)-induced part of the conductivity. The LFO contribution to the current density is (see Refs.~\cite{ips-rsw-ph,ips_tilted_dirac} for a detailed derivation)
\begin{align}
& \bs{J}^{s, \rm lf} =
-\,e^2 \,  \tau \int \frac{d^3 \bs{k}} {(2\,\pi)^3}
\left(  \bs{w}_s +   \bs{W}_s \right )
\, f_0^\prime (\mathcal{E}_s)\, \mathcal{Y}_s \,,
\text{ where } \check{L}
  = (\bs{w}_s \times \bs{B}) \cdot \nabla_{\bs{k}}\,,
\nn &
\bs{W}_s
= e \left  ( \bs{w}_s \cdot
  \bs{\Omega}_s \right  ) \bs{B}\,,
  \quad
\mathcal{Y}_s =
\sum_{n = 1}^{\infty}
\left (e \, \tau \, D_s \right )^n \check{L}^n
\left [    D_s \left
\lbrace \bs{w}_s
+  \bs{W}_s  \right
\rbrace \cdot \bs{E}   \right ] \,.
\end{align}
This part arises from the action of the LFO $\check{L}$. The nomenclature reflects the fact that it includes the classical Hall effect due to the Lorentz force. The corresponding components of the electric conductivity are
\begin{align}
\left(\sigma_s^{\rm lf} \right)_{i j} & =
-\,e^2 \,  \tau^2 \int \frac{d^3 \bs{k}} {(2\,\pi)^3}
\, \left[  (w_s)_i  + (W_s)_i \right ]
\, f_0^\prime (\mathcal{E}_s) \,
\frac{\partial \mathcal{Y}_s } {\partial E_j}\,.
\end{align}
The solution is obtained by taking the terms in the summation up to a chosen value of $n$ and, thereafter, expanding the expressions up to the desired power in $|\bs{B}|$.

The $n=1$ term leads to the current density
\begin{align}
& \bs{J}^{s, \rm lf} =
-\,e^3 \,  \tau^2 \int \frac{d^3 \bs{k}} {(2\,\pi)^3}
\, \left[ \bs{w}_s
+   \bs{W}_s \right] D_s\, f_0^\prime (\mathcal{E}_s)\left( t_1+ t_2  \right ), 
\quad
t_1 = D_s \,\check{L} \left [   \left ( \bs{w}_s
+ \bs{W}_s \right ) \cdot \bs{E}   \right ] ,\quad
%%%%%%%%%%%%%%%
t_2 =\left [ \left ( \bs{w}_s + \bs{W}_s \right ) \cdot \bs{E}  \right ] \check{L} \,D_s\,.
\end{align}
Expanding terms up to $\mathcal{O}(|\bs{B}|^3)$, we obtain
\begin{align}
 t_1 &=
 \lbrace 1 -e \left (\bs{\Omega}_s \cdot   \bs{B} \right)
 + e^2   \left (\bs{\Omega}_s \cdot   \bs{B}  \right)^2 \rbrace
  (\bs{v}_s \times \bs{B}) \cdot \nabla_{\bs{k}}
  \left( \bs{v}_s \cdot \bs{E} \right )
+ (\bs{v}_s \times \bs{B})
 \cdot \nabla_{\bs{k}}
 \left( \bs{u}_m   \cdot \bs{E} \right ) \nn
 & \quad  +\lbrace 1 -e \left (\bs{\Omega}_s
 \cdot   \bs{B}  \right) \rbrace
  (\bs{v}_s \times \bs{B}) \cdot
  \nabla_{\bs{k}}
  \left[ \left( \bs{u}_m  + \bs{V}_s \right) \cdot \bs{E} \right]
+ (\bs{u}_m  \times \bs{B}) \cdot
\nabla_{\bs{k}}\, \left[ \left( \bs{u}_m
 + \bs{V}_s \right) \cdot \bs{E} \right]  \nn
 & \quad
 +\lbrace 1 -e \left (\bs{\Omega}_s \cdot
 \bs{B}  \right) \rbrace
  (\bs{u}_m  \times \bs{B}) \cdot \nabla_{\bs{k}}
   \left(  \bs{v}_s  \cdot \bs{E} \right ),
\end{align}
and
\begin{align}
t_2 &=
\left ( \bs{v}_s \cdot \bs{E} \right)
 (\bs{v}_s \times \bs{B})  \cdot \nabla_{\bs{k}}
 \left[  -e \left (\bs{\Omega}_s \cdot   \bs{B}  \right)
 + e^2   \left (\bs{\Omega}_s \cdot   \bs{B}  \right)^2 \right]
 + \left ( \bs{v}_s \cdot \bs{E} \right)
 (\bs{u}_m  \times \bs{B})  \cdot \nabla_{\bs{k}}
 \left[  -e \left (\bs{\Omega}_s \cdot   \bs{B}  \right) \right] \nn
 & \quad
 +  \left ( \bs{u}_m + \bs{V}_s\right)\cdot \bs{E}
 \left(\bs{v}_s \times \bs{B}\right)  \cdot \nabla_{\bs{k}}
 \left[  -e \left (\bs{\Omega}_s \cdot   \bs{B}  \right) \right] .
\end{align}
Let us express the current density as
\begin{align}
& \bs{J}^{s, \rm lf} =
-e^3 \,  \tau^2 \int \frac{d^3 \bs{k}} {(2\,\pi)^3}
\sum_{\delta =1}^3
\bs{\mathcal{N}}_{1,\delta} \,,
\end{align}
where $\bs{\mathcal{N}}_{1, \delta} $ has a $|\bs{B}|^\delta$-dependence. The linear-in-$|\bs{B}|$ term is
\begin{align}
\bs{\mathcal{N}}_{1,1} &=
   \bs{v}_s\,  f^\prime_0  (\varepsilon_s )
   \left(\bs{v}_s \times \bs{B}\right)  \cdot \nabla_{\bs{k}}
   \left ( \bs{v}_s \cdot \bs{E} \right) .
\end{align}
The quadratic-in-$|\bs{B}|$ term is
\begin{align}
\bs{\mathcal{N}}_{1,2}  &=\bs{v}_s
   \left ( \bs{v}_s \cdot \bs{E} \right) f^\prime_0  (\varepsilon_s )
    (\bs{v}_s \times \bs{B})  \cdot \nabla_{\bs{k}}
  \left[
-e \left (\bs{\Omega}_s \cdot  \bs{B}  \right) \right]
   + \bs{v}_s \,  f^\prime_0  \left(\varepsilon_s  \right)
     \left(\bs{v}_s \times \bs{B}\right)  \cdot \nabla_{\bs{k}}
     \left[  \left(
     \bs{u}_m   + \bs{V}_s \right) \cdot \bs{E} \right]
   \nn
   & \quad
   +\left[ \left \lbrace -2 \, e \left( \bs{\Omega}_s \cdot  \bs{B}   \right) \bs{v}_s
   + \left(  \bs{u}_m +  \bs{V}_s \right) \right \rbrace
   f^\prime_0  (\varepsilon_s )
   + \varepsilon_m \, \bs{v}_s
   \,   f^{\prime \prime}_0  \left( \varepsilon_s  \right)
\right]
\left( \bs{v}_s \times \bs{B} \right)  \cdot \nabla_{\bs{k}}
\left (\bs{v}_s \cdot  \bs{E}   \right ) \nn
   & \quad +\bs{v}_s \, f^{ \prime}_0  (\varepsilon_s )
   \left( \bs{u}_m  \times \bs{B} \right)  \cdot \nabla_{\bs{k}}
    \left( \bs{v}_s \cdot  \bs{E}   \right ).
\end{align}
The cubic-in-$|\bs{B}|$ term is
\begin{align}
\bs{\mathcal{N}}_{1,3}  &=
\Big[
\left \lbrace -e \left(  \bs{\Omega}_s
\cdot  \bs{B}\right)  \bs{v}_s
\left( \bs{v}_s \cdot  \bs{E}  \right)
+ \bs{v}_s  \left( \bs{u}_m  +\bs{V}_s   \right)
 \cdot  \bs{E}
  + \left( \bs{v}_s \cdot  \bs{E}  \right)
     \left( \bs{u}_m  +\bs{V}_s   \right)
  \right \rbrace
      f^{ \prime}_0  (\varepsilon_s )
 + \varepsilon_m \, \bs{v}_s
       \left( \bs{v}_s \cdot  \bs{E}  \right)
  f^{\prime \prime}_0  (\varepsilon_s) \Big]      \nn
& \quad \times
 \left[   \left(\bs{v}_s \times \bs{B}\right)
    \cdot \nabla_{\bs{k}}
\left \lbrace -e \left( \bs{\Omega}_s
\cdot  \bs{B}  \right)\right \rbrace \right ] \nn
 & \quad
 + \bs{v}_s  \left( \bs{v}_s \cdot  \bs{E}  \right)
 f^{ \prime}_0  (\varepsilon_s )
     \left[ \left(\bs{v}_s \times \bs{B}\right)
    \cdot \nabla_{\bs{k}}
    \left[ e^2 \left( \bs{\Omega}_s \cdot  \bs{B}
  \right)^2 \right]
    + \left(\bs{u}_m  \times \bs{B}\right)
     \cdot \nabla_{\bs{k}}
    \left \lbrace -e \left( \bs{\Omega}_s \cdot  \bs{B}
  \right) \right \rbrace  \right] \nn
 & \quad  +
     \Big[ \left \lbrace 3\, e^2\left( \bs{\Omega}_s
  \cdot  \bs{B}    \right)^2  \bs{v}_s
     -2\, e \left( \bs{\Omega}_s \cdot  \bs{B}
     \right)  \left(\bs{u}_m +\bs{V}_s   \right)
     + \bs{u}_m \right \rbrace
      f^{\prime}_0  (\varepsilon_s )
 \nn & \hspace{ 1 cm }
 - \Big\{ 2\, e \left( \bs{\Omega}_s \cdot  \bs{B}  \right)  \bs{v}_s
- \left(\bs{u}_m +\bs{V}_s   \right)
       \Big\}   \varepsilon_m
        f^{ \prime \prime}_0  (\varepsilon_s )
     + \frac{ \left( \varepsilon_m \right)^2}{2}
\, \bs{v}_s\,
         f^{ \prime \prime \prime}_0  (\varepsilon_s )
  \Big]
      \left(\bs{v}_s \times \bs{B}\right)  \cdot
   \nabla_{\bs{k}} \left ( \bs{v}_s \cdot \bs{E} \right)\nn
& \quad +
 \left[ \left \lbrace-2\, e \left( \bs{\Omega}_s \cdot  \bs{B}    \right)
      \bs{v}_s
      + \left(\bs{u}_m  + \bs{V}_s    \right) \right \rbrace
      f^{ \prime }_0  (\varepsilon_s )
      + \varepsilon_m \, \bs{v}_s
 \, f^{ \prime \prime}_0  (\varepsilon_s )
      \right]  \nn
& \quad \times
\Big[ (\bs{v}_s \times \bs{B})
   \cdot \nabla_{\bs{k}}
     \left \lbrace
 \left (\bs{u}_m   + \bs{V}_s \right)
 \cdot \bs{E}  \right \rbrace
 +  (\bs{u}_m  \times \bs{B})
   \cdot \nabla_{\bs{k}}
     \left (  \bs{v}_s \cdot \bs{E} \right )
    \Big]  \nn
& \quad
+ \bs{v}_s \,f^{ \prime}_0  (\varepsilon_s )
    \left[ (\bs{v}_s \times \bs{B})
     \cdot \nabla_{\bs{k}}
\left (  \bs{u}_m  \cdot \bs{E} \right )
+  \left(\bs{u}_m  \times \bs{B}\right)
     \cdot \nabla_{\bs{k}}
 \left \lbrace
 \left  (\bs{u}_m
 + \bs{V}_s \right) \cdot
 \bs{E} \right \rbrace  \right].
\end{align}

The $n=2$ term leads to the current density
\begin{align}
\bs{J}^{s, \rm lf} =
-e^4 \,  \tau^3 \int \frac{d^3 \bs{k}} {(2\,\pi)^3}
 \lbrace \bs{w}_s
+   \bs{W}_s \rbrace
 D_s^2 f_0^\prime (\mathcal{E}_s)\,
\check{L}^2
\left [  D_s \left
\lbrace \bs{w}_s
+\bs{W}_s  \right
\rbrace \cdot \bs{E}   \right ] .
\end{align}
Due to the presence of $\check{L}^2$, there is no linear-in-$|\bs{B}|$ term here. Expressing the current density as
\begin{align}
& \bs{J}^{s, \rm lf} =
-\,e^4 \,  \tau^3 \int \frac{d^3 \bs{k}} {(2\,\pi)^3}
\sum_{\delta =2}^3 \bs{\mathcal{N}}_{2,\delta} \,,
\end{align}
the quadratic-in-$|\bs{B}|$ term is
\begin{align}
\label{eqb12}
\bs{\mathcal{N}}_{2, 2} &=
   \bs{v}_s\, f^\prime_0  (\varepsilon_s ) \left(\bs{v}_s \times \bs{B}\right)
   \cdot \nabla_{\bs{k}}
   \left[
   \left(\bs{v}_s \times \bs{B}\right)  \cdot \nabla_{\bs{k}}
   \left ( \bs{v}_s \cdot \bs{E} \right)\right],
\end{align}
and the cubic-in-$|\bs{B}|$ term is
\begin{align}
\bs{\mathcal{N}}_{2, 3} &=
     \bs{v}_s\, f^\prime_0  (\varepsilon_s )
   \nabla_{\bs{k}}
 \Big[
 -e \left(  \bs{\Omega}_s \cdot  \bs{B}\right)
     \left( \bs{v}_s \times  \bs{B}  \right)
     \cdot \nabla_{\bs{k}} \left ( \bs{v}_s \cdot \bs{E} \right)
+  \left( \bs{v}_s \times  \bs{B}  \right)
     \cdot \nabla_{\bs{k}} \left\lbrace \left ( \bs{u}_m
  +  \bs{V}_s \right)
     \cdot \bs{E} \right \rbrace
 \nn & \hspace{ 2.5 cm}
+ \left( \bs{u}_m  \times  \bs{B}  \right) \cdot
      \nabla_{\bs{k}} \left ( \bs{v}_s \cdot \bs{E} \right)\Big]
\cdot \left(\bs{v}_s \times \bs{B}\right)
%%%%%%%%%
+  \bs{v}_s\, f^\prime_0  (\varepsilon_s )
       \left( \bs{u}_m  \times  \bs{B}  \right) \cdot
      \nabla_{\bs{k}} \left[ \left( \bs{v}_s \times  \bs{B}  \right)
   \cdot \nabla_{\bs{k}}
   \left( \bs{v}_s \cdot \bs{E} \right)  \right]
\nn & \hspace{ 2.5 cm}
 + \bs{v}_s\, f^\prime_0  (\varepsilon_s )
       \left( \bs{v}_s \times  \bs{B}  \right) \cdot
      \nabla_{\bs{k}} \left[ \left( \bs{v}_s \cdot \bs{E}  \right)
      \left( \bs{v}_s \times  \bs{B}  \right) \cdot
      \nabla_{\bs{k}} \left \lbrace - e \left (  \bs{\Omega}_s \cdot
      \bs{B} \right) \right \rbrace \right]\nn
& \quad
+\Big[ \left \lbrace -2\, e\, \bs{v}_s \left (  \bs{\Omega}_s
\cdot \bs{B} \right) +
\bs{u}_m
+ \bs{V}_s \right \rbrace
f^\prime_0  (\varepsilon_s )
+ \varepsilon_m \, \bs{v}_s \,
 f^{\prime \prime}_0  (\varepsilon_s )\Big]
     \left(\bs{v}_s \times \bs{B}\right)
\cdot
\left [\nabla_{\bs{k}}
  \left \lbrace
\left( \bs{v}_s \times  \bs{B}  \right) \cdot
 \nabla_{\bs{k}}
 \left (  \bs{v}_s \cdot \bs{E} \right)
  \right \rbrace \right ].
\end{align}

The $n=3 $ term leads to the current density
\begin{align}
\bs{J}^{s, \rm lf} =
-e^5 \,  \tau^4 \int \frac{d^3 \bs{k}} {(2\,\pi)^3}
\lbrace \bs{w}_s
+   \bs{W}_s \rbrace  D_s^3\,f_0^\prime (\mathcal{E}_s)\,\check{L}^3
\left [    D_s \left \lbrace \bs{w}_s+\bs{W}_s  \right\rbrace \cdot \bs{E}   \right ] .
\end{align}
Due to the presence of $\check{L}^3 $, only a cubic-in-$|\bs{B}|$ term needs to be extracted here. We can express the current density as
\begin{align}
& \bs{J}^{s, \rm lf} = -\, e^5 \,  \tau^4
 \int \frac{d^3 \bs{k}} {(2\,\pi)^3} \,
 \bs{\mathcal{N}}_{3,3} \,,
 \text{ where }
\bs{\mathcal{N}}_{3, 3}  =
   \bs{v}_s\, f^\prime_0  (\varepsilon_s )
  \left(\bs{v}_s \times \bs{B}\right)
   \cdot \nabla_{\bs{k}} \left[
   \left(\bs{v}_s \times \bs{B}\right)  \cdot \nabla_{\bs{k}}
   \lbrace \left(\bs{v}_s \times \bs{B}\right) \cdot \nabla_{\bs{k}}
   \left ( \bs{v}_s \cdot \bs{E} \right)\rbrace \right].
   \label{eqb15}
\end{align}

%======================================================================

\section{Detailed steps for the LFO-induced conductivity in sVNR}
\label{app-lf-vnr}
%======================================================================

\subsection{Set-up I}
\label{appset1vnr}

At $n = 1$, three sub-parts give nonzero values for the out-of-plane conductivity. The term $\bs{\mathcal N}_{1,1}$ contributes to
\begin{align}
\sigma^{\text{lf, h}}_{zx} & = \frac{ -\tau^2 \, e^3 \, v_0^2 \,k_0\,B_y}
{8\,\pi }  .
\end{align}
The term $\bs{\mathcal N}_{1,2}$ contributes to
\begin{align}
\sigma^{\text{lf, bc}}_{yx} & =
\frac{\tau^2\,e^4\,v_0^7\,\left(k_0-k_\mu \right)^2\,
\left(B_x^2 + B_y^2 \right)\,
}
{64\,\pi\,\mu^5}
\left[3\,k_0^2+6\,k_0\,k_\mu+7\,k_\mu^2\right]
 \,,  \nn
\sigma^{\text{lf, m}}_{yx} & =
\frac{\tau^2 \,e^4\,v_0 \left(B_x^2 + B_y^2 \right)
}
{64\,k_\mu^2\,\pi\,\mu^5}
\left[
-8 \,k_0^6\, v_0^6
+ 8 \,k_0^5\, k_\mu \,v_0^6
+ 4 \,k_0^4 \,v_0^4 \,\mu^2
+ k_0^2 \,v_0^2\, \mu^4
+ 3 \,\mu^6
\right].
\end{align}
The term $\bs{\mathcal N}_{1,3}$ contributes to
\begin{align}
& \sigma^{\text{lf, bc}}_{zx}  =
\frac{3 \, \tau^2 \, e^5 \, k_0 \, v_0^6 \, B_y \,{\bs B}^2}
{512\,\pi \, \mu^8}
\left[72 \, k_0^4 \, v_0^4
- 76 \, k_0^2 \, v_0^2 \, \mu^2+ 11 \, \mu^4+ k_\mu \, k_0 \, v_0^2
\left(- 72 \, k_0^2 \, v_0^2
+ 40 \, \mu^2\right)\right], \nn
& \sigma^{\text{lf, m}}_{zx}  = \frac{\tau^2 \, e^5 \, v_0^2 \, B_y \,{\bs B}^2}
{512 \, k_\mu^3\,\pi \, \mu^8}\Big[
648 \, k_0^8 \, v_0^8- 1604 \, k_0^6 \, v_0^6 \, \mu^2
+ 1224 \, k_0^4 \, v_0^4 \, \mu^4
- 248 \, k_0^2 \, v_0^2 \, \mu^6+ 8 \, \mu^8 \nn
& \hspace{2.25 cm} 
+ k_\mu \, k_0 \, v_0^2
\left( 1280 \, k_0^4 \, v_0^4 \, \mu^2- 648 \, k_0^6 \, v_0^6
- 665 \, k_0^2 \, v_0^2 \, \mu^4 + 33 \, \mu^6\right)\Big], \nn &
%%%%%%%%%%%%%%%%
\sigma^{\text{lf, conc}}_{zx}  =
\frac{\tau^2 \, e^5 \, v_0^4 \, B_y \,{\bs B}^2 }
{256 \, k_\mu \, \pi \, \mu^8} \big[- 72 \, k_0^6 \, v_0^6
+ 80 \, k_0^4 \, v_0^4 \, \mu^2- 20 \, k_0^2 \, v_0^2 \, \mu^4+ 4 \, \mu^6
 + k_\mu \, k_0 \, v_0^2
\left(72 \, k_0^4 \, v_0^4- 44 \, k_0^2 \, v_0^2 \, \mu^2+ 7 \, \mu^4 \right)\big].
\end{align}

At $n = 2$, nonzero contributions appear in the in-plane response, arising from two sub-parts. The term $\bs{\mathcal N}_{2,2}$ contributes to
\begin{align}
\sigma^{\text{lf, bc}}_{xx} & =  \sigma^{\text{lf, m}}_{xx} =  \sigma^{\text{lf,  conc} }_{xx}  = 0\,,
\quad \sigma^{\text{lf, h}}_{xx}  =
\frac{\tau^3 \, e^4 \, k_0^2 \, v_0^6 \,
\left(B_x^2 + 3 \, B_y^2 \right)\,
\left(-k_0 + k_\mu \right)}
{16 \, \pi \, \mu^3} \, , \nn
\sigma^{\text{lf, bc}}_{yx} & =  \sigma^{\text{lf, m}}_{yx} =  \sigma^{\text{lf,  conc} }_{yx}  = 0\,,
\quad\sigma^{\text{lf, h}}_{yx} =
\frac{\tau^3 \, e^4 \, k_0^2 \, v_0^6 \, B_x \, B_y \,
\left(k_0 - k_\mu \right)}
{8 \, \pi \, \mu^3}.
\end{align}
The term $\bs{\mathcal N}_{2,3}$ contributes to
\begin{align}
\sigma^{\text{lf, h}}_{zx}  &= \sigma^{\text{lf,  conc} }_{zx}  = 0, \nn
\sigma^{\text{lf, bc}}_{zx} &=
\frac{\tau^3 \, e^5 \, v_0^5 \, B_x \, \left(B_x^2 + B_y^2 \right)}
{64 \, \pi \, \mu^6}
\left[
7 \, \mu^4
+ k_0^2 \,v_0^2 \left(-14 \,k_0^2 \,v_0^2 + 3 \, \mu^2\right)
+ k_0 \,k_\mu\, v_0^2 \left(14 \,k_0^2 \,v_0^2 + 4 \, \mu^2\right)
\right] ,\nn
\sigma^{\text{lf, m}}_{zx} & =
- \, \frac{
\tau^3 \, e^5 \, v_0^5 \, \left(k_0^2 \,v_0^2 - 3 \, \mu^2 \right)\, B_x \,
\left(B_x^2 + B_y^2 \right)}
{64 \, \pi \, \mu^6 \, k_\mu^2\, v_0^2 }
\left[-2 \,k_0^4 \,v_0^4 + 2 \,k_0^3\, k_\mu \,v_0^4 + k_0^2 \,v_0^2 \,\mu^2 + \mu^4\right].
\end{align}

At $n = 3$, the sole nonzero conductivity component is the $zx$ component, sourced from terms arising from $\bs{\mathcal N}_{3,3}$:
\begin{align}
\sigma^{\text{lf, h}}_{zx} &=
- \, \frac{3 \, \tau^4 \, e^5 \, k_0^2 \, v_0^8 \, B_y \,
\left(B_x^2 + B_y^2 \right)\,
\left(-k_0 + k_\mu \right)}
{16 \, \pi \, \mu^4}.
\end{align}
Notably, this receives no contribution from the BC or OMM.

\subsection{Set-up II}
\label{appset2vnr}

The individual expressions shown here feed into the final expressions summarised in Sec.~\ref{secset2}.

At $n = 1 $, the term $\bs{\mathcal N}_{1,1}$ does not contribute. Only the out-of-plane conductivity is nonzero, sourced by two sub-parts. The term $\bs{\mathcal N}_{1,2}$ contributes to
\begin{align}
& \sigma^{\text{lf, h}}_{yx}  = \sigma^{\text{lf,  conc} }_{yx}  = 0\,, 
%%%%%%%%%%%%%%%%%%%%
\nn &\sigma^{\text{lf, bc}}_{yx}  =\frac{\tau^2 \, e^4 \, v_0^3}
{64 \, \pi \, \mu^5} \left[
B_x^2 \left\{3 \, \mu^4+ 4 \, k_\mu^2 \, v_0^2 \left(
2 \, k_0^2 \, v_0^2- 2 \, k_0 \, k_\mu \, v_0^2- \mu^2\right)\right\}
+ 2 \, B_z^2 \left\{4 \, k_0^2 \, v_0^2 \left(-2 \, k_0^2 \, v_0^2+ \mu^2\right)
+ 8 \, k_0^3 \, k_\mu \, v_0^4+ \mu^4\right\}\right] \nn
%%%%%%%%%%%%%%%%%%%%%%%%%
& \sigma^{\text{lf, m}}_{yx} =
\frac{\tau^2 \, e^4 \, v_0}
{64 \, k_\mu^2 \, \pi \, \mu^5}\Bigg[2 \, B_z^2\Big(
9 \, \mu^6 + k_0^2 \, v_0^2
\left(8 \, k_0^4 \, v_0^4+ 20 \, k_0^2 \, v_0^2 \, \mu^2- 37 \, \mu^4\right)
- 8 \, k_0 \, k_\mu \, v_0^2\left(k_0^4 \, v_0^4
+ 3 \, k_0^2 \, v_0^2 \, \mu^2- 3 \, \mu^4\right)\Big)
\nn & \hspace{ 3.25 cm }
+ B_x^2 \left(8 \, k_0^5 \, k_\mu \, v_0^6
+ 3 \, \mu^6+ k_0^2 \, v_0^2\left(
- 8 \, k_0^4 \, v_0^4+ 4 \, k_0^2 \, v_0^2 \, \mu^2+ \mu^4\right)\right)\Bigg].
\end{align}
The term $\bs{\mathcal N}_{1,3}$ contributes to
\begin{align}
\sigma^{\text{lf, bc}}_{yx} & =
\frac{\tau^2 \, e^5 \, v_0^4 \, B_z}{128 \, k_\mu \, \pi \, \mu^8}
\Big[18 \, B_z^2 \, k_0 \, v_0^2
\left(8 \, k_0^5 \, v_0^4- 12 \, k_0^3 \, v_0^2 \, \mu^2
+ 4 \, k_0 \, \mu^4+ k_\mu \left(
- 8 \, k_0^4 \, v_0^4+ 8 \, k_0^2 \, v_0^2 \, \mu^2- \mu^4
\right)\right) \nn & \quad
+ B_x^2 \big\{- 216 \, k_0^6 \, v_0^6
+ 372 \, k_0^4 \, v_0^4 \, \mu^2- 184 \, k_0^2 \, v_0^2 \, \mu^4+ 24 \, \mu^6
+ k_\mu \left( 216 \, k_0^5 \, v_0^6- 264 \, k_0^3 \, v_0^4 \, \mu^2
+ 79 \, k_0 \, v_0^2 \, \mu^4\right)
\big\}\Big], \nn
\sigma^{\text{lf, m}}_{yx} & =
\frac{\tau^2 \, e^5 \, v_0^2 \, B_z}
{128 \, k_\mu^3 \, \pi \, \mu^8}
\Big[6 \, B_z^2
\big\{- 72 \, k_0^8 \, v_0^8+ 132 \, k_0^6 \, v_0^6 \, \mu^2
- 68 \, k_0^4 \, v_0^4 \, \mu^4+ 12 \, k_0^2 \, v_0^2 \, \mu^6- 2 \, \mu^8 \nn
& \quad + k_\mu
\left(
72 \, k_0^7 \, v_0^8
- 96 \, k_0^5 \, v_0^6 \, \mu^2
+ 29 \, k_0^3 \, v_0^4 \, \mu^4
- 5 \, k_0 \, v_0^2 \, \mu^6\right)\big\}
%\nn & \quad
+ B_x^2 \big\{
648 \, k_0^8 \, v_0^8
- 1300 \, k_0^6 \, v_0^6 \, \mu^2
+ 744 \, k_0^4 \, v_0^4 \, \mu^4
- 96 \, k_0^2 \, v_0^2 \, \mu^6- 4 \, \mu^8\nn & \quad
+ k_\mu\left(- 648 \, k_0^7 \, v_0^8
+ 976 \, k_0^5 \, v_0^6 \, \mu^2
- 337 \, k_0^3 \, v_0^4 \, \mu^4
+ 9 \, k_0 \, v_0^2 \, \mu^6
\right)
\big\}
\Big]
, \nn
\sigma^{\text{lf, conc}}_{yx} & =
\frac{\tau^2 \, e^5 \, B_z}
{64 \, \pi \, \mu^8}
\Bigg[
\frac{6 \, B_z^2 \, v_0^4}{k_\mu}
\left\{
8 \, k_0^6 \, v_0^6
- 36 \, k_0^4 \, v_0^4 \, \mu^2
+ 30 \, k_0^2 \, v_0^2 \, \mu^4- 4 \, \mu^6+ k_\mu
\left(- 8 \, k_0^5 \, v_0^6+ 32 \, k_0^3 \, v_0^4 \, \mu^2- 15 \, k_0 \, v_0^2 \, \mu^4
\right)\right\}\nn
& \quad + \frac{B_x^2 \, v_0^2}{k_\mu^3}
\big\{- 72 \, k_0^8 \, v_0^8+ 280 \, k_0^6 \, v_0^6 \, \mu^2- 306 \, k_0^4 \, v_0^4 \, \mu^4
+ 92 \, k_0^2 \, v_0^2 \, \mu^6+ 2 \, \mu^8\nn
& \quad+ k_\mu
\left(
72 \, k_0^7 \, v_0^8
- 244 \, k_0^5 \, v_0^6 \, \mu^2
+ 193 \, k_0^3 \, v_0^4 \, \mu^4
- 21 \, k_0 \, v_0^2 \, \mu^6
\right)
\big\}
\Bigg]
.
\end{align}

At $n =2 $, only the longitudinal components survive, arising from two sub-parts. The term $\bs{\mathcal N}_{2,2}$ contributes to
\begin{align}
&\sigma^{\text{lf, bc}}_{xx} =  \sigma^{\text{lf,  conc} }_{xx}  = \sigma^{\text{lf, m}}_{xx} = 0,
\quad
\sigma^{\text{lf, h}}_{xx} =
\frac{\tau^3 \, e^4 \, k_0 \, v_0^6}{8 \, k_\mu \, \pi \, \mu^3}
\left[ B_x^2 \left( \frac{k_0^2 \left(k_0 - k_\mu \right) }{2} 
- \frac{k_0 \, \mu^2}{2 \, v_0^2}\right)
+ B_z^2 \left(2 \, k_0^2 \left(-k_0 + k_\mu \right)
+ \frac{k_\mu \, \mu^2}{v_0^2}\right)\right].
\end{align}
The term $\bs{\mathcal N}_{2,3}$ contributes to
\begin{align}
\sigma^{\text{lf, h}}_{xx} & =\sigma^{\text{lf,  conc} }_{xx}=0   \,,\nn
\sigma^{\text{lf, bc}}_{xx}&  =
\frac{\tau^3 \, e^5 \,v_0^3 \, B_z}{64 \, k_\mu^2 \, \pi \, \mu^6}
\big[
B_x^2 \left(3 \, k_0^2\,  v_0^2 - \mu^2\right)
\left\{
-50 \, k_0^4 \, v_0^4
+ 63 \, k_0^2\,  v_0^2\,  \mu^2
- 13 \, \mu^4
+ k_\mu \left(
50 \, k_0^3 \, v_0^4
- 38 \, k_0\,  v_0^2 \, \mu^2
\right)
\right\}
\nn
& \quad
+ 3 \, B_z^2 \left\{
40 \, k_0^6 \, v_0^6
- 52\,  k_0^4\,  v_0^4 \, \mu^2
+ 11 \, k_0^2 \, v_0^2\,  \mu^4
+ \mu^6
+ k_\mu \left(
-40 \, k_0^5 \, v_0^6
+ 32\,  k_0^3 \, v_0^4 \, \mu^2
\right)
\right\}
\big]
, \nn
\sigma^{\text{lf, m}}_{xx} & =
\frac{\tau^3 \, e^5 \, v_0 \, B_z}{64 \, k_\mu^4 \, \pi \, \mu^6}
\big[
B_x^2 \big\{
122 \, k_0^8 \, v_0^8
- 207 \, k_0^6 \, v_0^6 \, \mu^2
+ 54 \, k_0^4 \, v_0^4 \, \mu^4
+ 25 \, k_0^2 \, v_0^2\,  \mu^6
+ 6 \, \mu^8
+ k_\mu \big(
-122 \, k_0^7\,  v_0^8
+ 146\,  k_0^5\,  v_0^6 \, \mu^2
\nn
& \quad
+ 4\,  k_0^3 \, v_0^4\,  \mu^4
- 10 \, k_0 \, v_0^2 \, \mu^6
\big)
\big\}
+ B_z^2 \big\{
-120\,  k_0^8 \, v_0^8
+ 108\,  k_0^6 \, v_0^6 \, \mu^2
+ 167 \, k_0^4 \, v_0^4\,  \mu^4
- 178\,  k_0^2 \, v_0^2\,  \mu^6
+ 23\,  \mu^8
\nn
& \quad
+ k_\mu \left(
120\,  k_0^7\,  v_0^8
- 48\,  k_0^5 \, v_0^6 \, \mu^2
- 176 \, k_0^3 \, v_0^4 \, \mu^4
+ 96 \, k_0 \, v_0^2 \, \mu^6
\right)
\big\}
\big]
\,,\nn
\sigma^{\text{lf, h}}_{zx}  &= \sigma^{\text{lf, conc}}_{zx} =0 , \quad \nn
\sigma^{\text{lf, bc}}_{zx}& =
\frac{\tau^3 \, e^5 \, v_0^3 \, B_x}{64 \, k_\mu^2 \, \pi \, \mu^6}
\big[
B_x^2 k_\mu^2\,  v_0^2
\left\{
-14 \, k_0^4 \, v_0^4
+ 3 \, k_0^2 \, v_0^2 \, \mu^2
+ 7 \, \mu^4
+ k_\mu \left(
14 \, k_0^3 \, v_0^4
+ 4 \, k_0\,  v_0^2\,  \mu^2
\right)
\right\}
\nn
& \quad
+ B_z^2 \left\{
104 \, k_0^6 \, v_0^6
- 156 \, k_0^4 \, v_0^4\,  \mu^2
+ 55 \, k_0^2 \, v_0^2\,  \mu^4
- 3 \, \mu^6
+ k_\mu \left(
-104 \, k_0^5 \, v_0^6
+ 104 \, k_0^3 \, v_0^4\,  \mu^2
- 16\,  k_0 \, v_0^2\,  \mu^4
\right)
\right\}
\big]
, \nn
\sigma^{\text{lf, m}}_{zx} &=
\frac{\tau^3 \, e^5 \, v_0 \, B_x}{64 \, \pi \, \mu^6 \, k_\mu^4 }
\big[
- B_x^2 \left\{
2\,  k_0^3 \, k_\mu\,  v_0^4
- k_\mu^2 \left(3 \, k_0^2 - k_\mu^2 \right) v_0^4
\right\}
\left(
k_0^4 \, v_0^4
- 4 \, k_0^2 \, v_0^2 \, \mu^2
+ 3\,  \mu^4
\right)
\nn
& \quad
+  B_z^2 \big\{
-40 \, k_0^8 \, v_0^8
+ 116 \, k_0^6 \, v_0^6\,  \mu^2
- 123 \, k_0^4 \, v_0^4 \, \mu^4
+ 58\,  k_0^2 \, v_0^2\,  \mu^6
- 11 \, \mu^8
+ k_\mu \big(
40 \, k_0^7 \, v_0^8
- 96 \, k_0^5 \, v_0^6 \, \mu^2
\nn
& \quad
+ 80 \, k_0^3 \, v_0^4\,  \mu^4
- 32 \, k_0 \, v_0^2\,  \mu^6
\big)
\big\}
\big].
\end{align}

At $n = 3 $, only the out-of-plane component of the response survives, sourced by $\bs{\mathcal N}_{3,3}$, which is given by
\begin{align}
\sigma^{\text{lf, h}}_{yx} &=
\frac{\tau^4 \, e^5 \, k_0 \, v_0^2 \, B_z}{8 \, k_\mu^4 \, \pi \, \mu^4}
\big[
- B_x^2 \, k_\mu^2 \, v_0^2
\left\{
8 \, k_0^4\,  v_0^4
- 9 \, k_0^2 \, v_0^2\,  \mu^2
+ \mu^4
+ k_\mu \left(
-8 \, k_0^3\,  v_0^4
+ 5 \, k_0 \, v_0^2 \, \mu^2
\right)
\right\}
\nn
& \quad
+ 2 \, B_z^2
\left\{
4 \, k_0^6 \, v_0^6
- 7 \, k_0^4\,  v_0^4\,  \mu^2
+ 2 \, k_0^2 \, v_0^2\,  \mu^4
+ \mu^6
+ k_\mu \left(
-4 \, k_0^5 \, v_0^6
+ 5\,  k_0^3 \, v_0^4\,  \mu^2
\right)
\right\}
\big]
\,.
\end{align}

\subsection{Set-up III}
\label{appset3vnr}

At $n = 1 $, the surviving ones are the out-of-plane components, which arise from three parts. The term $\bs{\mathcal N}_{1,1}$ contributes to
\begin{align}
\sigma^{\text{lf, h}}_{yz} & =
- \, \frac{\tau^2\, e^3 \,  k_0 \, v_0^2 \,  B_x}{8 \, \pi}\,.
\end{align}
The term $\bs{\mathcal N}_{1,2}$ contributes to
\begin{align}
\sigma^{\text{lf, bc}}_{yz} & =
\frac{\tau^2 \, e^4 \, v_0^3 \, B_x \, B_z}{32 \, \pi \, \mu^5}
\left[
-8 \, k_0^4\,  v_0^4
+ 8\,  k_0^3 \, k_\mu \, v_0^4
+ 4 \, k_0^2\,  v_0^2 \, \mu^2
+ \mu^4
\right]
, \quad \nn
\sigma^{\text{lf, m}}_{yz}  & =
\frac{\tau^2 \, e^4 \, v_0 \, B_x \, B_z}{32 \, k_\mu^2 \, \pi \, \mu^5}
\left[
8 \, k_0^6 \, v_0^6
- 12\,  k_0^4 \, v_0^4\,  \mu^2
+ 7 \, k_0^2 \, v_0^2\,  \mu^4
- 3 \, \mu^6
+ k_\mu\left(
-8 \, k_0^5 \, v_0^6
+ 8 \, k_0^3\,  v_0^4 \, \mu^2
- 4\,  k_0 \, v_0^2 \, \mu^4
\right)
\right]
.
\end{align}
The term $\bs{\mathcal N}_{1,3}$ contributes to
\begin{align}
\sigma^{\text{lf, bc}}_{yz} & =
\frac{\tau^2 \, e^5 \, k_0 \, v_0^6 \, B_x}{512 \, k_\mu \, \pi \, \mu^8}
\big[
4 \, B_z^2 \left\{
216 \, k_0^5 \, v_0^4
- 288\,  k_0^3 \, v_0^2 \, \mu^2
+ 80 \, k_0\,  \mu^4
+ k_\mu \left(
-216 \, k_0^4\,  v_0^4
+ 180 \, k_0^2 \, v_0^2 \, \mu^2
- 17 \, \mu^4
\right)
\right\}
\nn
& \quad
+ 3\,  B_x^2 \left\{
-72 \, k_0^5 \, v_0^4
+ 112 \, k_0^3 \, v_0^2\,  \mu^2
- 40 \, k_0 \, \mu^4
+ k_\mu \left(
72\,  k_0^4\,  v_0^4
- 76\,  k_0^2 \, v_0^2\,  \mu^2
+ 11 \, \mu^4
\right)
\right\}
\big]
,
\nn
\sigma^{\text{lf, m}}_{yz} &  =
\frac{\tau^2 \, e^5 \, v_0^2 \, B_x}{512 \, k_\mu^3 \, \pi \, \mu^8}
\big[
4 \, B_z^2 \, k_0 \, v_0^2
\big\{
-648 \, k_0^7 \, v_0^6
+ 1492 \, k_0^5\,  v_0^4 \, \mu^2
- 1092\,  k_0^3 \, v_0^2 \, \mu^4
+ 256 \, k_0\, \mu^6
+ k_\mu\big(
648 \, k_0^6 \, v_0^6
\nn
 & \quad
- 1168 \, k_0^4\,  v_0^4 \, \mu^2
+ 589 \, k_0^2 \, v_0^2 \, \mu^4
- 69 \, \mu^6
\big)
\big\}
+
B_x^2 \big\{
648 \, k_0^8 \, v_0^8
- 1604 \, k_0^6 \, v_0^6 \, \mu^2
+ 1224 \, k_0^4 \, v_0^4 \, \mu^4
\nn
 & \quad
- 248\,  k_0^2 \, v_0^2 \, \mu^6
+ 8 \, \mu^8
+ k_\mu\left(
-648 \, k_0^7\,  v_0^8
+ 1280 \, k_0^5 \, v_0^6\,  \mu^2
- 665 \, k_0^3\,  v_0^4 \, \mu^4
+ 33 \, k_0 \, v_0^2\,  \mu^6
\right)
\big\}
\big]
   ,\nn
\sigma^{\text{lf,  conc} }_{yz} &  =
\frac{\tau^2 \, e^5 \, v_0^4\, B_x}{256 \, k_\mu \, \pi \, \mu^8}
\big[
4 \, B_z^2 \big\{
72 \, k_0^6 \, v_0^6
- 196 \, k_0^4 \, v_0^4 \, \mu^2
+ 120 \, k_0^2 \, v_0^2 \, \mu^4
- 8 \, \mu^6
+ k_\mu\big(
-72 \, k_0^5 \, v_0^6
+ 160 \, k_0^3 \, v_0^4 \, \mu^2
\nn
& \quad
- 49 \, k_0 \, v_0^2 \, \mu^4
\big)
\big\}
 + B_x^2 \big\{
-72\,  k_0^6 \, v_0^6
+ 80 \, k_0^4 \, v_0^4 \, \mu^2
- 20\,  k_0^2\,  v_0^2\,  \mu^4
+ 4 \, \mu^6
+ k_\mu\big(
72 \, k_0^5 \, v_0^6
\nn
&  \quad
- 44 \, k_0^3 \, v_0^4\,  \mu^2
+ 7 \, k_0 \, v_0^2 \, \mu^4
\big)
\big\}
\big]
.
\end{align}

At $n =2 $, the sole surviving part is the longitudinal component captured by two parts. The term $\bs{\mathcal N}_{2,2}$ contributes to
\begin{align}
\sigma^{\text{lf, h}}_{zz}& =
- \, \frac{ \tau^3\,e^4 \,  k_0\, v_0^4\, B_x^2   }
{8 \, \pi \, \mu}
.
\end{align}
The term $\bs{\mathcal N}_{2,3}$ contributes to
\begin{align}
\sigma^{\text{lf, bc}}_{xz} & =
- \, \frac{\tau^3 \,e^5 \,v_0^5 \,B_x}{64 \,k_\mu\, \pi \,\mu^6}
\big[
B_x^2 \left\{
2 \,k_0^5 \,v_0^4
- 8 \,k_0^3\, v_0^2 \,\mu^2
+ 6 \,k_0 \,\mu^4
+ k_\mu\left(
-2 \,k_0^4 \,v_0^4
+ 7 \,k_0^2 \,v_0^2 \,\mu^2
- 6 \,\mu^4
\right)
\right\}
\nn
& \quad
+ B_z^2 \left\{
40 \,k_0^5\, v_0^4
- 32 \,k_0^3 \,v_0^2 \,\mu^2
+ k_\mu\left(
-40\, k_0^4 \,v_0^4
+ 12 \,k_0^2\, v_0^2 \,\mu^2
+ \mu^4
\right)
\right\}
\big]
,\nn
\sigma^{\text{lf, m}}_{xz} &  =
\frac{\tau^3 \, e^5 \, v_0^3\,  B_x}{64 \, k_\mu^3\,  \pi \, \mu^6}
\big[
B_x^2
\left(
k_0^4 \, v_0^4 - 4 \, k_0^2 \, v_0^2\,  \mu^2 + 3 \, \mu^4
\right)
\left\{
-2 \, k_0^3 \, v_0^2
+ k_\mu\left(
2 \, k_0^2\,  v_0^2 + \mu^2
\right)
\right\}
+
B_z^2
\big\{
40 \, k_0^7 \, v_0^6
- 96 \, k_0^5\,  v_0^4 \, \mu^2
\nn
& \quad
+ 80 \, k_0^3 \, v_0^2 \, \mu^4
- 32 \, k_0 \, \mu^6
+ k_\mu \left(
-40 \, k_0^6 \, v_0^6
+ 76 \, k_0^4 \, v_0^4 \, \mu^2
- 47 \, k_0^2 \, v_0^2 \, \mu^4
+ 11\,  \mu^6
\right)
\big\}
\big]
\,,
\end{align}
\begin{align}
\sigma^{\text{lf, bc}}_{zz} & =
\frac{\tau^3 \, e^5\,  v_0^5\,  B_z \,B_x^2}{64 \, \pi \, \mu^6}
\left[
-48\,  k_0^4 \, v_0^4
+ 32 \, k_0^2 \, v_0^2\,  \mu^2
+ \mu^4
+ k_\mu\left(
48 \, k_0^3\,  v_0^4
- 8 \, k_0 \, v_0^2 \, \mu^2
\right)
\right]
\,, \nn
\sigma^{\text{lf, m}}_{zz} &  =
- \, \frac{\tau^3 \, e^5 \, B_z\, B_x^2 }{64\,  k_\mu \, \pi \, \mu^6}
\left[
32 \, k_0^5\,  v_0^9
- 16 \, k_0^3 \, v_0^7 \, \mu^2
+ k_\mu\left(
-32 \, k_0^4 \, v_0^9
+ v_0^5 \, \mu^4
\right)
\right]
.
\end{align}

At $n = 3 $, an out-of-plane transverse component is the sole nonzero part, sourced by $\bs{\mathcal N}_{3,3}$, which is
\begin{align}
\sigma^{\text{lf, h}}_{yz} &=
\frac{\tau^4 \, e^5 \, k_0\,  v_0^6 \, B_x}{16\,  k_\mu \, \pi \, \mu^4}
\left[
3 \, B_x^2 \, k_0\,  \left(
- k_0^2 \, v_0^2
+ k_0 \, k_\mu \, v_0^2
+ \mu^2
\right)
- 2 \, B_z^2 \left\{
-2 \, k_0^3 \, v_0^2
+ k_\mu\left(
2 \, k_0^2 \, v_0^2
+ \mu^2
\right)
\right\}
\right]
.
\end{align}

%======================================================================
\section{Detailed steps for the LFO-induced conductivity in HSM}
\label{app-lf-hopf}
%======================================================================

\subsection{Set-up I}
\label{app-lf-hopf1}

At $n = 1$, three sub-parts give nonzero values for the out-of-plane conductivity. The term $\bs{\mathcal N}_{1,1}$ contributes to
\begin{align}
\sigma^{\text{lf, h}}_{zx} &= - \, \frac{\tau^2\,e^3\,v_0\, \mu\,B_y}{6\,\pi^2},
\end{align}
the term $\bs{\mathcal N}_{1,2}$ contributes to
\begin{align}
\sigma^{\text{lf, bc}}_{yx} &= \frac{2\,\tau^2 \, e^4\,v_0^3  {\boldsymbol B}^2}{15\,\pi^2\,\mu}, \quad
\sigma^{\text{lf, m}}_{yx} = - \, \frac{\tau^2\, e^4 \, v_0^3  {\boldsymbol B}^2}{20\,\pi^2\,\mu},
\end{align}
and the term $\bs{\mathcal N}_{1,3}$ contributes to
\begin{align}
\sigma^{\text{lf, bc}}_{zx} &= - \, \frac{\tau^2\, e^5\, v_0^5\, B_y  {\boldsymbol B}^2}{35\,\pi^2\,\mu^3}, \quad
\sigma^{\text{lf, m}}_{zx} = - \, \frac{9 \, \tau^2 \,e^5\, v_0^5 \, B_y  {\boldsymbol B}^2 }{280\,\pi^2\,\mu^3}, \quad
\sigma^{\text{lf,  conc} }_{zx}
= \frac{ \tau^2\, e^5\, v_0^5 \,B_y  {\boldsymbol B}^2 }{30\,\pi^2\,\mu^3}.
\end{align}
At $n = 2$, the term $\bs{\mathcal N}_{2,2}$ contributes to
\begin{align}
\sigma^{\text{lf, h}}_{xx} &= - \, \frac{\,\tau^3\,e^4\,v_0^3\, B_y^2}{6\,\pi^2}, \quad
\sigma^{\text{lf, h}}_{yx} = \frac{\tau^3\,e^4\,v_0^3\, B_x\,B_y}{6\,\pi^2},
\end{align}
and the term $\bs{\mathcal N}_{2,3}$ contributes to
\begin{align}
\sigma^{\text{lf, bc}}_{zx} &= \frac{2\,\tau^3 \,e^5\,v_0^5\, B_x  {\boldsymbol B}^2}{15\,\pi^2\,\mu^2}, \quad
\sigma^{\text{lf,  m} }_{zx}= - \, \frac{\tau^3\,e^5\,v_0^5\, B_x  {\boldsymbol B}^2}{20\,\pi^2\,\mu^2}.
\end{align}
At $n = 3$, the nonzero contribution is
\begin{align}
\sigma^{\text{lf, h}}_{zx} = \frac{\tau^4\,e^5\,v_0^5 \, B_y  {\boldsymbol B}^2}{6\,\pi^2\,\mu}.
\end{align}

%%%%%%%%%%%%%%%%%%%%%%%%%
\subsection{Set-up II}
\label{app-lf-hopf2}

At $n = 1$, the nonzero contributions are as follows. The term $\bs{\mathcal N}_{1,1}$ gives
\begin{align}
\sigma^{\text{lf, h}}_{yx} &= \frac{\tau^2 \, e^3 \,v_0 \, \mu\,B_z}{6\,\pi^2}.
\end{align}
The term $\bs{\mathcal N}_{1,2}$ contributes to
\begin{align}
\sigma^{\text{lf, bc}}_{yx} &= \frac{\tau^2 \, e^4 \,v_0^3 \,\left(2\,B_x^2 + B_z^2 \right)}{15\,\pi^2\,\mu}, \quad
\sigma^{\text{lf, m}}_{yx} = - \, \frac{\tau^2 \, e^4\,v_0^3 \,(B_x^2+ 3\,B_z^2)}{20\,\pi^2\,\mu}.
\end{align}
The term $\bs{\mathcal N}_{1,3}$ contributes to
\begin{align}
\sigma^{\text{lf, bc}}_{yx} &= \frac{3\, \tau^2 \, e^5 \, v_0^5 \,B_z \left(2\,B_x^2 + B_z^2 \right)}{70\,\pi^2\,\mu^3}, \quad
\sigma^{\text{lf, m}}_{yx} = \frac{\tau^2 \, e^5\, v_0^5 \,B_z \left(5 \, B_x^2+ 6\,B_z^2 \right)}{168\,\pi^2\,\mu^3},  \quad
\sigma^{\text{lf,  conc} }_{yx} = 
- \, \frac{\tau^2 \, e^5\, v_0^5 \,B_z \left(B_x^2+B_z^2 \right)\,}{10\,\pi^2\,\mu^3}.
\end{align}
At $n = 2$, the term $\bs{\mathcal N}_{2,2}$ contributes to
\begin{align}
\sigma^{\text{lf, h}}_{xx} &= - \, \frac{\tau^3 \, e^4\, v_0^3 \,B_z^2}{6\,\pi^2}, \quad
\sigma^{\text{lf, h}}_{zx} = \frac{\tau^3\, e^4\, v_0^3\,B_x\,B_z}{6\,\pi^2}.
\end{align}
The term $\bs{\mathcal N}_{2,3}$ contributes to
\begin{align}
\sigma^{\text{lf, bc}}_{xx}&= - \, \frac{\tau^3 \, e^5 \, v_0^5 \,B_z\left(7\,B_x^2 + 3\,B_z^2 \right)}{30\,\pi^2\,\mu^2}, \quad
\sigma^{\text{lf, m}}_{xx} = \frac{\tau^3 \,e^5\, v_0^5 \,B_z\,(6\,B_x^2 + 11\,B_z^2)}{60\,\pi^2\,\mu^2} \nn
\sigma^{\text{lf, bc}}_{zx} &= \frac{2\,\,\tau^3\, e^5 \, v_0^5 \,B_x^3}{15\,\pi^2\,\mu^2}, \quad
\sigma^{\text{lf, m}}_{zx} = - \, \frac{\tau^3\,e^5\, v_0^5 \,B_x\,(3\,B_x^2 + 8\,B_z^2)}{60\,\pi^2\,\mu^2}.
\end{align}
At $n = 3$, the nonzero contribution is
\begin{align}
\sigma^{\text{lf, h}}_{yx} = - \, \frac{\tau^4 \,e^5 \, v_0^5 \,B_z\, \left(B_x^2 + B_z^2 \right)}{6\,\pi^2\,\mu}.
\end{align}

%%%%%%%%%%%%%%%%%%%%
\subsection{Set-up III}
\label{app-lf-hopf3}

At $n = 1$, the term $\bs{\mathcal N}_{1,1}$ contributes to
\begin{align}
\sigma^{\text{lf, h}}_{yz} &= - \, \frac{\tau^2 \, e^3\, v_0 \,\mu\,B_x}{6\,\pi^2},
\end{align}
the term $\bs{\mathcal N}_{1,2}$ contributes to
\begin{align}
\sigma^{\text{lf, bc}}_{yz}&= \frac{\tau^2 \, e^4 \, v_0^3 \, B_x \, B_z}
{15\,\pi^2\,\mu}, \quad
\sigma^{\text{lf, m}}_{yz} = \frac{\tau^2\, e^4\, v_0^3 \,B_x\, B_z}
{10\,\pi^2\,\mu}\,,
\end{align}
and the term $\bs{\mathcal N}_{1,3}$ contributes to
\begin{align}
\sigma^{\text{lf, bc}}_{yz}&= \frac{\tau^2 \, e^5\, v_0^5\, B_x 
\left(-2 \, B_x^2+ B_z^2 \right)}{70\,\pi^2\,\mu^3}, \quad
\sigma^{\text{lf, m}}_{yz} =- \, \frac{\tau^2 \, e^5\, v_0^5\,B_x
 \left(27 \, B_x^2 +32 \,B_z^2 \right)}
{840\,\pi^2\,\mu^3}, \quad
\sigma^{\text{lf, conc}}_{yz} = \frac{\tau^2 \, e^5\,v_0^5\, B_x\,\,(B_x^2\,+ B_z^2)}
{30\,\pi^2\,\mu^3}.
\end{align}
At $n =2 $, the term $\bs{\mathcal N}_{2,2}$ contributes to
\begin{align}
\sigma^{\text{lf, h}}_{xz} &= \frac{\tau^3\,e^4\, v_0^3 \, B_x\,B_z}{6\,\pi^2}\,, \quad
\sigma^{\text{lf, h}}_{zz} = - \, \frac{\tau^3\, e^4\, v_0^3 \,B_x^2}{6\,\pi^2}\,,,
\end{align}
and the term $\bs{\mathcal N}_{2,3}$ contributes to
\begin{align}
& \sigma^{\text{lf, bc}}_{xz} = \frac{\tau^3\,e^5 \, v_0^5 \,B_x
\left(3\,B_x^2-B_z^2 \right)}{30\,\pi^2\,\mu^2}, \quad
\sigma^{\text{lf, m}}_{xz} = - \, \frac{\tau^3\, e^5\, v_0^5 \,B_x
\left(3\,B_x^2 + 8\,B_z^2 \right)}{60\,\pi^2\,\mu^2} \,,\nn &
\sigma^{\text{lf, bc}}_{zz} = \frac{2\,\tau^3\,e^5 \, v_0^5 \,B_x^2\,B_z}
{15\,\pi^2\,\mu^2}\,, \quad
\sigma^{\text{lf, m}}_{zz} = \frac{\tau^3 \, e^5 \, v_0^5 \,B_x^2\,B_z}{12\,\pi^2\,\mu^2}\,.
\end{align}
At $n =3 $, the nonzero contribution is
\begin{align}
\sigma^{\text{lf, h}}_{yz} &= \frac{\tau^4 \, e^5 \, v_0^5 \,B_x
\left(B_x^2 + B_z^2 \right)\,}{6\,\pi^2\,\mu} \,.
\end{align}

%%%%%%%%%%%%%%%%%%%%%%%%%%
\section{Expressions for internode scattering}
\label{app-boltz-internode}

We now outline the formalism needed to compute the internode-scattering conductivity of the HSM, following Ref.~\cite{ips-internode}. For band $s$ at node $\chi$, with crystal momentum $\bs k$ and dispersion $\varepsilon_s^\chi(\bs k)$, the distribution function $f_s^\chi(\bs r,\bs k,t)$ is defined such that
\begin{align}
dN_s^\chi=g_s\,f_s^\chi(\bs r,\bs k,t)\,\frac{d^3\bs k}{(2\,\pi)^3}\,d^3\bs r
\end{align}
is the number of particles occupying the phase-space volume $dV_p=\frac{d^3\bs k}{(2\, \pi)^3}\,d^3\bs r$ centred at $\{\bs r,\bs k\}$ at time $t$, where $g_s$ is the degeneracy of the band. In the presence of a nonzero OMM, the dispersion acquires the Zeeman-like correction $\varepsilon_m^\chi=-\,\bs m_s^\chi\cdot\bs B$, and the OMM-corrected dispersion and band velocity are
\begin{align}
\mathcal E_s^\chi(\bs k)=\varepsilon_s^\chi(\bs k)+\varepsilon_m^\chi(\bs k)\,,\qquad
\bs w_s^\chi(\bs k)=\nabla_{\bs k}\varepsilon_s^\chi(\bs k)+\nabla_{\bs k}\varepsilon_m^\chi(\bs k)=\bs v_s^\chi+\bs u_m^\chi\,.
\end{align}
The equations of motion of the quasiparticles in static electric and magnetic fields are
\begin{align}
&\dot{\bs r}=\nabla_{\bs k}\mathcal E_s^\chi-\dot{\bs k}\times\bs\Omega_s^\chi\,,\qquad
\dot{\bs k}=-\,e\left(\bs E+\dot{\bs r}\times\bs B\right)\nn
\Rightarrow\, &\dot{\bs r}=D_s^\chi\left[\bs w_s^\chi+e\left(\bs E\times\bs\Omega_s^\chi\right)+e\left(\bs\Omega_s^\chi\cdot\bs w_s^\chi\right)\bs B\right],\qquad
\dot{\bs k}=-\,e\,D_s^\chi\left[\bs E+\bs w_s^\chi\times\bs B+e\left(\bs E\cdot\bs B\right)\bs\Omega_s^\chi\right],
\end{align}
where $-e$ is the charge of the quasiparticles, and
\begin{align}
D_s^\chi=\frac{1}{1+e\left(\bs B\cdot\bs\Omega_s^\chi\right)}
\end{align}
modifies the phase-space volume element from $dV_p$ to $(D_s^\chi)^{-1}dV_p$.

The Fermi-Dirac distribution
\begin{align}
f_0\left(\mathcal E_s^\chi(\bs k),\mu_\chi,T\right)=\frac{1}{1+\exp\left[\left(\mathcal E_s^\chi(\bs k)-\mu_\chi\right)/T\right]}
\end{align}
describes a local equilibrium at the temperature $T$ with a spatially uniform chemical potential $\mu_\chi$. For a small static and uniform $\bs E$, we write
\begin{align}
f_s^\chi(\bs k)=f_0\left(\mathcal E_s^\chi(\bs k)\right)+\delta f_s^\chi(\bs k)\,,
\end{align}
suppressing the dependence of $f_0$ on $\mu_\chi$ and $T$. In the linearised approximation, $\nabla_{\bs k}f_0(\mathcal E_s^\chi)=\bs w_s^\chi\,f_0'(\mathcal E_s^\chi)$, and the linearised Boltzmann equation reads
\begin{align}
\label{eq-val-boltz}
-\,e\,D_s^\chi\left[\left\{\bs w_s^\chi
+e\left(\bs\Omega_s^\chi\cdot\bs w_s^\chi\right)\bs B\right\}\cdot\bs E\right]f_0'(\mathcal E_s^\chi)
+e\,D_s^\chi\,\bs B\cdot\left(\bs w_s^\chi\times\nabla_{\bs k}\right)\delta f_s^\chi=I_{\rm coll}\,,\qquad
I_{\rm coll}=-\,\frac{\delta f_s^\chi}{\tau}\,,
\end{align}
where $\tau$ is the phenomenological relaxation time. With the LFO $\check L=(\bs w_s^\chi\times\bs B)\cdot\nabla_{\bs k}$, Eq.~\eqref{eq-val-boltz} becomes
\begin{align}
e\,D_s^\chi\left[\left\{\bs w_s^\chi+e\left(\bs\Omega_s^\chi\cdot\bs
w_s^\chi\right)\bs B\right\}\cdot\bs E\right]\left[-f_0'(\mathcal E_s^\chi)\right]
-e\,D_s^\chi\,\check L\,\delta f_s^\chi=-\,\frac{\delta f_s^\chi}{\tau}\,.
\end{align}

%%%%
%%%
\subsection{Solution in the presence of internode scattering}
\label{app-internode-internode}

Initially, the two nodes share the chemical potential $\mu$. On applying the fields, each node acquires a local equilibrium chemical potential $\mu_\chi$, with the local equilibrium distribution
\begin{align}
f_{s,L}^\chi\simeq f_0(\mathcal E_s^\chi)+\left[-f_0'(\mathcal E_s^\chi)\right]\delta\mu_\chi\,,\qquad
%%%
\delta\mu_\chi\equiv\mu_\chi-\mu\,.
\end{align}
The intranode part of the collision integral drives the distribution towards this local equilibrium,
\begin{align}
I_{\rm coll}^{\rm intra}=-\,\frac{f_s^\chi(\bs k)-f_{s,L}^\chi}{\tau}\,,
\end{align}
whereas the internode part drives it towards the global equilibrium,
\begin{align}
f_{s,G}^\chi\simeq f_0(\mathcal E_s^\chi)+\left[-f_0'(\mathcal E_s^\chi)\right]\delta\mu_G\,,\qquad
%%%
\mu_G=\frac{\mu_\chi+\mu_{-\chi}}{2}\,,\qquad
%%%
\delta\mu_G\equiv\mu_G-\mu=\frac{\delta\mu_\chi+\delta\mu_{-\chi}}{2}\,,
\end{align}
such that
\begin{align}
I_{\rm coll}^{\rm inter}=-\,\frac{f_s^\chi(\bs k)-f_{s,G}^\chi}{\tau_G}\,,
\end{align}
where $\tau_G$ is the internode relaxation time. We parametrise the nonequilibrium distribution function as
\begin{align}
f_s^\chi(\bs k)=f_0(\mathcal E_s^\chi)+\left[-f_0'(\mathcal E_s^\chi)\right]\bar g_s^\chi(\bs k)\,,
\end{align}
where $\bar g_s^\chi$ quantifies the deviation of the chemical potential caused by the probe fields. The total collision integral then takes the form
\begin{align}
I_{\rm coll}=I_{\rm coll}^{\rm intra}+I_{\rm coll}^{\rm inter}\,,\qquad
\frac{I_{\rm coll}^{\rm intra}}{f_0'(\mathcal E_s^\chi)}=\frac{\bar g_s^\chi(\bs k)-\delta\mu_\chi}{\tau}\,,\qquad
%%%
\frac{I_{\rm coll}^{\rm inter}}{f_0'(\mathcal E_s^\chi)}=\frac{\bar g_s^\chi(\bs k)-\delta\mu_G}{\tau_G}\,.
\end{align}
The average of a physical observable $\mathcal O_s^\chi$ over the electron states of node $\chi$ is defined as
\begin{align}
\bar{\mathcal O}_\chi\equiv\left\langle\mathcal O_s^\chi\right\rangle=
%%%
\frac{\sum_s\int\frac{d^3\bs k}{(2\, \pi)^3}\left(D_s^\chi(\bs k)\right)^{-1}\left[-f_0'(\mathcal E_s^\chi(\bs k))\right]\mathcal O_s^\chi}
%%%%
{\sum_{\tilde s}\int\frac{d^3 \bs k}{(2\, \pi)^3}\left(D_{\tilde s}^\chi(\bs k)\right)^{-1}\left[-f_0'(\mathcal E_{\tilde s}^\chi(\bs k))\right]}\,,
\end{align}
and
\begin{align}
\label{eq-val-rho}
\rho_\chi\equiv
\sum_s\int\frac{d^3\bs k}{(2\, \pi)^3}\left(D_s^\chi(\bs k)\right)^{-1}\left[-f_0'(\mathcal E_s^\chi(\bs k))\right]
\end{align}
is the density of states at node $\chi$. The local equilibrium is captured by $\bar{\bar g}_\chi=\delta\mu_\chi$, where $\bar{\bar g}_\chi\equiv\langle\bar g_s^\chi\rangle$. For $\tau\ll\tau_G$, the system first reaches the local equilibrium through intranode scattering, which allows us to approximate $\bar g_s^\chi\simeq\bar{\bar g}_\chi$ in $I_{\rm coll}^{\rm inter}$. The global conservation of the net charge imposes
\begin{align}
\rho_\chi\,\delta\mu_\chi=-\,\rho_{-\chi}\,\delta\mu_{-\chi}\,,
\end{align}
%
%%%%%%%%%%
which gives
\begin{align}
\frac{I_{\rm coll}}{f_0'(\mathcal E_s^\chi)}\simeq
%%%%%%%%
%%%%%%%
\frac{\bar g_s^\chi(\bs k)-\delta\mu_\chi}{\tau}+\frac{\delta\mu_\chi-\delta\mu_{-\chi}}{2\,\tau_G}
%%%%%%%%%
=\frac{\bar g_s^\chi(\bs k)}{\tau}-\frac{\delta\mu_\chi}{\tau}\left[1-\frac{\tau}
%%%%%%%
{2\,\tau_G}\left(1+\frac{\rho_\chi}{\rho_{-\chi}}\right)\right].
\end{align}
The linearised Boltzmann equation is thus modified to
%%%%%%%%
%%%%%%%%
\begin{align}
&e\,D_s^\chi\left[\left\{\bs w_s^\chi+e\left(\bs\Omega_s^\chi\cdot\bs w_s^\chi\right)\bs B\right\}\cdot\bs E\right]
%%%%5
-e\,D_s^\chi\,\check L\,\bar g_s^\chi(\bs k)
%%%%%%5
%%%%%
=-\,\frac{\bar g_s^\chi(\bs k)}{\tau}+\frac{\delta\mu_\chi}{\tau}\left[1-\frac{\tau}{2\,\tau_G}\left(1+\frac{\rho_\chi}{\rho_{-\chi}}\right)\right]\nn
\Rightarrow\, 
%%%%%55
&\left(1-e\,\tau\,D_s^\chi\,\check L\right)\bar g_s^\chi(\bs k)
%%%%%%5
=-\,e\,\tau\,D_s^\chi\left[\left\{\bs w_s^\chi+e\left(\bs\Omega_s^\chi\cdot\bs w_s^\chi\right)\bs B\right\}\cdot\bs E\right]
%%%%%%%%
%%%%%%%%%%
+\delta\mu_\chi\left[1-\frac{\tau}{2\,\tau_G}\left(1+\frac{\rho_\chi}{\rho_{-\chi}}\right)\right].
\end{align}
Using that $\check L$ annihilates the $\delta\mu_\chi$-dependent term, we obtain
\begin{align}
\bar g_s^\chi(\bs k)=-
%%%%%%%%5
\,e\,\tau\sum_{n=0}^{\infty}\left(e\,\tau\,D_s^\chi\right)^n\check L^n
\left[D_s^\chi\left\{\bs w_s^\chi+e\left(\bs\Omega_s^\chi\cdot\bs w_s^\chi\right)\bs B\right\}\cdot\bs E\right]
%%%%%%%%%%%%%5
+\delta\mu_\chi\left[1-\frac{\tau}{2\,\tau_G}\left(1+\frac{\rho_\chi}{\rho_{-\chi}}\right)\right].
\end{align}
The terms with $n\geq1$ give the LFO-induced part treated in Appendix~\ref{appLF}. Excluding them, i.e., retaining the $n=0$ term,
\begin{align}
\label{eq-gbar}
\bar g_s^\chi(\bs k)=-
%%%%%%%%%
\,e\,\tau\,D_s^\chi\left\{\bs w_s^\chi+e\left(\bs\Omega_s^\chi\cdot\bs w_s^\chi\right)\bs B\right\}\cdot\bs E
+\delta\mu_\chi\left[1-\frac{\tau}
%%%%%%%5
{2\,\tau_G}\left(1+\frac{\rho_\chi}{\rho_{-\chi}}\right)\right].
\end{align}
Taking the average of both sides of Eq.~\eqref{eq-gbar} and using $\bar{\bar g}_\chi=\delta\mu_\chi$ determines $\delta\mu_\chi$ self-consistently:
\begin{align}
%%%%%5
&\delta\mu_\chi\left[1-\frac{\tau}{2\,\tau_G}\left(1+\frac{\rho_\chi}{\rho_{-\chi}}\right)\right]
%%%%%%
=\bar{\bar g}_\chi+e\,\tau\left\langle D_s^\chi\left\{\bs w_s^\chi+e\left(\bs\Omega_s^\chi\cdot\bs w_s^\chi\right)\bs B\right\}\right\rangle\cdot\bs E
\,\Rightarrow\,
%%%%%%5
\delta\mu_\chi=-\,\frac{2\,e\,\tau_G}{1+\frac{\rho_\chi}{\rho_{-\chi}}}\,\frac{\bs{\mathcal I}^\chi\cdot\bs E}{\rho_\chi}\,,\nn
&\text{where}\quad
%%%%%%%%5
\bs{\mathcal I}^\chi=\rho_\chi\left\langle D_s^\chi\left\{\bs w_s^\chi+e\left(\bs\Omega_s^\chi\cdot\bs w_s^\chi\right)\bs B\right\}\right\rangle
%%%%%%%%%5
%%%%%%5
=\sum_s\int\frac{d^3\bs k}{(2\, \pi)^3}\left[-f_0'(\mathcal E_s^\chi(\bs k))\right]\left\{\bs w_s^\chi+e\left(\bs\Omega_s^\chi\cdot\bs w_s^\chi\right)\bs B\right\}.
\end{align}
The non-AH contribution to the current, excluding the LFO-induced part, follows as
\begin{align}
&\bar{\bs J}_s^\chi
%%%%%5
=-\,e\int\frac{d^3\bs k}{(2\, \pi)^3}\left[-f_0'(\mathcal E_s^\chi)\right]\left[\bs w_s^\chi+e\left(\bs\Omega_s^\chi\cdot\bs w_s^\chi\right)\bs B\right]\bar g_s^\chi(\bs k)
%%%%5
=\bs J_s^{\chi,\rm intra}+\bs J_s^{\chi,\rm inter}\,,\nn
%%%%%%%%
&\bs J_s^{\chi,\rm intra}=e^2\,\tau\int\frac{d^3\bs k}{(2\, \pi)^3}\,D_s^\chi
%%%5
\left[-f_0'(\mathcal E_s^\chi)\right]
\left[\bs w_s^\chi+e\left(\bs\Omega_s^\chi\cdot\bs w_s^\chi\right)\bs B\right]\left\{\bs w_s^\chi+e\left(\bs\Omega_s^\chi\cdot\bs w_s^\chi\right)\bs B\right\}\cdot\bs E\,,\nn
%%%%%%%%%%%
%%%%%%5
&\bs J_s^{\chi,\rm inter}=\frac{2\,e^2}{1+\frac{\rho_\chi}{\rho_{-\chi}}}
%%%%
\left[\tau_G-\frac{\tau}{2}\left(1+\frac{\rho_\chi}{\rho_{-\chi}}\right)\right]
%%%%%%%
%%%%%%%
\int\frac{d^3\bs k}{(2\, \pi)^3}\left[-f_0'(\mathcal E_s^\chi)\right]\left[\bs w_s^\chi+e\left(\bs\Omega_s^\chi\cdot\bs w_s^\chi\right)\bs B\right]\frac{\bs{\mathcal I}^\chi\cdot\bs E}{\rho_\chi}\,.
\end{align}
The intranode part reproduces $\bar\sigma^s_{ij}$ of Eq.~\eqref{eq_elec}. The conductivity tensor corresponding to the internode-scattering-induced current is
\begin{align}
\label{eq-sig}
\left(\sigma_s^{\chi,\rm inter}\right)_{ij}
%%%%%%5
=\frac{e^2\,\rho_{-\chi}}{\rho_\chi\,\rho_G}\left[\tau_G-\frac{\tau\,\rho_G}{\rho_{-\chi}}\right]
%%%%%%%%%
\int\frac{d^3\bs k}{(2\, \pi)^3}\left[-f_0'(\mathcal 
%%%%5
E_s^\chi)\right]\left[(w_s^\chi)_i+e\left(\bs\Omega_s^\chi\cdot\bs w_s^\chi\right)B_i\right]\mathcal I_j^\chi\,,\qquad
\rho_G=\frac{\rho_\chi+\rho_{-\chi}}{2}\,.
\end{align}

%%%%5
%%%%%%5

\subsection{Terms expanded up to order \texorpdfstring{$|\bs B|^2$}{B^2}}
\label{sec-inter-expan}

Following Appendix~B of Ref.~\cite{ips-internode}, we expand the integrand of Eq.~\eqref{eq-sig} up to $\mathcal O(|\bs B|^2)$. Since the phase-space factor $(D_s^\chi)^{-1}=1+e\,\bs B\cdot\bs\Omega_s^\chi$ in Eq.~\eqref{eq-val-rho} is linear in $\bs B$, the expansion of $\rho_\chi$ contains no term quadratic in $\bs B\cdot\bs\Omega_s^\chi$ alone. For this purpose, we define
\begin{align}
\label{eq-rho-B1}
&\rho_\chi=\rho_\chi^{(0)}+\rho_\chi^{(1)}+\rho_\chi^{(2)}+\mathcal O(|\bs B|^3)\,,\nn
%%%%%%
&\rho_\chi^{(0)}=
\sum_s\int\frac{d^3 \bs k}{(2\, \pi)^3}\left\{-f_0'(\varepsilon_s^\chi(\bs k))\right\},\qquad
%
%%%%%%%%%%%
\rho_\chi^{(1)}=\sum_s\int\frac{d^3 \bs k}{(2\, \pi)^3}
%%5
\left[e\,\bs B\cdot\bs\Omega_s^\chi(\bs k)\left\{-f_0'(\varepsilon_s^\chi(\bs k))\right\}+\varepsilon_m^\chi(\bs k)\left\{-f_0''(\varepsilon_s^\chi(\bs k))\right\}\right],\nn
&\rho_\chi^{(2)}=\sum_s\int\frac{d^3 \bs k}{(2\, \pi)^3}\left[e\left\{\bs B\cdot\bs\Omega_s^\chi(\bs k)\right\}\varepsilon_m^\chi(\bs k)\left\{-f_0''(\varepsilon_s^\chi(\bs k))\right\}
%%%%%%%%%
+\frac{\left\{\varepsilon_m^\chi(\bs k)\right\}^2}{2}\left\{-f_0'''(\varepsilon_s^\chi(\bs k))\right\}\right],
\end{align}
\begin{align}
\label{eq-rhoG}
\rho_G=\rho_G^{(0)}+\rho_G^{(1)}+\rho_G^{(2)}+\mathcal O(|\bs B|^3)\,,\qquad
%%%%
\rho_G^{(0)}=\frac{\rho_\chi^{(0)}+\rho_{-\chi}^{(0)}}{2}\,,\qquad
\rho_G^{(1)}=\frac{\rho_\chi^{(1)}+\rho_{-\chi}^{(1)}}{2}\,,\qquad
\rho_G^{(2)}=\frac{\rho_\chi^{(2)}+\rho_{-\chi}^{(2)}}{2}\,,
\end{align}
and
\begin{align}
\label{eq-inter-I}
\mathcal I_j^\chi&=\mathcal I_j^{\chi,0}+\mathcal I_j^{\chi,1}+\mathcal I_j^{\chi,2}+\mathcal O(|\bs B|^3)\,,\qquad
%%
%%%%%%%%%
\mathcal I_j^{\chi,0}=\sum_s\int\frac{d^3 \bs k}{(2\, \pi)^3}\left(v_s^\chi(\bs k)\right)_j\left\{-f_0'(\varepsilon_s^\chi(\bs k))\right\},\nn
%%
%%%%%%%%%
\mathcal I_j^{\chi,1}&=\sum_s\int\frac{d^3 \bs k}{(2\, \pi)^3}\left[\left\{e\left(\bs\Omega_s^\chi(\bs k)\cdot\bs v_s^\chi(\bs k)\right)B_j+
\left(u_m^\chi(\bs k)\right)_j\right\}\left\{-f_0'(\varepsilon_s^\chi(\bs k))\right\}
+\varepsilon_m^\chi(\bs k)\left(v_s^\chi(\bs k)\right)_j\left\{-f_0''(\varepsilon_s^\chi(\bs k))\right\}\right],\nn
%%
%%%%%%%%%
\mathcal I_j^{\chi,2}&=\sum_s\int\frac{d^3 \bs k}{(2\, \pi)^3}\Big[e\left\{\bs\Omega_s^\chi(\bs k)\cdot\bs u_m^\chi(\bs k)\right\}B_j\left\{-f_0'(\varepsilon_s^\chi(\bs k))\right\}
+\varepsilon_m^\chi(\bs k)\left\{e\left(\bs\Omega_s^\chi(\bs k)\cdot\bs v_s^\chi(\bs k)\right)B_j+\left(u_m^\chi(\bs k)\right)_j\right\}\left\{-f_0''(\varepsilon_s^\chi(\bs k))\right\}\nn
&\hspace{2.6cm}+\frac{\left(\varepsilon_m^\chi(\bs k)\right)^2\left(v_s^\chi(\bs k)\right)_j}{2}\left\{-f_0'''(\varepsilon_s^\chi(\bs k))\right\}\Big].
\end{align}
The conductivity is then given by
\begin{align}
\label{eq-sig-inter}
\left(\sigma_s^{\chi,\rm inter}\right)_{ij}=
\frac{e^2\left[\tau\,\rho_G^{(0)}-\tau_G\,\rho_{-\chi}^{(0)}\right]}{\rho_G^{(0)}\,\rho_\chi^{(0)}}
\int\frac{d^3\bs k}
{(2\, \pi)^3}\left[\varsigma_{0,ij}^{\chi,s}+\varsigma_{1,ij}^{\chi,s}+\varsigma_{2,ij}^{\chi,s}\right]+\mathcal O(|\bs B|^3)\,,
\end{align}
where
\begin{align}
\label{eq-sig-inter-0}
\varsigma_{0,ij}^{\chi,s}(\bs k)=-\,\mathcal I_j^{\chi,0}\left(v_s^\chi\right)_i\left\{-f_0'(\varepsilon_s^\chi)\right\},
\end{align}
\begin{align}
\label{eq-sig-inter-1}
\varsigma_{1,ij}^{\chi,s}(\bs k)&=\left[\left[-\,e\,\bs\Omega_s^\chi\cdot\bs v_s^\chi\,B_i
+\left\{\frac{\tau_G}{\rho_G^{(0)}}\,\frac{\rho_G^{(0)}\,\rho_{-\chi}^{(1)}-\rho_G^{(1)}\,\rho_{-\chi}^{(0)}}{\tau\,\rho_G^{(0)}-\tau_G\,\rho_{-\chi}^{(0)}}+\frac{\rho_\chi^{(1)}}{\rho_\chi^{(0)}}\right\}\left(v_s^\chi\right)_i-
\left(u_m^\chi\right)_i\right]\mathcal I_j^{\chi,0}
-\left(v_s^\chi\right)_i\,\mathcal I_j^{\chi,1}\right]\left\{-f_0'(\varepsilon_s^\chi)\right\}\nn
&\quad-\,\varepsilon_m^\chi\left(v_s^\chi\right)_i\,\mathcal I_j^{\chi,0}\left\{-f_0''(\varepsilon_s^\chi)\right\},
\end{align}
and
\begin{align}
\label{eq-sig-inter-2}
\varsigma_{2,ij}^{\chi,s}(\bs k)&=\Big[\,e\,\rho_G^{(0)}\rho_\chi^{(0)}\,\bs\Omega_s^\chi\cdot\bs v_s^\chi\left\{\rho_G^{(0)}\left(\tau_G\,\rho_\chi^{(0)}\rho_{-\chi}^{(1)}+\tau\,\rho_G^{(0)}\rho_\chi^{(1)}\right)-\tau_G\,\rho_{-\chi}^{(0)}\left(\rho_G^{(1)}\rho_\chi^{(0)}+\rho_G^{(0)}\rho_\chi^{(1)}\right)\right\}B_i\nn
&\quad+e\left(\rho_G^{(0)}\rho_\chi^{(0)}\right)^2\bs\Omega_s^\chi\cdot\bs u_m^\chi\left(\tau_G\,\rho_{-\chi}^{(0)}-\tau\,\rho_G^{(0)}\right)B_i
-\tau_G\,\rho_{-\chi}^{(0)}\rho_G^{(0)}\rho_\chi^{(0)}\left\{\rho_G^{(1)}\rho_\chi^{(0)}+\rho_G^{(0)}\rho_\chi^{(1)}\right\}\left(u_m^\chi\right)_i\nn
%%%
%%
&\quad+\tau_G\left(\rho_G^{(0)}\rho_\chi^{(0)}\right)^2\rho_{-\chi}^{(1)}\left(u_m^\chi\right)_i
+\tau\left(\rho_G^{(0)}\right)^3\rho_\chi^{(0)}\rho_\chi^{(1)}\left(u_m^\chi\right)_i
+\tau_G\,\rho_{-\chi}^{(0)}\left(\rho_G^{(1)}\right)^2\left(\rho_\chi^{(0)}\right)^2\left(v_s^\chi\right)_i\nn
&\quad+\tau_G\,\rho_{-\chi}^{(0)}\left(\rho_G^{(0)}\right)^2\left(\rho_\chi^{(1)}\right)^2\left(v_s^\chi\right)_i
+\tau_G\,\rho_{-\chi}^{(0)}\rho_G^{(0)}\rho_G^{(1)}\rho_\chi^{(0)}\rho_\chi^{(1)}\left(v_s^\chi\right)_i
-
\tau_G\,\rho_G^{(0)}\rho_G^{(1)}\left(\rho_\chi^{(0)}\right)^2\rho_{-\chi}^{(1)}\left(v_s^\chi\right)_i\nn
&\quad-\tau\left(\rho_G^{(0)}\right)^3\left(\rho_\chi^{(1)}\right)^2\left(v_s^\chi\right)_i
-\tau_G\left(\rho_G^{(0)}\right)^2\rho_\chi^{(0)}\rho_{-\chi}^{(1)}\rho_\chi^{(1)}\left(v_s^\chi\right)_i
-\tau_G\,\rho_G^{(2)}\rho_{-\chi}^{(0)}\rho_G^{(0)}\left(\rho_\chi^{(0)}\right)^2\left(v_s^\chi\right)_i\nn
&\quad+\tau_G\,\rho_{-\chi}^{(2)}\left(\rho_G^{(0)}\right)^2\left(\rho_\chi^{(0)}\right)^2\left(v_s^\chi\right)_i
+\rho_\chi^{(2)}\left(\rho_G^{(0)}\right)^2\rho_\chi^{(0)}\left(\tau\,\rho_G^{(0)}-\tau_G\,\rho_{-\chi}^{(0)}\right)\left(v_s^\chi\right)_i\Big]
%%%%%%%%%5
\frac{\mathcal I_j^{\chi,0}\left\{-f_0'(\varepsilon_s^\chi)\right\}}{\left\{\rho_G^{(0)}\rho_\chi^{(0)}\right\}^2\left[\tau\,\rho_G^{(0)}-\tau_G\,\rho_{-\chi}^{(0)}\right]}\nn
%%%%%%%%%%%%%%%%%%%%%
&\quad+\left[-\,e\,\bs\Omega_s^\chi\cdot\bs v_s^\chi\,B_i-\left(u_m^\chi\right)_i
+\left\{\frac{\tau_G}{\rho_G^{(0)}}\,\frac{\rho_G^{(0)}\rho_{-\chi}^{(1)}-\rho_{-\chi}^{(0)}\rho_G^{(1)}}{\tau\,\rho_G^{(0)}-\tau_G\,\rho_{-\chi}^{(0)}}+\frac{\rho_\chi^{(1)}}{\rho_\chi^{(0)}}\right\}\left(v_s^\chi\right)_i\right]\mathcal I_j^{\chi,1}\left\{-f_0'(\varepsilon_s^\chi)\right\}
-\left(v_s^\chi\right)_i\,\mathcal I_j^{\chi,2}\left\{-f_0'(\varepsilon_s^\chi)\right\}\nn
%%%%%%%%%%%%%
&\quad+\varepsilon_m^\chi\left[\left[-\,e\,\bs\Omega_s^\chi\cdot\bs v_s^\chi\,B_i-\left(u_m^\chi\right)_i
+\left\{\frac{\tau_G}{\rho_G^{(0)}}\,\frac{\rho_G^{(0)}\rho_{-\chi}^{(1)}-\rho_G^{(1)}\rho_{-\chi}^{(0)}}{\tau\,\rho_G^{(0)}-\tau_G\,\rho_{-\chi}^{(0)}}+\frac{\rho_\chi^{(1)}}{\rho_\chi^{(0)}}\right\}\left(v_s^\chi\right)_i\right]\mathcal I_j^{\chi,0}
-\left(v_s^\chi\right)_i\,\mathcal I_j^{\chi,1}\right]\left\{-f_0''(\varepsilon_s^\chi)\right\}\nn
%%%%%%%
%%%%%%%%%%%%%%%
&\quad-\frac{\left(\varepsilon_m^\chi\right)^2\left(v_s^\chi\right)_i\,\mathcal I_j^{\chi,0}}{2}\left\{-f_0'''(\varepsilon_s^\chi)\right\}.
\end{align}
In the last three equations, we have suppressed the $\bs k$-dependence of the various quantities. For nodes whose BC and OMM are odd functions of the momentum components, such as those carrying BC monopoles, $\rho_\chi^{(1)}=\rho_G^{(1)}=\mathcal I_j^{\chi,0}=\int\varsigma_{0,ij}^{\chi,s}=\int\varsigma_{1,ij}^{\chi,s}=0$, and Eq.~\eqref{eq-sig-inter} reduces to Eqs.~(10) and (11) of Ref.~\cite{ips-internode}. For the BC-dipoles of Eq.~\eqref{eq-valley-BC}, $\bs\Omega_s^\chi$ and $\bs m_s^\chi$ are even in $\bs k$. The first-order corrections $\rho_\chi^{(1)}$ then need not vanish, and we evaluate the complete expressions of Eqs.~\eqref{eq-rho-B1}--\eqref{eq-sig-inter-2}.

%%%%%%%%%%%%%%%
\subsection{Internode scattering in a valley-Hopf semimetal}
\label{app-internode}

In this subsection, we consider two conjugate Hopf nodes and compute their internode-scattering-induced conductivity, following Ref.~\cite{ips-internode}, for the three set-ups of Sec.~\ref{secmethod}.
The Hamiltonian $H_h$ of Eq.~\eqref{eqbcd2} coincides with the continuum model of Ref.~\cite{graf-hopf} (say, $ H_3^{+1} $) for the valley index $\chi =+1$. In that reference, the valley index enters through $q_\pm^{\chi}=\chi \,q_x\pm i\,q_y$, which fixes the BC-dipole axis to $\bs \xi_\chi=(0,0,\chi )$. Lattice realisations of such nodes are of two kinds~\cite{graf-hopf}. Topological HSMs host an odd number of Hopf crossings, for instance a single crossing at the $\Gamma$ point described by $\chi = +1$. Valley-HSMs host the crossings in pairs with opposite dipole orientations: in the time-reversal-symmetric model built from stacked honeycomb layers, the two crossings sit at the $K$ and $K'$ points, and in the time-reversal-breaking model built from stacked square layers, at the $\Gamma$ and $M$ points. In both cases, their low-energy expansions are $H_3^{+1}$ and $H_3^{-1}$, respectively~\cite{graf-hopf}. The results of Secs.~\ref{secset1-hopf}--\ref{secset3-hopf} pertain to a single node and apply directly to a topological HSM with a single crossing. Internode scattering requires at least two nodes, and we therefore consider the valley-HSM here. Labelling the two valleys by $\chi = \pm1$, we have
%%%%%%%%%%%%%%%
\begin{align}
\label{eq-valley-H}
\mathcal H_3^\chi(\bs k)=v_0\left(\chi\,k_x\,\lambda_1+k_y\,\lambda_2+k_z\,\lambda_5\right).
\end{align}
Here, $\bs k$ is measured from the respective nodal point, and $\mathcal H_3^{+1}=H_h$. Since $\lambda_1$ is real while $\lambda_2$ and $\lambda_5$ are purely imaginary, the two valleys are related by
\begin{align}
\label{eq-valley-TR}
\mathcal H_3^{-\chi}(\bs k)=\mathcal H_3^\chi(-k_x,k_y,k_z)
=\left[\mathcal H_3^\chi(-\bs k)\right]^*.
\end{align}
The second equality shows that, in the time-reversal-symmetric realisation, the partner valley is not an independent input. For spinless time reversal, $\mathcal T=\mathcal K$, the Bloch Hamiltonians at $\bs K$ and $-\bs K$ are related by $H_{-\bs K}(\bs k)=[H_{\bs K}(-\bs k)]^*$, which yields $\mathcal H_3^{-1}$. Each valley individually breaks $\mathcal T$ (cf.\ Appendix~\ref{sec:AHeval}), whereas $\mathcal T$ maps one valley onto the other.

Both valleys have the eigenvalues $\varepsilon_s^\chi(\bs k)=(-1)^s\,v_0\,k$ for $s\in\{1,2\}$ and $\varepsilon_0^\chi=0$, and the band velocities $\bs v_s^\chi=(-1)^s\,v_0\,\bs k/k$. The BC and the OMM, obtained from the eigenstates of Eq.~\eqref{eq-valley-H}, are, for $\chi=+1$, those of Eq.~\eqref{eqbcommhopf}, and for general $\chi$
\begin{align}
\label{eq-valley-BC}
\bs\Omega_s^\chi(\bs k)=
\begin{cases}
\dfrac{\chi\,k_z\,\bs k}{k^4} & \text{for } s=1,2\\[8pt]
-\,\dfrac{2\,\chi\,k_z\,\bs k}{k^4} & \text{for } s=0
\end{cases}
\qquad\text{and}\qquad
\bs m_s^\chi(\bs k)=
\begin{cases}
\dfrac{(-1)^{s+1}\,\chi\,e\,v_0\,k_z\,\bs k}{2\,k^3} & \text{for } s=1,2\\[8pt]
 0 & \text{for } s=0
\end{cases}\,.
\end{align}
Hence,
\begin{align}
\label{eq-valley-rel}
\varepsilon_s^{\chi}=\varepsilon_s^{-\chi}\,,\qquad
\bs v_s^{\chi}=\bs v_s^{-\chi}\,,\qquad
\bs\Omega_s^{\chi}=-\,\bs\Omega_s^{-\chi}\,,\qquad
\bs m_s^{\chi}=-\,\bs m_s^{-\chi}\,,
\end{align}
which are the conditions satisfied by a pair of conjugate nodes of the same nature in Ref.~\cite{ips-internode}.

In the $T\to0$ limit, $\left\{-f_0'(\varepsilon_s^\chi)\right\}\to\delta(\varepsilon_s^\chi-\mu)$, and the integrals involving $\left\{-f_0''(\varepsilon_s^\chi)\right\}$ and $\left\{-f_0'''(\varepsilon_s^\chi)\right\}$ are evaluated by integrating the derivatives of the $\delta$-function by parts. For $\mu>0$, only the band $s=2$ of each node contributes, because $s=0$ and $s=1$ have no Fermi surface. Equations~\eqref{eq-rho-B1} and \eqref{eq-rhoG} give
\begin{align}
\label{eq-val-rho-res}
&\rho_\chi^{(0)}=\rho_{-\chi}^{(0)}=\rho_G^{(0)}=\frac{\mu^2}{2\,\pi^2\,v_0^3}\,,\qquad
\rho_\chi^{(1)}=-\,\rho_{-\chi}^{(1)}=\frac{\chi\,e\,B_z}{12\,\pi^2\,v_0}\,,\qquad
\rho_G^{(1)}=0\,,\nn
&\rho_\chi^{(2)}=\rho_{-\chi}^{(2)}=\rho_G^{(2)}=\frac{e^2\,v_0\left(B_x^2+B_y^2+3\,B_z^2\right)}{60\,\pi^2\,\mu^2}\,.
\end{align}
The integrands of $\rho_\chi^{(1)}$ and $\rho_\chi^{(2)}$ are even functions of $\bs k$, since $\bs\Omega_2^\chi$ and $\varepsilon_m^\chi$ are even under $\bs k\to-\bs k$. Consequently, $\rho_\chi^{(1)}$ is nonzero whenever $\bs B$ has a component along the dipole axis, i.e., in set-ups~II and III, and the conditions leading to Eqs.~(10) and (11) of Ref.~\cite{ips-internode} are not satisfied. In contrast, every integrand of $\mathcal I_j^{\chi,0}$, $\mathcal I_j^{\chi,1}$, and $\mathcal I_j^{\chi,2}$ in Eq.~\eqref{eq-inter-I} contains one factor that is odd under $\bs k\to-\bs k$, namely $\left(v^\chi\right)_j$, $\bs\Omega^\chi\cdot\bs v^\chi$, $\left(u_m^\chi\right)_j$, or $\bs\Omega^\chi\cdot\bs u_m^\chi$, multiplied by functions that are even, namely $\bs\Omega^\chi$, $\varepsilon_m^\chi$, and the derivatives of $f_0(\varepsilon^\chi)$. All these integrands are therefore odd functions of the momentum components and vanish upon integration, for arbitrary $(B_x,B_y,B_z)$:
\begin{align}
\label{eq-valley-I}
\mathcal I_j^{\chi,0}=\mathcal I_j^{\chi,1}=\mathcal I_j^{\chi,2}=0\,,\qquad j\in\{x,y,z\}\,,\qquad\chi=\pm1\,.
\end{align}
In particular, the BC term of $\mathcal I_j^{\chi,1}$, $e\,B_j\int\frac{d^3 \bs k}{(2\, \pi)^3}\left(\bs\Omega_2^\chi\cdot\bs v_2^\chi\right)\delta(\varepsilon_2^\chi-\mu)$, is proportional to the Berry flux through the Fermi surface. It is nonzero for a node carrying a BC monopole, where it gives rise to the chiral-anomaly-induced response, but it vanishes for the BC dipole, whose flux through any closed surface enclosing the node is zero.

Since every term of $\varsigma_{0,ij}^{\chi,s}$, $\varsigma_{1,ij}^{\chi,s}$, and $\varsigma_{2,ij}^{\chi,s}$ in Eqs.~\eqref{eq-sig-inter-0}--\eqref{eq-sig-inter-2} is proportional to $\mathcal I_j^{\chi,0}$, $\mathcal I_j^{\chi,1}$, or $\mathcal I_j^{\chi,2}$, Eq.~\eqref{eq-valley-I} gives
\begin{align}
\label{eq-valley-inter-zero}
\int\frac{d^3\bs k}{(2\, \pi)^3}\,\varsigma_{0,ij}^{\chi,s}=\int\frac{d^3\bs k}{(2\, \pi)^3}\,\varsigma_{1,ij}^{\chi,s}=\int\frac{d^3\bs k}{(2\, \pi)^3}\,\varsigma_{2,ij}^{\chi,s}=0\,,\qquad
\left(\sigma_s^{\chi,\rm inter}\right)_{ij}=0+\mathcal O(|\bs B|^3)\,,
\end{align}
for both nodes and for their sum, irrespective of the nonzero values of $\rho_\chi^{(1)}$ and $\rho_\chi^{(2)}$. The results for the three set-ups are summarised in Table~\ref{tab-valley}. The parity argument behind Eq.~\eqref{eq-valley-I} applies equally to the unexpanded $\bs{\mathcal I}^\chi$, which contains the factor $\bs w_s^\chi+e\left(\bs\Omega_s^\chi\cdot\bs w_s^\chi\right)\bs B$ and the weight $-f_0'(\mathcal E_s^\chi)$, with $\mathcal E_s^\chi$ even and $\bs w_s^\chi$ odd in $\bs k$. Hence, $\delta\mu_\chi=0$ and the internode current vanishes to all orders in $\bs B$, within the approximation adopted for $\bar g_s^\chi$.

%%%

\begin{table*}[t!]
\caption{\label{tab-valley}
Internode-scattering-induced conductivity of the valley-HSM in the three set-ups, up to $\mathcal O(|\bs B|^2)$, for $\chi=\pm1$ and for the sum over the two nodes.}
\begin{ruledtabular}
\begin{tabular}{lcccc}
Set-up & $\rho_\chi^{(1)}$ & $\rho_\chi^{(2)}$ & $\mathcal I_j^{\chi,0},\ \mathcal I_j^{\chi,1},\ \mathcal I_j^{\chi,2}$ & Internode conductivity\\
\colrule
I  & $0$ & $\dfrac{e^2v_0(B_x^2+B_y^2)}{60\pi^2\mu^2}$ & $0,\ 0,\ 0$ & $\sigma^{\rm inter}_{xx}=\sigma^{\rm inter}_{yx}=\sigma^{\rm inter}_{zx}=0$\\[8pt]
II  & $\dfrac{\chi eB_z}{12\, \pi^2v_0}$ & $\dfrac{e^2v_0(B_x^2+3B_z^2)}{60\pi^2\mu^2}$ & $0,\ 0,\ 0$ & $\sigma^{\rm inter}_{xx}=\sigma^{\rm inter}_{yx}=\sigma^{\rm inter}_{zx}=0$\\[8pt]
III  & $\dfrac{\chi eB_z}{12\, \pi^2v_0}$ & $\dfrac{e^2v_0(B_x^2+3B_z^2)}{60\pi^2\mu^2}$ & $0,\ 0,\ 0$ & $\sigma^{\rm inter}_{xz}=\sigma^{\rm inter}_{yz}=\sigma^{\rm inter}_{zz}=0$\\
\end{tabular}
\end{ruledtabular}
\end{table*}
%%%%%%%%%%%%%%%%%%%%%%%%%%55

The internode current is driven by the shift $\delta\mu_\chi\propto\bs{\mathcal I}^\chi\cdot\bs E$ of the local chemical potentials, i.e., by the charge pumped between the two nodes. Thus, internode-scattering-induced conductivity of Eq.~\eqref{eq-sig} is the transport signature of the chiral anomaly. Since $\bs{\mathcal I}^\chi=\bs 0$, no charge is pumped between conjugate Hopf nodes, and the valley-HSM does not exhibit the internode-scattering-induced magnetoconductivity of nodes carrying BC monopoles.
 Parallel components of $\bs E$ and $\bs B$ pump charge from one node to its conjugate partner and internode scattering (with the long relaxation time $\tau_G$) limits the resulting imbalance of the local chemical potentials. The present results show that this response is tied specifically to the monopole charge of the nodes, and not to the presence of a appreciable BC at the Fermi surface. 

The difference of the valley-HSM from a monopole system originates in the parity of the BC under $\bs k\to-\bs k$. For a BC monopole, $\bs\Omega_s^\chi\propto\bs k/k^3$ is odd. The BC term of $\mathcal I_j^{\chi,1}$, $e\,B_j\int\frac{d^3 \bs k}{(2\, \pi)^3}\left(\bs\Omega_s^\chi\cdot\bs v_s^\chi\right)\left\{-f_0'(\varepsilon_s^\chi)\right\}$, then has an even integrand, and it measures the BC flux through the Fermi surface, i.e., the monopole charge enclosed by it. As a consequence, $\delta\mu_\chi\neq0$, and Eq.~\eqref{eq-sig} yields $\left(\sigma_s^{\chi,\rm inter}\right)_{ij}\propto e^4\,v_0^3\,(\tau_G-\tau)\,B_i\,B_j/\mu^2$, with a prefactor set by the monopole charges of the nodes involved~\cite{ips-internode}. At the same time, the first-order correction $\rho_\chi^{(1)}$ to the density-of-states vanishes, because its integrand is odd. For a BC dipole, these roles are interchanged. Since $\bs\Omega_s^\chi$ and $\varepsilon_m^\chi$ are even under $\bs k\to-\bs k$, the density-of-states acquires a correction $\rho_\chi^{(1)}\propto\chi\,B_z$ that depends on the orientation of $\bs B$ relative to the dipole axis, whereas every term of $\mathcal I_j^{\chi,0}$, $\mathcal I_j^{\chi,1}$, and $\mathcal I_j^{\chi,2}$ has an odd integrand and vanishes. The field-induced charge pumping between the nodes is therefore absent, even though the BC itself is of the same order of magnitude, $|\bs\Omega_s^\chi|\sim1/k^2$, as for a monopole. 
This is consistent with Ref.~\cite{graf-hopf}, in which the linear-in-$\bs B$ magnetocurrents of the two valleys of a valley-HSM cancel because their dipoles are opposite.

%======================================================================
\section{Results for the NVNR}
\label{app_GNR}
%======================================================================

\subsection{Set-up I}

The non-AH part comprises
\begin{align}
\label{eqgnr1_d}
\sigma^{\rm d}_{xx} & = \frac{\tau \, e^2 \, k_0 \, v_0 }
{8 \, \pi \, v_z \, \mu}
\left(\mu^2 - \Delta^2 \right), \quad
\sigma^{\rm  bc}_{xx}  =
\frac{\tau \, e^4 \, v_z \, v_0^3  \, \Delta^2 \, k_0
\left( \mu^2 - \Delta^2 \right)}
{128 \, \pi \, \mu^7}
\left( B_x^2 + 3 \, B_y^2 \right), \nn
\sigma^{\rm  m}_{xx} & =
\frac{\tau \, e^4 \,v_z \, v_0^3 \, \Delta^2}
{128 \, \pi \, \mu^7}
\left[
-\left(B_x^2 + 3 \, B_y^2 \right) k_0
\left( 6\, \mu^2 - 5\, \Delta^2 \right)
+\frac{2 \left(B_x^2 + 3 \, B_y^2 \right) \mu^4}
{v_0\, \zeta}
\right], \quad
\sigma^{\rm conc}_{xx} =
\frac{-\,\tau \, e^4 \, v_z \, v_0^3  \, \Delta^4 \, k_0
\left( B_x^2 + 3 \, B_y^2 \right)}
{64 \, \pi \, \mu^7}\,.
\end{align}
\begin{align}
\label{eqgnr1_yx}
\sigma^{\rm d}_{yx} & = 0 \,, \quad
\sigma^{\rm  bc}_{yx}  =
\frac{ -\, \tau \, e^4 \, v_z \, v_0^3  \, \Delta^2 \, k_0
\left(  \mu^2 - \Delta^2 \right) B_x \, B_y}
{ 64 \, \pi \, \mu^7}, \nn
\sigma^{\rm  m}_{yx} & =
\frac{\tau \, e^4 \,v_z \, v_0^3 \, \Delta^2\,  B_x \, B_y}
{64 \, \pi \, \mu^7}
\left [  k_0 \left( 6\, \mu^2 - 5\, \Delta^2\right)
- \frac{2 \,\mu^4} {v_0\, \zeta} \right], \quad
\sigma^{\rm conc}_{yx}  =
\frac{\tau \, e^4 \,v_z \, v_0^3 \, \Delta^4\, k_0\, B_x \, B_y}
{32 \, \pi \, \mu^7}\,.
\end{align}

The AH part is
\begin{align}
\label{eqgnr1_ah}
\sigma^{\rm{ah}}_{zx} & =
\frac{ - \,e^3\, v_z \, v_0 \, k_0 \,\Delta^2 \, B_y}
{16 \, \pi\, \mu^4} \left (1 + \frac{ 9\, e^2  \,v_z^2 \, v_0^2  
\, \Delta^2\,{\bs B}^2 } { 4\, \mu^6 }  \right ).
\end{align}

The LFO-induced part comprises the in-plane components $\sigma^{\rm{lf}}_{xx}$ and $\sigma^{\rm{lf}}_{yx}$, together with the out-of-plane component $\sigma^{\rm{lf}}_{zx}$,
\begin{align}
\label{eqgnr1_lf_ip}
\sigma^{\rm{lf}}_{xx} & =
\frac{ -\,\tau^3 \, e^4\, v_0^4 \,v_z \, k_0^2  }
{ 16 \,\pi \,\mu^3 }
\left(B_x^2 + 3\, B_y^2 \right)
\left( k_0 \, v_0 -\zeta\right), \quad
\sigma^{\rm{lf}}_{yx}  =
\frac{ \tau^3  \, e^4 \,  v_z \,v_0^4 \, k_0^2 }
{8 \,\pi\, \mu^3}
\, B_x\, B_y
\left(k_0 \, v_0- \zeta\right). \nn
\sigma^{\rm{lf}}_{zx}
&= \frac{\tau^2 \, e^3 \, v_z \, v_0\, B_y}
{128\, \pi \, \mu^8 }\,
\Bigg[
16\, k_0 \, \mu^6  \left(\Delta^2 - \mu^2\right)
- 12 \, \Delta^4 \,e^2 \, k_0 \,v_0^2\, v_z^2
\left(B_x^2+B_y^2\right) 
+ 24  \, e^2 \,  k_0^2  \, \mu^4  \, \tau^2  \, v_0^3
\, v_z^2 \left(B_x^2+B_y^2\right)
\left(k_0  \, v_0 - \zeta \right)
 \nn
& \hspace{3 cm}
+ \frac{ 3  \, \Delta^2 \,  e^2 \,  v_0  \, v_z^2 \,\mu^2
\left(B_x^2+B_y^2\right)
\left(7  \, \zeta \, k_0  \,  v_0 - 2  \, \mu^2 \right) }
{\zeta}
\Bigg].
\end{align}

\subsection{Set-up II}

The non-AH part comprises
\begin{align}
\label{eqgnr2_long}
\sigma^{\rm d}_{xx} & = \frac{\tau \, e^2 \, k_0\, v_0 }
{8 \, \pi \,v_z \,\mu}
\left(\mu^2 - \Delta^2 \right), \quad
\sigma^{\rm  bc}_{xx} =
\frac{\tau \, e^4  \,v_z\, v_0^3 \, \Delta^2 \, k_0
\left( \mu^2 - \Delta^2 \right) }
{128 \, \pi \, \mu^7}\,
B_x^2 \,, \nn
\sigma^{\rm  m}_{xx} & =
\frac{\tau \, e^4 \,v_z \, v_0^3 \, \Delta^2}
{128 \, \pi \, \mu^7}
\left[
-\,B_x^2 \, k_0 \left( 6\, \mu^2 - 5\, \Delta^2 \right)
+\frac{2 \,B_x^2  \,\mu^4} {v_0\, \zeta}
\right], \quad
\sigma^{\rm conc}_{xx} =
\frac{ -\,\tau\, e^4\, v_z \, v_0^3 \, \Delta^4 \, k_0 \,B_x^2}
{64\, \pi\, \mu^7}\,.
\end{align}

As already noted in Sec.~\ref{secvnr-vs-nvnr}, the out-of-plane transverse component and the AH response vanish identically in this geometry,
\begin{align}
\label{eqgnr2_zero}
\bar\sigma_{zx} =  \sigma^{\rm{ah}}_{yx}  = 0\,.
\end{align}

The LFO-induced part is
\begin{align}
\label{eqgnr2_lf_ip}
\sigma^{\rm{lf}}_{xx} & =
\frac{\tau^3 \, e^4 \,v_0^3}{32\,\pi\, \mu^4}
\,\frac{2 \,k_0 \,v_0 \,\mu}{v_z \,\zeta}\,
\Bigg[
2 \,B_z^2\, v_0\,\Big(2 \,k_0^2 \,v_0^2
\left( \zeta- k_0 \, v_0\right)
-\zeta \left( \Delta^2- \mu^2 \right) \Big) 
+ B_x^2 \,k_0\, v_z^2 \left(k_0 \,v_0 (k_0\, v_0-\zeta)
+\Delta^2-\mu^2\right)
\Bigg], \nn
\sigma^{\rm{lf}}_{zx} & = 0\, \nn
\sigma^{\rm{lf}}_{yx} & =
\frac{\tau^2\, e^4 \, v_0^4 \, B_z }
{32\, \pi\, v_z\, \mu^6 \, \zeta^3 }
\Bigg[
2 \,\Delta^2  \, e  \, k_0 \,  v_0 \, B_x^2 \, v_z^2
\left(\Delta^2+k_0^2 \, v_0^2\right)
\left(k_0  \, v_0-\zeta  \right) 
+ \Delta^2  \,e  \,\mu^2 \, B_x^2  \,v_z^2
\left\lbrace 3\, (\Delta^2 -\mu^2 )
+ 2  \,\zeta \,  k_0 \, v_0\right\rbrace \nn
& \hspace{2 cm}
- 32 \,  e  \, \zeta^2  \, k_0^3  \, \mu^2
\, \tau^2 \, v_0^3 \, B_x^2 \, v_z^2
\left(\zeta -k_0 \,  v_0\right) 
-4  \, e \, \zeta^2  \, k_0  \, \mu^2  \, \tau^2
\, v_0 \,  B_x^2 \left(\Delta^2-\mu^2\right)
v_z^2 \left(\zeta -5 \,  k_0  \, v_0\right) \nn
& \hspace{2 cm}
-8  \,e \, k_0 \, \mu^2  \,\tau^2  \,v_0^3 \, B_z^2
\Big\lbrace \zeta  \left(\Delta^2-\mu^2\right)^2
+k_0^2 \, v_0^2 \left(\Delta^2-\mu^2\right)
\left(5  \, k_0  \, v_0-3  \, \zeta \right)
+ 4 \, k_0^4 \, v_0^4
\left(k_0  \,v_0-\zeta \right)
\Big\rbrace \Bigg]\,.
\end{align}

\subsection{Set-up III}

The non-AH part comprises
\begin{align}
\label{eqgnr3_long}
\sigma^{\rm d}_{zz} & = \frac{\tau \, e^2 \, v_z \, k_0}
{ 4 \, \pi \, v_0 \,\mu }
\left(\mu^2 - \Delta^2 \right), \quad
\sigma^{\rm  bc}_{zz} =
\frac{\tau \, e^4  \,v_z^3\, v_0 \, \Delta^2 \, k_0
\left( \mu^2 - \Delta^2 \right) }
{32 \, \pi \, \mu^7}\,
B_x^2 \,, \nn
\sigma^{\rm  m}_{zz} & =
\frac{\tau \, e^4  \,v_z^3\, v_0 \, \Delta^2 \, k_0}
{32 \, \pi \, \mu^7}
\, B_x^2 \left(5\, \Delta^2 -6\, \mu^2 \right), \quad
\sigma^{\rm conc}_{zz}  =
\frac{ -\,\tau\, e^4\, v_z^3\,k_0 \, v_0 \, \Delta^4 \, B_x^2}
{16\, \pi\, \mu^7}\,.
\end{align}

The component $\bar\sigma_{xz}$ vanishes identically, as already noted in Sec.~\ref{secvnr-vs-nvnr}, while the AH part is
\begin{align}
\label{eqgnr3_ah}
\bar\sigma_{xz} & = 0\,, \quad
\sigma^{\rm{ah}}_{yz} =
\frac{ -\, e^3\, v_z \, v_0 \, k_0 \,\Delta^2 \, B_x}
{16 \, \pi\, \mu^4}
\left[
1 +\frac{ 9\, e^2  \,v_z^2 \, v_0^2  \, \Delta^2 \, B_x^2 }
{ 4\, \mu^6 }
\right].
\end{align}

The LFO-induced part is
\begin{align}
\label{eqgnr3_lf_ip}
\sigma^{\rm{lf}}_{zz} & =\frac{ \tau^3\, e^4\,v_z^3\, v_0 \, k_0 \, \mu}{ 8 \,\pi\, \mu^4 }
\, B_x^2 \left(\Delta^2 - \mu^2\right), \quad
\sigma^{\rm{lf}}_{xz}  = 0\,, \nn 
\sigma^{\rm{lf}}_{yz} & =
\frac{\tau^2 \, e^3 \,  v_z \,  v_0 \,B_x }
{128 \,\pi\,\mu^8 \,\zeta } \Big[
16  \, \zeta  \,  k_0 \,  \mu^6
(\Delta^2 -\mu^2 )
-3  \, \Delta^2 \,  e^2  \, v_0^2 \,  v_z^2\,
\zeta  \,  k_0 \, B_x^2
\left(4  \, \Delta^2-7  \, \mu^2\right) 
-6  \, \Delta^2 \,  e^2  \, v_0 \,  v_z^2   \, \mu^4\, B_x^2\nn
& \hspace{3 cm} + 8  \, \Delta^2 \,  e^2 \,  k_0  \, \mu^2
\, v_0^4  \, B_z^2 \left(k_0 \,  v_0-\zeta \right) 
+ 24 \,  e^2 \,  \zeta  \,  k_0^2 \,  \mu^4
\, \tau^2 \,  v_0^3 \,  B_x^2  \, v_z^2
\left(k_0 \,  v_0-\zeta \right) \nn
& \hspace{3 cm}
+ 16 \,  e^2 \,  k_0 \,  \mu^4  \, \tau^2  \, v_0^4\, B_z^2 \left(\zeta -k_0 \,  v_0\right)^2
\left(\zeta +2  \, k_0 \,  v_0\right)\Big].
\end{align}

%======================================================================
\section{Results for the TSM}
\label{app-TSM}
%======================================================================

The Drude term $\sigma^{\rm d}_{xx}$ is unchanged from the HSM, as already given in Sec.~\ref{sechopfres}. In set-up~I, $B_z=0$, only $\bar\sigma_{xx}$ and $\bar\sigma_{yx}$ are nonzero. The out-of-plane response vanishes identically,
\begin{align}
\sigma^{\rm bc}_{zx}
=\sigma^{\rm m}_{zx}
=\sigma^{\rm conc}_{zx}=0\,,
\end{align}
as required by the monopolar $k_z$-parity structure. The remaining components are
\begin{align}
\sigma^{\rm bc}_{xx}
&=\frac{\tau\,e^4\,v_0^3\,(8\,B_x^2+B_y^2)}
{30\,\pi^2\,\mu^2}\,,
\quad
\sigma^{\rm m}_{xx}
=\frac{\tau\,e^4\,v_0^3\,B_x^2}
{24\,\pi^2\,\mu^2}\,,\quad
\sigma^{\rm conc}_{xx}
=-\,\frac{\tau\,e^4\,v_0^3\,(3\,B_x^2+B_y^2)}
{20\,\pi^2\,\mu^2}
\end{align}
and
\begin{align}
\sigma^{\rm bc}_{yx}
&=\frac{7\,\tau\,e^4\,v_0^3\,B_x \,B_y}
{30\,\pi^2\,\mu^2}\,,\quad
\sigma^{\rm m}_{yx} =\frac{\tau\,e^4\,v_0^3\,B_x \,B_y}
{24\,\pi^2\,\mu^2}\,,\quad
\sigma^{\rm conc}_{yx}
=-\,\frac{\tau\,e^4\,v_0^3\,B_x \,B_y}{10\,\pi^2\,\mu^2}\,.
\end{align}

In set-up~I, the in-plane AH response of the dispersive bands vanishes by the $k_z$-parity of the monopolar BC, $\sigma^{\rm{ah}}_{yx}=0$, whereas the out-of-plane AH response acquires a term linear in $B_y$ and a cubic-in-$\bs B$ correction, as already given in Eq.~\eqref{eq-tsm-ah}.

The LFO-induced part is organised by the iteration order $n$. At $n = 1$, three sub-parts give nonzero values for the out-of-plane conductivity:
\begin{align}
\sigma^{\rm lf, h}_{zx}
&=-\,\frac{\tau^2\,e^3\,v_0\,\mu\,B_y}{6\,\pi^2}\,.
\end{align}
While the term $\bs{\mathcal N}_{1,2}$ does not contribute, the term $\bs{\mathcal N}_{1,3}$ contributes to
\begin{align}
\sigma^{\rm lf, bc}_{zx}
&=
-\,\frac{\tau^2\,e^5\,v_0^5\,B_y\, {\boldsymbol B}^2}
{15\,\pi^2\,\mu^3}\, , \quad
\sigma^{\rm lf, m}_{zx}
=-\,\frac{13\,\tau^2\,e^5\,v_0^5\,B_y\, {\boldsymbol B}^2}
{120\,\pi^2\,\mu^3}\,,
\quad \sigma^{\rm lf, conc}_{zx}
= \frac{\tau^2\,e^5\,v_0^5\,B_y\, {\boldsymbol B}^2}
{5\,\pi^2\,\mu^3}\,.
\end{align}
At $n = 2$, the term $\bs{\mathcal N}_{2,2}$ contributes to
\begin{align}
\sigma^{\rm lf, h}_{xx}
=- \, \frac{\tau^3\,e^4\,v_0^3\,B_y^2}
{6\,\pi^2}\,, \quad
\sigma^{\rm lf, h}_{yx}
=
\frac{\tau^3\,e^4\,v_0^3\,B_x\, B_y}
{6\,\pi^2}\,.
\end{align}
At $n = 3$, the nonzero contribution is the geometry-independent term already given in Eq.~\eqref{eq-tsm-lf-n3}. All other LFO components in set-up~I vanish identically for the monopolar TSM.

In closing, we note that $\alpha_n$ is the spin-$1$ overlap coefficient appearing in the pseudospin-$J$ eigenvectors. For the isotropic TSM used throughout this comparison, $\alpha_n\to 1$. The flat-band contributes zero to the non-AH and LFO-induced parts at every order because $\bs v_{s=0}= 0$ in the TSM.

\bibliography{ref_nl}

\end{document}